\documentclass{article}

\usepackage{jheppub}
\usepackage[utf8]{inputenc}
\usepackage{amsmath}
\usepackage{amssymb}
\usepackage{amsfonts}
\usepackage{amsthm}
\usepackage{appendix}
\usepackage{physics}
\usepackage{upgreek}
\usepackage{slashed}
\usepackage{tensor}
\usepackage[most]{tcolorbox}
\tcbset{colback=yellow!10!white,colframe=red!50!black}
\usepackage{tikz-cd} 
\usepackage{xcolor}

 \makeatletter
\def\moverlay{\mathpalette\mov@rlay}
\def\mov@rlay#1#2{\leavevmode\vtop{%
   \baselineskip\z@skip \lineskiplimit-\maxdimen
   \ialign{\hfil$\m@th#1##$\hfil\cr#2\crcr}}}
\newcommand{\charfusion}[3][\mathord]{
    #1{\ifx#1\mathop\vphantom{#2}\fi
        \mathpalette\mov@rlay{#2\cr#3}
      }
    \ifx#1\mathop\expandafter\displaylimits\fi}
\makeatother

\def\LCL{\textit{LCL}}

\def\mrmI{\mathrm{I}}
\def\mrmII{\mathrm{II}}

\def \s= {\!\!\!=\!\!\!}

\def \dg {\delta}
\def \ds {\partial}

\def \ag {\alpha}
\def \bg {\beta}
\def \cg {\gamma}

\def \si {\sigma}

\def \ep {\epsilon}
\def \om {\omega}
\def \Om {\Omega}

\def \la {\lambda}

\def \xx {\mathcal{X}}

\def \oo {\mathcal{O}}

\def \Zs {\mathbb{Z}}
\def \Rs {\mathbb{R}}

\def \nn {\mathcal{N}}

\def \vt {\vartheta}

\def \Dg {\Delta}

\def \orm{\dutchcal{O}}
\def \DC{\mathrm{DC}}
\def \CC{\mathrm{CC}}

\DeclareMathAlphabet{\dutchcal}{U}{dutchcal}{m}{n}
\SetMathAlphabet{\dutchcal}{bold}{U}{dutchcal}{b}{n}
\DeclareMathAlphabet{\dutchbcal} {U}{dutchcal}{b}{n}

\newcommand{\hypg}[3] {\tensor[_2]{F}{_1} \begin{bmatrix} #1,#2 \\ #3  \end{bmatrix}}

\newcommand{\hygen}[4] {\tensor[_2]{#1}{_1} \begin{bmatrix} #2,#3 \\ #4  \end{bmatrix}}
\newcommand\vct[1]{\mathbf{#1}}

\newcommand{\ba}{\begin{align}}

\newcommand{\be}{\begin{equation}}
\newcommand{\ee}{\end{equation}}
\def\bd{\begin{tikzpicture}}
\def\ed{\end{tikzpicture}}

\title{Multipoint Lightcone Bootstrap from Differential Equations} 

\subheader{\normalsize
        \vspace*{-3.7em}
        \begin{flushright}
        ZMP-HH/22-21
        \end{flushright}
}

\author{Apratim Kaviraj$^1$,}
\author{Jeremy A. Mann$^{1,2}$,}
\author{Lorenzo Quintavalle$^{1,3}$,}
\author{Volker Schomerus$^{1,4}$}
\affiliation{$^1$ Deutsches Elektronen Synchroton DESY, Notkestrasse 85, 22603 Hamburg, Germany}
\affiliation{$^2$ Department of Mathematics, King’s College London, Strand, London, WC2R 2LS, UK}
\affiliation{$^3$ D\'epartement de Physique, de G\'enie Physique et d'Optique, Universit\'e Laval, Qu\'ebec, QC G1V0A6, Canada}
\affiliation{$^4$ II. Institut f\"ur Theoretische Physik, Universit\"at Hamburg, Luruper Chaussee 149, D-22761 Hamburg}
\affiliation{Zentrum f\"ur Mathematische Physik, Universit\"at Hamburg, Bundesstrasse 55, D-20146 Hamburg }

\emailAdd{apratim.kaviraj@desy.de}
\emailAdd{jeremy.mann@kcl.ac.uk}
\emailAdd{lorenzo.quintavalle.1@ulaval.ca}
\emailAdd{volker.schomerus@desy.de}

\date{December 2022}

\abstract{One of the most striking successes of the lightcone bootstrap has been 
the perturbative computation of the anomalous dimensions and OPE coefficients of double-twist operators 
with large spin. It is expected that similar results for multiple-twist families can
be obtained by extending the lightcone bootstrap to multipoint correlators. However, very little was known about multipoint lightcone blocks until now, in particular
for OPE channels of comb topology. Here, we develop a systematic theory of lightcone
blocks for arbitrary OPE
channels based on the analysis of Casimir and vertex differential
equations. Most of the novel technology is developed in the context 
of five- and six-point functions. Equipped with new expressions for lightcone blocks, 
we analyze crossing symmetry equations and compute OPE coefficients involving 
two double-twist operators that were not known before. In particular, for the 
first time, we are able to resolve a discrete dependence on tensor structures at large spin. 
The computation of anomalous dimensions for triple-twist families from six-point 
crossing equations will be 
addressed in a sequel to this work.}

\begin{document}
\addtolength{\baselineskip}{2mm}
\maketitle

\section{Introduction}

Understanding the non-perturbative dynamics of quantum field theories is a central
challenge of theoretical physics. This is particularly true for conformal field
theories which are, through Maldacena's celebrated AdS/CFT correspondence, related
to gravitational quantum theories in Anti-de Sitter space. 
The dynamics of conformal field theories turns out to be very strongly constrained by conformal symmetry. In
the case of 2-dimensional theories, this has been exploited for several decades now,
starting with the seminal work of Belavin, Polyakov, and Zamolodchikov \cite{Belavin:1984vu} which was the
first work to implement the conformal bootstrap program. But it took another 30
years until the constraining power of conformal symmetry was put to use in higher
dimensions \cite{Rattazzi:2008pe, Poland:2018epd}. The modern conformal bootstrap was originally based on a
numerical analysis of the bootstrap constraints. Analytic solutions of the higher-dimensional crossing symmetry equations seem out of reach for now, mostly due to
the complexity of the equations. But when pairs of insertion points in the
correlator are light-like separated, the operator products for higher-dimensional
conformal field theories are dominated by a finite set of leading-twist fields.
This was first exploited in \cite{Fitzpatrick:2012yx, Komargodski:2012ek} and initiated the development of the so-called 
lightcone or analytic bootstrap, see \cite{Alday:2015ewa, Alday:2017zzv, Kaviraj:2015cxa, Kaviraj:2015xsa, Li:2015itl,Li:2015rfa, Hofman:2016awc} and references therein. 

One of the most striking early results of the lightcone bootstrap was to show 
that any conformal field theory must contain infinite double-twist families in 
which the conformal weights approach the values of a free-field theory when the 
spin becomes large \cite{Fitzpatrick:2012yx, Komargodski:2012ek,Alday:2013cwa}.\footnote{The original analysis was based on some 
natural assumptions regarding the behavior of sums of conformal blocks in the double lightcone limit. A rigorous proof was only given very recently in \cite{Pal:2022vqc}.} 
Later, the analytic tools were sharpened such that it is now possible to 
systematically compute corrections to the free-field theory behavior for double-twist families in an expansion around large spin \cite{Caron-Huot:2017vep, Alday:2017vkk}. In terms of the AdS dual 
theories, these results allow in particular to compute the mass spectrum of 
two-particle bound states \cite{Cornalba:2006xm,Cornalba:2007zb}. 

It is certainly highly desirable to obtain similar results for multi-twist 
families in conformal field theory or multi-particle states in the dual gravitational theories on AdS, see \cite{Karlsson:2019txu, Karlsson:2020ghx, Fitzpatrick:2015qma} for some previous discussions. 
But not much is actually known yet. While double-twist families already 
appear in the operator product of two fields in a generalized free-field 
theory, it requires at least three fields to build a triple-twist family. 
Hence, to analyze the behavior of triple-twist fields in free-field theory 
and beyond, it seems natural to study the crossing symmetry of six-point correlators. This turns out to be a rather formidable task. Nevertheless,
it seems worth some effort. Indeed, in an interacting theory, triple- and higher-twist operators can be exchanged in the intermediate channel of a four-point 
function. As long as we remain agnostic about the behavior of higher-twist 
families, their exchange will contaminate four-point crossing symmetry 
equations and this contamination systematically limits the precision to which 
conformal data can be determined \cite{Simmons-Duffin:2016wlq,Liu:2020tpf}. 
Overcoming this limitation of existing bootstrap tools has recently motivated 
several groups to probe conformal field theories through multipoint correlation 
functions involving more than four scalar fields \cite{Bercini:2020msp,Bercini:2021jti,Antunes:2021kmm,Anous:2021caj,Dodelson:2022eiz}. 

The main bottleneck of the conformal bootstrap in general and of the multipoint
bootstrap in particular is our very limited knowledge of the relevant conformal
blocks. For $N>4$ insertion points, even partial results on conformal blocks are quite
recent, see e.g. \cite{Goncalves:2019znr,Hoback:2020pgj, Hoback:2020syd, Parikh:2019dvm, Fortin:2019zkm,  Fortin:2020yjz, Fortin:2020bfq, Fortin:2022grf, Poland:2021xjs} and references therein. Even though conformal blocks
are expected to simplify drastically when the insertion points lie on their respective lightcones (that is to say, the relevant regime in the lightcone bootstrap), 
there exist no universal results yet. The only cases that were understood 
recently concern blocks in which pairs of external fields can be expanded 
through the lightcone OPE of \cite{Ferrara:1971zy,Ferrara:1971tq,Ferrara:1972cq}. Through the use of this lightcone OPE, 
one can derive integral formulas for multipoint lightcone blocks for up to 
$N=6$ external points in the snowflake channel. By taking 
appropriate limits of these integral formulas, the authors of \cite{Bercini:2020msp,Bercini:2021jti,Antunes:2021kmm}
were able to analyze the crossing equation for a duality between two 
inequivalent snowflake channels in the lightcone limit and thereby derive new results on the leading large-spin behavior of OPE coefficients 
involving several double-twist families. But in order to effectively access 
triple-twist families, it is necessary to study crossing symmetry relations 
in which at least one channel has comb topology. Unfortunately, lightcone 
limits for general comb channel blocks cannot be accessed through the 
lightcone OPE formula of Ferrara et al. Some new tools are required. Here 
we shall advocate the use of differential equations as a universal tool to study 
lightcone blocks for any channel topology and in particular the comb channels 
that can give access to multi-twist families.  

The characterization of multipoint conformal blocks through differential equations 
is by now rather well understood, mostly through the integrability-based 
approach to conformal blocks that was initiated in \cite{Buric:2020dyz} and 
developed in \cite{Buric:2021kgy,Buric:2021ttm,Buric:2021ywo}.
In the case of four-point blocks, Dolan and Osborn first proposed to 
study them through a set of Casimir differential equations. While the 
original equations did not look very inviting, they indeed allowed to 
efficiently analyze four-point blocks in dimension $d \geq 2$ 
\cite{Dolan:2011dv}. Later, the Casimir differential operators of Dolan 
and Osborn were identified as the eigenvalue equations for the Hamiltonian 
of a two-particle integrable Calogero-Sutherland model \cite{Isachenkov:2016gim}. 
This insight initiated an integrability-based approach to four-point function 
that was advanced in \cite{Schomerus:2016epl,Schomerus:2017eny,Isachenkov:2017qgn,
Buric:2019dfk,Buric:2022ucg}.
Very remarkably, this relation between integrable models of quantum 
mechanics and conformal blocks persists for correlators involving 
more than four external fields. More specifically, as shown in 
\cite{Buric:2021ywo}, one can identify the relevant differential operators that 
characterize blocks for any number of external fields with the 
commuting Hamiltonians of an integrable Gaudin model, in a certain 
homogeneous limit. Even though the complexity of these differential
operators grows quickly with the number of cross-ratios, they can be
worked out and possess very nice properties. As we shall see below, 
these properties make it possible to evaluate the differential 
operators in the lightcone limit. After taking sufficiently many 
insertion points to be lightlike separated, these differential 
operators lose much of their complexity and one can often 
find explicit solutions in closed form. 

The main goal of this work is to develop a systematic theory of 
multipoint blocks in the lightcone limit, based on the study of the 
differential equations the blocks satisfy. In particular, our new 
approach does not rely on lightcone OPEs and hence applies to all 
channels including the important OPE channels highlighted above. For 
$N=5$ and $N=6$ external scalar fields, we will also apply our concrete 
results on blocks to the analysis of crossing symmetry equations, 
leading to a number of new results on OPE coefficients, see below.

\subsection{Lightcone differential equations - an overview}

In order not to get lost in technical details later on, we first 
want to give a bird's eye view of the new methods developed herein. 
We first describe the general setup, 
then explain some elements of constructing perturbative solutions 
to a set of differential equations, and finally turn to the 
multipoint lightcone bootstrap.
\subsubsection{The general setup}  
Our setup starts with $N$ external scalar fields that are inserted in
points $x_i \in \mathbb{R}^d$ with $i = 1, \dots, N$. As is common in
the conformal field theory literature, we will represent the insertions
points $x_i$ through projective null lines $X_i \in \mathbb{R}^{1,d+1}$
in embedding space. Given $X_i$ one can construct a certain number
$n_{\textrm{cr}}(N,d)$ of conformally-invariant cross-ratios. In the case
of $N=5$ points and $d \geq 3$, for example, there exist five independent
cross-ratios. For $N=6$ points, the number of independent cross-ratios
increases by three in $d=3$ and by four in $d \geq 4$. This number of 
independent cross-ratios is easy to count in general, see e.g.\
\cite[eqn.~(1.3)]{Buric:2021ywo}.

Let us now consider some correlation function $G$ of $N$
scalar fields. After some appropriate factor $\Omega$ is split off from
$G = \Omega g$, the remaining function $g$ depends on cross-ratios only.
In principle, one can evaluate $G$ by performing $N-2$ operator products
between fields until one ends up with a two-point function that is fixed
by conformal symmetry. The precise sequence of operator products that
are performed is known as the OPE channel. We shall denote OPE channels 
by $\mathcal{C}$. For $N=6$ points, for example, there exist $90$ different 
OPE channels with two different topologies, known as snowflake and comb 
topology, respectively, see the left-hand and right-hand sides of 
Fig.~\ref{fig:6pt_cse_outlook}, respectively. Once all operator product 
expansions (OPEs) are performed, the correlator is written as a product 
of $N-2$ constant OPE coefficients and the kinematically-determined 
conformal blocks that carry all the dependence on the cross-ratios. 
These conformal blocks depend on certain quantum numbers that parametrize 
the conformal weights and spins of intermediate fields appearing in 
the operator products, as well as the choice of tensor structures. There 
are as many such quantum numbers as there are independent cross-ratios. 
In the case of $N=5$ external scalars, for example, one has two intermediate 
fields that transform in symmetric traceless tensor representations (STTs)
of the conformal group and hence carry two quantum numbers each, the
conformal weight $\Delta$ and rank $J$ of the tensor. In addition, there 
is a fifth label that determines the choice of the tensor structure in 
the OPE between a scalar and an STT field. So, indeed, the total number 
of labels coincides with the number of cross-ratios. 

According to the approach advocated in \cite{Buric:2020dyz}, scalar $N$-point
conformal blocks may be characterized by a system of $n_\textrm{cr}(N,d)$
linear, higher-order differential equations. These arise as eigenvalue
equations of a system of commuting operators $\mathcal{D}_A, A = 1, \dots,
n_\textrm{cr}(N,d)$ that measure the quantum numbers of the 
conformal blocks. An explicit construction of such operators for any given 
OPE channel $\mathcal{C}$ was proposed in \cite{Buric:2021ywo}. Even though these 
operators are quite complicated in general, it is possible to introduce a 
special set of coordinates in the space of cross-ratios such that all the 
coefficients in the differential operators become polynomials, see 
\cite{Buric:2021kgy} for OPE channels with comb topology. These coordinates 
have been dubbed \textit{polynomial cross-ratios}. The  set of differential 
operators in polynomial cross-ratios is the starting point for the lightcone
analysis in the present work.

\subsubsection{Lightcone differential operators} 
In order to describe our analysis in a bit more detail, let us denote
the polynomial cross-ratios for an $N$-point function by $w_Q, Q =1\,
\dots, n_\textrm{cr}(N,d)$. The lightcone bootstrap analyzes crossing
symmetry constraints in the limit where some pairs of insertion
points $x_i$ become lightlike separated, i.e. in which $X_{ij} := X_i 
\cdot X_j$ tends to zero for some set of pairs $(i,j)$. How many and 
which pairs of points are selected to become lightlike depends on the 
pair of OPE channels that one relates through crossing symmetry, see 
below. We will denote the total number of independent lightlike pairs 
by $m$ and enumerate the
individual pairs $(i_\nu, j_\nu)$ by $\nu= 1, \dots, m$. For each
lightlike pair we shall introduce a formal parameter $\epsilon_\nu,
\nu = 1, \dots, m$. The parameters $\epsilon_\nu$ control the
approach to the lightcone of the products $X_{i_\nu j_\nu}$.
Since our cross-ratios $w_Q$ are multi-homogeneous in each of the 
products $X_{ij}$, the substitution rules $X_{i_\nu j_\nu} 
\rightarrow \epsilon_\nu X_{i_\nu j_\nu}$ determine a map 
$\gamma_\epsilon$ that attaches formal variables to the 
coordinates $w_Q$, 
\begin{equation} \label{eq:scalingw}
\gamma_\epsilon (w_Q) = \epsilon^{s_Q} w_Q\ .
\end{equation}
Here, $s = (s_\nu) \in \mathbb{Z}^m$ are multi-indices and we used
the notation $\epsilon^s = \prod_{\nu=1}^m \ep_\nu^{s_\nu}$ as usual.
We shall refer to the multi-index $s_Q$ as the $\ep$-scaling of the
variable $w_Q$ and to eq.~\eqref{eq:scalingw} as scaling laws of the
variables $w_Q$.

Next, let us consider the algebra $\mathcal{D}^{(n)}$ of differential
operators $\mathcal{D}$ in the $n$ variables $w_{Q}$. We can extend our
scaling law \eqref{eq:scalingw} to a homomorphism $\gamma_{\epsilon}$
on differential operators by imposing the additional rule
\begin{equation}  \label{eq:scalingdw}
 \gamma_{\epsilon} (\partial_{w_Q})
 =  \epsilon^{-s_Q} \partial_{w_Q}\
\end{equation}
where we use  $\epsilon^{-s}= \prod_{\nu=1}^m \ep_\nu^{-s_\nu}$. After
application of $\gamma_{\epsilon}$ to some differential operator
$\mathcal{D}$ we obtain an expression in the formal variables
$\epsilon_\nu$ that takes values in the space of differential
operators. All operators we shall work with below have finite order
and polynomial coefficients in some set of variables~$w_Q$. Given
any such operator, we can apply $\gamma_\ep$ and expand the result
in a formal series in the variables~$\ep_\nu$
\begin{equation}
 \gamma_\epsilon (\mathcal{D}) \equiv \mathcal{D}^{\epsilon}
= \sum_{k \in \mathbb{Z}^m} \epsilon^{\, k }\
\mathcal{D}^{(k)} \ \ .
\label{eq:epsilonexp}\end{equation}
Note that for differential operators of finite order with polynomial
coefficients, the sum on the right-hand side involves a finite number
of terms. From time to time we shall refer to the index $k$ as the
$\ep$-scaling of the differential operator $\mathcal{D}^{(k)}$.

We will be studying differential operators in a limiting regime in
which pairs of points become lightlike separated in some particular
order, with the distance between the first pair becoming lightlike
much faster than the second, which in turn becomes lightlike much
faster than the third, etc. As we remarked above, the pairs are
enumerated by our index $\nu = 1, \dots, m$. Because of this
hierarchy of lightcone limits, the ordering of the formal
parameters $\ep_\nu$ does matter. Whenever we want to stress this
ordering, we shall write
$$ \vec{\ep} = (\ep_1, \dots, \ep_m) \ . $$
Similar notations are used for the multi-indices $s,k, \dots$. The
ordering of the formal variables $\ep_\nu$ introduces a lexicographic
order among the scalings, i.e. we write $\vec{k} < \vec{k}'$ if the
first non-vanishing entry in $\vec{k}' - \vec{k}$ is positive. Since 
the differential operators we are dealing with throughout this work 
have finite order, their expansion \eqref{eq:epsilonexp} contains a 
term that contains the leading term. The associated grade vector will be denoted 
by $\vec{k}_0$. We can then rewrite the expansion \eqref{eq:epsilonexp} 
in the form
 \begin{equation}
 \gamma_\epsilon (\mathcal{D}) \equiv \mathcal{D}^{\vec{\epsilon}}
= \vec{\ep}^{\,\, \vec{k}_0} \sum_{\vec{p} \in \mathbb{N}^m}
\vec{\epsilon}^{\,\, \vec{p}}\
\mathcal{D}^{(\vec{k}_0 + \vec{p})} = \vec{\ep}^{\,\, \vec{k}_0}
\mathcal{D}^{(\vec{k_0})} + \tilde{\mathcal{D}}
\label{eq:epsilonexpp}
\end{equation}
where the sum runs over a finite set of vectors $\vec{p}$ with 
non-negative integer components only. On the right-hand side we 
have split $\mathcal{D}$ into its leading term and the subleading 
remainder.  

As we explained above, we will be interested in eigenfunctions of the 
differential operator $\mathcal{D}$. In this work we will only look at 
the leading terms of the eigenvalue equation, i.e.\ at eigenfunctions 
of the leading singular term $\mathcal{D}^{(\vec{k}_0)}$ that is 
selected by our choice of $\vec{\ep}$. For the associated eigenfunctions
there are two cases that will occur, depending on the scaling behavior
of the eigenvalues $\lambda$. In order to describe the scaling behavior 
of eigenvalues, we extend the map $\gamma$ to the eigenvalues, i.e. we 
introduce $\lambda^{\vec{\ep}} = \gamma_{\vec{\ep}}(\lambda)$. The first 
case that turns out to be relevant for us is when the eigenvalue does not 
scale, i.e.\ when $\lambda^{\vec{\ep}} = \lambda$. If that is the case and
if the grade vector $\vec{k}_0$ of the leading singular term is negative, 
then the differential equation forces the leading behavior of the 
eigenfunction $\psi$ to lie in the kernel of the differential operator 
$\mathcal{D}^{(\vec{k}_0)}$. The other relevant case that we will 
encounter below is when the eigenvalue $\lambda$ scales in the same 
way as the leading term of the differential operators, i.e. when 
$\lambda^{\vec{\ep}} = \vec{\ep}^{\, \, \vec{k}_0} \lambda + \tilde \lambda$. 
With this scaling of the eigenvalue, we are led to consider eigenfunctions
of the differential operator $\mathcal{D}^{(\vec{k}_0)}$ for 
eigenvalue $\lambda$.  While beyond the scope of this paper, this 
expansion of the differential equations actually defines a perturbative 
expansion of the eigenfunctions that can be used to calculate corrections
to lightcone blocks away from the strict lightcone limit. 

\subsubsection{Multipoint lightcone bootstrap}

After these more general comments on the perturbative study of
differential eigenvalue equations, we now return to the study 
of lightcone blocks and lightcone bootstrap. The crossing 
equations we want to analyze are associated with a pair of OPE 
channels. We shall refer to one of these channels as the 
\textit{direct channel} and denote it as $\mathcal{C}^\DC$. The 
second channel is referred to as \textit{crossed channel} and 
denoted by $\mathcal{C}^\CC$. The choice of direct and crossed 
channel largely determines the choice of the vector $\vec{\ep}$ 
that featured in the previous subsection. In particular, the 
first $N-3$ entries of $\vec{\ep}$ must ensure that the lightcone 
limit of the direct channel receives its leading contributions 
from leading-twist operators. In order to remove higher-twist 
exchange from the crossed-channel contributions as well, one 
needs to take at least $N-3$ additional lightcone limits. Which 
pairs of points are required to become lightlike depends on the 
crossed OPE channel. Along with the first $N-3$ lightlike 
limits that removed direct-channel higher twists, the minimal number 
of lightlike limits thus amounts to $2N-6$. As a result, the vector $\vec{\ep}$ 
contains at least $2N-6$ components, of which the first $N-3$ 
entries are determined by the direct channel while the latter 
depends on the crossed channel. Obviously, taking additional 
lightcone limits further reduces the complexity of the crossing 
equations, but it also comes at the price of reducing 
resolution. 

Once the vector $\vec{\ep}$ has been chosen, we can apply the 
general discussion from the previous subsection to the differential 
operators $\mathcal{D}_A$ that characterize multipoint conformal 
blocks, see above. As we have explained, this set of differential 
operators depends on the OPE channel. Since we have singled out a pair 
of such channels, namely the direct and the crossed channel, we shall 
denote the dependence on the channel by $\mathcal{D}^\DC_A$ and 
$\mathcal{D}^\CC_A$, respectively. With a sufficiently large number 
$m \geq N-3$ of lightcone limits performed, the original complexity 
of the differential operators is very much reduced and one obtains 
rather simple expressions for the leading terms 
$\mathcal{D}^{(\vec{k}_{A0})}_A$, as well as the subleading corrections. 
Many examples will be spelled out explicitly throughout the following 
sections. 

Given explicit formulas for the leading terms of direct-channel
differential operators, one can construct expressions for the lightcone 
limit of direct-channel blocks. In this case, the eigenvalues are chosen 
to scale trivially with the lightcone limit. Following our
brief comments at the end of the previous section, the lightcone limit 
of direct-channel blocks must lie in the kernel of the leading terms 
of the differential operators $\mathcal{D}_A^\DC$. Given the simple 
expressions of the leading singular terms, the kernel can be 
constructed explicitly. The lightcone limit of the direct-channel 
blocks is some particular vector in this linear space. We will 
comment on how to identify this vector within the kernel in a 
moment. 

Before we do so, we want to take a first look at the crossed channel.
It turns out that, in order to reproduce the leading terms in the 
direct-channel expansion within the crossed channel, 
we need to take an appropriate scaling limit in the eigenvalues of the crossed-channel differential operators.  
To determine the scaling laws for crossed-channel eigenvalues, we adopt 
a procedure that was first described by David Simmons-Duffin in the 
context of the four-point lightcone bootstrap \cite{Simmons-Duffin:2016wlq}. 
It involves applying the crossed-channel differential operators to the 
leading terms in the direct channel expansion. So, given the explicit 
expression for the crossed-channel differential operators, we are able 
to determine the scaling of the crossed-channel quantum numbers as soon 
as we have sufficient information about the direct-channel lightcone 
blocks. The scaling of crossed-channel eigenvalues in effect drives
contributions from the crossed channel to large spins. 

Given the relevant scaling of the quantum number/eigenvalues in the 
crossed-channel differential equations, we can now determine the 
lightcone limit of the crossed-channel blocks by solving the 
associated differential equations. In all the cases we have looked at, 
these solutions can be found explicitly in terms of exponentials and 
Bessel functions. Furthermore, the linear differential equations obtained 
from the complete set of commuting differential operators possess a finite-
dimensional space of solutions. Once again, the lightcone limit of the 
crossed-channel conformal blocks corresponds to one particular vector 
within this space.    

The identification of the relevant vector turns out to be challenging.  
Even before taking any lightcone limits, the differential equations for 
conformal blocks possess finite-dimensional solution spaces. Within 
these spaces, the blocks are selected by boundary conditions that 
fix their behavior in the OPE limit, which we define in Lorentzian signature as the limit of two coincident points within lightlike separation. However, the OPE limit 
is not part of the lightcone regime for which we solve the lightcone 
differential equations. So, in order to select the limiting behavior 
of the blocks, both in direct and crossed channel, we must somehow 
connect the OPE limit with the lightcone limit. At least for the 
cases we have studied, it is indeed possible to obtain sufficient 
control along some curve in the space of cross-ratios that connects 
the OPE with the lightcone limit. To obtain this control for $N > 4$ 
is the main challenge within the approach we advance. Once this is 
overcome, the gate is open to analyze and solve the crossing equation in the 
lightcone limit. 
Before we describe more concretely the setups in which the above procedure 
is implemented, let us add one more 
comment. As we have reviewed above, the differential operators that 
are used to characterize conformal blocks fall in two classes. Casimir 
differential operators measure the weight and spin of the intermediate 
fields in a given OPE channel, while vertex differential operators 
measure the choice of tensor structure. But in order for these 
quantum numbers to be measurable simultaneously, we had to introduce 
a new basis of tensor structures that has not yet been well explored. 
It is certainly quite different from the choice of tensor structure 
that is commonly used in the conformal field theory literature. In 
our construction of lightcone blocks we will initially focus on the 
Casimir differential operators and stick to a more conventional basis 
of tensor structures. This also simplifies the comparison with 
previous work, most notably the results in lightcone blocks in 
\cite{Antunes:2021kmm}. Nevertheless, we shall have a look at the 
lightcone limit of the vertex differential operator once the lightcone 
blocks are constructed. This will provide interesting novel insight 
into the relation between eigenfunctions of vertex differential 
operators and the more conventional choice of tensor structures. 
There is an immediate payoff in the analysis of the crossing 
equations: crossed-channel lightcone vertex operators can be used much 
in the same way as the Casimir operators to determine the scaling of 
crossed-channel spins in the lightcone limit. Given our new
understanding of the relation between eigenvalues of vertex operators and 
conventional tensor structure labels in the lightcone limit, we can 
apply the lightcone vertex operators to determine the scaling of 
tensor structures in the lightcone limit.

\subsection{Summary of new results}

Even though the main focus of this work is to sharpen universal tools for 
the multipoint lightcone bootstrap, and in particular for the analysis of 
multi-twist operators, see section~\ref{sect:six_pt_outlook} and~\cite{In_preparation}, this paper does contain a few new results already, 
both on lightcone blocks and OPE coefficients. The purpose of this short 
subsection is to highlight these concrete new results before we dive into 
a broader outline of the paper. 
\smallskip 

Let us begin with the new results on five-point lightcone blocks. The most 
novel aspect of our analysis is that we study blocks in a partial lightcone 
limit where only four of the five cross ratios are sent to limiting values. That is, we only insist that four of the five distances between 
neighboring points become lightlike. In this restricted lightcone limit, we 
obtain new explicit formulas for the associated blocks, both in the direct and the crossed channel. For the direct channel, these are spelled out in eqs.~(\ref{blocks_dc_v1zero_final},~\ref{eq:sumgfin}). In these expressions, it is also straightforward to send the fifth cross ratios to zero  
and thereby obtain formulas for direct channel blocks in the full 
lightcone limit, see eqs.~(\ref{eq:5ptLCLdc},~\ref{eq:5ptLCLdcresult}). We stress that even the latter are new for the regime in which the 
half-twists $h_1, h_2$ of the exchanged fields satisfy $h_1,h_2 > h_\phi$. 

For the crossed channel, the lightcone blocks in the restricted lightcone limit 
can be found in eqs.\ (\ref{blocks_case_II},~\ref{norm_block_case_II}). The 
limiting behavior in the full lightcone limit is stated in eq.\ \eqref{sol_5pt_I}. 
In both the restricted and full lightcone limit, we obtain closed-form expressions 
for the blocks that were not known before. The results we  have listed so far work 
with the usual basis of tensor structures at the central vertex of the five-point 
blocks. In section~\ref{sect:VertexOp}, we also derive the analogous results for 
the crossed channel blocks in the vertex operator eigenbasis, see 
eq.~\eqref{eq:Case1_Vop_eigenf} and eqs.~(\ref{eq:vert_eig_caseII_decomp},
~\ref{t_of_N},~\ref{ctm}). 
\smallskip 

By exploiting our new results on five-point blocks in the restricted lightcone limit, we obtain new formulas for several OPE coefficients. These concern the OPE 
decompositions 
$$ [\phi \phi] \times \phi \ \sim \ [\phi\phi] \quad \mathit{ and } \quad 
[\phi \oo_\star] \times \phi \ \sim \ [\phi \phi]  \ . $$
In the second decomposition, the field $\oo_\star$ is assumed to appear in the OPE 
of $\phi$ with itself and to have lower twist than $\phi$, i.e. $h_\star  < h_\phi$. In the limit of large tensor structures (and large spins, of course), the 
relevant OPE coefficients 
$$C_{[\phi\phi]_{0,J_1} \phi [\phi\phi]_{0,J_2}}^{(\eta)} 
\quad \textit{ and } \quad C_{[\phi\oo_\star]_{0,J_1} \phi [\phi\phi]_{0,J_2}}^{(\eta)}$$ 
were first spelled out by Antunes et al.~\cite{Antunes:2021kmm}.  Here, we recover the same 
expressions, see eqs.~\eqref{PJphiJ} and~\eqref{PJstarphiJ}, despite relying on different 
methods and assumptions. But we can do better. Our new results on five-point blocks in the 
restricted lightcone limit enable us to analyze crossing symmetry without sending the fifth 
cross ratio to zero. Using this technology, we bootstrap the above OPE coefficients 
in a new regime with large spins but discrete tensor structure, i.e. in a regime where the 
tensor structure label $\eta$ takes values $\eta = J_1 - \dg n$ with $\dg n = 0,1,2, \dots, 
\infty$. For the OPE coefficient that involves both $[\phi\phi]$ and $[\phi\oo_\star]$, the relevant formula for the OPE coefficient is 
presented in eq.~\eqref{caseII_h*<hphi}. For the other OPE coefficient that involves two 
double twist fields of the form $[\phi\phi]$, our new results are spelled out in eqs.\ 
\eqref{caseII_1phiexch} and \eqref{ope_coeffs_discrete_ts}. If the field $\phi$ 
appears in the OPE of $\phi$ with itself, eq.\ \eqref{ope_coeffs_discrete_ts} represents  
a subleading correction to the leading term \eqref{caseII_1phiexch}.

\subsection{Detailed plan of part I } 

Let us now outline the plan of this work in some detail. In order to introduce our 
analytic tools and explore how to use them, we shall begin with the case of scalar 
four-point functions in section~\ref{sect:four_pt_boot}. 
In contrast to the usual treatment, however, we will derive all required results 
in lightcone blocks directly from the Casimir equations in the (restricted) 
lightcone limit. In the direct channel we first look at the Casimir equation in 
the restricted lightcone limit in which only one pair of points becomes lightlike 
separated, see section~\ref{ssec:4pt-dc-blocks}. This regime contains both the full lightcone 
limit and the OPE limit. It is well known that the associated Casimir equation 
can be solved in closed form. Then we look at the Casimir equations for both the 
direct and the crossed channel in the full lightcone limit and determine the 
finite-dimensional space of solutions. The existence of a closed-form solution
in the restricted lightcone limit makes it easy to select the relevant 
solutions that are associated with the OPE boundary condition, including 
the overall normalization of the lightcone blocks. Once the lightcone 
blocks are constructed, we review the familiar lightcone bootstrap analysis
in section~\ref{4pointlightcone}. Some emphasis is put on how to determine the lightcone 
scaling behavior of the spin quantum numbers in the crossed channel by acting 
with crossed-channel Casimir operators on terms in the direct-channel expansion.

After this warm-up, we then turn to $N=5$ in section~\ref{sect:lightcone_blocks_five_pt}. Scalar five-point functions
provide an ideal framework to develop our new analytic tools. Indeed, in this case, 
lightcone blocks can be studied through the lightcone OPE, as was exploited in 
\cite{Bercini:2020msp,Bercini:2021jti,Antunes:2021kmm}. Hence, all the results that we shall obtain through the mere use of 
Casimir differential equations can be cross-checked with those standard 
techniques. In the full lightcone limit, five pairs of points become lightlike 
separated. Once again, we shall start in section~\ref{ssec:5pt_blocks_dc} by looking at direct-channel 
blocks in a restricted lightcone limit where only two pairs become lightlike separated, and 
which still contains both the OPE and the full lightcone limit. The only difference 
with $N=4$ is that lightcone blocks in the restricted lightcone limit are not 
written down in closed form but rather through an integral formula. As we shall 
show, the integral representation we obtain by solving lightcone Casimir 
equations is equivalent to the formula one obtains from the lightcone OPE. Then, 
we solve the direct- and crossed-channel Casimir equations in the (full) lightcone 
limit where additional pairs of points become lightlike separated. The correct 
solutions of the lightcone Casimir equations can be selected and normalized with 
the help of the integral formula for direct channel blocks in the restricted 
lightcone limit. As we shall show later in the text, see section~\ref{sec:5pt_bootstrap}, the 
analysis of crossing symmetry requires the construction of crossed-channel lightcone 
blocks for two different scaling laws of the eigenvalues. Our analysis of 
lightcone blocks results in a set of relatively simple formulas for direct-
and crossed-channel blocks that are clearly marked, see
section~\ref{sect:five_pt_crossed_blocks}.

Section \ref{sect:VertexOp} is devoted to the study of the unique 
non-trivial vertex operator that exists for five external points. We 
find that this operator, being quite complicated for 
generic kinematics \cite{Buric:2021ttm}, simplifies considerably in the 
lightcone limit --- so much so that we are able to find its joint eigenfunctions
with the Casimir operators explicitly. This provides important new 
insight into the relation between the conventional tensor structure 
labels which are used in section \ref{sect:lightcone_blocks_five_pt}
and the eigenvalues of vertex differential operators. In analogy to 
Simmons-Duffin's notion of Casimir singular terms we can use vertex 
differential operators to introduce the new notion of vertex singular 
behavior. In this context, the results of section \ref{sect:VertexOp}
allow us to determine the scaling behavior of tensor structure labels 
in the lightcone regime, see section \ref{sec:5pt_bootstrap}.
 
The explicit formulas for lightcone blocks and differential operators 
from sections \ref{sect:lightcone_blocks_five_pt} and \ref{sect:VertexOp}
prepare us for the analysis of the crossing symmetry equation for five 
external scalars in the lightcone limit, see section~\ref{sec:5pt_bootstrap}. 
There we shall show how our results on blocks allow us to recover all of the 
known OPE coefficients and go beyond. In section~\ref{ssec:5ptBS DC} we 
spell out the leading terms within the direct channel. Then we address each 
of these terms in section~\ref{ssec:5ptBS_CC} and show how to reproduce 
them from the crossed-channel expansion. While most of this analysis just 
reproduces results from \cite{Antunes:2021kmm}, there is one important 
direct-channel contribution 
for which we can go much beyond the previous status. Our progress is 
related to the fact that we will be able to construct the lightcone 
blocks in both the direct and crossed channel with the minimal number 
$2N-6 = 4$ of lightcone limits performed. This is one limit less than in 
\cite{Antunes:2021kmm} and will allow us for the first time to resolve 
the dependence of the OPE coefficients on the tensor structure. This 
motivates us to reanalyze the crossing equation in the fourfold lightcone 
limit in section~\ref{ssec:discrete_tensor_structures}, where we obtain the 
OPE coefficients in the new scaling regime of large spin and discrete tensor 
structures. Finally, we review some applications and checks of our results in
section~\ref{ssec:applications}.

Section~\ref{sect:six_pt_outlook} contains an outlook to the second 
part \cite{In_preparation} which is devoted to the 
lightcone bootstrap for six-point functions. As we shall explain, the crossing 
equation we will address in the second part relates a direct snowflake 
channel to a crossed comb channel. While direct channel lightcone blocks 
in the snowflake channel can be studied through lightcone-OPE integral 
formulas, such tools no longer exist for the crossed comb channel, and 
here is where our new technology comes to shine. Indeed, it turns out 
that the program we described in section~\ref{sect:lightcone_blocks_five_pt} 
carries over to six-point 
comb-channel blocks and for the first time provides explicit formulas
for the relevant lightcone blocks. We shall show a few examples and 
also showcase some of the simplest results that can be extracted from 
the bootstrap analysis. Additional formulas for relevant lightcone blocks, 
detailed derivations, as well as applications to the six-point lightcone 
bootstrap will be given in the second part. This includes a discussion 
of the anomalous corrections to the conformal weights of triple-twist 
operators.

\section{Lightcone Bootstrap for Four Points}
\label{sect:four_pt_boot}

The purpose of this section is to provide a smooth introduction to the multipoint
lightcone bootstrap. In order to do so we shall review the standard lightcone
bootstrap analysis for scalar four-point functions. The lightcone bootstrap program
requires a good knowledge of conformal blocks in certain limits. When dealing with
four-point functions, the blocks and their limiting behavior are well known. Our
discussion here will put some stress on the derivation of these properties from
Casimir differential equations. We explain the usefulness of polynomial cross-ratios,
the way we take (scaled) limits, and the resulting expressions for lightcone blocks.
Special attention will be paid to the normalization of the blocks in the limit.

\subsection{Preliminaries on blocks and lightcone limits}

There are many ways to parametrize conformal invariants of four points. The
most basic choice is the two cross-ratios $(u,v)$ that are defined
by
\begin{equation}
u=\frac{X_{12}X_{34}}{X_{13}X_{24}} = z \bar z,
\quad v=\frac{X_{14}X_{23}}{X_{13}X_{24}} = (1-z)(1-\bar z).
\end{equation}
Here, $X_i$ denote the embedding space insertion points of the four scalar fields
and $X_{ij} =X_i \cdot X_j$, as usual. Note that we have also introduced a second
set of cross-ratios, $(z,\bar z)$. For reasons we shall discuss below, we shall
refer to $u,v$ as \textit{polynomial cross-ratios} and use the term \textit{OPE
cross-ratios} when dealing with $(z,\bar z)$.

The usual conformal partial wave or conformal block expansions for correlation
functions of four identical scalar fields $\phi$ with weight $\Delta_\phi$ are
given by
\begin{align} \label{eq:4ptcrossing1}
\langle \phi(X_1)\dots \phi(X_4)\rangle &=
(X_{14}X_{23})^{-\Dg_\phi} \sum_{\oo} C_{\phi\phi\oo}^2
\psi_\oo^{(14)(23)}(u,v)\\[2mm]
&= (X_{12}X_{34})^{-\Dg_\phi}
\sum_{\oo} C_{\phi\phi\oo}^2 \psi_\oo^{(12)(34)}(u,v)
\end{align}
The central objects in these equations are the conformal blocks $\psi$. The
second line is an expansion in terms of $s$-channel blocks $\psi^{(12)(34)}$
while in the first line we expanded the same correlator in terms of $t$-%
channel blocks $\psi^{(14)(23)}$. The two sets of blocks are related by a simple
exchange of the two cross-ratios, $\psi^{(14)(23)}(u,v)=\psi^{(12)(34)}(v,u)$.
The famous crossing symmetry equations for scalar four-point functions,
\begin{equation} \label{eq:4ptcrossing2}
 \sum_\oo C_{\phi\phi\oo}^2 \psi_\oo^{(14)(23)}(u,v) = v^{\Dg_\phi} u^{-\Dg_\phi}
 \sum_\oo C_{\phi\phi\oo}^2 \psi_\oo^{(12)(34)}(u,v) \ .
\end{equation}
are obtained by equating the two expansions in terms of $t$- and $s$-channel
blocks, respectively.

As emphasized by Dolan and Osborn~\cite{Dolan:2011dv}, the conformal blocks can best be characterized through the differential equations they satisfy. For $s$-channel blocks, these are
given by
\begin{align}\mathcal{D}_{12}^2 \psi_\oo^{(12)(34)} = \left\{ h_\oo(h_\oo-d+1) +\bar h_\oo(\bar h_\oo-1)
 \right\} \psi_\oo^{(12)} =: \lambda_\oo \psi_\oo^{(12)}, \\[2mm]
\mathcal{D}_{12}^2 := (X_{12}X_{34})^{\Dg_\phi} \frac{1}{4} \mathrm{tr}\,
\left(\mathcal{T}_1+\mathcal{T}_2\right)^2 (X_{12}X_{34})^{-\Dg_\phi},
\end{align}
To write the eigenvalue $\lambda_\oo$ on the right-hand side we defined the half-twist
$h:= \frac{\Dg-J}{2}$ and the half-anti-twist $\bar h := \frac{\Dg+J}{2}$ in terms of
the weight $\Dg$ and the spin $J$ of the intermediate field $\oo$. In the second line
we have introduced $\mathcal{T}_i$ to denote the usual action of the generators of the
conformal Lie algebra on the primary field $\phi(X_i)$. The operator $\mathcal{D}_{12}^2$
was first computed by Dolan and Osborn~\cite{Dolan:2003hv}. When expressed in terms of the cross-ratios $u,v$,
it takes the form
\begin{equation} \label{eq:uvCas}
\mathcal{D}_{12}^2 = (1-u-v) \partial_v v \partial_v +
u \partial_u\left(2u\partial_u - d\right) - (1+u-v)
\left( u \partial_u + v\partial_v\right)^2.
\end{equation}
We note that the coefficients of this second-order differential operator are polynomials
in the cross-ratios $u,v$. This is why we refer to them as polynomial cross-ratios.
Clearly, a similar discussion applies to the $t$-channel blocks, only that $u$ and $v$
are exchanged in passing from one channel to the other.

While the crossing symmetry equations of $d$-dimensional conformal field theories
are too difficult to solve analytically for now, at least if $d > 2$, there exist
certain limits in which they simplify drastically. The most interesting of these
is the so-called lightcone limit that is reached after continuation to Lorentzian
signature when $x_4$ is light-like separated from $x_1$, i.e. $X_{14} \sim 0$.
In Lorentzian signature, this limit can be performed without making $x_1 = x_4$
and hence without imposing $u=1$.\footnote{\textbf{Caution:} Contrary to the usual
conventions of the four-point bootstrap and of \cite{Antunes:2021kmm}, we will take
the channel containing the $(12)$ OPE as the \emph{crossed channel}. This will simplify
the computations of limits of blocks in the polynomial cross-ratios later on.}
To perform the relevant limit, we shall assign appropriate orders to the cross
ratios. Let us note that all the cross-ratios depend on the insertion points only
through $X_{ij}$. We can keep track of how a given cross-ratio behaves as we make
a pair of points $x_i$ and $x_j$ light-like separated by introducing a parameter
$\epsilon_{ij}$ and performing the substitution $X_{ij} \mapsto \epsilon_{ij}
X_{ij}$. In the case at hand, we want to make $x_1$ and $x_4$ light-like separated
and hence introduce a parameter $\epsilon_{14}$. Upon the substitution $X_{14}
\rightarrow \epsilon_{14} X_{14}$, the cross-ratio $v$ behaves as $v \rightarrow
\epsilon_{14} v$ while the second cross-ratio $u$ is invariant. We can therefore
think of $v$ as a quantity of $\epsilon_{14}$-order $\orm(\epsilon_{14})$ while the
cross-ratio $u$ has order $\orm(\epsilon_{14}^0)$. Later in the analysis, we will
furthermore make $x_1$ and $x_2$ light-like separated. This motivates us to study
the order also with respect to $\epsilon_{12}$. It is easy to see that $u$ is of
$\epsilon_{12}$ order $\orm(\epsilon_{12}^1)$ while $v$ is $\orm(\epsilon^0_{12})$. We
note that the order extends from the polynomial cross-ratios to the Casimir
differential operators. The operator \eqref{eq:uvCas}, for example, has terms of
$\epsilon_{12}$-order $\orm(\epsilon_{12}^n)$ with $n=0,1$. With respect to the
$\epsilon_{14}$-order, on the other hand, it contains terms of order
$\orm(\epsilon_{14}^m)$ with $m=-1,0,1$, i.e.\ there are also $\epsilon_{14}$-singular
terms. Often we want to keep track of several of these orders at the same time.
In the case at hand, we introduce $\vec{\epsilon} = (\epsilon_{14},\epsilon_{12})$
and then denote the $\vec{\epsilon}$-order by $\orm(\epsilon_{14}^n \epsilon_{11}^m)$
or simply $(n,m)$. With respect to this $\vec{\epsilon}$-order, our cross-ratios
$u$ and $v$ have order $(0,1)$ and $(1,0)$, respectively. This implies that
$\vec{\epsilon}$ is associated with the regime
$$ \textit{\LCL}_{\vec{\epsilon}}: \ v \ll u \ll 1 \  $$
in the space of cross-ratios. Note that the vector $\vec{\epsilon}\ ' =
(\epsilon_{12},\epsilon_{14})$ is associated with a different regime in which
we take $u$ and $v$ to zero in the opposite order, i.e. in which $u \ll v \ll 1$.

\subsection{Lightcone blocks in the direct channel}
\label{ssec:4pt-dc-blocks}
In order to study the behavior of the $t$-channel blocks in the lightcone regime
$\LCL_{\vec{\ep}}$, we shall study the limit of the Casimir eigenvalue equation. Let
us note that the $t$-channel Casimir operator $\mathcal{D}_{14}^2$ is obtained from
the $s$-channel operator we have displayed in eq.~\eqref{eq:uvCas} by exchanging
the cross-ratios $u$ and $v$, i.e.\
\begin{equation} \label{eq:vuCas}
\mathcal{D}_{14}^2 = (1-u-v) \partial_u u \partial_u +
v \partial_v\left(2v\partial_v - d\right) - (1+v-u)
\left( u \partial_u + v\partial_v\right)^2.
\end{equation}
For reasons that will become clear later, we will implement the lightcone limit 
of the direct channel blocks in stages, starting with the restricted limit in 
which we only send $v$ to zero while keeping $u$ finite at first. For the associated regime, we shall write 
$$ \LCL^{(1)}_\ep:  \  v \ll 1 \ .  $$
The superscript $(1)$ reminds us that we only consider the first component $\ep_{14}$
of the order $\vec{\ep}$. The dependence of the Casimir operator $\mathcal{D}_{14}^2$ 
on $\epsilon_{14}$ gives rise to a split into a sum of two terms, i.e.\ 
\begin{equation}
\mathcal{D}_{14}^{2, \ep_{14}} = \ep_{14}^0 \mathcal{D}_{14}^{(0)}+ \ep_{14}
                \mathcal{D}_{14}^{(1)}
\end{equation}
where the leading term is given by
\begin{equation}
\mathcal{D}_{14}^{(0)} = (1-u)\tensor[_2]{\mathcal{D}}{_1}(v\ds_v,v\ds_v;2 v\ds_v;
1-u;-\ds_u)+v\ds_v(2 v\ds_v-d)
\end{equation}
Here, we find it useful to introduce the symbol $\tensor[_2]{\mathcal{D}}{_1}
(a,b;c;z,\ds)$ that will be used in later discussions:
\begin{equation}
\tensor[_2]{\mathcal{D}}{_1}(a,b;c;w,\ds):= \ds(w\ds+c-1)-(w\ds+a)(w\ds+b).
\label{eq:2D1def}
\end{equation}
In general, the parameters $a,b,c$ can be differential operators that commute with $w,\partial$, while $w$ is some function of the cross-ratios. Since only
$v$ scales with $\ep_{14}$ and $\mathcal{D}_{14}^{(0)}$ is homogeneous
of degree zero in $\ep_{14}$, this operator commutes with the Euler operator 
$\vartheta_v \equiv v\ds_v$.

The $t$-channel conformal blocks are eigenfunctions of the Casimir differential operator $\mathcal{D}_{14}^2$. Hence,
if we assume the eigenvalues not to scale with $\ep_{14}$, their limits in the 
regime $\LCL^{(1)}$ where we send $v$ to zero must solve the differential 
equation
\begin{equation} \label{D14eq}
\mathcal{D}_{14}^{(0)} \psi^{(14)(23);(0)}_\oo(u,v) =  \left\{ h_\oo(h_\oo-d+1) +
\bar h_\oo(\bar h_\oo-1) \right\} \psi_\oo^{(14)(23);(0)}(u,v)\ .
\end{equation}
Since the differential operator $\mathcal{D}_{14}^{(0)}$ commutes with the
Euler operator $v\partial_v$, we can make the following Ansatz for a complete
set of solutions
\begin{equation}\label{4-ptdirect}
\psi^{(14)(23);(0)}_{(h,\bar h)}(u,v) = v^h g_{h,\bar h}(u)\ .
\end{equation}
We insert this Ansatz into the leading order eigenvalue equation using
\begin{equation}
v^{-h}\left(\mathcal{D}_{14}^{(0)}-h(h-d+1) \right) v^h = (1-u)\,
\tensor[_2]{\mathcal{D}}{_1}(h,h;2h;1-u,-\ds_u) .
\end{equation}
It is now easy to see that eigenfunctions take the form
\begin{equation}
\psi^{(14)(23)}_{(h,\bar h)}(u,v) \stackrel{\LCL^{(1)}}{=}
\psi^{(14)(23);(0)}_{(h,\bar h)}(u,v) = v^h (1-u)^{J}
\tensor[_2]{F}{_1}(\bar h, \bar h; 2\bar h;1-u).
\label{lightcone_block_4pt}
\end{equation}
with $J = \bar h - h$, as before. The formula we wrote requires some more comments. 
Obviously, the differential equation \eqref{D14eq} is of second order and, hence, 
possesses two linearly independent solutions. But in the regime $\LCL^{(1)}$ it is 
not difficult to determine which one is relevant. Let us recall that the relevant 
solution of the four-point Casimir equation is uniquely determined by its  behavior 
in the OPE limit, which corresponds to $ v \sim 0$ and $u \sim 1$\footnote{Recall that our direct channel is what is often referred to as $t$-channel 
and hence the cross-ratios $u$ and $v$ are exchanged with respect to the usual 
$s$-channel discussion.} with $v \ll (1-u)$, in accordance with its Lorentzian definition. In this limit, the conformal block behaves as 
\begin{equation}
\psi^{(14)(23)}_{(h,\bar h)}(u,v) = v^h (1-u)^{J} \left( 1+\orm(v,1-u)\right)\ .
\label{norm_OPE_12}
\end{equation}
With this normalization condition for the blocks stated, we now note a fortunate
fact: the regime $\LCL^{(1)}$ in which we were able to solve the Casimir equation
through eq.~\eqref{lightcone_block_4pt} does contain the OPE limit point at which 
blocks are selected and normalized. So, all we need to do is verify that the solution we proposed in eq.~\eqref{lightcone_block_4pt} satisfies the boundary condition \eqref{norm_OPE_12} at $u \sim 1$. This is obviously the case. 
\medskip 

After discussing conformal blocks in the partial lightcone limit, i.e.\ in the 
regime $\LCL^{(1)}$, we now want to address the full lightcone limit in which $u$ is sent to zero as well. Obviously, we could easily find the limiting behavior of the blocks from our formula \eqref{lightcone_block_4pt}, using standard properties
of hypergeometric functions. But here we want to go back to the study of differential equations instead and see how much we can deduce from them. In order to study the full lightcone limit we reinstate the parameter $\ep_{12}$ that determines the behavior of $u$ as we go in the lightcone regime $\LCL$.
The leading term of the Casimir operator \eqref{eq:vuCas} in the order $\vec{\ep} = (\ep_{14},\ep_{12}$) 
reads 
\begin{equation}
\mathcal{D}^{2, \vec{\ep}}_{14} = \ep_{12}^{-1} \mathcal{D}_{14}^{(0,-1)} + 
\orm(\ep_{12}^0),
\end{equation}
where the superscript reminds us that the leading term of the quadratic Casimir operator
is of order $\orm(\epsilon^0_{14}\epsilon_{12}^{-1})$. Explicitly one finds that
\begin{equation} \label{eq:Cas4ptLCL}
\mathcal{D}_{14}^{(0,-1)}= \ds_u u \ds_u.
\end{equation}
Let us also note that $\orm(\ep_{12}^0)$ consists of two terms, one that is constant
in $\ep_{12}$ and one that is linear. Their precise form is easily found but irrelevant for
us. If we now assume that the eigenvalue $\bar h(\bar h-1)+h(h-d+1)$ of the Casimir operator
does not depend on $\ep_{12}$, we conclude that the leading contribution of our conformal
blocks in the limit in which $\ep_{12}$ is sent to zero must be in the kernel of the
operator $\mathcal{D}_{14}^{0,-1}$,
\begin{equation} \label{eq:ker4pt}
\mathrm{ker}(\mathcal{D}_{14}^{(0,-1)}) =\mathrm{Span}(v^h, v^h \log u).
\end{equation}
Here the eigenvalue $h$ of the Euler operator $v \partial_v$ can assume any non-negative
real value. This behavior of the lightcone blocks is indeed consistent with our previous
equation \eqref{lightcone_block_4pt} for the normalized lightcone blocks. As one can verify
with the help of standard results in the behavior of hypergeometric functions, lightcone
blocks behave as
\begin{equation}\label{eq:4ptlcbdc}
\psi^{(14)(23)}_{(h,\bar h)}(u,v) \stackrel{\LCL_{\vec{\ep}}}{\sim} \begin{cases} 
\ \  v^h \quad \quad 
\mathrm{if} \,  \bar h=0 \\[2mm] \ \ 
-\mathrm{B}_{\bar h}^{-1} v^h \log u+\orm(1) 
\quad \mathrm{if}\,
\bar h >0
\end{cases},
\end{equation}
where $\mathrm{B}_{\bar h} :=\Gamma(2 \bar h)^{-1}\Gamma(\bar h)^2$. In the first line 
we have kept the prefactor $v^h$ even though $\bar h =0$ actually implies $h=0$ in a 
unitary theory. The solution we displayed is consistent with the statement 
\eqref{eq:ker4pt}. But the analysis of the Casimir equation in the full lightcone 
limit $\LCL$ is not sufficient to fully determine the solution. Recall that the 
solution is determined through the behavior of the blocks in the OPE limit. But the 
latter is not contained in $\LCL$. The only way in which we can determine the 
precise vector in the kernel \eqref{eq:ker4pt} that is chosen by the OPE boundary 
condition is to go back to our solution \eqref{lightcone_block_4pt} which we argued 
to satisfy the OPE boundary condition. We can then take the limit $u \rightarrow 0$ 
to obtain our results in eq.~\eqref{eq:4ptlcbdc} with the correct normalization. 
This follows from standard limiting formulas for hypergeometric functions.
\smallskip 

The analysis of direct channel blocks we presented here serves as a good model for 
our discussion of lightcone limits for multipoint blocks. It will turn out that in 
all the relevant cases we will be able to solve for blocks in a partial lightcone 
regime which contains both the OPE and the full lightcone limit. Though the asymptotics of these solutions are not well studied in comparison with standard hypergeometric functions, we will obtain integral formulas that provide access to the limiting 
behavior in both the OPE and the lightcone limit. In the case at hand, for 
example, we can also write the solution to the Casimir equation \eqref{D14eq}
in the form 
\begin{equation} 
\psi^{(14)(23)}_{(h,\bar h)}(u,v) \stackrel{\LCL^{(1)}}{\sim} 
\frac{v^h (1-u)^J}{\mathrm{B}_{\bar h}} \int_0^1 \dd t\,  
(t (1-t))^{\bar h-1} (1-(1-u)t)^{-\bar h} \ .
\label{eq:LCint4ptcr}
\end{equation}
It is easy to see that the integral in the last expression has a logarithmic 
divergence for $u \sim 0$ which stems from the integration near $t=1$. In order 
to determine the leading term of the integral as we send $u$ to zero, we can write
\begin{equation}
\int_0^1 \dd t  t^{\bar h-1} (1-t)^{\bar h-1} (1-(1-u)t)^{-\bar h} \stackrel{u \ll 1}{\sim}
\int_0^1 \dd t \frac{1-u}{1-(1-u)t} = \int_u^{1} \frac{\dd v}{v} \sim -\log u \ .
\end{equation}
In sum, we indeed reproduce the expected $\log u$ divergence with a coefficient 
$\mathrm{B}_{\bar h}^{-1}$ of the integral in eq.~\eqref{eq:LCint4pt}. This 
matches our result in eq.~\eqref{eq:4ptlcbdc}.
\smallskip 

While the integral formula \eqref{eq:LCint4ptcr} can be found through the analysis of the Casimir differential equations, it can also be obtained from the lightcone OPE. 
Let us note that even in generic kinematics, i.e. before sending any of the cross-ratios to zero, conformal blocks possess integral representations. 
But these are difficult to evaluate. They simplify significantly, however, when some of the external insertion points are lightlike separated so that one can use the 
lightcone OPE. In the case of $N=4$ external points, the integral formula for blocks becomes
\begin{equation} \label{eq:4ptLCOPE}
\psi^{(14)(23)}_{(h,\bar h)}(u,v) = (X_{14}X_{23})^h (X_4 \wedge X_1 \cdot X_2 
\wedge X_3)^J \int_{\Rs_+^2} \frac{\dd s_1\dd s_4 (s_1s_4)^{\bar h-1}}{\Rs^\times
\mathrm{B}_{\bar h}} X^{-\bar h}_{a2} X^{-\bar h}_{a3} \left(1+\orm(X_{14}) \right).
\end{equation}
In this expression, $X_a := s_1 X_1+s_4X_4$ is a point on the projective lightcone 
when $X_{14}=0$, $\mathrm{B}_{\bar h} := \Gamma(\bar h)^2\Gamma(2\bar h)^{-1}$ is 
the diagonal of the Euler Beta function, and $\Rs^\times := \int_0^\infty r^{-1} 
\dd r$ is the volume of the dilation group $ \{(s_1,s_2)\rightarrow (r s_1,
r s_2) \}$. Formula \eqref{eq:4ptLCOPE} is obtained by inserting the lightcone 
OPE --- see eq.~\eqref{eq:lOPE} of appendix~\ref{app:5pt_integral_formula} --- into the four-point function, before evaluating the 
resulting three-point function in the integrand using standard formulas for the 
three-point function of two scalar fields $\phi$ and one spinning field with 
weight $\Delta_\oo = h + \bar h $ and spin $J = \bar h - h$. Since the integral 
is homogeneous of degree zero in 
$X_1,X_2,X_3,X_4$ and manifestly $\mathrm{SO}(1,d+1)$ invariant, it can be written as a function of the cross-ratios $(u,v)$.
After the change of variables $(s_1,s_4)=(r t, r(1-t))$, it indeed takes 
the form
\begin{eqnarray}
\psi^{(14)(23)}_{(h,\bar h)}(u,v) &\stackrel{v \ll 1}{\sim}&
  \frac{\left(x^2_{14} x_{23}^2\right)^{h}}{\mathrm{B}_{\bar h}} \int \dd t \,  \frac{\left(x^2_{24} x_{13}^2 - x_{34}^2
x_{12}^2\right)^J} {\left(x_{12}^2t +(1-t)x_{24}^2\right)^{\bar h} \left( x_{13}^2t +
(1-t)x_{34}^2 \right)^{\bar h}}\nonumber \\[2mm]
& = & \frac{v^h (1-u)^J}{\mathrm{B}_{\bar h}} \int_0^1 \dd t\,  (t (1-t))^{\bar h-1} (1-(1-u)t)^{-\bar h} \ .
\label{eq:LCint4pt}
\end{eqnarray}
The expression in the second line gives the integral representation of the 
hypergeometric solution \eqref{lightcone_block_4pt} that we used in the previous paragraph to evaluate 
the limiting behavior in the full lightcone limit. 
With these comments stated, we close the discussion of the direct channel and turn to 
the analysis of lightcone blocks in the crossed channel.

\subsection{Lightcone blocks in the crossed channel}

Let us now discuss the lightcone limit of conformal blocks in the crossed $s$-channel.
We recall that the Casimir differential operator $\mathcal{D}^2_{12}$ in the crossed
channel has been displayed in eq.~\eqref{eq:uvCas}. To analyze its eigenfunctions 
in the full lightcone limit $\LCL$, we expand the Casimir operator
$\mathcal{D}_{12}^2$ as  
\begin{align}
\mathcal{D}_{12}^{2, \vec{\ep}} = 
\ep_{14}^{-1} \left( \ds_v v \ds_v - \ep_{12}u\ds_v v \ds_v
\right) + \orm(\ep_{14}^0).
\end{align}
Since the leading contribution commutes with the Euler operator $\vt_u \equiv 
u\ds_u$, we can without loss of generality choose a basis of functions whose 
leading behavior takes the form
\begin{equation}
\psi^{(12)(34)}_{(h,\bar h)}(u,v)  \sim u^h g_{(h,\bar h)}(v)(1+\orm(u^{h+1}))
\end{equation}
as we send the cross-ratio $u$ to zero. The leading term eigenvalue equation 
then becomes
\begin{equation}
\ds_v v \ds_v g_{(h,\bar h)}(v) = \la_{(h,\bar h)} g_{(h,\bar h)}(v),
\quad \la_{(h,\bar h)} = h(h-d+1) + \bar h(\bar h-1)\ .
\end{equation}
Contrary to the direct $t$-channel, the $v \rightarrow 0$ limit does not
restrict the twist of blocks appearing in the $s$-channel. Thus, the Casimir
eigenvalue can vary over all positive numbers, and we can distinguish three
regimes
\begin{equation}
g(v) \stackrel{v \ll 1}{\sim} \nn\ \begin{cases} 1 \,\, \mathrm{or}\, \log v
\quad \mathrm{if} \, \la = \orm(\ep_{14}^0), \\[2mm]
K_0(2 \sqrt{\lambda v}) \quad \mathrm{if} \, \la = \orm(\ep_{14}^{-1}),
\\[2mm] 0 \quad \mathrm{otherwise}. \end{cases}
\end{equation}
In the first case, the eigenfunctions must be in the kernel of $\partial_v v
\partial_v$. If the eigenvalue $\lambda^{\vec{\ep}} = \ep_{14}^{-1} \lambda$ 
scales with $\ep_{14}$ as stated in the second line, the right-hand side of 
the eigenvalue equation contributes and we obtain the equation
\begin{equation}
\left(\partial_v v \partial_v - \lambda \right) g = 0\
\end{equation}
which can be transformed into Bessel's differential equation by a simple
change of variables. This goes a long way toward explaining the behavior 
of $g(v)$ we have displayed. But as in the direct channel, we still need 
to ensure that the blocks satisfy the correct boundary conditions and in 
particular determine the normalization $\mathcal{N}$.

In order to so, we exploit the fact that the leading term $\ds_v
v \ds_v$ in the Casimir operator $\mathcal{C}_{12}^2$ at order
$\orm(\ep_{14}^{-1} \ep_{12}^0)$ is the same regardless of whether 
we take $\ep_{14}$ or $\ep_{12}$ to zero first, i.e. 
\begin{equation} 
\lim_{\ep_{14} \rightarrow 0} \lim_{\ep_{12}\rightarrow 0} 
\ep_{14} \left( \mathcal{D}^{2,\vec{\ep}}_{12}-\lambda^{\vec{\ep}} \right) = 
\lim_{\ep_{12} \rightarrow 0} \lim_{\ep_{14}\rightarrow 0} 
\ep_{14} \left( \mathcal{D}^{2,\vec{\ep}}_{12}-\lambda^{\vec{\ep}} \right)\ . 
\end{equation} 
We can thus retrieve these asymptotics of the crossed channel lightcone 
blocks by taking $\ep_{12}$ to zero first. In this case, the analysis is 
identical to the one we described in our discussion of the direct channel. 
Consequently, in the limit $\ep_{12} \rightarrow 0$ (with $v$ kept finite), our 
blocks take the exact form of \eqref{lightcone_block_4pt} with 
$u\leftrightarrow v$ and with $h$ corresponding to a fixed half-twist,
\begin{equation} \label{eq:4ptpsiF}
\psi^{(12)(34)}_{(h,\bar h)}(u,v) \stackrel{u \ll 1}{\sim}
u^h (1-v)^J\tensor[_2]{F}{_1}( \bar h, \bar h; 2 \bar h;v)\ .
\end{equation}
Here we have already inserted the normalization that we had determined
previously. Consequently, given that the eigenvalue is $\lambda
= h(h-d+1)+\bar h(\bar h-1)$, and that we have fixed $h$ already, our three
regimes correspond to
\begin{equation}\label{eq:4ptlcbcc}
\psi^{(12)(34)}_{(h,\bar h)}(u,v)  \stackrel{\LCL_{\vec{\ep}}}{\sim} \
\begin{cases}\  1 \, \,
\mathrm{or}\, \mathrm{B}_{\bar h}^{-1} \log v^{-1} \quad \mathrm{if} \, \bar
h = \orm(\ep^0_{14}) \\[2mm]
\ \nn^{\mathrm{LS}}_{(h,\bar h)} u^h K_0( 2 \bar h \sqrt{v}), \quad \mathrm{if} \,
\bar h = \orm(\ep_{14}^{-1/2}) \\[2mm]
\ 0 \quad \mathrm{otherwise},\end{cases}
\end{equation}
with
\begin{equation} \label{eq:NLS}
\nn^{\mathrm{LS}}_{(h,\bar h)}= 4^{\bar h} \sqrt{\bar h/\pi}
\end{equation}
and $\mathrm{B}_{x} := \mathrm{B}(x,x)=\Gamma(x)^2/\Gamma(2x)$ given by the
diagonal of the beta function as before. In order to determine the normalization
$\nn$ in the second case we have used that $K_0 (2x) \sim -\log (x)$ near
$x \sim 0$ along with the Stirling formula to evaluate $\mathrm{B}_{\bar h}^{-1}$ 
as $\bar h$ tends to infinity. This concludes our discussion of the
lightcone limit of four-point blocks for identical scalars in the direct and
the crossed channel. We stress once again that in order to obtain the two
key results \eqref{eq:4ptlcbdc} and \eqref{eq:4ptlcbcc} for the lightcone
limits of blocks in the direct and crossed channel, respectively, we only
needed the Casimir differential equations. This applies in particular to
the dependence of the lightcone blocks on the cross-ratios. The
normalizations could also be determined from the Casimir equations, though
for this purpose we had to solve them outside the lightcone limit to connect
with the OPE limit. 

\subsection{Review of the four-point lightcone bootstrap}\label{4pointlightcone}
We can now exploit the results on lightcone blocks to analyze the crossing
symmetry equation \eqref{eq:4ptcrossing2}. Here, our discussion will closely
follow some of the original literature \cite{Fitzpatrick:2012yx,Komargodski:2012ek}. 
Let us first evaluate the direct
channel, i.e. the left-hand side of the crossing equation, in the regime
$v \ll u \ll 1$. Using the result \eqref{eq:4ptlcbdc} for the limit of the
blocks one finds
\begin{equation}
\sum_{\oo} C_{\phi\phi\oo} \psi_\oo^{(14)(23)}(u,v) \sim
1 + C_{\phi\phi\oo_*}^2 \frac{\Gamma(2 \bar h_*)}{\Gamma(\bar h_*)^2}
\left[ v^{h_*} \left(\log u^{-1} + \orm(u^0) \right) +\orm(v^{h_*+1})\right]
+ \orm(v^{h>h_*}).
\label{eq:4ptdc}
\end{equation}
Here, we denote the leading-twist operator in the operator product of $\phi$ with
itself by $\oo_\star$, and we assume it is unique. Note that uniqueness does not
apply to generalized free-field (GFF) theory, where the leading twist is infinitely degenerate and the operator product takes the form
\begin{equation} \label{eq:phi2ope}
\phi(X_1) \phi(X_4) = 1+ \sum_{n=0}^{\infty} X_{14}^{n} \sum_{J \in 2\Zs_+}^\infty
C_{\phi\phi [\phi\phi]_{n,J}} f_{\phi\phi [\phi\phi]_{n,J}}(X_1-X_4,\ds_{X_4})
[\phi\phi]_{n,J}(X_4).
\end{equation}
Here we have introduced the standard notation $[\phi\phi]_{n,J}$ for the double-twist operators of the form
\begin{equation}
[\phi\phi]_{n,J;\mu_1, \dots \mu_J}(X) = \phi(X) \, \square^n \,
\partial_{\mu_1} \cdots \partial_{\mu_J} \phi(X)\ .
\end{equation}
The label $J$ denotes the spin of the field, while $\tau_{n,J}= 2 \Dg_\phi + 2n$
is the twist in GFF theory. In terms of our parameters $h$ and $\bar h$
this means
\begin{equation}\label{eq:doubletwists}
h_{[\phi\phi]_{n,J}}= \tau_{n,J}/2 =\Dg_\phi+n \quad , \quad
\bar h_{[\phi\phi]_{n,J}}=\Dg_\phi+n+J\ .
\end{equation}
 Note that if $\phi$ is a real scalar field, then only \emph{even} spins will appear in the $\phi\times\phi$ OPE, see e.g. \cite[footnote~12]{Rattazzi:2008pe}. As one can read off from these formulas, the operator product \eqref{eq:phi2ope}
contains an infinite tower of operators $\{[\phi\phi]_{n,J}\}_{n,J\in 2\Zs_+}$ at
leading twist. There are degeneracies that may be lifted by interactions. 

Now, analytic bootstrap teaches us that for a general conformal field theory, with an isolated operator contribution like $\oo_\star$ in eq.\ \eqref{eq:4ptdc},  there exist towers of operators whose OPE coefficients and twists approach those of GFF double-twist operators at large spin. The main goal is to find precise expressions for how their OPE data is corrected. Note
also that  $C_{\phi\phi\oo_*}^2 \orm(v^{h_*+1})$ in eq.\ \eqref{eq:4ptdc} corresponds to the
contributions of descendants of the leading-twist operator, whereas the
next-to-leading-twist operator $\oo'$ would give $C_{\phi\phi\oo'}^2 \orm
(v^{h'})$. In many cases of interest, we can expect several $h'<h_\star +1$
primaries to dominate over the first descendants. In the 3D Ising model, for
example, $\epsilon$ and  $T_{\mu\nu}$  give important contributions \cite{Simmons-Duffin:2016wlq}.
\smallskip

We will now re-derive two of the important results of the lightcone bootstrap, 
the computation of the OPE coefficients $C_{\phi\phi [\phi\phi]_{0,J}}$ and the 
anomalous dimensions
$h_{[\phi\phi]_{0,J}}\!-\Dg_\phi$, to leading order in the large $J$ limit. Here $[\phi\phi]_{0,J}$ is the family of operators whose twists approach the GFF double twist tower for $n=0$.
Given our results \eqref{eq:4ptlcbcc} on the lightcone limit of crossed-channel
blocks, we can write the $s$-channel sum on the right-hand side of the crossing
symmetry equation \eqref{eq:4ptcrossing2} in the $v\ll u \ll 1$ limit as
\begin{align}
&\sum_\oo C_{\phi\phi\oo}^2 \psi^{(12)(34)}(u,v) =\nonumber \\[2mm]
&1 + \int_{\orm(\ep_{14}^0)}^{\infty} \frac{\dd \la_\oo}{4 \sqrt{\la_\oo}} \nn^{\mathrm{LS}}
C_{\phi\phi\oo}^2  u^{h_{\oo}^{\mathrm{min}}} K_0(2 \sqrt{\la_\oo v})
\left(1+ \orm(v^{\frac{\tau}{2}}) \right) + \sum_{\la_\oo = \orm(\ep_{14}^0)}
\frac{\log v^{-1}}{\mathrm{B}_{\bar h}}\left(1+ \orm(v^0) \right).
\label{eq:4ptcc}
\end{align}
In the integral over large Casimir eigenvalues, $h_{\oo}^{\mathrm{min}}$ is the minimal twist that appears in the operator product of $\phi$
with itself for which the Casimir $\la_\oo$ is not bounded above. If this
$h_\oo^{\mathrm{min}}$ is finite (which it usually is), then this implies
that $\oo$ has large spin
\begin{equation}
\bar h_\oo = J_\oo + h_\oo^{\mathrm{min}} = \sqrt{\la_\oo}+\orm(\ep_{12}^0).
\end{equation}
Comparing the two expressions \eqref{eq:4ptdc} and \eqref{eq:4ptcc} for the limiting
behaviors of the two sides of the crossing equation \eqref{eq:4ptcrossing2} we deduce to
leading order
\begin{equation}
1+\orm(v^{h_\star}) = v^{\Dg_\phi}u^{-\Dg_\phi}
\sum_\oo C_{\phi\phi\oo}^2 \psi^{(12)(34)}(u,v).
\end{equation}
One immediately observes that the $s$-channel sum must yield a divergent power law
$v^{-\Dg_\phi}$ to leading order --- this eliminates the finite sum over primaries
with $\la_\oo=\orm(v^0)$, since these contributions diverge \emph{logarithmically}
at most. We thus need to solve
\begin{equation}\label{eq:4ptboot}
1+\dots = v^{\Dg_\phi} \int_{\orm(1)}^\infty \frac{\dd \la_\oo }{4 \sqrt{\lambda_\oo}}\nn^{\mathrm{LS}}C_{\phi\phi\oo}^2
u^{h_\oo^{\mathrm{min}}-\Dg_\phi} K_0(2 \sqrt{\la_\oo v}) +\dots
\end{equation}
This equation requires us to impose $h_\oo^{\mathrm{min}}=\Dg_\phi=2h_\phi$, from which we deduce that the left-hand side of the
crossing symmetry equations is obtained from the exchange of the double-twist
operators $[\phi\phi]_{0,J}$ with spins $J$ that scale as
$$\sqrt{\la}=J+\orm(1)= \orm(v^{-1/2})\ . $$
If we make such a scaling Ansatz, we then find
\begin{equation}
\nn^{\mathrm{LS}}C_{\phi\phi\oo}^2 = c_0 \frac{8 J^{\bg-1}}{\Gamma(\bg/2)^2}
\left(1+\orm(J^{-\tau}) \right),
\end{equation}
and using the auxiliary formula
\begin{equation}
\int_{\orm(\sqrt{v})}^\infty \frac{\dd y}{y} \left(\frac{y}{2}\right)^\bg K_{\ag}(y)
= \frac{1}{4} \Gamma\left(\frac{\bg+\ag}{2}\right)\Gamma\left(\frac{\bg-\ag}{2}\right)
+\orm(v^{\frac{\bg-\ag}{2}}), \quad \bg > \ag,
\end{equation}
we reduce the leading order terms of the crossing symmetry equation to
\begin{equation}
1+\dots = c_0v^{\Dg_\phi-\frac{\bg}{2}}+... \Rightarrow \nn^{\mathrm{LS}}
C_{\phi\phi\oo}^2 = \frac{8 J^{2\Dg_\phi-1}}{\Gamma(\Dg_\phi)^2}
\left(1+\orm(J^{-\tau}) \right).
\end{equation}
This is the same as the large spin limit of the OPE coefficients of double-twist operators in a GFF theory \cite{Heemskerk:2009pn}, obtained for a more general conformal field theory using the lightcone bootstrap.

Moving on to the next-to-leading order in the crossing equation, we have
\begin{equation}\label{eq:4ptboot-gamma}
1+\frac{C_{\phi\phi\oo}^2}{\mathrm{B}(\bar h_\star)}v^{h_\star}
\left(\log u^{-1}\! +\orm(u^0)\right)+\orm(v^{h>h_\star}) = v^{\Dg_\phi}\!\int_{\orm(1)}^\infty
\frac{\dd J}{J} \frac{2 J^{2\Dg_\phi}}{\Gamma(\Dg_\phi)^2} u^{h_{[\phi\phi]_{0,J}}-
\Dg_\phi} K_0(2 J \sqrt{v}) (1+\orm(J^{-\tau})).
\end{equation}
Here we once again discarded the finite spin contributions with $\orm(\log v)$ divergence
at most. To reproduce $v^{h_\star}\log u$ asymptotics, one makes the following ansatz for
leading anomalous dimension corrections to double-twist operators at large spin:
\begin{equation}
h_{[\phi\phi]_{0,J}}= \Dg_\phi + \frac{\gamma}{2 J^{2 h_\star}} +
\orm\left( J^{-\tau}\right), \quad \tau>2h_\star.
\end{equation}
Expanding $v^{h_\star}$ in the $t$-channel, we obtain
\begin{equation}
\frac{C_{\phi\phi\oo_\star}^2}{\mathrm{B}_{\bar h_\star}} = -\frac{\cg}{2}
\frac{\Gamma(\Dg_\phi-h_\star)^2}{\Gamma(\Dg_\phi)^2}.
\end{equation}
Note that the anomalous dimension correction vanishes when $h_\star = \Dg_\phi$. This is
the signature of GFF: the leading-twist operators are $[\phi\phi]_{0,J}$,
and their scaling dimensions are not anomalous. Analogous results for the $n>0$ families in eq.~\eqref{eq:doubletwists} can be obtained by focusing on higher powers of $u$ in eqs.~\eqref{eq:4ptboot} and \eqref{eq:4ptboot-gamma} \cite{Kaviraj:2015cxa, Kaviraj:2015xsa}.
\smallskip

In the previous analysis, we have seen that we can only reproduce the leading terms on the left-hand side of the crossing symmetry equations by summing over large spin double-twist operators on the right-hand side, i.e.\ with the help of crossed channel lightcone
blocks \eqref{eq:4ptlcbcc} for which the Casimir eigenvalue $\lambda$ scales as $\lambda
\sim \ep_{14}^{-1}$. In \cite{Simmons-Duffin:2016wlq}, Simmons-Duffin gave a simple proof
of this fact. The argument is based on the observation that the Casimir operator
$\mathcal{D}_{12}^2$ has a stable action on any finite spin subspace in the $(12)$ channel,
whereas its action on $v^{-\Dg_\phi}$ produces higher-order divergences in $v$. To see this in more detail, let us first define the vector space
\begin{equation}
\mathbb{V}_{L} := \mathrm{Span}\left\{ \psi_\oo^{(12)(34)}:h_\oo = \Dg_\phi,
\bar h_\oo < L \ \right\}, \quad L<\infty.
\end{equation}
We rewrite the crossing equation as
\begin{equation}
u^{\Dg_\phi} v^{-\Dg_\phi} +\dots = \sum_{\oo} C_{\phi\phi\oo}^2 \psi_\oo^{(12)(34)}(u,v).
\end{equation}
Acting repeatedly with $\mathcal{D}_{12}^2$ on the identity contribution, we find
\begin{align*}
&\ \mathcal{D}_{12}^2 u^{\Dg_\phi} v^{-\Dg_\phi} = \Dg_\phi(\Dg_\phi+1)u^{\Dg_\phi}
v^{-(\Dg_\phi+1)}+\orm(v^{-\Dg_\phi}) \\[2mm]
\longrightarrow &\ \mathcal{D}_{12}^2 u^{\Dg_\phi} v^{-(\Dg_\phi+1)} = (\Dg_\phi+1)
(\Dg_\phi+2)u^{\Dg_\phi} v^{-(\Dg_\phi+1)}+\orm(v^{-(\Dg_\phi+1)}) \\[2mm]
\longrightarrow &\ \dots
\end{align*}
As long as $\Dg_\phi \notin \Zs_{<0}$, this series never truncates and $\mathcal{D}^2_{14}$
stabilizes only on an infinite dimensional vector space $\mathrm{Span}(u^{\Dg_\phi}
v^{-(\Dg_\phi+n)})_{n \in \Zs_{>\geq 0}}$ --- we say that
$(u/v)^{\Dg_\phi}$ is $\mathcal{D}_{12}^2$-singular. It follows that the expansion of $(u/v)^{\Dg_\phi}$
in eigenvectors of $\mathcal{D}_{12}^2$ has \emph{infinite} support. In short,
\begin{equation}
(u/v)^{\Dg_\phi} \notin \mathbb{V}_L, \quad \forall L < \infty.
\label{inf_support}
\end{equation}
In general, a direct channel contribution is reproduced by large spin operators in the crossed
channel if and only if it is $\mathcal{D}$-singular for some Casimir differential operator
$\mathcal{D}$ in the crossed channel. In five- and six-point crossing equations, where we obtain
more unconventional asymptotics in the lightcone limit of the direct channel, we will demonstrate
that they yield large spin conformal field theory data by virtue of being Casimir-singular.

\section{Lightcone Blocks for Five Points}
\label{sect:lightcone_blocks_five_pt}

 The aim of this section is to develop the technology that is needed to determine
the lightcone limits of multipoint conformal blocks from the differential equations
they satisfy. Our constructions are carried out in the case of $N=5$, for which 
lightcone blocks have been studied through the lightcone OPE, so that we can compare 
many of our findings with previous results, most notably those obtained in~\cite{Antunes:2021kmm}. On the other hand, the approach we follow here can be 
carried out for arbitrary topology of the OPE channel, and in particular for comb 
channels where no other approach exists. Even for $N=5$ insertion points, our approach 
adds to the study of \cite{Antunes:2021kmm} in two respects. On the one hand, for any exchanged twists $h_1,h_2$, we are able to construct lightcone blocks explicitly in a limit where only four of the five 
cross-ratios are taken to zero. This will give us a much higher resolution in the 
bootstrap analysis later on. In addition, we can put to a stringent test and provide evidence for one of the central assumptions of  \cite{Antunes:2021kmm}, namely that it is possible to compute crossed-channel lightcone blocks by taking limits of 
an integral formula that is valid only in the direct channel.  

In the first subsection, we review some basic notations that are needed in
dealing with five-point functions and their blocks. We will also introduce the
crossing symmetry equation to be studied in the next section. This equation
provides us with a precise understanding of the relevant limit and a list
of blocks whose lightcone behavior we need to determine. The relevant
lightcone blocks are then computed in the remainder of the section, first
for the direct channel and then for the crossed one. The special case of three-dimensional lightcone blocks with parity-odd tensor structures is treated in appendix~\ref{app:parity-odd_blocks}, and is based on the results present in the main text. As we already saw in our
discussion of four insertions, the differential equations are a
powerful tool to determine the dependence of the lightcone blocks on the
cross-ratios, but they leave an overall normalization undetermined. To
find the normalizations, we need to connect the lightcone limit to the
OPE limit. In order to do so, we shall use a strategy that is closely 
modeled after the strategy we explained for four-point blocks in the 
previous section. 

\subsection{Preliminaries on blocks and lightcone limits}
Five-point functions in $d \geq 3$
are parametrized by five cross-ratios. The following polynomial cross-ratios for five-point
functions were first constructed in \cite{Buric:2021kgy},
\begin{equation}
		\begin{gathered}
		u_1=\frac{X_{12} X_{34}}{X_{13} X_{24}}\,,
		\qquad v_1=\frac{X_{14} X_{23}}{X_{13} X_{24}}\,,\\[2mm]
		u_2=\frac{X_{23} X_{45}}{X_{24} X_{35}}\,, \qquad
		v_2=\frac{X_{25} X_{34}}{X_{24} X_{35}}\,,
		\end{gathered}
		 \qquad U^{(5)}_1=\frac{X_{15} X_{23}X_{34}}{X_{24} X_{13}X_{35}}\,.
\label{eq:five_point_polynomial}
\end{equation}
It will often be convenient to use the ``snowflake" cross-ratios $u_{si},i=1, \dots, 5$ used
in \cite{Bercini:2020msp,Antunes:2021kmm}. In terms of our polynomial cross-ratios, the
cross-ratios $u_{si}$ read
\begin{eqnarray}
u_{s1} & = & u_1/v_2 = \frac{X_{12}X_{35}}{X_{13}X_{25}}, \quad
u_{s2} = v_1 =\frac{X_{14} X_{23}}{X_{13} X_{24}} , \\[2mm]
u_{s3} & = & v_2 = \frac{X_{25} X_{34}}{X_{24} X_{35}}, \quad
u_{s4} = u_2/v_1 = \frac{X_{45} X_{13}}{X_{35}X_{14}},\\[2mm]
u_{s5} & = & \frac{U^5}{v_1 v_2} = \frac{X_{15}X_{24}}{X_{14}X_{25}}.
\end{eqnarray}
These cross-ratios have the simple shift properties $u_{s(i+1)} = u_{si}
\vert_{X_i\rightarrow X_{i+1}}$ as we shift the label $i$ of the external scalar
fields by one unit.

While we shall mostly work with the snowflake cross-ratios, there is one more set
of cross-ratios that will play some role below, namely the so-called OPE cross-ratios
$z_1,z_2,\bar z_1,\bar z_2$ and $\xx:=1-w$ that were introduced in \cite{Buric:2021kgy}.
These are the five-point analog of the cross-ratios $z$ and $\bar z$ that Dolan and Osborn introduced to study the OPE limit. In the case of five-point functions, the
OPE limit corresponds to the regime
\begin{equation}
\mathrm{OPE}^{(12)3(45)}: \bar z_1,\bar z_2 \ll z_1,z_2 \ll 1,
\label{ope_lim}
\end{equation}
For the purposes of this paper, it suffices to record the map from the snowflake cross 
ratios to the OPE cross-ratios near the region where $\bar z_{1,2}$ are small, 
\begin{eqnarray}
u_{s1} & = & \frac{\bar z_1 z_1}{1-z_2}(1+\orm(\bar z_2)), \qquad 
u_{s2} = 1-z_1+\orm(\bar z_1) , \\[2mm]
u_{s4} & = & \frac{\bar z_2 z_2}{1-z_1} (1+\orm(\bar z_1)), \qquad 
u_{s3}=1-z_2+\orm(\bar z_2), \\[2mm]
 u_{s5} & = & 1- \frac{z_1 z_2(1-w)}{(1-z_1)(1-z_2)}+\orm(\bar z_1,\bar z_2).
\label{snowflake_to_ope_lc}
\end{eqnarray}
The behavior of conformal blocks in this OPE limit provides boundary conditions that
are used to select a particular solution of the differential equations.
\medskip

\begin{figure}[ht]
\centering
\includegraphics[scale=0.1]{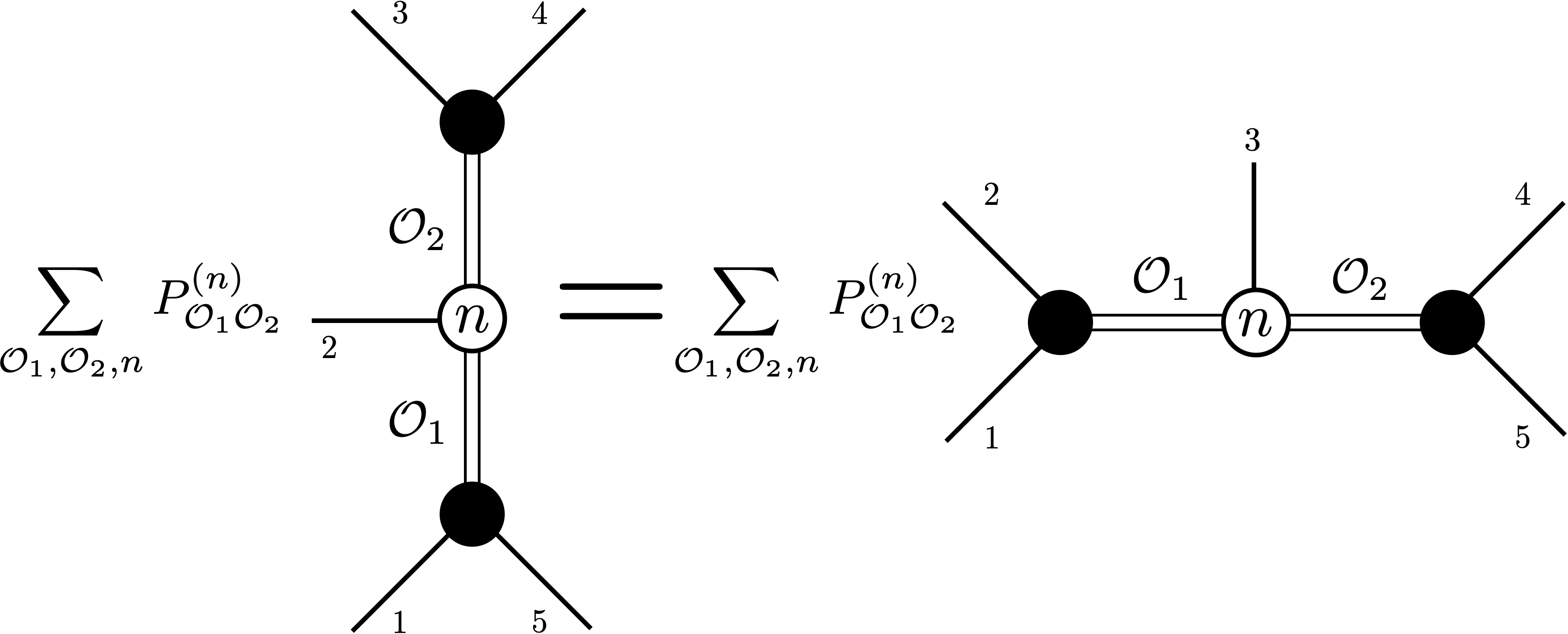}
\caption{Graphical representation of the planar five-point crossing symmetry equation studied
in this paper. The crossed channel is obtained from the direct channel by cyclic permutation
of points, $i \rightarrow i+1$ mod $5$. }
\label{fig:5pt_cse_general}
\end{figure}
For five points there exist a total number of 15 different channels which are all of a unique
(comb-channel) topology. Here we shall look at one particular pair of channels that is obtained
by shifting the label of the external fields $i \mapsto 1+i\,  \text{mod}\,  5$ by one unit, see
fig.~\ref{fig:5pt_cse_general}. Let us stress that in the case of five points, this duality can
be used to generate all others by iterated application since all OPE channels have the same
topology, as in the case of four points. We will fix the cyclic enumeration of external points
to be such that
\begin{equation}
\DC = (51) 2 (34) \quad \textit{and} \quad \CC = (12) 3 (45)
\end{equation}
as shown in fig.~\ref{fig:5pt_cse_general}. To fix our conventions for the conformal blocks
let us display the expansion in the crossed channel
\begin{equation} \label{eq:5ptcorrblock}
\langle\phi(X_1)\dots \phi(X_5)\rangle = \left( X_{12} X_{45} 
\sqrt{\frac{X_{23}X_{34}}{X_{24}}}
\right)^{-\Dg_\phi} \!\sum_{\oo_1,\oo_2,n} P_{\oo_1\oo_2}^{(n)} \psi^{\CC}_{\oo_1\oo_2;n}
 (u_{si}),
\end{equation}
where the coefficients $P$ are determined by the coefficients $C$ of the involved 
operator products as 
\begin{equation} \label{eq:PCCC}
P_{\oo_1 \oo_2}^{(n)}:= C_{\phi\phi\oo_1} C_{\oo_1 \phi\oo_2}^{(n)} C_{\phi\phi\oo_2},
\end{equation}
Here $\oo_a, a=1,2,$ denote two STT primary fields of weight $\Delta_a = h_a + \bar h_a$
and spin $J_a = \bar h_a -  h_a$. Given the two spin labels $J_1$ and $J_2$, the parameter
$n=0,\dots,\mathrm{min}(J_1,J_2)$ labels the basis of tensor structures of the three-point
function in the expansion
\begin{equation}
\langle \oo_1(X_1,Z_1) \phi(X_3) \oo_2(X_2,Z_2)\rangle = \sum_{n=0}^{\mathrm{min}(J_1,J_2)}
C_{\oo_1 \phi\oo_2}^{(n)} \prod_{i<j} X_{ij}^{n_{ij}} J_{1,32}^{J_1} J_{2,31}^{J_2}
\mathcal{X}^n, \quad \mathcal{X} := \frac{H_{12} X_{13}X_{23}}{J_{1,32}J_{2,31}},
\label{3pt_stt_scalar_stt}
\end{equation}
where $\mathcal{X}$ is a three-point cross-ratio that corresponds precisely to the OPE limit reduction of the five-point cross-ratio $(1-w)$ when projecting to exchanged STT primaries. The objects $H_{ij}$ and $J_{i,jk}\equiv V_{i,jk} X_{jk}$ are tensor structures used to parametrize three-point functions, see \cite{Costa:2011mg}. 
This choice of basis in the space of three-point tensor structures implies that the 
five-point blocks in eq.~\eqref{eq:5ptcorrblock} with expansion coefficients 
\eqref{eq:PCCC} satisfy the normalization condition 
\begin{equation}
\lim_{z_1,z_2\rightarrow 0}\lim_{\bar z_1,\bar z_2\rightarrow 0}
\frac{\psi^{(12)3(45)}_{\oo_1\oo_2;n}(z_a,\bar z_a,w)}
{\prod_{a=1}^2 \bar z_a^{h_a} z_a^{\bar h_a} \, (1-w)^n} =1, 
\label{ope_bc_exact}
\end{equation}
see \cite{Buric:2021kgy} for notations and the derivation.  With all our conventions fixed, the crossing equation we want to analyze takes the form
\begin{equation} \label{eq:5ptcrossing}
\sum_{\oo_1,\oo_2,n} P_{\oo_1\oo_2}^{(n)} \psi_{\oo_1\oo_2;n}^{\DC} =
 \left(\frac{u_{s5}\sqrt{u_{s3}}}{u_{s4} \sqrt{u_{s1}}}\right)^{\Dg_\phi} \sum_{\oo_1,\oo_2,n}
 P_{\oo_1\oo_2}^{(n)} \psi^{\CC}_{\oo_1\oo_2;n}.
\end{equation}
Up to relabeling, this is the same planar crossing equation as in \cite[Sec.~(3.1)]{Antunes:2021kmm}.
To evaluate the constraints that arise from this crossing symmetry equation we shall consider
a regime in which
\begin{equation} \label{eq:5ptLCL}
X_{15} \ll X_{34} \ll X_{12} \ll X_{45}\ll X_{23} \ll 1
\end{equation}
i.e.  we make pairs of neighboring points light-like separated in the order that is given by 
reading the previous line from left to right. While the first two limits favor internal leading-twist exchanges in the direct channel, the following two have the same effect on crossed-channel
exchanges. The last limit in which we place $x_3$ on the lightcone of $x_2$ is not that 
fundamental and in fact it will be important for some of our results to include corrections 
to limiting behavior as we send $X_{23}$ to zero. The limit we consider here amounts to introducing the following order, see previous section, 
$$\vec{\epsilon} = (\epsilon_{15},\epsilon_{34},\ep_{12},\ep_{45},\ep_{23})$$
With respect to this prescription, the cross-ratios $u_{si}$ are of order $\orm(\ep_{i,i+1})$.
At the same time, the transition between direct and crossed channel is also easily expressed
in terms of these cross-ratios $u_{si}$.
\begin{equation}
\psi^{(12)3(45)}(u_{si}) = \psi^{(51)2(34)}(u_{s,i-1})
\label{cyclicity}
\end{equation}
We conclude that for the crossing equation we are about to study,  the cross-ratios $u_{si}$
are the perfect generalization of the cross-ratios $u,v$ we used in the discussion of four-point
functions. Once again, the duality between the two channels corresponds to a cyclic permutation
of the cross-ratios and each of the cross-ratios has unit order with respect to exactly one of
the order parameters $\ep_{ij}$ we are taking to zero. On the other hand, the cross-ratios
$u_{si}$ are simple rational functions of our polynomial cross-ratios \eqref{eq:five_point_polynomial}. Hence, when expressed in
terms of the $u_{si}$, all terms in the Casimir differential operators possess some definite
$\vec{\ep}$-order.

\subsection{Lightcone blocks in the direct channel}
\label{ssec:5pt_blocks_dc}

As in the four-point case that began this discussion,  we shall perform the lightcone 
limit of the five-point blocks for the direct channel in two stages, starting with
only that part of the limit that exposes leading-twist contributions in the direct
channel
\begin{equation} \label{eq:5ptLCLds1}
X_{15} \ll X_{34} \ll 1\ .
\end{equation}
As we shall discuss in the first subsection, going to the partial lightcone 
regime is sufficient for the blocks to possess a relatively simple integral
representation that is amenable to further evaluation. Since the OPE-limit point at which we normalize our blocks lies within the regime \eqref{eq:5ptLCLds1},
see discussion in the previous subsection, the limiting blocks are also easy to
normalize. For our bootstrap analysis we will only need the lightcone blocks in
the (almost) full lightcone limit \eqref{eq:5ptLCL}. These are derived in the second and third subsections by sending the remaining three cross-ratios to zero one by one.

\subsubsection{Partial lightcone limit}

In terms of the cross-ratios $u_{si}$, the partial lightcone limit
\eqref{eq:5ptLCLds1} amounts to considering the order $\vec{\ep}^{\,(2)}
= (\ep_{15},\ep_{34})$. In terms of cross-ratios, the associated regime 
is given by 
$$ \textit{LCL}_{\vec{\ep}}^{(2)}: \quad  u_{s5} \ll u_{s3} \ll 1 \ . $$
Here, we use the superscript $(2)$ to signal that we perform only the first two 
limits. The following analysis and in particular the normalization of the limiting 
blocks turn out to be simplest when written in terms of OPE cross-ratios. In general, 
the relation between the snowflake and OPE cross-ratios is somewhat complicated, but in 
the regime $\textit{LCL}^{(2)}$, i.e.\ after sending $u_{s5}$ and $u_{s3}$ to zero, 
the relation simplifies. We had seen this for the crossed channel already in the 
eqs.~\eqref{snowflake_to_ope_lc}. In order to treat the direct channel we recall 
that after the shift of the index $i$ that brings us from the crossed to the direct 
channel, the snowflake cross-ratios are related to the OPE cross-ratios of the direct channel via the relation
\begin{equation}
(u_{s5},u_{s1},u_{s2},u_{s3},u_{s4}) = \left(\frac{z_1}{1-z_2} \bar z_1,1-z_1,1-z_2,\frac{z_2}{1-z_1}\bar z_2,1-\frac{(1-w) z_1z_2}{(1-z_1)(1-z_2)}\right)+\orm(\bar z_a).
\label{usi_to_zbzw}
\end{equation}
The inverse map is 
\begin{equation}
(\bar z_1,\bar z_2,z_1,z_2,w)= \left(\frac{u_{s5}u_{s2}}{(1-u_{s1})},\frac{u_{s3}u_{s1}}{(1-u_{s2})}, 1-u_{s1},1-u_{s2},1-\frac{u_{s1}u_{s2}(1-u_{s4})}{(1-u_{s1})(1-u_{s2})}\right)+\orm(u_{s5,3}).
\label{zbzw_to_usi}
\end{equation}
In comparison to the OPE cross-ratios we have discussed in the previous subsection, 
the index of the snowflake cross-ratios is now shifted by one unit, i.e. the OPE 
cross-ratios in this subsection are the ones relevant for the OPE limit in 
the direct channel.

After these brief comments on coordinates, we are now ready to display and study 
the Casimir differential equations in the lightcone regime $\textit{LCL}^{(2)}$. 
If we use the OPE cross-ratios, the Casimir operators take the form
\begin{align}
&\mathcal{D}_{15}^{2,\vec{\ep}^{(1)}} = \ep_{15}^0 \ep_{34}^0
\left(  z_1\, \tensor[_2]{\mathcal{D}}{_1}\left(\hat{A}_1,0;0;z_1,\ds_{z_1}\right)
+\bar z_1\ds_{\bar z_1}(\bar z_1\ds_{\bar z_1}-d+1) \right) + \orm(\ep_{15}), \\[2mm]
&\hat{A}_1= \vt_{z_2}-h_\phi+w(\vt_{\bar z_2}-\vt_{z_2}+\vt_w)\\[2mm] 
&\mathcal{D}_{34}^{2,\vec{\ep}^{(1)}}= \ep_{15}^0 \ep_{34}^0
\left(z_2\, \tensor[_2]{\mathcal{D}}{_1}\left(\hat{A}_2,0;0;z_2,\ds_{z_2}\right)+
\bar z_2\ds_{\bar z_2}(\bar z_2\ds_{\bar z_2}-d+1) \right) + \orm(\ep_{34}), \\[2mm]
&\hat{A}_2= \vt_{z_1}-h_\phi+w(\vt_{\bar z_1}-\vt_{z_1}+\vt_w) ,
\end{align}
where the operator ${}_2\mathcal{D}_1$ was defined in eq.\ \eqref{eq:2D1def}.
We will now study the associated Casimir eigenvalue equations and determine their 
solutions in order to show that 
\begin{equation} \label{eq:DCLCL2}
\psi^{\DC}_{(h_a,\bar h_a;n)}(u_{si}(\bar z_a,z_a,w)) \stackrel{\LCL^{(2)}}{\sim} 
\prod_{a=1}^2 \bar z_a^{h_a}z_a^{\bar h_a} (1-w)^n 
\tilde{F}_{(h_a,\bar h_a;n)}(z_1,z_2,w), 
\end{equation} 
where $\tilde{F}$ is given by the following double integral over two integration 
variables $t_1$ and $t_2$
\begin{align}
\tilde{F}_{(h_a,\bar h_a;n)}(z_1,z_2,w) = \prod_{a \neq b} \int_0^1 
\frac{\dd t_a}{\mathrm{B}_{\bar h_a}} \frac{(t_a(1-t_a))^{\bar h_a-1} 
(1-w z_a t_a)^{J_b-n}}{ (1-z_1 t_1-z_2t_2+wz_1z_2t_1t_2)^{\bar h_{12;\phi}}}.
\label{int_5pt_from_cas}
\end{align}
Recall that the cross-ratios $z_i,w$ can be considered as functions of the 
snowflake cross-ratios $u_{s1},u_{s2}$ and $u_{s4}$ that are not sent to zero 
in the lightcone regime we consider. Let us note that indeed the right hand 
side of eq.~\eqref{eq:DCLCL2} is correctly normalized since one infers from 
eq.~\eqref{int_5pt_from_cas} that 
\begin{equation}
\lim_{z_1,z_2\rightarrow 0}  \tilde{F}_{(h_a,\bar h_a;n)}(z_1,z_2,w) = 1
\label{OPE_f_5pt}
\end{equation}
and the factor in front of $\tilde{F}$ gives the usual behavior of the block in the 
OPE limit. The remainder of this subsection is devoted to the derivation of 
eqs.~\eqref{eq:DCLCL2}, \eqref{int_5pt_from_cas}. 
\smallskip

Intuitively, it is rather clear that the two second-order Casimir equations 
suffice to reconstruct the dependence of the blocks on $z_1,z_2$ from the 
behavior in the OPE limit where the $z_a$ are sent to zero. In order to give a
formal proof, we first note that our second-order Casimir equations for the 
lightcone blocks imply 
\begin{align} \label{eq:DCcas15}
\left(\mathcal{D}_{15}^{2\, \vec{\ep}} -\bar h_1(\bar h_1-1)-h_1(h_1-d+1)\right)
\psi_{(h_a,\bar h_a;n)}(u_{si}) 
\propto  \, \mathcal{D}_{1} \cdot \tilde{F}_{(h_a,\bar h_a;n)}(z_1,z_2,w) = 0  \\[2mm]
\left(\mathcal{D}_{45}^{2\, \vec{\ep}} -\bar h_2(\bar h_2-1)-h_2(h_2-d+1)\right)
\psi_{(h_a,\bar h_a;n)}(u_{si}) \propto 
\mathcal{D}_{2}  \cdot \tilde{F}_{(h_a,\bar h_a;n)}(z_1,z_2,w) = 0 ,
\label{eq:DCcas34}
\end{align}
where $\propto$ means that we dropped some non-vanishing overall prefactor. The operators $\mathcal{D}_a, a=1,2$ that act on the function
$\tilde{F}$ are given by 
\begin{align}
\mathcal{D}_{a}= \vt_{z_a}(2\bar h_a+
\vt_{z_a}-1)&-z_a(\bar h_a+\vt_{z_a})(\bar h_{12;\phi}+\vt_{z_1}+
\vt_{z_2}-\vt_w) \nonumber \\[1mm]
& \quad \quad - w z_a(\bar h_a+\vt_{z_a})(n-J_b-\vt_{z_b}+\vt_w),
\label{eq:Da_for_Ftilde}
\end{align}
where $b \in \{1,2\}$ is chosen such that $b \neq a$. 
We will now determine a unique solution to this system of differential equations 
subject to the OPE limit boundary condition \eqref{OPE_f_5pt}.
 
Our strategy will be to solve these differential equations for some special 
2-dimensional submanifolds first and then to reconstruct the whole function
using the various solutions as boundary condition. Let us first note that 
the two differential equations can be solved very explicitly if we set 
either $w=0$ or $w=1$. For $w=1$, the solution can easily be obtained in 
terms of Gauss' hypergeometric functions as 
\begin{align} \label{eq:solw=1}
\tilde{F}_{(h_a,\bar h_a;n)}(z_1,z_2,w=1)= \prod_{a\ne b} 
\hypg{\bar h_a}{\bar h_a+h_{b\phi}+n}{2\bar h_a}(z_a). 
\end{align}
When we set $w=0$, the set of differential equations may be seen to coincide 
with the equations that characterize Appell's hypergeometric function $F_2$. 
After matching all the parameters we find 
\begin{align} \label{eq:solw=0}
\tilde{F}_{(h_a,\bar h_a;n)}(z_1,z_2,w=0) =  
F_2\left(\bar h_{12;\phi}; \bar h_1,\bar h_2;2\bar h_1,2\bar h_2;z_1,z_2\right).
\end{align} 
Similarly, we can solve our system of differential equations if we set one of 
the variables $z_b=0$. Once again, we can construct the solution in terms of 
Appell's hypergeometric functions, only this time it is in terms of $F_1$,  
\begin{equation} \label{eq:solz=0}
\tilde{F}_{(h_a,\bar h_a;n)}(z_a,w) \vert_{z_b=0} = 
F_1\left(\bar h_c;\bar h_{12;\phi},n-J_b;2 \bar h_c;z_c,w z_c\right).
\end{equation}
Here $c \in \{1,2\}$ with $c \neq b$. In order to be completely explicit about 
our conventions concerning the Appell functions $F_1$ and $F_2$ we used in the 
last two formulas, let us state their series expansions, 
\begin{align}
F_1(b;a_1,a_2;c;z_1,z_2):= \sum_{m_1,m_2=0}^\infty
\frac{z_1^{m_1}}{m_1!}\frac{z_2^{m_2}}{m_2!} (a_1)_{m_1}
(a_2)_{m_2}\frac{(b)_{m_1+m_2}}{(c)_{m_1+m_2}} \label{eq:ps_F1} \\[2mm] 
F_2(a;b_1,b_2;c_1,c_2;z_1,z_2):= \sum_{m_1,m_2=0}^\infty
\frac{z_1^{m_1}}{m_1!}\frac{z_2^{m_2}}{m_2!}
(a)_{m_1+m_2}\frac{(b_1)_{m_1}}{(c_1)_{m_1}}\frac{(b_1)_{m_1}}{(c_1)_{m_1}}.
\end{align}
Note that all of these solutions may be obtained by rewriting the action of the
Casimir operators on a power series in $z$ and/or $w z$ as a recursion relation 
with initial condition governed by the OPE limit. 

To find a general expansion of $\tilde{F}_{(h_a,\bar h_a;n)}$ outside of the $z_b=0$ 
limit,  we can make use of the relations 
\begin{equation}
\mathcal{D}_c \vt_{z_b}=\vt_{z_b} \mathcal{D}_c, \quad 
\mathcal{D}_c (w z_b)^k = (w z_b)^k \mathcal{D}_c,  
\quad \mathcal{D}_c z_b^k = z_b^k \mathcal{D}_c \vert_{J_b\rightarrow J_b+k},
\end{equation}
where we assumed $c \neq b$ and in the last relation we take $\bar h_b=h_b+J_b$ while keeping $h_b$ fixed. In the vicinity of $z_b=0$ we can conclude that 
the general solution admits an expansion of the form 
\begin{equation}
\tilde{F}_{(h_a,\bar h_a;n)}(z_1,z_2,w) = \sum_{m=0}^\infty f_{b,m}(w z_b) z_b^{m} 
F_1 \left(\bar h_c;\bar h_{12;\phi}+m,n-J_b-m;2\bar h_c;z_c, w z_c \right)
\label{expansion_wzb}
\end{equation}
with some coefficients $f_{b,m}$ that need to be determined. Note that we 
have introduced two different expansions here which depend on which of the 
two variables $z_b, b=1,2$ we set to zero in the leading term. Each of the 
two distinct expansions for $b=1,2$ are explicitly in the kernel of 
$\mathcal{D}_c$. At this stage, the functions $f_{b,m}(w z_b)$ can be solved
explicitly by recasting $\mathcal{D}_b f=0$ as a recursion relation in $m$.  
However, we can find the general solution more efficiently by making use 
of the integral representation of the Appell $F_1$:
\begin{equation}
F_1 \left(\bar h_c;\bar h_{12;\phi}+m,n-J_b-m;2\bar h_c;z_c, w z_c \right) 
=\int_0^1 \frac{\dd t_c}{\mathrm{B}_{\bar h_c}} \frac{(t_c(1-t_c))^{\bar h_c-1}} 
{(1-z_at_c)^{\bar h_{12;\phi}+m}} \, (1-w z_ct_c)^{m+J_b-n},
\end{equation}
which is equivalent to the convergent power series \eqref{eq:ps_F1} for $0 \leq z_c,wz_c <1$. Assuming we can commute the sum over $m$ and the integral over $t_c$, we can 
rewrite the expansion \eqref{expansion_wzb} as
\begin{equation}
\tilde{F}_{(h_a,\bar h_a;n)}(z_1,z_2,w) = \int_0^1 \frac{\dd t_c}{\mathrm{B}_{\bar h_c}}
\frac{(t_c(1-t_c))^{\bar h_c-1}}{(1-z_ct_c)^{\bar h_{12;\phi}}}\,  
(1-w z_ct_c)^{J_b-n} f_b\left(\frac{1-w z_ct_c}{1-z_ct_c} z_b, wz_b \right), 
\end{equation}
where $f_b(x,y)=\sum_m f_{b,m}(y)x^m$ is a power series in each variable.  
At $z_c=0$,  where we know that $\tilde{F}_{(h_a,\bar h_;n)}$ is an Appell $F_1$, the
above integral simplifies to $f_b(z_b,w z_b)$, such that 
$$ f_b(x,y) = F_1(\bar{h}_b;\bar h_{12;\phi}, n-J_c;2\bar h_b;x,y)\ . $$  
Plugging this in the integral representation for the Appell function $F_1$ 
we obtain the formula  \eqref{int_5pt_from_cas}. We have thus established
that in the lightcone regime $\LCL^{(2)}$, the direct channel blocks are given 
by eqs.~\eqref{eq:DCLCL2} and \eqref{int_5pt_from_cas}. Note that all we used
were the limiting expressions \eqref{eq:DCcas15} and \eqref{eq:DCcas34} for the
Casimir operators. 

\paragraph{Generalized Euler transformation.} In our derivation, the double integral of eq.~\eqref{int_5pt_from_cas} appeared merely as a tool to make the power series in $z_1,z_2,w$ that solves the Casimir equations more compact.  To write the latter, one can expand the integrand explicitly into a power series in $z_a,w z_a$ in the domain $0\leq z_a,w z_a<1$ before integrating each summand to a product of Euler Beta functions.  However,  in our applications to the lightcone bootstrap, we must analyze the asymptotics of blocks near certain edges $z_a,w\rightarrow 1$ of this domain,  where the formal hypergeometric series need not converge. Here, the integral provides a useful tool to analyze the convergence properties near the singular regions, as is well-known from the classical theory of hypergeometric functions.   In particular, two singular regions relevant to lightcone bootstrap take the form of a double scaling limit in OPE cross-ratios:
\begin{equation}
z_a=1+\orm(\ep_a), \quad w=1+\orm(\ep_a), \quad \ep_a \rightarrow 0,
\label{eq:double_scaling_lc}
\end{equation}
for $a=1$ or $2$.  When restricted to two-dimensional submanifolds $z_b=0$ or $w=1$, the asymptotics of the blocks ($\,\tensor[_2]{F}{_1}$ or Appell $F_1$) in the above limit depend on the parameters $h_b,h_\phi,n$.  In particular, on the $w=1$ submanifold,  the parameter dependence of $z_a\rightarrow 1$ asymptotics is captured by the Euler transformation of the Gauss hypergeometric function,
\begin{equation}
\hypg{\bar h_a+h_{b\phi}+n}{\bar h_a}{2\bar h_a}(z_a) = (1-z_a)^{-(h_{b\phi}+n)} \hypg{\bar h_a-h_{b\phi}-n}{\bar h_a}{2\bar h_a}(z_a).
\label{euler_2F1_w=1}
\end{equation} 
The series on the left-hand side diverges while the series on the right-hand side converges if $h_{b\phi+n}>0$, and vice versa when $h_{b\phi}+n<0$\footnote{In the case where $h_{b\phi}+n=0$ we saw in section~\ref{sect:four_pt_boot} that the hypergeometric function admits a logarithmic divergence at $z_a=1$.}. In the integral representation \eqref{int_5pt_from_cas} evaluated at $w=1$, this relation is obtained from the change of variables 
\begin{equation*}
\tilde{t}_a = \frac{1-t_a}{1-z_a t_a}. 
\end{equation*}
We would now like to find an appropriate generalization of the Euler transformation \eqref{euler_2F1_w=1} away from the $w=1$ submanifold that controls the parameter-dependence of the lightcone limits $(z_a,w)=(1,1)+\orm(\ep_a)$,  $a=1,2$. To this end, we first introduce a new set of variables $(v_a(z_a),x(z_1,z_2,w))$ defined as
\begin{equation}
v_a(z_a):=1-z_a,
\quad x(z_1,z_2,w):= \frac{z_1 z_2 (1-w) }{(1-z_1)(1-z_2)}. 
\label{vx_to_zw}
\end{equation}
In $v,x$ cross-ratios, the double scaling limit 
of OPE cross-ratios defined in eq.~\eqref{eq:double_scaling_lc} is equivalent to the limit $v_a\rightarrow 0$ at $v_b,x$ fixed. The change of variables is simple to invert,
\begin{equation}
z_a(v_a)=1-v_a, \quad w(v_1,v_2,x) = 1-\frac{v_1v_2 x}{(1-v_1)(1-v_2)}.
\label{zw_to_vx}
\end{equation}
Now, if we apply the same change of variables $\tilde{t}_a=(1-z_at_a)^{-1}(1-t_a)$ to the double integral in eq.~\eqref{int_5pt_from_cas} and express $z_a,w$ in terms of $v_a,x$ using eq.~\eqref{zw_to_vx}, then we obtain the desired ``generalized Euler transformation" of the form
\begin{equation}
\tilde{F}_{(h_a,\bar h_a;n)}(1-v_1,1-v_2,w(v_1,v_2,x)) = \prod_{a=1}^2 v_a^{-(h_{b\phi}+n)} F_{(h_a,\bar h_a;n)}(v_1,v_2,x),
\label{block_to_int_ope}
\end{equation}
where the function $F$ on the right-hand side is given by the integral 
\begin{equation} \label{eq:Fint} 
F_{(h_a,\bar h_a;n)}(v_1,v_2,x) = \prod_{b\neq c} \int_0^1 
\frac{\dd t_b}{\mathrm{B}_{\bar h_b}} \frac{(t_b(1-t_b))^{\bar h_b-1} 
(1-(1-v_b)t_b)^{h_{c\phi}+n-\bar h_b}}{ (1+\frac{v_{b} x}{1-v_{b}} 
(1-t_c))^{n-J_b}(1-x(1-t_1)(1-t_2))^{\bar h_{12;\phi}}}. 
\end{equation}
For $h_{b\phi}+n>0$ and $0\leq x<1$, this expression evaluates at $v_a=0$ to a convergent integral that can be expanded into a power series in $x$. 
\medskip 

We conclude this subsection with a couple of important comments. One nice feature of the integral formulas \eqref{int_5pt_from_cas}
and \eqref{eq:Fint} we have derived is that the two integrations 
decouple in the limit in which we send $w$ to one (see eq.~\eqref{eq:solw=1} for the function $\tilde{F}$) or, equivalently, $x$ to zero. For the function $F$, one has 
\begin{equation}
\lim_{x\rightarrow 0}F_{(h_a,\bar h_a;n)}(v_1,v_2,x)
= f_1^{\mathrm{dec}}(v_1) f_2^{\mathrm{dec}}(v_2)  
\end{equation}
where each factor $f_a^{\mathrm{dec}}$ depends only on a single cross-ratio and is given by 
\begin{equation} \label{eq:deffdec}
f_a^{\mathrm{dec}}(v) = \hypg{\bar h_a}{\bar h_a - h_{b\phi} - n}
{2 \bar h_a} (1-v) \quad a \neq b = 1,2. 
\end{equation} 
Finally, recall that we derived these integrals with the help 
of the Casimir equations. It is actually possible to obtain the 
same integral formulas through the lightcone OPE, in close analogy 
to what we explained in our discussion of the four-point
lightcone blocks. The details for five external points can be 
found in appendix~\ref{app:5pt_integral_formula}.  In order to obtain 
the first integral representation \eqref{int_5pt_from_cas} one 
needs to choose the gauge,  
\begin{equation}
(s_1,s_2,s_4,s_5)=(r_1t_1,r_1(1-t_1),r_2(1-t_2),r_2t_2), 
\quad r_1,r_2 \in \Rs^\times,  
\end{equation}
see  appendix~\ref{app:5pt_integral_formula} for notations. The derivation of 
the second integral formula \eqref{eq:Fint} from the lightcone OPE is 
carried out in some detail in the same appendix. 

\subsubsection{The full lightcone limit}

From the integral formulas for blocks in the regime $\LCL^{(2)}$ that we obtained in the previous subsection, we now want to move towards the full lightcone limit by 
sending the remaining cross-ratios to zero, one by one. 

\paragraph{Sending $u_{s1}$ to zero.} Let us start with the cross 
ratio $u_{s1}$ so that we pass from $\LCL^{(2)}$ to the new regime 
\begin{equation}
\LCL_{\vec{\ep}}^{(3)}: u_{s5} \ll u_{s3} \ll u_{s1} \ll  1.
\end{equation} 
Our plan is to derive the asymptotics of the blocks by evaluating the integral 
expressions in eqs.\ \eqref{int_5pt_from_cas} or \eqref{eq:Fint} at $u_{s1}=0$.  
In the case of the expression \eqref{eq:DCLCL2}, the blocks formally  
evaluate to
\begin{align} 
\frac{\psi^{\DC}_{(h_a,\bar h_a;n)}(u_{si})}{u_{s5}^{h_1} u_{s2}^{h_1+n} 
(1-u_{s2})^{J_2-n}} &\stackrel{{\LCL}^{(3)}}{\sim}  u_{s1}^{h_2+n} u_{s3}^{h_2}(1-u_{s4})^n  
\tilde{F}_{(h_a,\bar h_a;n)}(1,1-u_{s2},1)\nonumber \\[2mm]
& \hspace*{-3cm} = u_{s1}^{h_2+n} u_{s3}^{h_2}(1-u_{s4})^n  
\hypg{\bar h_1}{\bar h_1+h_{2\phi}+n}{2\bar h_1}(1) \ 
\hypg{\bar h_2}{\bar h_2+h_{1\phi}+n}{2\bar h_1}(1-u_{s2}). 
\label{eq:limit1}
\end{align}
Note, however, that the first Gauss hypergeometric series with argument $z_1 = 
1-u_{s1} = 1$ does not converge when $h_\phi \leq h_2 + n$. On the other hand, it evaluates to the following finite combination of $\Gamma$ functions in the complementary case:
\begin{equation}
\hypg{\bar h_1}{\bar h_1+h_{2\phi}+n}{2\bar h_1}(1)  = \frac{\Gamma(2\bar h_1)
\Gamma(h_\phi-h_2-n)}{\Gamma(\bar h_1-h_{2\phi}-n)\Gamma(\bar h_1)} \quad 
\mathit{ for } \quad h_\phi>h_2+n \ .  
\end{equation}
Let us now look at the second integral formula \eqref{block_to_int_ope}. Evaluating 
the limit $u_{s1}=0$ yields
\begin{align}
\frac{\psi^\DC_{(h_a,\bar h_a;n)}(u_{si})}{u_{s5}^{h_1} u_{s2}^{h_\phi} (1-u_{s2})^{J_2-n}} 
&\stackrel{\LCL^{(3)}}{\sim} u_{s1}^{h_\phi} u_{s3}^{h_2} (1-u_{s4})^n 
F_{(h_a,\bar h_a;n)}(0,u_{s2},1-u_{s4}).
\label{eq:limit2}
\end{align}
In this case, one can show that $F_{(h_a,\bar h_a;n)}(0,u_{s2},x)$ evaluates to a 
convergent power series in $1-u_{s2}$ and $x = 1 - u_{s4}$ \emph{only} if $h_2+n>h_\phi$, see eq.~\eqref{eq:Fint_one_v_to_0} in appendix~\ref{app:direct_channel_5pt}. This is precisely the regime in which the functions $\tilde{F}$ from the first integral expression were singular. We can understand this parameter dependence of the $u_{s1}\rightarrow 0$ asymptotics in terms of the 
Casimir differential operators. More specifically, for the second-order Casimir $\mathcal{D}^2_{15}$ we find
\begin{align}
&\mathcal{D}^{2,\vec{\ep}^{(3)}}_{15} = \ep_{15}^0 \left\{ \ep_{12}^{-1} \mathcal{D}_{15}^{(0,0,-1)} +\orm(\ep_{12}^0)\right\}+\orm(\ep_{15}), 
\\[2mm]
&\mathcal{D}_{15}^{(0,0,-1)} = \left(\ds_{u_{s1}}-\frac{\vt_{u_{s3}}+\vt_{1-u_{s4}}}{u_{s1}} \right)(\vt_{u_{s1}}-h_\phi).\label{eq:D1500m1}
\end{align}
Note that the operator contains a single singular term. As long as we are interested in direct channel contributions that arise from fields with finite weight and spin, the eigenvalues of the quadratic Casimir operators do not scale and hence, the leading term of the relevant lightcone blocks must lie in the kernel of the singular term.  At the same time, any second-order differential equation admits two independent solutions to  
the kernel condition. But our equation is second order only in the derivative 
$\ds_{u_{s1}}$ with respect to the variable that we are sending to zero. Hence, 
depending on the parameter $h_\phi$ in the right factor the operator
\eqref{eq:D1500m1}, there is a unique solution that dominates the behavior 
at small $u_{s1}$. For $h_\phi>h_2+n$ we see that the function in the second 
line of eq.~\eqref{eq:limit1} is indeed annihilated by the linear operator 
in the left factor of the Casimir operator \eqref{eq:D1500m1} (after commuting it to the right) and hence it 
lies in the kernel. On the other hand, the function on the right-hand side
of equation \eqref{eq:limit2} lies in the kernel of $\vt_{u_{s1}}-h_\phi$
and hence also in the kernel of the Casimir operator. 

It remains to discuss the case $h_{\phi}= h_2+n$. When this equality holds, $\tilde{F}_{(h_a,\bar h_a;n)}(1-u_{s1},1-u_{s2},1-u_{s4})=F_{(h_a,\bar h_a ;n)}(u_{s1},u_{s2},1-u_{s4})$ and we expect both expressions to diverge as $u_{s1}$ is sent 
to zero.  More specifically, in the case where $\vt_{u_{s3}} \psi 
= h_2 \psi$ and $\vt_{1-u_{s4}} \psi = n\psi$,  the Casimir equation admits 
a logarithmic solution 
\begin{equation}
 \left(\ds_{u_{s1}}-\frac{\vt_{u_{s3}}+\vt_{1-u_{s4}}}{u_{s1}} \right)(\vt_{u_{s1}}-h_2-n) u_{s3}^{h_2}(1-u_{s4})^n u_{s1}^{h_2+n} \log u_{s1}=0.
\end{equation} 
In accordance with this solution, a power series expansion of $F_{\oo_1\oo_2;n}(u_{s1},u_{s2},x)$ around $x=0$ shows that only the $x^0$ term exhibits a logarithmic divergence, while all higher order terms $x^m$ are $\orm(1)$ at most, i.e.
\begin{align}
F_{\oo_1\oo_2;n}(u_{s1},u_{s2},0) = \frac{\log u_{s1}^{-1}}{\mathrm{B}_{\bar h_1}} f_2^{\mathrm{dec}}(u_{s2})+\orm(u_{s1}^0), \quad \ds_{x}^m F_{\oo_1\oo_2;n}(u_{s1},u_{s2},x) \vert_{x=0} = \orm(u_{s1}^0),  
\end{align}
for all integers $m > 0$. Here we have used the symbol $f_2^\mathrm{dec}$ as a 
shorthand for the 
hypergeometric function \eqref{eq:deffdec} that we introduced above. A more detailed derivation of this result can be found in appendix~\ref{app:direct_channel_5pt}. In summary, we obtain the following asymptotics in the $u_{s1} \ll 1$ 
limit of our partial lightcone blocks,
\begin{equation}
\frac{\psi^{\DC}_{(h_a,\bar h_a;n)}(u_{si})}{u_{s5}^{h_1} u_{s3}^{h_2} (1-u_{s4})^n 
u_{s2}^{h_\phi}(1-u_{s2})^{J_2-n}} \stackrel{\LCL^{(3)}}{\sim}  \begin{cases} 
\ \ \mathrm{B}_{\bar h_1}^{-1} \mathrm{B}_{\bar h_1,\abs{h_{2\phi}+n}} 
u_{s1}^{h_2+n}f_2^{\mathrm{dec}}(u_{s2}), \quad & h_2+n<h_\phi, \\[2mm]
\ \ \mathrm{B}_{\bar h_1}^{-1}u_{s1}^{h_\phi} \log u_{s1} f_2^{\mathrm{dec}}(u_{s2}), \quad 
& h_2+n=h_\phi, \\[2mm]
\ \  F_{(h_a,\bar h_a;n)}(0,u_{s2},1-u_{s4}), \quad & h_2+n>h_\phi.
\end{cases}
\label{blocks_dc_v1zero}
\end{equation}
 
\paragraph{Sending $u_{s4}$ to zero.} We can now move on the next level 
by sending the cross-ratio $u_{s4}$ to zero, i.e. we study lightcone blocks 
in the regime  
\begin{equation}
\LCL^{(4)}_{\vec{\ep}}: u_{s5} \ll u_{s3} \ll u_{s1} \ll u_{s4} \ll 1\ . 
\label{eq:LCL4}
\end{equation}    
Unlike the $u_{s1}\rightarrow 0$ limit, there is no sign of a divergence 
when evaluating each expression in our formula \eqref{blocks_dc_v1zero} at $u_{s4}=0$. Once again, this convergent behavior is corroborated by the 
analysis of Casimir differential operators. Indeed, in the regime 
$\LCL^{(4)}_{\vec{\ep}}$,  we find the leading order singular terms to be
\begin{align}
\mathcal{D}_{15}^{(0,0,-1,-1)} = \left(\ds_{u_{s1}}-\frac{h_\phi}{u_{s1}} \right)\ds_{u_{s4}}, \quad \mathcal{D}_{34}^{(0,0,0,-1)}=\ds_{u_{s4}}\left(\vt_{u_{s4}}-\vt_{u_{s1}} - \vt_{1-u_{s2}}-\frac{1-u_{s2}}{u_{s2}}h_\phi\right).
\end{align}
The lightcone block in our regime $\LCL^{(4)}_{\vec{\ep}}$ must lie in the 
kernel of both operators. As one can immediately notice, any function that is 
independent of $u_{s4}$ lies in the common kernel of the two leading order 
singular terms.  We conclude that the asymptotics of blocks in the lightcone limit is given by
\begin{equation}
\tcboxmath{
\frac{\psi^{\DC}_{(h_a,\bar h_a;n)}(u_{si})}{u_{s5}^{h_1} u_{s3}^{h_2} (1-u_{s4})^n u_{s2}^{h_\phi}(1-u_{s2})^{J_2-n}} \stackrel{\LCL^{(4)}_{\vec{\ep}}}{\sim}  \begin{cases} 
\ \  \mathrm{B}_{\bar h_1}^{-1} \mathrm{B}_{\bar h_1,\abs{h_{2\phi}+n}} u_{s1}^{h_2+n}f_2^{\mathrm{dec}}(u_{s2}), \quad & h_2+n<h_\phi, \\[2mm]
\ \  \mathrm{B}_{\bar h_1}^{-1}u_{s1}^{h_\phi} \log u_{s1}^{-1} f_2^{\mathrm{dec}}(u_{s2}), \quad & h_2+n=h_\phi, \\[2mm]
\ \  (1-u_{s2})^{n-J_2}g_{(h_a,\bar h_a;n)}^{\mathrm{fin}}(1-u_{s2}), \  & h_2+n>h_\phi,
\end{cases}}
\label{blocks_dc_v1zero_final}
\end{equation}
where the new functions $g^{\mathrm{fin}}(z)$ can be written as the following 
double integral 
\begin{align}
g_{(h_a,\bar h_a;n)}^{\mathrm{fin}}(z) & := 
z^{J_2-n} F_{(h_a, \bar h_a;n)}(0,1-z,1) =  
\int_0^1 \frac{\dd t_2}{t_2(1-t_2)}  
\frac{(t_2(1-t_2))^{\bar h_2}}{\mathrm{B}_{\bar h_2}} 
(1-z t_2)^{h_{1\phi}+n-\bar h_2} \cdot 
\nonumber \\[2mm]
&  \int_0^1 \frac{\dd t_1}{t_1(1-t_1)} 
\frac{t_1^{\bar h_1} (1-t_1)^{h_{2\phi}+n}}{\mathrm{B}_{\bar h_1}} (1-(1-z)t_1)^{J_2-n} \label{eq:intgfin}
 (1- (1-t_1)(1-t_2))^{-\bar h_{12;\phi}}. 
\end{align}
While the convergence of this integral is not immediately clear, note that it converges at $z=0,1$ to the following quantities:
\begin{align*}
g^{\mathrm{fin}}_{(h_a,\bar h_a;n)}(0) &= \frac{\Gamma(2\bar h_1)\Gamma(2\bar h_2) \Gamma(\bar h_2-h_\phi)\Gamma( h_\phi)}{\Gamma(\bar h_1)\Gamma(\bar h_2) \Gamma(\bar h_2+h_\phi) \Gamma(\bar h_{12;\phi})}, \\[2mm]
g^{\mathrm{fin}}_{(h_a,\bar h_a;n)}(1) &=\!\!\!\prod_{a\neq b=1}^2\! \frac{\Gamma(2\bar h_a)\Gamma(h_{b\phi}+n)}{\Gamma(\bar h_a)\Gamma(\bar h_a+h_{b\phi}+n)} \, \tensor[_3]{F}{_2}\left(\bar h_{12;\phi},h_{1\phi}+n,h_{2\phi}+n;h_{1\phi}+n+\bar h_2,h_{2\phi}+n+\bar h_1;1\right).
\end{align*}
Consequently, we can expand the integrand of eq.~\eqref{eq:intgfin} and integrate each term to obtain a convergent power series for $0\leq z\leq 1$ of the form
\begin{align}
g_{(h_a,\bar h_a;n)}^{\mathrm{fin}}(z) =& \frac{\Gamma(2\bar h_2)\Gamma(h_{1\phi}+n)}{\Gamma(\bar h_2)\Gamma(\bar h_2+h_{1\phi}+n)} \sum_{k=0}^\infty \frac{(\bar h_{12;\phi})_k}{k!} \frac{(\bar h_1)_k}{(2\bar h_1)_k} \frac{(h_{1\phi}+n)_k}{(h_{1\phi}+n+\bar h_2)_k} \nonumber \\[2mm]
& \hypg{\bar h_1-h_{2\phi}-n}{\bar h_1}{2\bar h_1+k}(z) \, \,\hypg{n-J_1}{\bar h_2}{\bar h_2+h_{1\phi}+n+k}(1-z). \label{eq:sumgfin}
\end{align}
\paragraph{Sending $u_{s2}$ to zero.}
Finally,  to reach the full lightcone regime $\LCL_{\vec{\ep}}$, we take the leading $u_{s2}\rightarrow 0$ asymptotics of eq.\ \eqref{blocks_dc_v1zero_final} 
and the expansion \eqref{eq:sumgfin} of $g^{\mathrm{fin}}$. Before we spell 
out the final result, let us introduce the following family of functions
\begin{equation} \label{eq:deff}
f_{\beta}(v) = \begin{cases} \ \ v^{\beta}, \quad \ \ \mathrm{if}
\quad \beta < 0, \\[2mm]
\ \ -\log v, \quad \mathrm{if} \quad \beta=0, \\[2mm]
\ \ 1, \quad \ \quad \mathrm{if} \quad \beta > 0. \\[2mm]
\end{cases}
\end{equation}
It is defined such that we are able to write the dependence of the 
direct channel blocks on the cross-ratios $u_{s1},u_{s2}$ in a single 
line 
\begin{equation}
\tcboxmath{\psi^{\DC\, (0)}_{(h_a,\bar h_a;n)}(u_{si}) \
\stackrel{{\LCL}_{\vec{\ep}}}{\sim} \nn^{\DC\, (0)}_{(h_a,\bar h_a;n)}\,
(u_{s1} u_{s2})^{h_\phi} u_{s5}^{h_1} u_{s3}^{h_2}\,
f_{n+h_{2\phi}}(u_{s2})f_{n+h_{1\phi}}(u_{s1}).}
\label{eq:5ptLCLdc}
\end{equation}
Note that $f_\beta$ depends on whether $\beta$ is negative, zero,
or positive and hence the single line we have displayed is indeed 
capable of summarizing the limiting behavior of the three lines in 
eq.~\eqref{blocks_dc_v1zero_final}. It remains to work out the 
the normalization $\mathcal{N}$. From the normalizations we 
listed in eq.~\eqref{blocks_dc_v1zero_final} as well as the 
evaluation of $g_{(h_a,\bar h_a;n)}^{\mathrm{fin}}(1)$ we 
obtain 
\begin{align} \label{eq:5ptLCLdcresult}
\tcboxmath{
\nn^{\DC\, (0)}_{(h_a,\bar h_a;n)}
= \begin{cases}\  \prod\limits_{a\neq b}
\frac{\Gamma(2\bar h_a)}{\Gamma(\bar h_a) \Gamma(\bar h_a+\abs{h_{b\phi}+n})}
\mathring{\Gamma}(\abs{h_{a\phi}+n}),
\quad \mathrm{iff} \quad   h_{1\phi}+n\leq 0\ \mathit{or}\ h_{2\phi} +n \leq 0 \\[6mm]
\ \prod\limits_{a\neq b} \frac{\Gamma(2\bar h_a) \Gamma(h_{b\phi}+n)}{\Gamma(\bar h_a)
\Gamma(\bar h_a+h_{b\phi}+n)}\  \tensor[_3]{F}{_2}\begin{bmatrix}
\bar h_{12;\phi}, & h_{1\phi}+n, &  h_{2\phi}+n \\
& \bar h_1+h_{2\phi}+n, & \bar h_2+h_{1\phi}+n
\end{bmatrix}(1) \ .
\end{cases}}
\end{align}
The second line applies whenever the conditions in the first line are not satisfied, i.e.\ if $h_{1\phi} + n$ and $h_{2\phi} + n$ are both positive. There, note that the ${}_3 F_2(1)$ converges for any $h_\phi>0$. In the first line, we used the `regularized' $\Gamma$ function 
$\mathring{\Gamma}$ which coincides with $\Gamma(x) = \mathring{\Gamma}(x)$ 
for $x > 0$ but is defined as $ \mathring{\Gamma}(0):= \mathrm{Res}_{x=0} 
\Gamma(x) = 1$ at $x=0$ where $\Gamma(x)$ has a pole.
As a final comment, we would like to emphasize the improved analytical control we have built through the differential equations in comparison with previous works on five-point lightcone bootstrap \cite{Bercini:2020msp,Antunes:2021kmm}. First, we extend the asymptotics of leading-twist blocks to the regime of twists $h_1,h_2\geq h_\phi$ on the second line of eq.\ \eqref{eq:5ptLCLdcresult}. Second, we are granted access to less restrictive limits such as $\LCL^{(4)}_{\vec{\ep}}$ in eq.\ \eqref{blocks_dc_v1zero_final}. The latter will allow us to bootstrap double-twist OPE coefficients with finite tensor structure labels in section~\ref{ssec:discrete_tensor_structures}.
This concludes our 
discussion of the direct channel lightcone blocks.

\subsection{Lightcone blocks in the crossed channel}
\label{sect:five_pt_crossed_blocks}

Our goal in the remainder of this section is to compute those crossed 
channel blocks that we will later need for the analysis of the crossing 
symmetry equation. In order to do so, we shall study the crossed-channel 
Casimir equations in the lightcone limit. As we have seen before, however,
these equations cannot fully determine the lightcone blocks since they are 
linear second-order equations with multiple independent solutions. So some 
work is required to select solutions that describe the 
lightcone limit of actual conformal blocks, which are uniquely 
determined by their behavior in the OPE limit. In the first subsection, 
we will take a first look at the leading terms in the Casimir equations
and sketch the strategy we use in order to find the relevant solution
that describes the limiting behavior of blocks.
This strategy is then 
implemented in detail in the two subsequent subsections. 

\subsubsection{Crossed channel Casimir equations}
\label{ssec:Crossed_Casimir_eqs}
Let us now use the Casimir operators $\mathcal{D}_{12}^2$ and $\mathcal{D}_{45}^2$
to study the behavior of crossed channel blocks in the lightcone limit. To leading
$\vec{\epsilon}$-order these operators read
\begin{align}
&\mathcal{D}_{12}^{2,\vec{\ep}} = \frac{1}{\ep_{15}} \left(\ep_{34}^0
\left( \frac{1}{\ep_{23}} \left(\ep_{12}^0\ep_{45}^0
\mathcal{D}_{12}^{(-1,0,-1)}+\orm(\ep_{12}) \right)+\orm(\ep_{23}^0) \right)
+\orm(\ep_{15}^0)\right), \\[1mm]
&\mathcal{D}_{12}^{(-1,0,-1)} = \left( \ds_{u_{s2}}- \frac{h_\phi}{u_{s2}} \right) 
\ds_{u_{s5}}, \label{D12_m10m1} \\[2mm]
&\mathcal{D}_{45}^{2,\vec{\ep}} = \frac{1}{\ep_{15}} \left(\frac{1}{\ep_{34}} \ep_{23}^0
\left(\ep_{12}^0\ep_{45}^0 \mathcal{D}_{12}^{(-1,-1,0)}+\orm(\ep_{56}) \right)+\orm(\ep_{34}^0)\right)
+\orm(\ep_{15}^0), \\[1mm]
&\mathcal{D}_{45}^{(-1,-1,0)} = \left( \ds_{u_{s3}} - \frac{h_\phi}{u_{s3}} \right)\ds_{u_{s5}} .\label{D45_m1m10}
\end{align}
Since all the terms we spelled out have vanishing grade with respect to the formal
variables $\ep_{12},\ep_{45}$ we shortened the label $(\vec{k})$ and only displayed
three entries, i.e. the superscript on the right-hand side only gives the grades 
with respect to $(\ep_{15},\ep_{34},\ep_{23})$.

We are going to analyze the Casimir eigenvalue equations for the crossed channel in
two different cases that will turn out to be relevant below. But as we explained in
the introduction, the analysis depends crucially on whether the eigenvalue of the operator scales in the same way as the most singular term or not. Correspondingly, we
are going to study two very different cases that will turn out to be relevant for
the discussion of the bootstrap constraints. We call these case~I and case~II
respectively.

\begin{itemize}
\item[\ ] \textbf{Case~I:}\ In this case, we assume the eigenvalues $\la_1$ and $\la_2$
    to scale such that the products $\la_1 u_{s2} u_{u_{s5}}$ and $\la_1 u_{s3} u_{s5}$
    stay finite in the lightcone limit. This means that
    \begin{equation}
    \mathrm{LS}_\mrmI: \quad \la^{\vec{\epsilon}}_1 = 
    \ep_{15}^{-1}\ep_{23}^{-1} \lambda_1, \quad
    \la^{\vec{\epsilon}}_2 = \ep_{15}^{-1}\ep_{34}^{-1} \lambda_2,
    \label{case_I_regime}
    \end{equation}
    i.e. both eigenvalues scale in the same way as the most singular term of the
    associated Casimir operator.
\item[\ ] \textbf{Case~II:}\ In this case, we assume that only one of the eigenvalues
    scales like the singular term. For definiteness, let this be the eigenvalue $\lambda_2$
    and assume that we scale $\lambda_1$ such that $\lambda_1 u_{s5}$ remains finite, i.e.
    \begin{equation}
    \mathrm{LS}_\mrmII: \quad \la_1^{\vec{\ep}} = \ep_{15}^{-1} \lambda_1, \quad
    \la_2^{\vec{\ep}} = \ep_{15}^{-1}\ep_{34}^{-1} \lambda_2.
     \label{case_II_regime}
    \end{equation}
    \end{itemize}
This list is by no means complete. Clearly, we could also consider the case in which
the scaling behavior of both eigenvalues $\lambda_1$ and $\lambda_2$ is subleading.
Of course, we could change the scaling behavior of the subleading eigenvalue, e.g. keep them
finite, etc. But fortunately, case~I and case~II are (almost) all we need.

Before we dive into the computation of these wave functions, first for case~I 
and then for case~II, we want to explain our general strategy for how to select 
and normalize solutions of the Casimir equations to describe the 
lightcone limit of the actual blocks. The procedure we shall use mimics to 
a certain extent the strategy we outlined for four points, but the additional 
fifth variable $u_{s5}$ (or $w$ in terms of OPE cross-ratios) that arises 
from the need to parametrize the choice of tensor structures at the middle 
vertex causes some new challenges, in particular when it comes to normalizing 
our lightcone blocks.

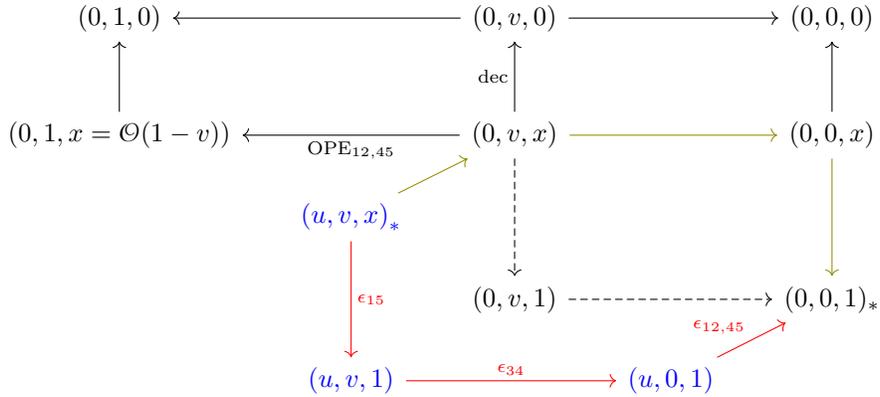
\begin{figure}[ht]
\centering
\begin{equation*}
\begin{tikzcd}[row sep=scriptsize, column sep=scriptsize]
(0,1,0)& & \arrow[ll](0,v,0) \arrow[rr] &  & (0,0,0) \\
& & &  & \\
(0,1,x=\orm(1-v))\arrow[uu] & & \arrow[ll,"\mathrm{OPE}_{12,45}"] (0,v,x)\arrow[uu,"\mathrm{dec}"] \arrow[rr,olive] \arrow[dd,dashed] & & (0,0,x) \arrow[uu] \arrow[dd,olive] \\
& \color{blue}{(u,v,x)}_\ast \arrow[ur,olive]  \arrow[dd,red,"\ep_{15}"] & &    \\
& & (0,v,1) \arrow[rr,dashed] & & (0,0,1)_\ast \\
&\color{blue}{(u,v,1)}\arrow[rr,red,"\ep_{34}"] & & \color{blue}{(u,0,1)} \arrow[ur,red,"\ep_{12,45}"] \\
\end{tikzcd}
\end{equation*}
\caption{Diagrammatic representation of the space of five-point cross-ratios. To make this plot 3-dimensional, we have combined $u_{s1}$ and $u_{s4}$ into a single letter $u$ and we omitted $u_{s2}$. The variable $v = u_{s3}$ increases from right 
to left, while $x = 1-u_{s5}$ increases from top to bottom. Our letter $u$ corresponds to the direction that leaves the plane of the paper. From the generic point $(u,v,x)_\ast$ in the center of the diagram we can reach the lightcone limit $(0,0,1)_\ast$ in the bottom right corner in the direct (green) and the crossed (red) channel. The top half of the figure contains the OPE limit $(0,1,x)$ on the left which provides boundary conditions for blocks. }
\label{fig:lim_diag_5pt_gen}
\end{figure}

In order for the discussion to apply to both case~I and case~II blocks, 
we will keep the variable $u_{s2}$ separate and group the remaining four 
variables into three groups $u,v,x$, where $u$ denotes the pair $u = \{u_{s1},u_{s4}\}$ while  $v = u_{s3}$ and $x = 1-u_{s5}$. Once we have
formed these three groups, we can draw a three-dimensional map for the 
space of cross-ratios, see fig.~\ref{fig:lim_diag_5pt_gen}. The black 
vertices in this diagram label points in the $u=0$ plane while the blue 
vertices have $u \neq 0$. Moreover, $v$ varies along the horizontal direction
while $x$ varies along the vertical direction. We have marked the generic 
point with cross-ratios $(u,v,x)$ by an index $\ast$. The lightcone limit 
we explore below corresponds to the vertex $(0,0,1)_\ast$ in the figure. 
In the crossed channel, we reach this point from generic cross-ratios by 
going along the red arrows. We want to compute the lightcone blocks by 
studying the Casimir differential equations at the limit point. As we 
shall see, it is actually not that difficult to solve the limiting 
differential equations. The only issue is, however, that we have to 
pick the solution that satisfies the OPE boundary conditions. Now, the 
OPE limit is actually the point with coordinates $(0,1,0)$ in the
upper left corner of fig.~\ref{fig:lim_diag_5pt_gen}, very far away 
from the lightcone limit. The only thing the two points have in common 
is that they both lie in the $u=0$ plane.  

Fortunately, within this $u=0$ plane, the blocks are under relatively good 
control --- one can infer this from our discussion of the direct channel blocks
in the previous subsection. It implies that we can actually write down
an integral representation that is quite amenable to evaluation. By 
adapting our formulas \eqref{eq:DCLCL2} and \eqref{block_to_int_ope} to the crossed-channel 
setup, we have 
\begin{equation} \label{block_to_int_ope_CC}
\psi^{\CC}_{(h_a,\bar h_a;n)}(u_{si}) \stackrel{u=0}{\sim} 
u_{s1}^{h_1}u_{s4}^{h_2}(u_{s2} u_{s3})^{h_\phi-n} z_1^{\bar h_1-h_1} z_2^{\bar h_2-h_2} 
(1-w)^n F_{(h_a,\bar h_a;n)}(u_{s2},u_{s3},1-u_{s5}),
\end{equation} 
where the function $F$ is given by the integral \eqref{eq:Fint}, only 
that now the arguments are $v_1 = u_{s2}, v_2 = u_{s3}$ and $x = 
1-u_{s5}$. In our regime $u = \{u_{s1},u_{s4}\} = 0$ the variables 
$z_i,w$ are computed from the non-zero snowflake cross-ratios through
\begin{equation} 
z_1 = 1 - u_{s2}\ , \quad z_2 = 1- u_{s3}\ , 
\quad 1-w = \frac{u_{s2}u_{s3}(1-u_{s5})}{(1-u_{s2})(1-u_{s3})}
\end{equation} 
Compared to the previous formula, we have just shifted the index of the 
snowflake cross-ratios by one to account for the different labeling of 
external points we use in the crossed channel. 

As we have seen before, we can write a closed-form expression 
for the function $F$ all along the upper edge $x=0$ of 
fig.~\ref{fig:lim_diag_5pt_gen}: 
\begin{equation} \label{5ptblockdecoupling}
\lim_{x\rightarrow u}  \lim_{u \rightarrow 0} F_{(h_a,\bar h_a;n)}
(u_{si}) = f_1^{\mathrm{dec}}(u_{s2}) f_2^{\mathrm{dec}}(u_{s3})
\end{equation}
where
\begin{equation}
f_a^{\mathrm{dec}}(v) =  
\hypg{\bar h_a}{\bar h_a-h_{b\phi}-n}{2\bar h_a}(1-v), 
\quad a\neq b=1,2.
\label{dec_f_a}
\end{equation}
In some sense, this formula is the analog of eq.~\eqref{eq:4ptpsiF} 
in our discussion of four-point blocks, only that now it holds merely for 
$x=0$. In order to move from the upper edge of fig.~\ref{fig:lim_diag_5pt_gen}
down to the lightcone limit, we can start from the decoupling limit in 
the region where $v$ is small (upper right corner) and then expand the integral formula in the variable $x$. Up to subleading corrections in the small $u_{s3}$, large $J_2$ limit (see appendix~\ref{app:cross_channel}), this expansion can be written in 
the form 
\begin{align}
&F_{(h_a,\bar h_a;n)}(u_{si}) \ \sim \ \mathcal{Q}\left(
\left(1- x \mathcal{S}_1(\vt_{u_{s2}})\mathcal{S}_2(\vt_{u_{s3}})
\right)^{h_\phi-\bar h_1-\bar h_2}\right) f_1^{\mathrm{dec}}(u_{s2})
f_2^{\mathrm{dec}}(u_{s3}),  \label{block_formula_decsum} \\[2mm]
&\mathit{where} \quad \mathcal{Q}\left(\mathcal{S}_a(\vt)^k\right) :=
\frac{(h_{b\phi}+n+\vt)_k}{(h_{b\phi}+n+\bar h_a)_k}, \quad a\neq b=1,2.
\label{Sop_formula}
\end{align}
and $\vartheta_u = u \partial_u$ is the Euler operator, as before. In order to
evaluate the first line we first evaluate the argument of $\mathcal{Q}$ as a power
series in $x\mathcal{S}_1 \mathcal{S}_2$ and then apply the `quantization map'
$\mathcal{Q}$ defined in the second line in order to express powers of the
commuting objects $\mathcal{S}_1$ and $\mathcal{S}_2$ as differential operators
that act on functions of $v_1 = u_{s2}$ and $v_2 = u_{s3}$. It is this formula that will 
allow us to determine the normalization of the crossed channel lightcone 
blocks below. 

\subsubsection{Lightcone blocks for case~I} 
We will begin our analysis of crossed
channel lightcone blocks with the case in which the eigenvalues scale as in eq.\
\eqref{case_I_regime}. Looking back at the expressions \eqref{D12_m10m1} and
\eqref{D45_m1m10}, we note that both leading terms commute with the Euler
operators $\vartheta_{u_{s1}}$ and $\vartheta_{u_{s4}}$. Consequently, we can 
expand the lightcone blocks in eigenfunctions of these two Euler operators. 
But there is another simple differential operator that commutes with the operators \eqref{D12_m10m1}
and \eqref{D45_m1m10} as well as the two Casimir operators $\vartheta_{u_{s1}}$ and
$\vartheta_{u_{s4}}$: The operator $\partial_{u_{s5}}$. Hence we can expand the
lightcone limit of the crossed channel blocks as\footnote{The summation over $\eta$
is symbolic. Of course, in general one might expect a continuum of eigenvalues
$\eta$ to contribute in which case one would replace the sum with an integral. As
we shall see, however, in this case the leading contribution arises from a single
term, namely $\eta = n$.}
\begin{equation}
\psi^{\CC,\mrmI}_{(h_a,\bar h_a;n)}(u_{si}) \stackrel{\LCL_{\vec{\ep}}}{\sim}
u_{s1}^{h_1} u_{s4}^{h_2} (u_{s2} u_{s3})^{h_\phi}
\sum_{\eta} \bra{\eta}\ket{n}
e^{-\eta u_{s5}} g_{h_a,\bar h_a;\eta}(u_{s2},u_{s3}).
\label{expansion_eta}
\end{equation}
Here $\bra{\eta}\ket{n}$ is just a symbol for the unknown expansion coefficients.
As in our discussion of the direct channel blocks, the OPE boundary condition
$$ \psi_{(\bar h_a,h_a;n)}^{\CC}(u_{si}) \stackrel{\mathrm{OPE}}{\propto}
u_{s1}^{h_1} u_{s4}^{h_2} + \dots $$
implies that the eigenvalues of the Euler operators $\vartheta_{u_{s1}}$ and
$\vartheta_{u_{s4}}$ are fixed and coincide with the quantum numbers $h_1$
and $h_2$ of the exchanged fields in the crossed channel. Giving an interpretation of the eigenvalue $\eta$ and finding the precise coefficients $\bra{\eta} \ket{n}$ is however a bit more difficult.

Before we go into the details of this, let us provide a rather suggestive
argument. To begin with, let us recall that the operator that measures the quantum number $n$ that labels tensor structures in the decoupling limit is given
by the Euler operator
$$  \vartheta_x = \vartheta_{1-u_{s5}} = (u_{s5}-1) \partial_{u_{s5}}
\stackrel{\LCL_{\vec{\epsilon}}}{\sim} - \partial_{u_{s5}}\ .
$$
this may lead us to suspect that the eigenvalue $\eta$ of the operator
$\partial_{u_{s5}}$ in the lightcone limit is directly related to the
quantum number $n$. In addition, let us also observe that the argument
$\eta u_{s5}$ of the exponential can only be finite when $u_{s5}$ goes to zero if the eigenvalue $\eta$ scales as
\begin{equation}
\eta^{\vec{\epsilon}}  = \ep_{15}^{-1} \eta
\end{equation}
If we combine this fact with the expectation that $\eta = n$, then we are led to conclude that the lightcone limit of our blocks can only have a nontrivial dependence on the cross-ratio $u_{s5}$ if we scale $n$ such that $nu_{s5}$ remains finite. Note that this scaling ansatz $n=\orm(\ep_{15}^{-1})$ is consistent with the scaling of the vertex operator at leading order in the lightcone limit, see section~\ref{sect:VertexOp} and the argument after equation~\eqref{eq:vop_on_id_phi}.
As an attentive reader may have noticed, the regime we are probing here is such that the tensor structure label $n=\orm(\epsilon_{15}^{-1})$ scales faster than the spins $J_i=\sqrt{\lambda_i}=\orm\!\left(\epsilon_{15}^{-\frac{1}{2}}\epsilon_{(i+1)(i+2)}^{-\frac{1}{2}}\right)$. This seemingly unphysical regime can nonetheless be defined in the continuum limit for the spins $J_i$ that is used when replacing the sum over large spins with an integral, see appendix~\ref{app:tensor_structure_larger} for its construction based on the vertex differential operator.

Let us now turn to discuss the non-trivial functions $g(u_{s2},u_{s3})$ of the two
remaining cross-ratios $v_1 = u_{s2}$ and $v_2 =u_{s3}$.  For these functions, the
two Casimir equations reduce to the following set of linear differential equations
\begin{equation}
\eta \ds_{u_{s2,3}} g_{h_a,\bar h_a;\eta}(u_{s2},u_{s3}) = - \la_{1,2}
g_{h_a,\bar h_a;\eta}(u_{s2},u_{s3}) \ .
\end{equation}
These equations are solved by
\begin{equation}  \label{eq:gIsol}
g_{h_a,\bar h_a;\eta}(u_{s2},u_{s3})
= e^{-\frac{\la_1 u_{s2}+\la_2u_{s3}}{\eta}}.
\end{equation}
Let us stress that, by our assumption on the scaling behavior \eqref{case_I_regime}
of the eigenvalues $\lambda_a$, the argument of the exponential function is finite
as we send the cross-ratios $u_{s5}, u_{s2}$ and $u_{s3}$ to zero.
\smallskip

Putting things together we can now conclude that the lightcone limit of our
crossed channel blocks in case~I is given by
\begin{equation}
\tcboxmath{\psi^{\CC,I }_{(h_a,\bar h_a;n)}(u_{si})
\stackrel{\LCL_{\vec{\ep}}}{\sim}
\nn^{\CC,\mrmI}_{(h_a,\bar h_a;n)}
 u_{s1}^{h_1} u_{s4}^{h_2} (u_{s2}u_{s3})^{h_\phi}
 e^{-n u_{s5}-\frac{\bar h_1^2 u_{s2}+ \bar h_2^2 u_{s3}}{n}}.}
\label{sol_5pt_I}
\end{equation}
Let us stress again that the validity of this formula requires scaling the quantum numbers $\bar h_a$ and $n$ such that $\bar h^2_1 u_{s2}/n, \bar h^2_2
u_{s3}/n $ and $nu_{s5}$ stay finite as we sent the cross-ratios into the
regime $\LCL_{\vec{\ep}}$.

We still need to determine the normalization $\nn^{\CC,\mrmI}$ and to justify the expectation $\eta=n$.
In order to do so we note that the exponential term in the crossed channel
lightcone blocks \eqref{sol_5pt_I} drops out in the limit in which $\bar
h_1^2 u_{s2},\bar h_2^2 u_{s3} \ll n  \ll u_{s5}^{-1}$. After taking the
limit, the remaining dependence on cross-ratios coincides with that of
the direct channel blocks \eqref{eq:5ptLCLdc} in case $h_{a\phi}+n > 0$
for both $a=1,2$, of course up to the obvious replacement $u_{si}\rightarrow
u_{s,i-1}$. This suggests that we can simply compute the normalization of
our crossed channel lightcone blocks by taking an appropriate limit of the
direct channel normalization, i.e.
\begin{equation}
\nn^{\CC,\mrmI}_{(h_a,\bar h_a;n)}  = \lim_{n\rightarrow \infty}
\lim_{\bar h_a \rightarrow \infty} \nn^{\DC\, (0)}_{(h_a,\bar h_a;n)}
= \lim_{n\rightarrow \infty} \lim_{\bar h_a \rightarrow \infty}
\prod\limits_{a\neq b} \frac{\Gamma(2\bar h_a) \Gamma(h_{b\phi}+n)}
{\Gamma(\bar h_a)\Gamma(\bar h_a+h_{b\phi}+n)} .
\end{equation}
The direct channel normalizations can be found in eq.~\eqref{eq:5ptLCLdcresult},
and $\lim$ is a shorthand for taking the leading term in the Stirling formula
for the Gamma functions\footnote{In particular, one can check that
$\tensor[_3]{F}{_2}(1) \sim 1$ in the large $\bar h_1, \bar h_2$ limit of the
normalization formula \eqref{eq:5ptLCLdcresult}.}.
The evaluation with the help of Stirling's formula gives
\begin{equation}
\tcboxmath{
\nn^{\CC,\mrmI}_{(h_a,\bar h_a;n)} =
\frac{1}{2n} \left(\frac{n}{e}\right)^{2n} n^{h_1+h_2-2h_\phi}
\prod_{a\neq b} 4^{\bar h_a} \bar h_a^{\frac{1}{2}-(h_{b\phi}+n)}.}
\label{norm_case_I}
\end{equation}
The two formulas \eqref{sol_5pt_I} with the normalization \eqref{norm_case_I}
are indeed correct, even though our derivation was based on the identification
that we did not derive rigorously yet.
\medskip

In order to establish eqs.~\eqref{sol_5pt_I} and \eqref{norm_case_I}
rigorously, we go back to the expansion \eqref{expansion_eta} and note
that the formula \eqref{eq:gIsol} completely determines the dependence
of each term in the expansion on the cross-ratios $u_{s2}$ and $u_{s3}$.
Hence, in order to determine the coefficients $\bra{\eta}\ket{n}$ and
thereby the relation between $\eta$ and $n$, it suffices to look at the
limit where $u_{s2}$ and $u_{s3}$ tend to zero. In this limit, we now
want to compare with the formula \eqref{block_formula_decsum}. Recall,
that this formula computes the lightcone behavior from the decoupling
limit in the direct channel.

In the case of four-point functions, we pointed out that the leading
terms of the crossed channel Casimir operators in the limit where
$u \ll v$ coincide with the leading terms of the direct channel Casimir
operators with $u$ and $v$ exchanged. A closely related statement holds
true for five-point Casimirs. More specifically, the leading terms in 
the differential equations are the same regardless of whether we use 
the usual ordering prescription with $\vec{\ep}=(\ep_{15},\ep_{34},
\ep_{12},\ep_{45},\ep_{23})$ or the one that is obtained by shifting 
all indices,  $\vec{\ep}^{\,\prime}=(\ep_{12},\ep_{45},\ep_{23},\ep_{15},
\ep_{34})$,
\begin{equation}
\lim_{\vec{\ep}\rightarrow 0 } \ep_{15}\ep_{23,34}
\left(\mathcal{D}_{12,45}^{2,\vec{\ep}}-\la_{1,2}^{\vec{\ep}}\right)
=\lim_{\vec{\ep}^{\,\prime} \rightarrow 0} \ep_{15}\ep_{23,34}
\left(\mathcal{D}_{12,45}^{2,\vec{\ep}}-\la_{1,2}^{\vec{\ep}}\right).
\end{equation}
The left-hand side of this equation represents the red path in the diagram
\ref{fig:lim_diag_5pt_gen}, while the right-hand side represents the green
path, with the same scaling of $\la_1,\la_2,n$ in both cases.

This suggests that we can indeed determine the matrix elements $\bra{\eta}
\ket{n}$ in the expansion \eqref{expansion_eta} from the formula
\eqref{block_formula_decsum}. After re-instating the factor
that was removed in order to pass from $\psi$ to $F$, eq.\
\eqref{block_formula_decsum} reads
$$
\psi_{(h_a,\bar h_a;n)}^{\CC}(u_{si}) \sim u_{s1}^{h_1} u_{s2}^{h_2}
(u_{s2} u_{s3})^{h_\phi} (1-u_{s5})^n  \mathcal{Q}\left(
\left(1- x \mathcal{S}_1(\vt_{u_{s2}})\mathcal{S}_2(\vt_{u_{s3}})
\right)^{h_\phi-\bar h_1-\bar h_2}\right) f_1^{\mathrm{dec}}(u_{s2})
f_2^{\mathrm{dec}}(u_{s3}),
$$
For definitions of the various objects, see eqs.~\eqref{dec_f_a},\eqref{Sop_formula}. In the limit where
$u_{s2,3}\rightarrow 0$, we see that the only $x$-dependence comes from the factor
$x^n = (1-u_{s5})^n$.  In fact, since in case~I the eigenvalues $\lambda_a$ are sent to
infinity with a scaling law \eqref{case_I_regime}, the operators
$\mathcal{S}_a^k \sim \bar h_a^{-k}$ vanish in this regime. Hence, our task is
to reproduce the factor $\exp (-\eta u_{s5})$ in eq.~\eqref{expansion_eta} from
$x^n$. This requires that we scale $n$ such that $nu_{s5}$ remains finite. Then,
in the limit of large $n$,  we find
$$
\lim_{n\rightarrow \infty} x^n =
\lim_{n \rightarrow \infty} \left(1- \frac{nu_{s5}}{n}\right)^n = e^{- n u_{s5}}
\ .
$$
Comparing with the expansion \eqref{expansion_eta} we obtain agreement provided
that
$$ \bra{\eta}\ket{n} = \dg_{\eta n}\nn_{(h_a,\bar h_a;n)}^{\CC,\mrmI}.$$
In addition, we can compute the normalization $\nn$ from the limiting
behavior of the functions $f^\mathrm{dec}(v)$ as we send $v$ to zero. The result
coincides with our formula \eqref{norm_case_I}. This concludes our rigorous
derivation of formulas \eqref{sol_5pt_I} and \eqref{norm_case_I}.

\subsubsection{Lightcone blocks for case~II}
In case~II, one of the two eigenvalues is subleading and hence, in order to
find eigenfunctions depending on the subleading eigenvalues, we must
expand the corresponding Casimir operator to the desired order at least.
Let us recall from eq.\ \eqref{case_II_regime} that the eigenvalue $\lambda_1$
of the Casimir $\mathcal{D}^2_{12}$ is subleading in the last entry 
$\epsilon_{23}$ of our order $\vec{\ep}$ and we should include the 
next-to-leading order term in $u_{23}$ which reads
\begin{equation}
\mathcal{D}_{12}^{(-1,0,0)} = \ds_{u_{s5}}\left(\vt_{u_{s5}} - \vt_{u_{s2}}-\vt_{u_{s3}}+h_{\phi} \right) \label{D12_m100}
\end{equation}
Once again, our notation suppresses the entries $\ep_{12},\ep_{45}=0$
in the superscript so that we only have three entries $(\ep_{15},\ep_{34},
\ep_{23})$. We notice that this term still commutes with the Euler operators
$\vartheta_{u_{s1}}$ and $\vartheta_{u_{s4}}$ for the cross-ratios $u_{s1}$
and $u_{s4}$. As we explained before, when acting on lightcone blocks the value of these
two Euler operators is fixed to $h_1$ and $h_2$, respectively. Hence, the
lightcone blocks have the form
\begin{equation}
\psi^{\CC, \mrmII}_{(h_a,\bar h_a;n)}(u_{si}) = u_{s1}^{h_1} u_{s4}^{h_2}
g(u_{s2},u_{s3},u_{s5})\ .
\end{equation}
Before we continue our discussion of differential equations, let us pass
from the blocks $\psi$ to the functions $F$ by splitting off the simple
prefactor  
\begin{equation} \label{eq:omegaII} 
\om^{\CC, \mrmII}_{(h_a,\bar h_a;n)} = 
u_{s1}^{h_1} u_{s4}^{h_2} (u_{s2} u_{s3})^{h_\phi}
 (1-u_{s2})^{J_1 - n}\ .
\end{equation} 
In passing from $\psi$ to $F$, the simple dependence of the lightcone
block on $u_{s1}$ and $u_{s4}$ is removed and one find
\begin{equation}
F^{\CC, \mrmII}_{(h_a,\bar h_a;n)} (u_{si}) =
F^{\CC, \mrmII}_{(h_a,\bar h_a;n)}(u_{s2},u_{s3},u_{s5}).
\end{equation}
To fix the dependence of the functions on the remaining variables we need
to consider the eigenvalue equations
\begin{align}
& \mathcal{D}_{12}^{(-1,0,*)} : \ds_{u_{s5}}\left(\vt_{u_{s5}}-
\vt_{1-u_{s2}}-\vt_{u_{s3}}
-J_1+n-h_\phi\right) F_{(h_a,\bar h_a;n)} =  \la_1 F_{(h_a,\bar h_a;n)} \label{D12m10_eq} \\[2mm]
& \mathcal{D}_{45}^{(-1,-1,0)} : \ds_{u_{s5}}\ds_{u_{s3}} F_{(h_a,\bar h_a;n)}
= \la_2 F_{(h_a,\bar h_a;n)}. \label{D45m1m10_eq}
\end{align}
The $*$ in the label of the first Casimir operator reminds us that the
equation includes the term that is subleading in $u_{s2}$. Note that we have
written these terms as differential operators for the function $F$, i.e.
after removing $\om$. This needs to be taken into account when comparing
with our expressions \eqref{D12_m10m1}, \eqref{D12_m100} and \eqref{D45_m1m10}
which were written for the action on $\psi$.

We observe that the two operators on the left-hand side of the differential
equations \eqref{D12m10_eq} and \eqref{D45m1m10_eq} commute with the Euler
operator $\vartheta_{1-u_{s2}}= (1-u_{s2})\ds_{u_{s2}}$. This suggests to
expand the functions $F$ as
\begin{equation} \label{eq:CCIIAnsatz}
F_{(h_a,\bar h_a;n)}(u_{s2},u_{s3},u_{s5}) = \sum_\mu \bra{\mu}\ket{n}
(1-u_{s2})^\mu F_{(h_a,\bar h_a;n)}^\mu(u_{s3},u_{s5})\ .
\end{equation}
Inserting this Ansatz into our eigenvalue equations and using the notation
$J_1-n:=\dg n$ we obtain the following differential equations
\begin{align}
& \mathcal{D}_{12}^{(-1,0,*)} : \ds_{u_{s5}}\left(\vt_{u_{s5}}-\vt_{u_{s3}}-h_\phi-\dg n-\mu\right) F_{(h_a,\bar h_a;n)}^\mu(u_{s3},u_{s5}) =
\la_1 F_{(h_a,\bar h_a;n)}^\mu(u_{s3},u_{s5}) \\[2mm]
& \mathcal{D}_{45}^{(-1,-1,0)} : \ds_{u_{s5}}\ds_{u_{s3}} F_{(h_a,\bar h_a;n)}^{\mu} (u_{s3},u_{s5})(u_{s3},u_{s5})= \la_2 F_{(h_a,\bar h_a;n)}^\mu (u_{s3},u_{s5}).\label{eq:int_k_II}
\end{align}
The solution can be derived by expanding $F^\mu$ into a basis of common eigenfunctions of $\mathcal{D}_{45}^{(-1,-1,0)}$ and $\ds_{u_{s5}}$,
\begin{equation*}
F^\mu_{(h_a,\bar h_a;n)}(u_{s3},u_{s5}) = \int_{0}^\infty \dd k\, f^\mu_{(h_a,\bar h_a;n)}(k) \, e^{-\left(k u_{s5}+k^{-1}\la_2 u_{s3}\right)}.
\end{equation*}
Here, we assume that only the positive part of the $\ds_{u_{s5}}$ spectrum contributes to 
the superposition and that said spectrum is continuous. Acting with the Casimir operator $\mathcal{D}_{12}^{(-1,0,\ast)}$ on the integral and integrating by parts (assuming appropriate 
falloff at large $k$), we obtain a first-order differential equation for the function 
$f^\mu_{(h_a,\bar h_a;n)}(k)$ that fixes it up to a multiplicative constant $f_0$:
\begin{equation}
f^\mu_{(h_a,\bar h_a;n)}(k) = \frac{f_0}{2} \frac{e^{-\la_1/k}}{k^{1+h_\phi+\dg n+\mu}}.
\end{equation}
Plugging this back into eq.~\eqref{eq:int_k_II} yields the integral representation \eqref{bessel_clifford_integral} of the modified Bessel-Clifford function $\mathcal{K}_{\ag}(x)$, see appendix~\ref{app:bessel_clifford} for its definition and properties. Thus, up to a multiplicative constant, the solution with boundary conditions suitable for our problem is given by 
\begin{equation}
F^{\mu}_{(h_a,\bar h_a;n)}(u_{s3},u_{s5}) = u_{s5}^{h_\phi+\dg n+\mu} \mathcal{K}_{h_{\phi}+\dg n+\mu}(\la_1u_{s5}+\la_2u_{s3}u_{s5}).
\label{eq:Case2Fmu}
\end{equation} 
Inserting the solution~\eqref{eq:Case2Fmu} into the ansatz \eqref{eq:CCIIAnsatz}, we obtain
\begin{equation}
F_{(h_a,\bar h_a;n)}(u_{s2},u_{s3},u_{s5}) = \sum_\mu \bra{\mu}\ket{n}
(1-u_{s2})^\mu u_{s5}^{h_\phi+\dg n+\mu} \mathcal{K}_{h_\phi+\dg n+\mu}
(\la_1 u_{s5}+ \la_2 u_{s3} u_{s5}).
\label{F_mu_basis}
\end{equation}
To summarize, we have shown that any function of the form \eqref{F_mu_basis}
satisfies our two Casimir equations to the desired order, regardless of the coefficients $\bra{\mu}\ket{n}$.

To determine the domain of the summation index $\mu$, the scaling of the
tensor structure label $n$ and the coefficients $\bra{\mu}\ket{n}$, we follow
the strategy we outlined in section~\ref{ssec:Crossed_Casimir_eqs} and executed in our discussion of
case~I blocks already. The procedure rests on the observation that the
relevant terms of the two Casimir operators are the same if we take the limits with
the ordering $\vec{\ep}$ (red path in fig.~\ref{fig:lim_diag_5pt_gen})
or choose the direct channel ordering $\vec{\ep}\,'$ instead (green path
in fig.~\ref{fig:lim_diag_5pt_gen}),
\begin{align*}
\lim_{\vec{\ep}\rightarrow 0}  \ep_{15}\left(\mathcal{D}_{12}^{2,\vec{\ep}}-\la_{1}^{\vec{\ep}}\right) &=\lim_{\vec{\ep}^{\,\prime}\rightarrow 0} \ep_{15}\left(\mathcal{D}_{45}^{2,\vec{\ep}}-\la_{1}^{\vec{\ep}}\right),
\\[2mm]
\lim_{\vec{\ep} \rightarrow 0} \ep_{15} \ep_{34}
\left(\mathcal{D}_{45}^{2,\vec{\ep}}-\la_{2}^{\vec{\ep}}\right) &=
\lim_{\vec{\ep}^{\,\prime}\rightarrow 0} \ep_{15}\ep_{34}
\left(\mathcal{D}_{45}^{2,\vec{\ep}}-\la_{2}^{\vec{\ep}}\right).
\end{align*}
Thus, in direct analogy to case~I blocks, we should retrieve the form \eqref{F_mu_basis} from the expansion \eqref{block_formula_decsum} of blocks around the decoupling limit which is defined here at the $(0,0,x)$ 
node of fig.~\ref{fig:lim_diag_5pt_gen} for $v=u_{s3}$. 

Our first goal is to show that the only contribution to the sum 
\eqref{F_mu_basis} comes from $\mu=0$. In order to do this, we start by observing 
that the sum runs over $\mu \geq 0$ only. This can be shown by writing the leading contribution to $F_{\oo_1\oo_2;n}$ in 
eq.~\eqref{block_formula_decsum} as a power series in $1-u_{s2}$. Now, to deduce the scaling of $n$, we make
the simple observation
\begin{equation}
\mathcal{D}_{12}^{(-1,0,*)} -J_1^2 = \ds_{u_{s5}} \vt_{1-u_{s2}} +\orm(\ep_{15}^{-1}).
\end{equation}
On the left-hand side, the Casimir operator and its eigenvalue 
$J_1^2+\orm(J_1)$ scale like $\orm(\ep_{15}^{-1})$ at leading order, while on the 
right-hand side, $\ds_{u_{s5}}$ also scales like $\orm(\ep_{15}^{-1})$ 
and $\vt_{1-u_{s2}}$ has eigenvalues $\dg n+\mu$, $\mu\geq 0$.  Assuming $n<\mathrm{min}(J_1,J_2)=J_1$, there 
is only one way to ensure that the right-hand side does not scale 
faster than the $\ep_{15}^{-1}$ to infinity, namely to keep 
$\dg n$ finite,  
\begin{equation}
\dg n = J_1-n =\orm (1).
\end{equation}
More specifically, we will assume from now on that $\dg n=0,1,2,\dots$ is a positive integer. In section~\ref{sect:VertexOp}, we will further demonstrate that this discrete tensor structure labeling is consistent with the spectrum and eigenbasis of the vertex operator.

Let us now analyze the scaling behavior of the coefficients 
$\bra{\mu}\ket{n}$.  Looking at eq.~\eqref{F_mu_basis}, we see 
that the function multiplied by $\bra{\mu}\ket{n}$ scales like 
$\ep_{15}^\mu$ at order $(1-u_{s2})^\mu$.  As a result,  
the $\mu$-th term in the sum can only contribute to the leading behavior if its coefficient compensates for the aforementioned suppression, i.e. $\bra{\mu}\ket{n} \sim \ep_{15}^{-\mu} \bra{\mu=0}\ket{n}$. 
Whether this is the case or not can be deduced from our formula 
\eqref{block_formula_decsum}. Upon expansion of the $x$-dependent 
part into a binomial sum, we find 
\begin{equation}
(1-(1-u_{s5}) \mathcal{S}_1\mathcal{S}_2)^{-\bar h_{12;\phi}} = 
\sum_{k=0}^\infty \frac{(1-u_{s5})^k}{k!}(\bar h_{12;\phi})_k 
\mathcal{S}_1^k \mathcal{S}_2^k.
\end{equation}
In the limit where $u_{s5}$ is of order $\orm(\ep_{15})$,  the 
region in which $k$ is of order $\orm(\ep_{15}^{-1})$ dominates 
this sum\footnote{In analogy to the crossed channel of a crossing equation, one can think of $k$ as the ``spin" and 
$(u_{s5}-1)\ds_{u_{s5}}$ as the ``crossed channel Casimir". }.  
Now, we can expand each contribution $\mathcal{S}_1(\vt_{u_{s2}})^kf_1^{\mathrm{dec}}(u_{s2})$ explicitly as a hypergeometric in $1-u_{s2}$ by equating it with the right-hand side of eq.~\eqref{Ska} in appendix~\ref{app:cross_channel} for $\nu=k$, $a=1$, $v_1=u_{s2}$. In this case, we find:
\begin{equation}
\mathcal{S}_1(\vt_{u_{s2}})^kf_1^{\mathrm{dec}}(u_{s2}) = \frac{(\bar h_1)_k}{(2\bar h_1)_k} \left( 1+ (\dg n+h_\phi-h_1-h_2) \bar h_1 \frac{1-u_{s2}}{k}+\orm\left( \bar{h}_1^2 k^{-2} (1-u_{s2})^2\right) \right).
\end{equation}
Given $\bar h_1^2,k=\orm(\ep_{15}^{-1})$, we can read off from this equation that $\bra{\mu}\ket{n}\sim \ep_{15}^{\mu/2}
\bra{\mu=0}\ket{n}$. Consequently, the coefficients in the sum 
\eqref{F_mu_basis} do not diverge fast enough in the $u_{s5}\rightarrow 0$ limit for $\mu \neq 0$ terms to compete with the $\mu=0$ term, such that only the $\mu=0$ term survives. Putting all this together, we finally arrive at the following 
expression for the lightcone blocks
\begin{equation}
\tcboxmath{\hspace*{-3mm}\psi^{\CC,\mrmII}_{(h_a,\bar h_a;J_1-\dg n)}(u_{si}) \stackrel{{\LCL}_{\vec{\ep}}^{(4)}}
{\sim} \nn^{\CC, \mrmII}_{(h_a,\bar h_a;n)} u_{s1}^{h_1} u_{s4}^{h_2}
(u_{s2}u_{s3})^{h_\phi} (1-u_{s2})^{\dg n} u_{s5}^{h_\phi + \dg n}
\mathcal{K}_{h_\phi+\dg n} \left(\bar h^2_1 u_{s5} +
\bar h_2^2 u_{s3} u_{s5}\right).\hspace*{-5mm}}
\label{blocks_case_II}
\end{equation}
Here, the superscript $(4)$ on the $\LCL$ reminds us that we keep
the cross-ratios $u_{s2}$ finite, i,e, we only send four of the five 
cross-ratios to zero. As we shall demonstrate below, the normalization 
is given by 
\begin{equation}
\tcboxmath{
\nn^{\CC, \mrmII}_{(h_a,\bar h_a;J_1-\dg n)}= 4^{\bar h_1+\bar h_2}
\sqrt{\frac{\bar h_2}{2\pi}} \, e^{-\bar h_1}\bar h_1^{\bar h_1} \,
\bar h_2^{\bar h_1-h_\phi-\dg n}.}
\label{norm_block_case_II}
\end{equation}
The derivation of the normalization here is less intuitive than in 
case~I --- this is because even at all orders in $u_{s2}$, only the 
first term of the power series in $(1-u_{s2})$ survives the case~II 
limit.  In analogy with a derivative expansion, the sum over 
descendants $\mu >0$ truncates,  and this sum is precisely what is 
counted by the second factor in the normalization 
\eqref{eq:5ptLCLdcresult} coming from $f_1^{\mathrm{dec}}(0)$.  
At the same time,  the descendants $\eta >0$ coming from higher 
powers of $(1-u_{s5})$ counted in the $\tensor[_3]{F}{_2}(1)$
in eq.~\eqref{eq:5ptLCLdcresult} is modified,  because it depends 
itself the contribution of $\mu\geq 0$ descendants. With a more 
careful tracking of the sum over $(\mu,\eta)$ descendants, that 
is to say, powers of $1-u_{s2}$ and $x$ in eq.~\eqref{block_formula_decsum}, 
we can obtain the normalization via the limit
\begin{equation}
F_{(h_a,\bar h_a;n)} \stackrel{\bar h_2^2 u_{s2} u_{s5}\ll \bar h_1^2u_{s5}\ll 1}{\sim}
\nn^{\CC, \mrmII}_{(h_a,\bar h_a;n)} \bar h_1^{-2(h_\phi+\dg n)}\frac{\Gamma(h_\phi+\dg n)}{2}.
\label{norm_caseII_limit}
\end{equation}
In the formula \eqref{block_formula_decsum}, this is equivalent to setting
\begin{equation}
(u_{s2},u_{s3})=(1,0), \quad \mathcal{S}_1^k = \frac{(\bar h_1)_k}{(2\bar h_1)_k}, \quad \mathcal{S}_2^k(\vt)=\mathcal{S}_2^k(0).
\end{equation}
From this we obtain 
\begin{equation}
\frac{1}{2}\nn^{\CC, \mrmII}_{(h_a,\bar h_a;J_1-\dg n)} = \lim_{\bar h_1\rightarrow \infty}
\bar h_1^{2(h_\phi+\dg n)} \frac{\Gamma(2\bar h_1)}{\Gamma(\bar h_1)\Gamma(\bar h_1+h_\phi+\dg n)}  \lim_{\bar h_2\rightarrow \infty}
\frac{\Gamma(2\bar h_2) \Gamma(\bar h_1-h_\phi-\dg n)}{\Gamma(\bar h_2)\Gamma(\bar h_2+\bar h_1-h_\phi-\dg n)}.
\end{equation}
The evaluation of the limits using Stirling's formula gives the explicit simple 
expression for the normalization of the case~II lightcone blocks that we 
anticipated in eq.~\eqref{norm_block_case_II}. We have thus completed all 
the goals we set ourselves for this section: computing all the 
lightcone blocks we need for our analysis of the crossing symmetry equation. 
We will not go there right away, however, but instead pause for a moment to 
comment on the behavior of vertex differential operators in the lightcone 
limit.

\section{Interlude: Lightcone limit of vertex operators} 
\label{sect:VertexOp}

In the last few sections, we have analyzed second-order Casimir equations 
in the lightcone limit. But these are not the only differential equations 
that are needed to characterize conformal blocks. On the one hand, there 
are also higher-order Casimir operators. For the scalar four- and five-point functions we considered here, only the fourth-order 
Casimir equations are needed to characterize the blocks. It is clear from 
our discussion that in the lightcone limit, these can be expressed in terms of 
the second-order Casimir and Euler operators. So, there was no need to include 
them in our discussion. On the other hand, starting from $N=5$ external scalars, 
Casimir operators no longer provide a complete set of differential equations. 
As was explained in \cite{Buric:2020dyz,Buric:2021ywo}, additional vertex differential operators 
are needed in order to fully characterize multipoint conformal blocks. These 
have to do with the appearance of non-trivial tensor structures at the vertices. 
Just as the eigenvalues of the Casimir operators measure the weight and spin 
of exchanged fields in a given OPE channel, the eigenvalues of the vertex 
operators can be used to measure the choice of tensor structures. In the case 
of five-point blocks, a single such vertex operator needs to be taken into 
account. 

As we have emphasized in the introduction already, vertex differential 
operators are relevant for the lightcone bootstrap in that they can 
be used to determine the scaling of tensor structures much in the same way 
as Casimir operators allow to extract the scaling of spin labels. The main 
obstacle to overcome in setting up a theory of vertex singular behavior is 
to understand how the eigenvalues of vertex operators are actually related 
to more conventional tensor structure labels. In general, very little is 
known. But in the lightcone limit, we can now add some significant new 
insight based on the results of the previous section. This new insight 
will suffice to determine the scaling of tensor structures in the next 
section. 

Below we shall analyze the vertex differential operator for five-point functions,
both for the case~I and the case~II lightcone regimes studied in the previous
section. The resulting expressions for the vertex differential operator are remarkably simple. This simplicity will enable us to construct their 
eigenfunctions from a linear combination/integral transform of the 
lightcone blocks $\psi_n$ we constructed in the previous section.  

\subsection{Vertex operator for case~I}
\label{ssect:VertexOp_I}

To begin with, let us first recall that in \cite{Buric:2021ywo}, the vertex 
operator  was constructed out of generators $\mathcal{T}$ of the conformal 
Lie algebra as follows:
\begin{equation}
	    \mathcal{V}\equiv \mathcal{D}_{\rho,(12)3}^{4,3}=\frac{1}{2}
	    \textit{str}\left[\left(\mathcal{T}_1+\mathcal{T}_2\right)^3
	    \left(\mathcal{T}_3\right)\right] \,. 
	    \label{fivepointsvertexop}
\end{equation}
All relevant notations can be found in \cite{Buric:2021ywo} along with a 
Mathematica notebook that expresses $\mathcal{V}$ as a differential operator 
acting on the five cross-ratios of a five-point function. Starting from these
expressions, we can now go to the lightcone limit $\LCL$, see eq.~\eqref{eq:5ptLCL}. In doing so, it is straightforward to notice that the 
leading divergences of the operator $\mathcal{V}$ can be expressed in terms 
of the second-order Casimir operators $\mathcal{D}_{12}$ and $\mathcal{D}_{45}$. 
The precise relation is
\begin{equation}
    \mathcal{V}^{\vec{\ep}} = 
    -\epsilon_{15}^{-2}\epsilon_{23}^{-2} 
    \left(\mathcal{D}_{12}^{(-1,0,-1)}\right)^2+
     \epsilon_{15}^{-2}\epsilon_{23}^{-1}
    \epsilon_{34}^{-1} \mathcal{D}_{12}^{(-1,0,-1)}\mathcal{D}_{45}^{(-1,-1,0)}+ 
    \orm\!\left(\epsilon_{15}^{-1}\right).
\end{equation}
As in our discussion in section~\ref{sect:lightcone_blocks_five_pt}, we have 
suppressed the components $\ep_{12},\ep_{45}=0$ from the superscript on the 
right-hand side. In other words, the superscript only lists $(\ep_{15},\ep_{34},
\ep_{23})$. To expose those terms in the vertex operator that are actually independent 
of the quadratic Casimir operators we have already studied, we propose to 
subtract $(\mathcal{D}_{12}\mathcal{D}_{45}-\mathcal{D}_{12}^2)/4$. Indeed, 
after this subtraction one  finds 
\begin{equation}
    \left[\mathcal{V}+\mathcal{D}_{12}^2-\mathcal{D}_{12}
    \mathcal{D}_{45}\right]^{\vec{\ep}} =
    \epsilon_{15}^{-1}\left(\epsilon_{34}^{-1}\left(\epsilon_{23}^{-1} \,
    \mathcal{V}^{(-1,-1,-1)} +\orm\!\left(\epsilon_{23}^{0} \right)\right)
    +\orm\!\left(\epsilon_{34}^{0} \right)\right)+\orm\!
    \left(\epsilon_{15}^{0}\right),
    \label{eq:Vert_op_CaseI_subtracted}
\end{equation}
where 
\begin{equation}
\mathcal{V}^{(-1,-1,-1)} = 2\, \left(\partial_{u_{s2}}- \frac{h_\phi}{u_{s2}}\right) 
\left(\partial_{u_{s3}}-\frac{h_\phi}{u_{s3}}\right) 
\partial_{u_{s5}} \left(\vartheta_{u_{s2}}+\vartheta_{u_{s3}}-\vartheta_{u_{s5}} - \vartheta_{u_1} - \vartheta_{u_4} +\frac{d-2}{2}\right).
\label{eq:vertex_op_caseI}
\end{equation} 
In this form, the operator $\mathcal{V}^{(-1,-1,-1)}$ cannot be expressed entirely in 
terms of the second-order Casimir operators, i.e.\ it is an independent operator. 
On the other hand, it is also worth noticing that we can factor out the lightcone 
limit \eqref{D12_m10m1} or \eqref{D45_m1m10} of the second-order Casimir operator
$\mathcal{D}_{12}$ or $\mathcal{D}_{45}$ from the lightcone limit of the fourth-order vertex operator. Hence, when applied to eigenfunctions of the lightcone
Casimir operators, the lightcone vertex operator gives effectively a second 
order eigenvalue equation. This makes it rather easy to solve the combined 
system of Casimir and vertex differential equations in the lightcone limit. 

In order to construct the eigenfunctions of the vertex operator in the lightcone 
limit explicitly, we note  that it also commutes with the Euler operators 
$\vartheta_{u_{s1}}$ and $\vartheta_{u_{s4}}$, which implies we can diagonalize 
these and work in a basis of functions of the form 
\begin{equation} \label{eq:psitog}
    \psi^{\CC}_{(h_a,\bar{h}_a; \mathfrak{t})}(u_{si})=
    u_{s1}^{h_1}u_{s4}^{h_2}\left(u_{s2} u_{s3}\right)^{h_\phi} 
    g_{(h_a,\bar{h}_a; \mathfrak{t})}(u_{s2},u_{s3},u_{s5})\ . 
\end{equation}
Here we have also extracted a prefactor $(u_{s2}u_{s3})^{h_\phi}$ as suggested by 
previous experience and the explicit form of $\mathcal{V}$. Note that we have also 
denoted the tensor structure label by $\mathfrak{t}$ in place of the label $n$ we 
used above. This label $\mathfrak{t}$, that is to say the eigenvalue of the vertex operator, is now used to label eigenfunctions of the 
vertex operator. After conjugation with the prefactor in eq.~\eqref{eq:psitog}
the relevant leading contribution of the subtracted vertex differential operator 
becomes 
\begin{equation}
     \widetilde{\mathcal{V}}^{(-1,-1,-1)}=2\,  \partial_{u_{s2}}\partial_{u_{s3}}
     \partial_{u_{s5}}
     \left(\vartheta_{u_{s2}}+\vartheta_{u_{s3}}-\vartheta_{u_{s5}}+
     2h_\phi-h_1-h_2+\frac{d-2}{2}\right).
     \label{eq:lead_Vert_op}
\end{equation}
By definition, the functions $g$ in eq.~\eqref{eq:psitog} are eigenfunctions of 
$\tilde{\mathcal{V}}$ with an eigenvalue that is determined by $\mathfrak{t}$. To 
diagonalize $\tilde{\mathcal{V}}$, let us start from the basis of functions that 
diagonalize the leading Casimirs as well as $\partial_{u_{s5}}$, 
see section~\ref{sect:lightcone_blocks_five_pt}, 
\begin{equation}
    \psi^{\CC}_{(h_a,\bar{h}_a; \eta)}(u_{si}) \sim 
    \nn^{\CC,\mrmI}_{(h_a,\bar h_a;n)} u_{s1}^{h_1}u_{s4}^{h_2}
    \left(u_{s2} u_{s3}\right)^{h_\phi} e^{-\eta u_{s5}-\frac{\lambda_1 u_{s2}
    +\lambda_2 u_{s3}}{\eta}}.
    \label{eq:CaseI_eta_blocks}
\end{equation}
Here $\sim$ implies that we display the leading term in the lightcone limit up 
to some numerical prefactor. As we outlined in the introductory paragraphs to 
this section, a joint eigenfunction 
of the Casimir operators and the vertex operator~\eqref{eq:lead_Vert_op} is a 
superposition of the blocks~\eqref{eq:CaseI_eta_blocks}, and can be thus written 
as
\begin{equation}
    \psi^{\CC}_{(h_a,\bar{h}_a; \mathfrak{t})}(u_{si}) \sim 
    u_{s1}^{h_1}u_{s4}^{h_2}\left(u_{s2} u_{s3}\right)^{h_\phi} 
    \int_0^{\infty} f(\eta) e^{-\eta u_{s5}-\frac{\lambda_1 u_{s2}+\lambda_2 
    u_{s3}}{\eta}}\dd \eta\,. 
    \label{eq:integral_Vop_blocks}
\end{equation}
Here we assumed that only the positive part of the $\eta$ spectrum contributes to 
the physical blocks and that said spectrum is continuous. Acting with the vertex 
differential operator $\mathcal{V}$ on the integral~\eqref{eq:integral_Vop_blocks}
and integrating by parts (assuming appropriate 
falloff at large $\eta$), we obtain a differential equation for the function 
$f(\eta)$, 
\begin{equation}
    -2\lambda_{1}\lambda_2\left[f'(\eta)+\frac{p+1}{\eta}f(\eta)\right]=
    \mathfrak{t}f(\eta), \qquad p:=2h_\phi-h_1-h_2+\frac{d-2}{2}. 
\end{equation}
This equation is easy to solve. As one can see immediately, the solution is determined up to a multiplicative constant $f_0$ by 
\begin{equation} \label{eq:feta} 
    f(\eta)= \frac{f_0}{2} e^{-\eta \frac{\mathfrak{t}}{2\lambda_1\lambda_2}} 
    \eta^{-(1+p)}.
\end{equation}
In writing this solution, we assumed that the eigenvalue $\mathfrak{t}$ of the 
vertex operator $\mathcal{V}$ scales with $\epsilon_{15}^{-1}\epsilon_{23}^{-1}
\epsilon_{34}^{-1}$, just as the leading term of the vertex operator itself. 
Plugging this form of the function $f(\eta)$ into eq.~\eqref{eq:integral_Vop_blocks} provides us with the blocks
\begin{equation}
   \psi^{\CC,I}_{(h_a,\bar{h}_a; \mathfrak{t})}(u_{si}) \stackrel{\LCL_{\vec{\ep}}}{=} 
    f_0\,  u_{s1}^{h_1}u_{s4}^{h_2}\left(u_{s2} u_{s3}\right)^{h_\phi}  \left(u_{s5}+\frac{\mathfrak{t}}{2\lambda_1\lambda_2}\right)^p\mathcal{K}_{p}\left[\left(u_{s2}\lambda_1+u_{s3}\lambda_2\right)
    \left(u_{s5}+\frac{\mathfrak{t}}{2\lambda_1\lambda_2}\right)\right],
    \label{eq:Case1_Vop_eigenf}
\end{equation}
where we used the integral representation of the modified Bessel-Clifford function $\mathcal{K}_p(x)$ in eq.~\eqref{bessel_clifford_integral}. Thereby we have indeed 
obtained an explicit expression for the lightcone limit of the joint eigenfunctions of Casimir 
and vertex differential operators. The formula \eqref{eq:feta} for the coefficients of 
the eigenfunctions \eqref{eq:CaseI_eta_blocks} in the $\eta$ basis has a simple interpretation: 
In order to go from the $\eta$ basis we used in section~\ref{sect:lightcone_blocks_five_pt} to 
the $\mathfrak{t}$ basis in which the vertex differential operator is diagonal, we apply a Laplace transform, after multiplying the argument of the Laplace transform with 
$\eta^{-1-p}$. Recall from our discussion in
section~\ref{sect:lightcone_blocks_five_pt} that the variable $\eta$ coincides 
with the usual tensor structure label $n$ in the lightcone limit. Hence, in the limit, the tensor 
structure label $\mathfrak{t}$ that was introduced in \cite{Buric:2021ywo} within the integrable systems 
approach to conformal blocks is essentially the Laplace transform of the standard basis labeled by $n$. 
\subsection{Vertex operator for case~II}
\label{ssect:VertexOp_II} 

In our analysis of the eigenfunctions of Casimir operators, we have actually gone a little further 
and evaluated the eigenfunctions to all orders in $u_{s2}$. Quite remarkably, we can also include
all these finite-order contributions in the analysis of the vertex operator. Once we consider all 
orders in $\epsilon_{23}$, we obtain a more complicated expression for the vertex operator which 
reads 
\begin{multline}
    \widetilde{\mathcal{V}}^{(-1,-1,\ast)}= 2\Bigl[ \left(1-z_1\right) z_1^2 \partial _{z_1}^2+ z_1 \left(h_{\phi }+h_1-h_2\right) \left(\vt_{u_{s5}}-\vt_{u_{s3}}\right) \\[2mm] 
    -  
    \left(z_1\left(2 h_{\phi } +h_1 -h_2 +1+ \vt_{u_{s3}} -  \vt_{u_{s5}}\right)-2h_1+\frac{d-2}{2}\right)\vt_{z_1}\\[2mm] 
      -\left(h_{\phi }+h_1-h_2\right) h_{\phi }z_1 
    +h_1^2+\frac{d^2}{12}- 
    \frac{d}{4} \left(2 h_1+1\right)\Bigr]
    \partial_{u_{s3}} \partial _{u_{s5}}\,,
\end{multline}
where we introduced $z_1:=1-u_{s2}$. In our discussion of the leading order, we constructed eigenfunctions
of the vertex operators as linear combinations of case~I blocks in eq.~\eqref{eq:CaseI_eta_blocks}. Now 
we need to work with the case~II blocks \eqref{blocks_case_II} instead, i.e. we write the 
eigenfunctions $\psi_{\mathfrak{t}}$ of the vertex operators as a superposition of case~II blocks, 
\begin{eqnarray}
    g^{\CC,\mrmII}_{(h_a,\bar h_a;\mathfrak{t})}(u_{s2},u_{s3},u_{s5}) & =  & 
    \sum_{m\geq 0} c^{\mathfrak{t}}_m g^{\CC,\mrmII}_{(h_a,\bar h_a;m)}
    (u_{s2},u_{s3},u_{s5}) \nonumber 
    \\[2mm] & & \quad = \sum_{m\geq 0} c^{\mathfrak{t}}_{m} 
    u_{s5}^{h_\phi+m}z_1^m 
    \mathcal{K}_{h_\phi+m}(\bar h_1^2 u_{s5}
    +\bar h^2_2 u_{s3} u_{s5})\,.
    \label{eq:vert_eig_caseII_decomp}
\end{eqnarray}
In writing this expansion, we have not explicitly displayed the dependence of the expansion 
coefficients $c^{\mathfrak{t}}_m$ on the quantum numbers $(h_a,\bar h_a)$ of the exchanged 
fields. The eigenvalue equation for the vertex operator acting on these blocks can be recast 
as a recurrence  relation for the coefficients $c$, 
\begin{equation*}
    \lambda _2 c^{\mathfrak{t}}_m 
    \left(d^2-3 d \left(2 h_1+2 m+1\right)+12 \left(h_1+m\right){}^2\right)-
    6 \mathfrak{t} \, c^{\mathfrak{t}}_m-12 \lambda _1 \lambda _2 
    c^{\mathfrak{t}}_{m-1} \left(h_{\phi }+h_1-h_2+m-1\right)=0\ ,
\end{equation*}
where $\lambda_i= \orm(\bar{h}_i^2)$, as usual.
We can solve this case~II lightcone recursion relation through the 
following explicit expression involving Pochhammer symbols $(\cdot)_{m-1}$, 
\begin{equation}
    c^{\mathfrak{t}}_m=\mathcal{N} \frac{\lambda _1^{m-1} 
    \left(h_1-h_2+h_{\phi }+1\right)_{m-1}}{\left(-\frac{d}{4}+h_1-\frac{\sqrt{8 \mathit{t} 
    \lambda _2-\frac{1}{3} d(d-12)  \lambda _2^2}}{4 \lambda _2}+2\right){}_{m-1} 
    \left(-\frac{d}{4}+h_1+\frac{\sqrt{8 \mathit{t} \lambda _2-\frac{1}{3} d(d-12)  
    \lambda _2^2}}{4 \lambda _2}+2\right){}_{m-1}}\,.
\end{equation}
The square roots that appear in the solution suggest the following parametrization of 
the vertex operator eigenvalues
\begin{equation}
    \mathfrak{t}_N=2 \lambda_2 \left(\left(-\frac{d}{4}+h_1+N\right)^2-\frac{1}{4} d 
    \left(1-\frac{d}{12}\right)\right), 
    \label{t_of_N}
\end{equation}
for some integer $N\in\Zs_{\geq 0}$. Once this parametrization is adopted, we can simplify the expressions and make the 
coefficients $c$ vanish for $m<N$, assuming $N$ is a positive integer. By reabsorbing some constants in the normalization factor, we are then left with
\begin{equation}
c^{\mathfrak{t}}_m=\mathcal{N}^{\,\prime}\frac{ \lambda _1^{m} \left(h_1-h_2+h_{\phi }+N\right)_{m-N}}
    {(m-N)! \left(2 h_1-\frac{d-2}{2}+2N\right)_{m-N}}
    \label{ctm}
\end{equation}
which, together with eq.~\eqref{eq:vert_eig_caseII_decomp}, provide a representation for 
the vertex operator eigenfunctions at all orders in $\epsilon_{23}$.
\paragraph{Alternative representations and consistency with previous results.} There are two ways that we can check the consistency of the case~II vertex operator spectrum \eqref{t_of_N} and eigenfunctions \eqref{eq:vert_eig_caseII_decomp}, \eqref{ctm} with other computations. First, we can reproduce the case~I vertex operator eigenfunctions starting from the following representation of the case~II eigenfunctions:
\begin{equation}
g_{(h_a,\bar h_a;\mathfrak{t}_N)}^{\CC,\mathrm{II}}(u_{s2},u_{s3},u_{s5}) = \mathcal{N}' \int_0^\infty \frac{\dd k}{k^{1+h_\phi}} \left(\frac{\la_1 z_1}{k}\right)^N \tensor[_1]{F}{_1}\begin{bmatrix}
   a+N\\ b+2N 
\end{bmatrix} \left(\frac{\la_1 z_1}{k}\right) e^{-\left(k u_{s5}+\frac{\la_1}{k}+\frac{\la_2 u_{s3}}{k}\right)},
\end{equation}
where $a:=h_1+h_\phi-h_2$ and $b:=2h_1-(d-2)/2$. In the scaling limit
\begin{equation}
1-z_1=u_{s2}=\orm(\ep_{23}), \quad \la_1=\orm(\ep_{15}^{-1}\ep_{23}^{-1}), \quad N^2=\orm(\ep_{23}^{-1}),
\end{equation}
we can approximate the confluent hypergeometric function (see \cite[eq.~(13.7.1)]{dlmf}) by
\begin{equation}
\tensor[_1]{F}{_1}\begin{bmatrix}
   a+N\\ b+2N 
\end{bmatrix} \left(\frac{\la_1 z_1}{k}\right) = \left(\frac{\la_1 z_1}{k}\right)^{a-b-N} e^{\frac{\la_1}{k}(1-u_{s2}) -\frac{k N^2}{\la_1}} \left(1+\orm(\ep_{23}^{1/2})\right).
\end{equation}
In this case, the limiting form of the case~II eigenfunctions is
\begin{equation}
g_{(h_a,\bar h_a;\mathfrak{t}_N)}^{\CC,\mathrm{II}} (u_{s2},u_{s3},u_{s5})\stackrel{\ep_{23}\rightarrow 0}{\sim} \frac{\mathcal{N}'}{\la_1^{b-a}} \left(u_{s5}+\frac{N^2}{\la_1}\right)^{h_\phi+a-b} \mathcal{K}_{h_\phi+a-b}\left([\la_1 u_{s2}+\la_2 u_{s3}]\left(u_{s5}+\frac{N^2}{\la_1}\right)\right).
\end{equation}
This expression coincides with eq.~\eqref{eq:Case1_Vop_eigenf} as $p=h_\phi+a-b$ and $\mathfrak{t}=\mathfrak{t}_N = 2\la_2 N^2$. The second check is the case of scalar exchange $\la_1=0+\orm(1)$, for which there is only one single tensor structure which we expect to be at $N=0$. The corresponding vertex operator eigenfunction must agree with case~II blocks in the $n$-basis when $J_1=\dg n=0$. We can check this from the following Euler integral representation of the case~II vertex operator eigenfunctions:
\begin{multline}
g_{(h_a,\bar h_a;\mathfrak{t}_N)}^{\CC,\mathrm{II}}(u_{s2},u_{s3},u_{s5}) =\\
\mathcal{N}'(\la_1 u_{s5}z_1)^Nu_{s5}^{h_\phi} \int_0^1 \frac{\dd t}{t(1-t)} \frac{t^{a+N}(1-t)^{b-a+N}}{\mathrm{B}_{a+N,b-a+N}} \mathcal{K}_{h_\phi+N}\left(\la_1 u_{s5}(1-z_1 t)+\la_2 u_{s3} u_{s5}\right).
\end{multline}
For scalar exchange, where $\la_1 u_{s5}\rightarrow 0$, the integral over $t$ factorizes. We are thus left with
\begin{align}
g_{(h_a,\bar h_a;\mathfrak{t}_N)}^{\CC,\mathrm{II}}(u_{s2},u_{s3},u_{s5}) \stackrel{\la_1 u_{s5}\rightarrow 0}{\sim} \mathcal{N}'(\la_1 u_{s5}z_1)^Nu_{s5}^{h_\phi}\mathcal{K}_{h_\phi+N}\left(\la_2 u_{s3} u_{s5}\right).
\end{align}
We indeed retrieve the asymptotics of blocks with scalar exchange for $N=0$, see appendix~\ref{app:cross_channel} for a direct computation. 

This concludes our discussion of case~I and II solutions of the vertex differential 
equations. In this work, we shall stick to the more conventional basis of tensor 
structures that is labeled by the integer $n$, also for comparison of our 
bootstrap analysis in the next section with previous work. But it is very 
promising that the vertex operator basis can be calculated explicitly in the 
lightcone limit and that, at least in the leading order case~I lightcone blocks,
is related to the $n$ basis through a simple integral transform. 

\section{Five-point lightcone bootstrap}
\label{sec:5pt_bootstrap}

In this section, we will apply the results on five-point lightcone blocks we derived
in the previous section to the analysis of the crossing symmetry equation depicted
in fig.~\ref{fig:5pt_cse_general}.
In the first subsection, we shall address the
leading contributions in the direct channel. The remaining subsections are then
devoted to the crossed channel. There we shall explain how to reproduce the various
direct channel terms from the crossed channel block expansion. Along the way we
will determine various OPE coefficients involving two double-twist operators, see
eqs.~\eqref{PJphiJ}, \eqref{PJstarphiJ} and \eqref{ope_coeffs_discrete_ts}.  The first two of these are known from the work \cite{Antunes:2021kmm}, so our derivation of these results simply represents a confirmation of their validity via a more rigorous treatment of the order of limits. Our formula \eqref{ope_coeffs_discrete_ts}, instead, is new. It applies in particular to theories for which the leading-twist field $\oo_\star$ that appears in the OPE
of $\phi$ with itself has twist $h_\star > h_\phi$. The derivation of our new
formula requires going beyond the leading terms in the lightcone limit. This is
explained in the section~\ref{ssec:discrete_tensor_structures}, where we derive the large spin expansion of OPE coefficients at discrete tensor structures that reproduce all leading-twist exchanges in the direct channel. Finally, in section~\ref{ssec:applications}, we go over the applications of our results to specific models and check their validity against an independent computation of tree-level OPE coefficients in $\phi^3$ theory. 

\subsection{Direct channel contributions to lightcone limit}
\label{ssec:5ptBS DC}

We will expand both channels of the crossing equation \eqref{eq:5ptcrossing} near two different sequences of lightcone limits: first the null polygon limit $X_{i(i+1)}=0$, with the specific hierarchy stated in eq.
\eqref{eq:5ptLCL}, and then the limit $LCL_{\vec{\ep}}^{(4)}$ in eq.~\eqref{eq:LCL4}, obtained by relaxing the $X_{23}=0$ limit. Starting with the direct channel, we observe that the hierarchy of lightcone limits ensures an expansion whose first contributions are the leading-twist fields in the two operator products of the direct
channel. More precisely, taking both $X_{15}\ll1$ and $X_{34}\ll 1$, we
obtain
\begin{align}
\sum_{\oo_1,\oo_2,n} & P_{\oo_1\oo_2}^{(n)} \psi_{\oo_1\oo_2;n}^{\DC} =
 \ C_{\phi\phi\phi} \left(\left(u_{s1}u_{s3}\right)^{\frac{\Dg_\phi}{2}}+
\left(u_{s5}u_{s2}\right)^{\frac{\Dg_\phi}{2}} \right) +
\nonumber \\[2mm]
& + \sum_{n=0}^{J_\star} P_{\oo_\star\oo_\star}^{(n)}
\,   \psi_{\oo_\star \oo_\star;n}(u_{s1},u_{s2},u_{s4})+
\orm(X_{15}^{h>h_\star}). \label{eq:5ptdirect}
\end{align}
Here, we have used the conventions that were stated in eq.~\eqref{eq:5ptcorrblock},
i.e. we multiplied the five-point correlation function by a factor
$$ \Omega_{\DC}(X_i) = \left(X_{15}X_{34} \sqrt{\frac{X_{12}X_{23}}{X_{13}}}
\right)^{\Delta_\phi}\ . $$
The three terms that appear on the right-hand side correspond to intermediate
exchange of $[\oo_1\vert\oo_2] = [\textbf{1}\vert\phi], [\phi\vert \textbf{1}]$ and
$[\oo_1\vert\oo_2] = [\oo_\star\vert\oo_\star]$ at the top and bottom line respectively. Note that there cannot
be identity exchange in both intermediate channels simultaneously. Furthermore,
a single identity exchange forces the second intermediate exchange to coincide
with the external operators, i.e. it forces $\phi$ exchange in the other
operator product. The two terms with a single identity exchange are the first
two terms on the right-hand side of the previous equation. The third term,
which may involve a sum over tensor structures $n$, is associated with the
leading-twist field $\oo_\star$ in the operator product of $\phi$ with itself, under the assumption that this field is unique.
We parameterize its weight and spin by $h_\star$ and $J_\star$, and we require $h_\star<2h_\phi$ in order to avoid the infinite summation over the double-twist fields~$[\phi\phi]_{n,J}$. For this term,
there are a few mutually exclusive cases to distinguish. The leading-twist
field $\oo_\star$ may coincide with the external field $\phi$ itself, i.e.
$[\oo_1\vert\oo_2]= [\oo_\star\vert\oo_\star] = [\phi\vert\phi]$. But this is not always
realized, especially when the appearance of $\phi$ in the operator product
of $\phi$ with itself is excluded by some selection rule. It turns
out that $\oo_\star \neq \phi$ falls again into two subcases, depending on
whether $h_\star < h_\phi$ or $2h_\phi> h_\star > h_\phi$. Depending on which of the
three scenarios is realized, the blocks $ g_{\oo_\star\oo_\star;n}(u_{s1},u_{s2},
u_{s4})$ possess the following asymptotic behavior in the fivefold lightcone limit \eqref{eq:5ptLCL}, see section~\ref{ssec:5pt_blocks_dc}.
\begin{itemize}
     \item[$(-)$] in case $\oo_\star \neq \phi$ and $h_\star <h_\phi$, then for all $0 \leq n < h_\phi-h_\star$ the blocks possess the following     power law behavior in the lightcone limit,
     \begin{equation} \label{eq:case-}
     \psi^{\DC\, (0)}_{\oo_\star\oo_\star;n}(u_{si}) =
     \nn^{\DC\, (0)}_{\oo_\star\oo_\star;n}
     \left(u_{s1}u_{s2} u_{s3}u_{s5}\right)^{h_\star}
     \left(u_{s1} u_{s2}\right)^n \left(1+\orm(u_{s2})\right).
     \end{equation}
     Note that even when $h_\phi-h_\star>1$, such that the direct channel sum includes $n>0$ contributions, the latter will be subleading of relative order $(X_{12}X_{23})^n$ compared to the $n=0$ block.
    \item[$(0)$] In case $\oo_\star = \phi$ the tensor structure is trivial, i.e.
    $n=0$, and the blocks possess a logarithmic divergence of the form
    \begin{equation}\label{eq:case0}
    \psi^{\DC\, (0)}_{\oo_\star \oo_\star;0} (u_{si})=
     \nn^{\DC\, (0)}_{\phi\phi;0}\, \left(u_{s1}u_{s2} u_{s3}u_{s5}\right)^{h_\phi}
    \log u_{s1}\log u_{s2} + \orm(u_{s2} \log u_{s1}).
    \end{equation}
    \item[$(+)$] In case $\oo_\star \neq \phi$ and $2h_\phi>h_\star > h_\phi$, then for all $n=0,1,\dots,J_\star$, the blocks possess the following power law behavior in the lightcone limit,
    \begin{equation} \label{eq:case+}
     \psi^{\DC\, (0)}_{\oo_\star \oo_\star;n}(u_{si}) =  \nn^{\DC\, (0)}_{\oo_\star\oo_\star;n}
     \left(u_{s1}u_{s2} u_{s3}u_{s5}\right)^{h_\star}
     \left(u_{s1}u_{s2}\right)^{h_\phi-h_\star}\left(1+\orm(u_{s2})\right).
     \end{equation}
     Contrary to the case $(-)$, the $n>0$ blocks are not subleading.
\end{itemize}

In the fourfold lightcone limit \eqref{eq:LCL4}, the blocks present the same asymptotics in the first four cross-ratios but acquire a non-trivial dependence in the remaining finite cross-ratio $u_{s2}:=1-z$, see eq.~\eqref{blocks_dc_v1zero_final}.

\subsection{Reproducing direct channel terms from crossed channel}
\label{ssec:5ptBS_CC}

We will now proceed to the term-by-term analysis of the direct channel \eqref{eq:5ptdirect}
expanded up to order $(\ep_{15}\ep_{34})^{h_\star}$, reproducing it term by term in the crossed
channel. 
Similarly to the standard works~\cite{Fitzpatrick:2012yx, Komargodski:2012ek}, in all of the following computations we assume that the direct-channel contributions are reproduced uniquely by fields whose twist at large spin matches the leading power of the left-hand side. This has been proven in the context of four-point functions in~\cite{Pal:2022vqc}, and is thus a natural assumption to make in a higher-point setting.
The third term that does not involve identity exchange in the direct channel will
require a separate discussion for each of the three alternative cases $(-,0,+)$ listed above.
We shall indeed find that all direct channel terms can be reproduced based on what is known
about the behavior of anomalous
dimensions and operator product coefficients from previous work on four- and five-point lightcone bootstrap, see in particular
\cite{Fitzpatrick:2012yx} and \cite{Antunes:2021kmm}. But there is one notable exception. In
order to reproduce the terms without identity exchange for case $(+)$, the operator
product coefficients involving two double-twist families must satisfy a novel sum rule
that involves a summation/averaging over tensor structures. We will actually be able to
resolve this sum rule and derive the full dependence of these operator product
coefficients on the tensor structure in the next subsection, but this requires leaving the strict lightcone limit \eqref{eq:5ptLCL}.

\subsubsection{Single internal identity exchange in direct channel}
\label{ssec:5ptBS_CCsingle}

\begin{figure}[ht]
		\centering 
  \includegraphics[scale=0.1]{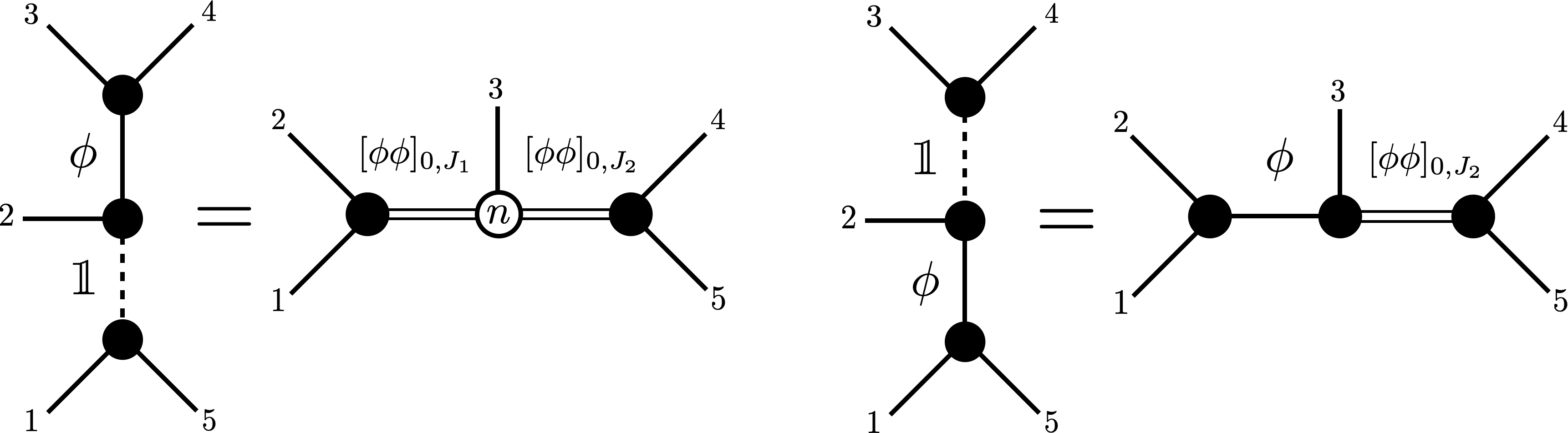}
		\
		\caption{
			Lightcone bootstrap of five-point correlator (with an identity exchange).
			\label{fig:case2_identity}
		}	
\end{figure}

Let us begin with the two direct channel terms that involve a single identity exchange,
either in the operator product $(15)$ or in the operator product $(34)$. Let us stress
again that double identity exchange is excluded so that the terms we consider here
represent the leading contributions to the lightcone limit of the direct channel. A
graphical representation of the relevant crossing equations is given 
in fig.~\ref{fig:case2_identity}.

\paragraph{$[\mathbf{1}\vert\phi]$-exchange in the direct channel.}\label{identity15}
As we explained above, the first term on the left-hand side of the direct channel
decomposition \eqref{eq:5ptdirect} is associated with the exchange of $[\oo_1\vert\oo_2]
= [\oo_{15}\vert\oo_{34}] = [\textbf{1}\vert\phi]$ in the direct channel. Such a term can
contribute non-trivially to the direct channel whenever the field $\phi$ appears in the
operator product of $\phi$ with itself. If this is the case, the leading term in the
direct channel is the first term in eq.~\eqref{eq:5ptdirect}. Taking into account the
prefactor on the right-hand side of the crossing symmetry equation \eqref{eq:5ptcrossing}
we deduce that
\begin{equation}\label{cross-15}
C_{\phi\phi\phi} \left(\frac{u_{s1} u_{s4}}{u_{s5}} \right)^{2 h_\phi} + ...
=  \sum_{\oo_1,\oo_2,n} P_{\oo_1\oo_2}^{(n)} \psi_{\oo_1\oo_2;n}^{\CC}.
\end{equation}
Here, $\dots$ denotes other (subleading) terms in the direct channel. In our lightcone
regime \eqref{eq:5ptLCL}, we first take $u_{s5} \ll u_{s3} \ll u_{s2}\ll 1$. Then, in
order to match the behavior of the leading direct channel term on the remaining cross
ratios $u_{s1}$ and $u_{s4}$, we conclude that the limiting crossed channel blocks must
obey
\begin{equation}
\psi_{\oo_1\oo_2;n}^{\CC}(u_{si}) \sim (u_{s1} u_{s4})^{2h_\phi}
\widetilde{\psi}^{\CC}_{h_1h_2}(u_{s2},u_{s3},u_{s5}),
\label{double_twist}
\end{equation}
This implies double-twist exchange in both operator products of the crossed channel, i.e.
we conclude that the direct channel terms must be reproduced from terms in the crossed
channel that involve $\oo_1, \oo_2 = [\phi\phi]_{0,J}$.

Next, we observe that the leading contribution $u_{s5}^{-2h_\phi}$ is Casimir singular
with respect to the relevant terms of both crossed-channel Casimir operators. Indeed,
using the explicit expressions we spelled out in eqs.~\eqref{D12_m10m1} and
\eqref{D45_m1m10} we obtain
\begin{align}
\big[	\mathcal{D}_{12}^{(-1,0,-1)} \big]^n  u_{s5}^{-2 h_\phi}  &= (2h_\phi)_n
(h_\phi)_n u_{s5}^{-2h_\phi-n} u_{s2}^{-n}\,,\\[2mm]
\big[	\mathcal{D}_{45}^{(-1,-1,0)} \big]^n  u_{s5}^{-2 h_\phi}&= (2h_\phi)_n
(h_\phi)_n u_{s5}^{-2h_\phi-n} u_{s3}^{-n}\label{D45id15}\,.
\end{align}
More specifically, the actions of the leading Casimir operators $\mathcal{D}_{12}$
and $\mathcal{D}_{45}$ increase the singularities by an order $\orm(\epsilon_{15}^{-1}
\epsilon_{23}^{-1})$ and $\orm(\epsilon_{15}^{-1}\epsilon_{34}^{-1})$ respectively. In
direct analogy to the four-point discussion above eq.~\eqref{inf_support}, this
implies that the direct channel has infinite support when expanding in a basis of
eigenvectors of the crossed-channel differential operators. This may be understood
by acting with one of the leading Casimirs (say $\mathcal{D}_{12}^{(-1,0,-1)}$) on
the crossing equation \eqref{cross-15}: the order of $\ep_{15}^{-1}\ep_{23}^{-1}$
only matches on both sides when the crossed channel blocks have $\la_{1}=
\orm(\ep_{15}^{-1}\ep_{23}^{-1})$. Similarly, acting with  $\mathcal{D}_{45}^{(-1,-1,0)}$
yields a match only for $\la_2= \orm(\ep_{15}^{-1}\ep_{23}^{-1})$. 
We can analogously understand the scaling of the tensor structure label by acting with the vertex operator we discussed in section~\ref{sect:VertexOp}. Acting with the leading vertex operator~\eqref{eq:vertex_op_caseI} on the left-hand side of eq.\ \eqref{cross-15}, we get:
\begin{equation}
    \mathcal{V}^{(-1,-1,-1)} \left(\frac{u_{s1} u_{s4}}{u_{s5}} \right)^{2 h_\phi}= \frac{h_\phi^3}{u_{s2}u_{s3}u_{s5}}\left(2h_\phi-\frac{d-2}{2}\right)\left(\frac{u_{s1} u_{s4}}{u_{s5}} \right)^{2 h_\phi}.
    \label{eq:vop_on_id_phi}
\end{equation}
Unless the field $\phi$ saturates the unitarity bound $2h_\phi=\frac{d-2}{2}$, which corresponds to it being a free scalar and having a vanishing five-point function, we see that the action of $\mathcal{V}^{(-1,-1,-1)}$ increases the singularities by an order $\orm(\ep_{15}^{-1}\ep_{23}^{-1}\ep_{34}^{-1})$, which requires its eigenvalue to scale as in eq.~\eqref{eq:Case1_Vop_eigenf}. For the superposition~\eqref{eq:integral_Vop_blocks} with coefficients~\eqref{eq:feta} to make sense, we see that the tensor structure label $\eta$ needs to scale as $\orm(\ep_{15}^{-1})$. We can thus conclude that, to
reproduce the correct behavior in the crossed channel, we must sum over the case
I blocks of eq.~\eqref{case_I_regime}. 

Having identified the correct regime, we can now address our goal to reproduce the
direct channel from a sum over crossed channel lightcone blocks in the case~I regime.
Using the limits of blocks computed in the previous section (see eq.~\eqref{sol_5pt_I}),
we can indeed reproduce the correct asymptotics from a large spin and tensor structure integral
\begin{equation}\label{cross5ptlargeJ}
C_{\phi\phi\phi} u_{s5}^{-2h_\phi}=(u_{s2}u_{s3})^{h_\phi} \int_{\left[\orm(1),\infty \right)^3}
 \dd \eta \frac{\dd \la_1}{4 \sqrt{\la_1}} \,\frac{\dd \la_2}{4 \sqrt{\la_2}}
 \,\nn_{(h_a,\bar h_a;\eta)}^{\CC,\mrmI}
 P_{[\phi\phi]_{0,J_1}[\phi\phi]_{0,J_2}}^{(\eta)}
  \, e^{-\eta u_{s5}-\frac{\la_1 u_{s2}+\la_2 u_{s3}}{\eta}},
\end{equation}
where $\la_a = \bar h_a^2+\orm(\bar h_a)$ and $\bar h_a=J_a+\orm(1)$ at leading order. In writing
the crossing equation we have divided both sides by $(u_{s1}u_{s4})^{2h_\phi}$. Through
explicit computation of the integrals it is not difficult to verify that the crossing
equation is satisfied if the coefficients $P$ in the crossed channel obey
\begin{equation}\label{NPJphiJ}
\frac{\nn_{(h_a,\bar h_a;\eta)}^{\CC,\mrmI}}{16 \sqrt{\la_1\la_2}} \,
P_{[\phi\phi]_{0,J_1}[\phi\phi]_{0,J_2}}^{(\eta)} = C_{\phi\phi\phi}
\frac{\left(\la_1\la_2\right)^{\Delta_\phi/2-1}}{\eta\,\Gamma^2(\Delta_\phi/2)\Gamma(\Delta_\phi)}.
\end{equation}
Using our formula \eqref{norm_case_I} for the normalization $\nn^{\CC,\mrmI}$
of the case~I lightcone blocks we obtain
\begin{equation}\label{PJphiJ}
\, P_{[\phi\phi]_{0,J_1}[\phi\phi]_{0,J_2}}^{(\eta)} = C_{\phi\phi\phi} 4^{-J_1 - J_2}
\frac{32 \times 16^{-\Delta_\phi}}{\Gamma^2(\Delta_\phi/2) \Gamma(\Delta_\phi)}
\eta^{-2\eta-\Delta_\phi} e^{2\eta} (J_1J_2)^{\eta + \frac32(\Delta_\phi - 1)} \ .
\end{equation}
Here we have used that $h_a = \Delta_\phi = 2 h_\phi, \bar h_a = h_a + J_a$ and
$\lambda_a \sim J_a^2$. Let us note that $\bar h_a$ can be replaced by $\bar h_a
\rightarrow J_a$ except if $\bar h_a$ appears in the exponent, i.e. in the factor
$4^{\bar h_a}$ of the normalization \eqref{norm_case_I}. Our result coincides with
the findings of \cite{Antunes:2021kmm}. In order to compare the two expressions one
should observe that our normalization conventions for blocks differ slightly from
those used by Antunes et al., see \cite[footnote~3]{Antunes:2021kmm}. This difference
of conventions gives rise to an additional factor $2^{J_1+J_2}$.

\paragraph{$[\phi\vert\textbf{1}]$-exchange in the direct channel.}
Let us now look at the second term $\left(u_{s2}u_{s5} \right)^{h_\phi} $ in the direct
channel that arises from the exchange of an identity field in the $(34)$ OPE.  Compared
to the leading term in eq.~\eqref{eq:5ptdirect}, the second term is subleading in
$X_{15}$, but higher order in $X_{12}$. In the presence of this second term, the
crossing equation \eqref{cross-15} is modified to
\be\label{cross-15-2}
C_{\phi\phi\phi} \Big[\left(\frac{u_{s1} u_{s4}}{u_{s5}} \right)^{2h_\phi} +
\left(\frac{u_{s1}u_{s2}u_{s4}^2 }{u_{s3}u_{s5}} \right)^{h_\phi}\Big]+... =
\sum_{\oo_1,\oo_2,n} P_{\oo_1\oo_2}^{(n)} \psi_{\oo_1\oo_2;n}^{\CC} .
\ee
Once again, $\dots$ denote other (subleading) terms in the direct channel. Since we have
already reproduced the first term on the right-hand side, we would now like to see how
the second term emerges from the crossed-channel sum on the right-hand side.

As a first step, we observe that the subleading direct channel term contains
$u_{s1}^{h_\phi}$ rather than the $u_{s1}^{2h_\phi}$ that appears in the first term.
Consequently, we expect to reproduce the subleading term from an exchange of the
field $\phi$ itself in the $(12)$ OPE of the crossed channel. For the $(45)$ OPE
the situation is unchanged with respect to the leading term, i.e.\ the crossed
channel must involve the exchange of double-twist operators in the $(45)$ OPE.

At the same time, we now find that the direct channel term under consideration is
in the kernel of the crossed channel Casimir operator $\mathcal{D}_{12}$ at both
leading $\orm(\ep_{15}^{-1}\ep_{23}^{-1})$ and subleading $\orm(\ep_{15}^{-1}
\ep_{23}^0)$ order --- these two properties are easily verified using the explicit formulas \eqref{D12_m10m1}
and \eqref{D12_m100} spelled out above. In fact, outside of any lightcone limit, the Casimir operator $\mathcal{D}_{12}$ is exactly diagonalized by the direct channel power law with eigenvalue $\frac{1}{2}\Dg_\phi(\Dg_\phi-d)$. We conclude that the
second term of the direct channel is associated with $\phi$ exchange in the
$(12)$ OPE of the crossed channel. On the other hand, the action of the second crossed-channel Casimir operator $\mathcal{D}_{45}$ is just as singular as in eq.\ \eqref{D45id15}.
Hence, we continue to have an infinite support of twist $h_2=2h_\phi$ operators in the
$(45)$ OPE with $\lambda_2= \orm(\ep_{15}^{-1}\ep_{34}^{-1})$.

According to these remarks, the lightcone blocks we need on the right-hand side satisfy
the same differential equations as case~II of our analysis in the previous section, see
eq.~\eqref{blocks_case_II}. But in contrast to what we discussed above, the parameters $J_1$
is now fixed to $J_1 = 0$ (which also implies $n=0$) and hence the normalization prescription must be modified. In
appendix~\ref{app:cross_channel}, we determine the correct normalization for the case at
hand and obtain
\begin{equation}
\psi_{\phi, [\phi\phi]_{0,J_2};0}^{\CC, \mrmII^\ast}(u_{si}) \sim
\frac{\Gamma(2h_\phi)}{\Gamma(h_{\phi})}
4^{\bar h_2} \sqrt{\frac{\bar h_2}{\pi}} \, (u_{s1} u_{s2} u_{s3} u^2_{s4} u_{s5})^{h_{\phi}}
\mathcal{K}_{h_\phi} (u_{s3}u_{s5}J_2^2).
\label{eq:blocks_caseIIstar}
\end{equation}
where the notation $\mrmII^\ast$ reminds us that $J_1=0$ and hence one should not copy
the normalization factor from the expression \eqref{norm_block_case_II}. Note that the asymptotics in eq.~\eqref{eq:blocks_caseIIstar} can also be determined from the vertex differential operator of section~\ref{sect:VertexOp}: these blocks are in fact eigenfunctions of the case~II vertex operator with eigenvalue~\eqref{t_of_N} for $N=0$, in the limit where $\la_1u_{s5}\rightarrow 0$.

We now insert these results into the crossing symmetry equation. After dividing both
sides of the equation by $C_{\phi\phi\phi}(u_{s1}u_{s2} u_{s4}^2)^{h_\phi}$, we
conclude
\begin{equation}
\left(u_{s3}u_{s5}\right)^{-h_\phi}= \frac{\Gamma(2h_\phi)}{\Gamma(h_{\phi})}
\int \frac{\dd J_2}{2} C_{\phi\phi [\phi\phi]_{0,J_2}}^2 \,   4^{2h_\phi+J_2}
\sqrt{\frac{J_2}{\pi}} \, (u_{s3} u_{s5})^{h_{\phi}} \mathcal{K}_{h_\phi}
(u_{s3}u_{s5}J_2^2).
\label{claim_phi_exch}
\end{equation}
Here, we have also factorized the coefficients in the crossed channel as
$P_{\mathcal{O}_1\mathcal{O}_2}^{(n)}=C_{\phi\phi\phi}C_{\phi\phi[\phi\phi]_{0,J_2}}^2$.
This integral equation is easily verified using the following expression for the square of
the OPE coefficient:
\begin{equation} \label{eq:CphiphidtphiJ}
 C_{\phi\phi[\phi\phi]_{0,J}}^2 = \frac{8}{\Gamma(2h_\phi)^2} \frac{\sqrt{\pi}}{4^{2 h_\phi+J}}
J^{4h_\phi - 3/2},
\end{equation}
which can be found e.g.~in \cite[eq.~(12)]{Fitzpatrick:2012yx}, with $\ell \leftrightarrow J$
and $\Dg_\phi\leftrightarrow 2h_\phi$. In conclusion, we have determined how the second term
in the direct channel can be reproduced from the crossed channel. In this case, however, the analysis does not provide access to any new dynamical data that goes beyond the four-point
lightcone bootstrap.

\subsubsection{No internal identity exchange in the direct channel}
\label{ssec:5ptBS_CCno}

In this subsection, we address the direct channel contributions that appear in the second
line of the expansion in eq.~\eqref{eq:5ptdirect}, in which there is no internal identity
operator exchanged. As we explained in section~\ref{ssec:5ptBS DC}, there are three
alternatives to consider which we denoted by $(-), (0)$ and $(+)$, respectively. Here we
shall address case $(0)$ first before discussing $(-)$ and $(+)$.
\begin{figure}[ht]
	\centering 
                    \includegraphics[scale=0.1]{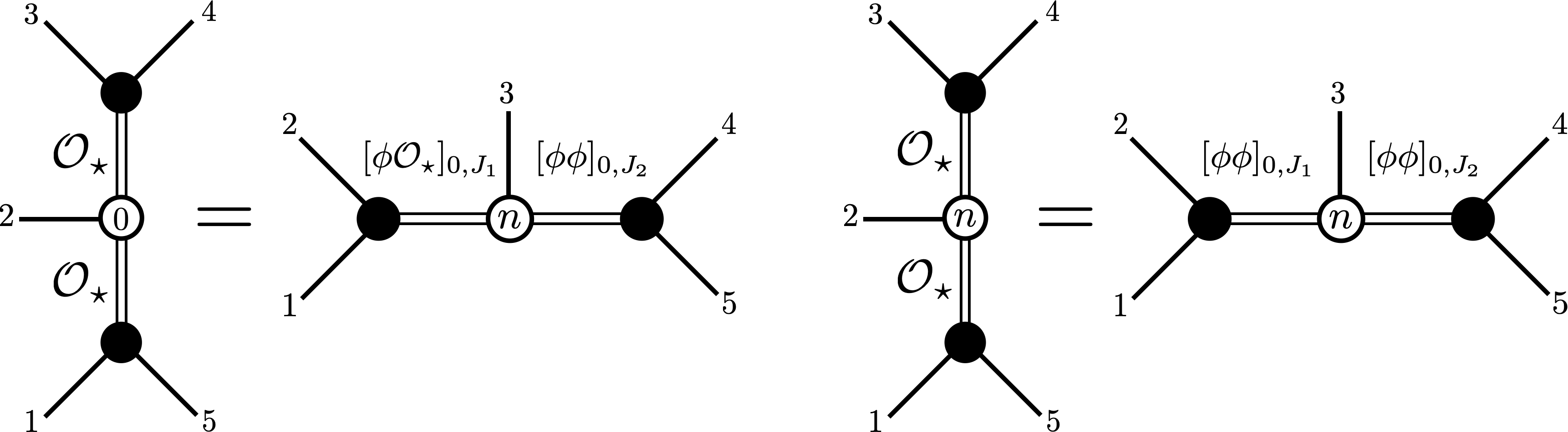}
	\
	\caption{
		Lightcone bootstrap of five-point correlator with non-identity leading-twist exchange in the direct channel. Left: lighter exchange $h_\star < h_\phi$. Right: heavier exchange $h_\star>h_\phi$. 
		\label{fig:case2_non_identity}
	}
\end{figure}

\paragraph{$[\phi\vert\phi]$-exchange in the direct channel.} Let us first consider the
case in which the field $\phi$ is exchanged in both the $(15)$ and $(34)$ OPE. In
particular, we shall assume that the field $\phi$ does appear in the operator product
of $\phi$ with itself. The associated direct channel contribution  is given by eq.\ 
\eqref{eq:5ptdirect}. The crossing symmetry equation is
\begin{equation}\label{cross15-2ext}
C_{\phi\phi\phi}^3 \nn^{\DC\, (0)}_{\phi\phi;0}(u^2_{s1}u_{s2}u^2_{s4} u_{s5}^{-1})^{h_\phi}\log u_{s1} \log u_{s2} +...
=  \sum_{\oo_1,\oo_2,n}P_{\oo_1\oo_2}^{(n)} \psi_{\oo_1\oo_2;n}^{\CC}.
\end{equation}
Here, we have not shown explicitly the leading terms involving identities, which were
accounted for earlier, i.e. the $\dots$ denote these leading terms as well as all the
subleading ones. With all the experience we have from the previous subsections, we can now skip the arguments that are mere repetitions of previous examples.

Let us first note that the $u_{s1}, u_{s4}$ asymptotics impose double-twist exchange in both
intermediate exchanges of the crossed channel, i.e. $\oo_a = [\phi\phi]_{0,J_a}$ by
the same argument as in section \eqref{double_twist}. Unlike the claim made in
\cite{Antunes:2021kmm}, however, the asymptotics of this direct channel contribution
\emph{cannot} be reproduced in every single block: The direct channel term that
results from $[\phi\vert\phi]$-exchange is still Casimir singular for the three
variables $u_{s2},u_{s3},u_{s5}$,
\begin{align}
\big[	\mathcal{D}_{12}^{(-1,0,-1)} \big]^n u_{s2}^{h_\phi}u_{s5}^{-h_\phi}\log u_{s2}
&= (n-1)!(h_\phi)_n u_{s2}^{h_\phi-n}u_{s5}^{-h_\phi-n}\\[2mm]
\big[	\mathcal{D}_{45}^{(-1,-1,0)} \big]^n u_{s3}^{0}u_{s5}^{-h_\phi}
&=(h_\phi)_n^2u_{s3}^{-n}u_{s5}^{-h_\phi-n}\,.
\end{align}
Therefore, as argued in section~\ref{identity15}, the crossed channel blocks must have the
eigenvalues $\la_1$ of order  $\orm(\ep_{15}^{-1}\ep_{23}^{-1})$ and $\la_2$ of order
$\orm(\ep_{15}^{-1}\ep_{34}^{-1})$, respectively. Similarly, one can use again the vertex operator argument of the same subsection to argue the scaling of the tensor structure label. Acting with the operator \eqref{eq:vertex_op_caseI} on the direct channel contribution, one gets
\begin{equation}
    \mathcal{V}^{(-1,-1,-1)}\!\left[(u^2_{s1}u_{s2}u^2_{s4} u_{s5}^{-1})^{h_\phi}\log u_{s1} \log u_{s2}\right]=-\frac{2(u^2_{s1}u_{s2}u^2_{s4} u_{s5}^{-1})^{h_\phi}}{u_{s2}u_{s3}u_{s5}}h_\phi^2\left(\!1\!+\!\left(2h_\phi-\frac{d-2}{2}\right)\log u_{s1}\!\right),
\end{equation}
which implies $\mathfrak{t}=\orm(\ep_{15}^{-1}\ep_{23}^{-1}\ep_{34}^{-1})$ and $\eta=\orm(\ep_{15}^{-1})$.
Consequently, the relevant crossing
equation is similar to eq.~\eqref{cross5ptlargeJ} studied in the previous
subsection, and we can use the case~I blocks of eq.~\eqref{sol_5pt_I} for the crossed-channel conformal block decomposition. The main new feature is of course the logarithmic factors
in $u_{s1}$ and $u_{s2}$. The case of the $\log u_{s1}$ factor is well known from the
four-point lightcone bootstrap: It can be reproduced in the crossed channel via
an anomalous correction to the weight of double-twist families that takes the form
\begin{equation}
h_1=h_{[\phi\phi]_{0,J_1}} = 2h_\phi + \frac{\gamma}{2 J_1^{2h_\phi}}+ ...
= 2h_\phi + \frac{\gamma}{2 \lambda_1^{h_\phi}} + \dots  \,.
\end{equation}
Indeed, from eq.~\eqref{sol_5pt_I}, we see that expanding the exponent $h_1$ of the
factor $u_{s1}^{h_1}$ around $2h_\phi$ gives the desired $\log u_{s1}$ at leading
order.

In order to understand the factor $\log u_{s2}$, we now study the sum (now approximated by an integral) on the right-hand side of the crossing symmetry equation \eqref{cross15-2ext}. After inserting the relevant lightcone blocks \eqref{sol_5pt_I}, we obtain
\begin{align}
\sum_{\oo_1,\oo_2,n}P_{\oo_1\oo_2}^{(n)} & \psi_{\oo_1\oo_2;n}^{\CC} = \\[2mm]
& = (u^2_{s1} u_{s2} u_{s3} u^2_{s4})^{h_\phi} \ \int \dd \eta \frac{\dd \la_1}{4 \sqrt{\la_1}}
 \frac{\dd \la_2}{4 \sqrt{\la_2}} \frac{\gamma \log u_{s1}}{2 \la_1^{h_\phi}}
 \nn^{\CC,\mrmI}  P_{[\phi\phi]_{0,J_1}[\phi\phi]_{0,J_2}}^{(\eta)}
  \, e^{-\eta u_{s5}-\frac{\la_1 u_{s2}+\la_2 u_{s3}}{\eta}}  \nonumber \\[2mm]
&   = - \frac{\gamma}{2} \frac{C_{\phi\phi\phi}}{\Gamma(\Dg_\phi)} \left(u_{s1}^2 u_{s2} u_{s4}^2 u_{s5}^{-1} \right)^{h_\phi} \log  u_{s1} \log  u_{s2}.
\end{align}
In order to reach the third line, we inserted the formula \eqref{NPJphiJ} for
the product $\nn P$ and performed the integration over $\lambda_1$ using
\begin{equation}
\int_{L_1^2}^\infty \frac{\dd \la_1}{\la_1} e^{-\la_1 u_{s2}\eta^{-1}}
= -\log u_{s2}+ \orm(1), \quad \forall L_1=\orm(1).
\end{equation}
Here, $L_1$ denotes a lower limit of spins above which the large spin approximation of the
blocks is valid. This same kind of large spin integral with logarithmic asymptotics appears in the four-point bootstrap at subleading orders --- see the discussion below \cite[eq.~(3.5)]{Fitzpatrick:2015qma}. The rest of the crossing equation is similar to eq.~\eqref{cross5ptlargeJ},
i.e. the integrals over large $\la_2$ and $\eta$ reproduce the correct powers law behavior
on the various cross-ratios. Comparison with the right-hand side of the crossing symmetry
equation yields
\begin{equation}
\frac{\Gamma(\Dg_\phi)^2}{\Gamma(\Dg_\phi/2)^4}\, C^3_{\phi\phi\phi} = - \frac{\gamma}{2} \frac{C_{\phi\phi\phi}}{\Gamma(\Dg_\phi)} \ .
\label{eq:gamma_from_phiphi}
\end{equation}
The result for the factor $\gamma$ in the anomalous dimension of double-twist families is
the same as for $\phi$ exchange in the direct channel of the four-point crossing equation,
see e.g. \cite[Eq.~(41)]{Fitzpatrick:2012yx} after setting the
quantum numbers involved to $\tau_m=\Delta_\phi$ and $\ell_m=0$. In addition, one identifies the operator product coefficients
of the four-point expansion as $P_m = C_{\phi\phi[\phi\phi]_{0,J_1}}^2$.

\paragraph{$[\mathcal{O}_\star\vert\mathcal{O}_\star]$-exchange in the direct channel,
$h_\star< h_\phi$}
\label{sssec:OO_light_star_lower}
We will now consider the case where the two exchanged operators in the direct channel are
identical and possess leading twist $h_\star <h_\phi$. The direct channel blocks for this
case $(-)$ were given in eq.~\eqref{eq:case-}, and we had already stressed that the
leading contribution comes solely from one block, namely the block with $n=0$.
Hence, the relevant term we need to reproduce from the crossed channel sum is
\begin{equation}\label{cross15-4}
\dots + P_{\oo_\star\oo_\star}^{(0)} \nn^{\DC\, (0)}_{\oo_\star\oo_\star;0}
u_{s1}^{h_\star + h_\phi} u_{s2}^{h_\star}  u^{h_\star - h_\phi}_{s3}
u_{s4}^{2h_\phi} u_{s5}^{h_\star - 2 h_\phi}
 +\dots =
\sum_{\oo_1,\oo_2,n} P_{\oo_1\oo_2}^{(n)} \psi^{\CC}_{\oo_1\oo_2;n}.
\end{equation}
In this case, the power law $u_{s1}^{h_\phi+h_\star}u_{s4}^{2h_\phi}$ implies that
operators $\oo_1 = [\phi\oo_\star]_{0,J_1}$  and $\oo_2 = [\phi\phi]_{0,J_2}$ are
exchanged in the two intermediate channels of the crossed channel. Furthermore, just
as for case $(0)$, the power law prefactor is Casimir singular with respect to the 
two Casimir operators $\mathcal{D}_{12}^{(-1,0,-1)}$ and 
$\mathcal{D}_{45}^{(-1,-1,0)}$,
	\begin{align*}
	\big[\mathcal{D}_{12}^{(-1,0,-1)}\big]^nu_{s2}^{h_\star}   u_{s5}^{{h_\star}-2 h_{\phi }}
&= (h_\phi-h_\star)_n  (2h_\phi-h_\star)_n \,u_{s2}^{h_\star} u_{s5}^{{h_\star}-2 h_{\phi }}, \\[2mm]
	\big[\mathcal{D}_{45}^{(-1,-1,0)}\big]^n   u_{s3}^{{h_\star}-h_{\phi }}  u_{s5}^{{h_\star}-2 h_{\phi }}
&=   (2h_\phi-h_\star)_n^2\,   u_{s3}^{{h_\star}-h_{\phi }}  u_{s5}^{{h_\star}-2 h_{\phi }}.
	\end{align*}
Finally, the direct channel power law is singular with respect to the vertex operator 
$\mathcal{V}^{(-1,-1,-1)}$ given in eq.~\eqref{eq:vertex_op_caseI}: 
 \begin{equation}
     \frac{\mathcal{V}^{(-1,-1,-1)}\left(u_{s1}^{h_\star + h_\phi} u_{s2}^{h_\star}  u^{h_\star - h_\phi}_{s3}
u_{s4}^{2h_\phi} u_{s5}^{h_\star - 2 h_\phi}\right)}{u_{s1}^{h_\star + h_\phi} u_{s2}^{h_\star}  u^{h_\star - h_\phi}_{s3}
u_{s4}^{2h_\phi} u_{s5}^{h_\star - 2 h_\phi}}=-2\frac{(h_\star-h_\phi)(h_\star-2h_\phi)^2\left(2h_\phi-\frac{d-2}{2}\right)}{u_{s2}u_{s3}u_{s5}}\,.
 \end{equation}
Our standard reasoning from previous subsections implies that, in order to reproduce the direct channel term, we need to sum over crossed channel lightcone blocks with eigenvalue
$\lambda_1$ of order $\orm(\ep_{15}^{-1}\ep_{23}^{-1})$, eigenvalue $\lambda_2$ of
order $\orm(\ep_{15}^{-1}\ep_{34}^{-1})$, and tensor structure label $\eta$ of order $\orm(\ep_{15}^{-1})$. Consequently, we use the case~I lightcone
blocks that were given in eq.~\eqref{sol_5pt_I}, in which case the crossed channel
sum reduces to
	\begin{align}
	\sum_{\oo_1,\oo_2,n} &P_{\oo_1\oo_2}^{(n)} \psi_{\oo_1\oo_2;n}^{\CC} =
\nonumber\\[2mm]
	&u_{s1}^{h_\star+h_\phi}u_{s4}^{2h_\phi}u_{s2}^{h_\star} u_{s3}^{2h_\phi-h_\star}
\int_{\Rs_+^3} \frac{\dd \eta\, \dd \la_1\dd\la_2}{16\sqrt{\la_1\la_2}} \,
\nn_{(h_a,\bar h_a;\eta)}^{\CC,\mrmI} P_{[\phi\oo_\star]_{0,J_1}
[\phi\phi]_{0,J_2}}^{(\eta)} e^{-\eta u_{s5}-\frac{\la_1 u_{s2}+\la_2u_{s3}}{\eta}}\ .
	\end{align}
In order to reproduce the left hand side of the crossing equation \eqref{cross15-4},
we must impose the following asymptotic behavior of the product
	\begin{equation}
	\frac{\nn_{(h_a,\bar h_a;\eta)}^{\CC,\mrmI}}{16\sqrt{\la_1\la_2}}
	 P_{[\phi\oo_\star]_{0,J_1}[\phi\phi]_{0,J_2}}^{(\eta)} =
P_{\oo_\star\oo_\star}^{(0)}\nn^{\DC\, (0)}_{\oo_\star\oo_\star;0}
\frac{(\eta^{-1}\la_1)^{h_\phi-h_\star-1}}{\Gamma(h_\phi-h_\star)}
\frac{(\eta^{-1}\la_2)^{2h_\phi-h_\star-1}}{\Gamma(2h_\phi-h_\star)}
\frac{\eta^{2 h_\phi-h_\star-3}}{\Gamma(2 h_\phi-h_\star)}.
	\end{equation}
at large quantum numbers $\eta,\lambda_1,\lambda_2$. After plugging in our
formulas \eqref{norm_case_I} and \eqref{eq:5ptLCLdcresult} for the relevant
normalizations $\nn$, we deduce
\begin{equation}\label{PJstarphiJ}
\, P_{[\phi\oo_\star]_{0,J_1}[\phi\phi]_{0,J_2}}^{(\eta)}
= P_{\oo_\star \oo_\star}^{(0)} \nn^{\DC}_{\oo_\star\oo_\star;0}
\frac{32 \times 4^{3h_\phi - h_\star}4^{-J_1 - J_2}e^{2\eta}}
{\Gamma(h_\phi-h_\star)\Gamma^2(2h_\phi-h_\star)}
 \eta^{-2\eta-2h_\phi}
J_1^{\eta + 3 h_\phi -2 h_\star-\frac32}
J_2^{\eta +4h_\phi- h_\star - \frac32},
\end{equation}
with
\begin{equation}
\nn^{\DC\, (0)}_{\oo_\star\oo_\star;0} = \frac{\Gamma^2(2h_\star) \Gamma^2(h_\phi-h_\star)}
{\Gamma^2(h_\star) \Gamma^2(h_\phi)}\ .
\end{equation}
To evaluate the crossed channel quantities we have used that $h_1 \rightarrow h_\phi+h_\star,
h_2 \rightarrow 2 h_\phi$ and $\bar h_a= h_a + J_a \rightarrow \sqrt \lambda_a$. In the direct channel,
on the other hand, we have $h_a = \bar h_a = h_\star$.  Our result coincides with the
findings of \cite{Antunes:2021kmm}, see eq.~(48) therein. In order to compare the two expressions,
one should observe that our normalization conventions for blocks differ slightly from those
used by Antunes et al., see \cite[footnote~3]{Antunes:2021kmm}. This difference of conventions
gives rise to an additional factor $2^{J_1+J_2}$. In addition, the labels $J_1$ and $J_2$
are exchanged. 

With this result, we have reproduced all computations of \cite{Antunes:2021kmm} pertaining to the scalar five-point function. In this context, our derivation explicitly shows that these OPE coefficients lie in a case~I regime where the tensor structure label $n$ scales to infinity faster than the spins $J_1,J_2$. In contrast, the derivation of \cite{Antunes:2021kmm} was based on a permutation of the order of limits in the crossed channel from $(\ep_{15},\ep_{34},\ep_{12},\ep_{45},\ep_{23})$ to $(\ep_{12},\ep_{45},\ep_{23},\ep_{34},\ep_{15})$, in which case $J_1,J_2\rightarrow \infty$ before $n\rightarrow \infty$. In general, it is not clear that these results will continue to predict the leading asymptotics of OPE coefficients in this permuted order of limits. Nonetheless, in the explicit example \cite[Sec.~4.2]{Antunes:2021kmm} of $\phi^3$ theory in $d=6-\varepsilon$ dimensions, which we will review in section~\ref{ssec:applications} in more detail, the authors show that the case~I asymptotics predicted from the lightcone bootstrap indeed coincide with the leading $J_1,J_2 \gg n \gg 1$ behavior of the OPE coefficients. To avoid the permutation of limits, it would be interesting to clarify the physical interpretation of this case~I regime and, more generally, to better understand the analyticity properties of OPE coefficients in the tensor structure label.

\paragraph{$[\mathcal{O}_\star\vert\mathcal{O}_\star]$-exchange in the
direct channel, $2h_\phi> h_\star> h_\phi$}
\label{sssec:OO_light_star_greater}

Let us now address the last case in which the leading-twist contribution $\oo_\star$
to the internal exchange in the crossed channel has $2h_\phi>h_\star > h_\phi$. As we have
pointed out in the first subsection, an infinite number of crossed
channel blocks can contribute in this case  at the same order, since the leading asymptotics
do not depend on the quantum number $n$. Using our formula \eqref{eq:case+}
for crossed channel lightcone blocks, the crossing equation becomes
\begin{equation}\label{caseIIcross0}
	\dots +  u_{s1}^{2h_{\phi }}u_{s2}^{h_\phi} u_{s3}^{{h_\star}-h_{\phi }}
u_{s4}^{2 h_{\phi }} u_{s5}^{{h_\star}-2 h_{\phi }}
\sum_{n} P_{\oo_\star\oo_\star}^{(n)} \nn^{\DC\, (0) }_{\oo_\star\oo_\star;n}+\dots =
\sum_{\oo_1,\oo_2,n} P_{\oo_1\oo_2}^{(n)} \psi^{\CC}_{\oo_1\oo_2;n}(u_{si}).
	\end{equation}
The familiar power law $(u_{s1}u_{s4})^{2h_\phi}$ for the dependence on the
cross-ratios $u_{s1}$ and $u_{s4}$ implies double-twist exchange of
$[\phi\phi]_{0,J_i}$ in the $(12)$ and $(45)$ OPE of the crossed channel.

The first interesting new feature of the case under consideration arises
from the kernel condition. In fact, using our explicit formula \eqref{D12_m10m1}
for the leading term of the crossed channel Casimir operator $\mathcal{D}_{12}$, we find
\begin{equation}
\mathcal{D}_{12}^{(-1,0,-1)} u_{s2}^{h_\phi} u_{s1}^{2h_{\phi }}
u_{s3}^{{h_\star}-h_{\phi }} u_{s4}^{2 h_{\phi }} u_{s5}^{{h_\star}-2 h_{\phi }}= 0.
\end{equation}
On the other hand, when we apply the next-to-leading term \eqref{D12_m100} in the Casimir operator
$D_{12}$ to the same power law on the left-hand side of the crossing equation
\eqref{caseIIcross0}, we find
\begin{align*}
	\big[\mathcal{D}_{12}^{(-1,0,0)}\big]^n u_{s2}^{h_\phi} u_{s1}^{2h_{\phi }}
u_{s3}^{{h_\star}-h_{\phi }} u_{s4}^{2 h_{\phi }} u_{s5}^{{h_\star}-2 h_{\phi }}
&=  (2h_\phi-h_\star)_n(h_\phi)_n u_{s2}^{h_\phi} u_{s1}^{2h_{\phi }}
u_{s3}^{{h_\star}-h_{\phi }} u_{s4}^{2 h_{\phi }} u_{s5}^{{h_\star}-2 h_{\phi }-n}  \ ,
	\end{align*}
i.e.\ the direct channel is $\mathcal{D}_{12}^2$-singular at order $\orm(\ep_{15}^{-1})$
but not at leading order.  At the same time, the action of the leading term \eqref{D45_m1m10}
in the second crossed channel Casimir operator $\mathcal{D}_{45}$ gives
	\begin{equation}\label{D45action}
		\big[\mathcal{D}_{45}^{(-1,0,-1)}\big]^n u_{s2}^{h_\phi}
u_{s1}^{2h_{\phi }} u_{s3}^{{h_\star}-h_{\phi }} u_{s4}^{2 h_{\phi }}
u_{s5}^{{h_\star}-2 h_{\phi }} = ((2h_\phi-h_\star)_n)^2 u_{s2}^{h_\phi}
u_{s1}^{2h_{\phi }} u_{s3}^{{h_\star}-h_{\phi }} u_{s4}^{2 h_{\phi }}
u_{s5}^{{h_\star}-2 h_{\phi }-n}\,.
	\end{equation}
In comparison with the discussion in section~\ref{identity15}, the crossed channel is now
dominated by lightcone blocks with eigenvalue $\lambda_1$ of order $\orm(\ep_{15}^{-1})$ and
eigenvalue $\lambda_2$ of order $\orm(\ep_{15}^{-1}\ep_{34}^{-1})$. To see the way the tensor structure label scales, we can use the case~II vertex operator of section~\ref{sect:VertexOp}. Its action on the direct channel contribution reads
\begin{equation}
    \frac{\mathcal{V}^{(-1,-1,0)}\left(u_{s2}^{h_\phi}
u_{s1}^{2h_{\phi }} u_{s3}^{{h_\star}-h_{\phi }} u_{s4}^{2 h_{\phi }}
u_{s5}^{{h_\star}-2 h_{\phi }}\right)}{u_{s2}^{h_\phi}u_{s1}^{2h_{\phi }} u_{s3}^{{h_\star}-h_{\phi }} u_{s4}^{2 h_{\phi }}
u_{s5}^{{h_\star}-2 h_{\phi }}}=\frac{\left(h_\star-2h_\phi\right)^2\left(36 h_\phi^2-12 d h_\phi+d^2-3d\right)}{6 u_{s3}u_{s5}}\,,
\end{equation}
which shows that its eigenvalue scales like $\lambda_2$, that is, precisely like in eq.\ \eqref{t_of_N} 
with $N$ finite.
This regime corresponds
precisely to the case~II blocks \eqref{blocks_case_II} we studied at the end of the
previous section. Consequently, the crossed-channel sum on the right-hand side of
equation \eqref{caseIIcross0} becomes
	\begin{align}\label{caseIIcross}
	&\sum_{\oo_1,\oo_2,n} P_{\oo_1\oo_2}^{(n)} \psi_{\oo_1\oo_2;n}^{\CC}(u_{si}) =
\\[2mm]
	&\left({u_{s1}^2 u_{s4}^2 u_{s2}u_{s3} u_{s5}}\right)^{h_\phi}\!\! \int_{\Rs_+^2}
\!\frac{\dd \la_1\dd \la_2}{16 \sqrt{\la_1\la_2}} \sum_{\dg n=0}^\infty
\nn_{[\phi\phi]_{0,J_1}[\phi\phi]_{0,J_2};J_1-\dg n}^{{\CC,\mrmII}}P_{[\phi\phi]_{0,J_1}
[\phi\phi]_{0,J_2}}^{(J_1-\dg n)} \mathcal{K}_{h_\phi+\dg n}(\la_1 u_{s5}+\la_2u_{s3}u_{s5}).
\nonumber
	\end{align}
In order to match this crossed-channel integral with the direct-channel term on the
left-hand side of the crossing equation \eqref{caseIIcross0}, we propose the following
Ansatz:
\begin{equation}
\frac{\nn_{[\phi\phi]_{0,J_1}[\phi\phi]_{0,J_2};J_1-\dg n}^{{\CC,\mrmII}}}
{16\sqrt{\la_1\la_2}} P_{[\phi\phi]_{0,J_1}[\phi\phi]_{0,J_2}}^{(J_1-\dg n)} =
b_{\dg n }\frac{2 \la_1^{h_\phi+\dg n-1}}{\Gamma(h_\phi+\dg n) \Gamma(2h_\phi-h_\star)}
\frac{\la_2^{2h_\phi-h_\star-1}}{\Gamma(2h_\phi-h_\star)},
	\label{ope_coeffs_case_II}
	\end{equation}
where $b_{\dg n}$ is a set of constants that depend on the choice of tensor structure.
As we shall see, these constants cannot be determined by our equation \eqref{caseIIcross0}.
After plugging the Ansatz \eqref{ope_coeffs_case_II} into the right-hand side of eq.~\eqref{caseIIcross},
we evaluate the two integrations over $\lambda_1$ and $\lambda_2$ with the help of the following
integral formula:
	\begin{equation}
	  \int_{\Rs_+^2} \frac{\dd x}{x} \frac{\dd y}{y} x^{\ag} y^\bg \mathcal{K}_\ag(x+y)=
\frac{1}{2} \Gamma(\ag) \Gamma(\bg)^2,
	  \label{eq:double_integral}
	\end{equation}
for $\ag=h_\phi+\dg n$ and $\bg = 2 h_\phi-h_\star$. This identity is proven in appendix~\ref{app:double_integral}. Once the dust settles, we see that the crossing equation
\eqref{caseIIcross0} is indeed satisfied, provided that the undetermined constants
$b_{\dg n}$ satisfy the following sum rule:
	\begin{equation}
	\sum_{n=0}^{J_\star} P_{\oo_\star\oo_\star}^{(n)} \nn^{\mathrm{DC}}_{\oo_\star\oo_\star;n}
=  \sum_{\dg n=0}^\infty b_{\dg n }
	\label{sum_rule}
	\end{equation}
In other words, while the crossing symmetry equation \eqref{caseIIcross0} suffices to
determine the dependence of the operator product coefficients in eq.~\eqref{ope_coeffs_case_II}
on the spins $J_1$ and $J_2$ of the intermediate fields, it only constrains the `average' over
the choice $\delta n$ of tensor structure. It is clear that we could not have done better here
because the lightcone blocks from case~II depend nontrivially on only two of the five
cross-ratios, namely on $u_{s3}$ and $u_{s5}$. The dependence of the other three cross-ratios
is through a simple power law that matches directly the dependence of the lightcone blocks in the
direct channel and moreover does not depend on $\delta n$. Hence, at the order we have
considered here, crossing symmetry is unable to resolve the $\delta n$ dependence of the
operator product coefficients beyond the sum rule \eqref{sum_rule}. In the next subsection,
we will explain how to determine the coefficients $b_{\dg n}$ in eq.~\eqref{ope_coeffs_case_II}
by solving the crossing equation at subleading orders in $u_{s2}$.

\subsection{Solving OPE coefficients for discrete tensor structures}
\label{ssec:discrete_tensor_structures}
In the last subsection, we determined the operator product coefficients 
involving two double-twist families in the limit of large tensor structure 
label $\eta$, see eqs.~\eqref{PJphiJ} and \eqref{PJstarphiJ}. Through the analysis 
of the five-point crossing equation \eqref{caseIIcross0} in theories for which 
the leading-twist field $\oo_\star$ that appears in the operator product of 
$\phi$ with itself has weight $h_\star > h_\phi$, we were able to at least 
constrain the operator product coefficients with finite tensor structure 
label $\delta n$ through the sum rule \eqref{sum_rule}. Our goal in this 
subsection is to do better and to fully determine the operator product 
coefficients for finite tensor structure label $\delta n = J_1-n$.  Our 
comments in the final paragraph of the previous subsection suggest that our 
goal can be reached if we manage to analyze corrections to the crossing 
symmetry constraint \eqref{caseIIcross0} that are subleading in the 
cross-ratio $u_{s2}$, i.e. we should look at the crossing equation in the 
regime
\begin{equation}
\LCL^{(4)}_{\vec{\ep}}: X_{15} \ll X_{34} \ll X_{12} \ll X_{45} \ll 1,
\end{equation}
which imposes no condition on $X_{23}$. Fortunately, we have already determined 
both direct channel and case~II crossed channel blocks in this regime. For the 
direct channel, we constructed the full $u_{s2}$ dependence of the  relevant blocks 
in eq.~\eqref{blocks_dc_v1zero_final}, on the way from the partial lightcone limit 
$\LCL^{(2)}$ to the full lightcone limit.  By relaxing the last lightcone limit and 
summing over crossed channel blocks in a different scaling regime of spins,  we can 
also bootstrap the leading contributions with $h_\star \leq h_\phi$ to derive OPE 
coefficients in the less stringent scaling limit $J_2\gg J_1,n \gg 1$ with 
$J_1-n=\dg n=0,1,2,\dots,\infty$.

\subsubsection{Single internal identity exchange in the direct channel}
\label{sssec:singleid_CaseII}
\paragraph{$[\mathbf{1}\vert\phi]$-exchange in the direct channel.} Relaxing the 
$u_{s2}\rightarrow 0$ limit does not change the leading contribution:
\begin{equation}\label{cross-15-repeated}
C_{\phi\phi\phi} \left(\frac{u_{s1} u_{s4}}{u_{s5}} \right)^{2 h_\phi} + ...
=  \sum_{\oo_1,\oo_2,n} P_{\oo_1\oo_2}^{(n)} \psi_{\oo_1\oo_2;n}^{\CC}.
\end{equation}
We have constructed the full $u_{s2}$ dependence of the case~II crossed channel 
blocks,  see eq.~\eqref{blocks_case_II} in section~\ref{sect:five_pt_crossed_blocks}:
\begin{eqnarray}
& & \psi_{([\phi\phi]_{0,J_1},[\phi\phi]_{0,J_2};J_1-\dg n)}^{(12),(45)} 
(u_{si}) \\[2mm]
& & \quad \quad \stackrel{\LCL^{(4)}}{\sim}
\nn_{( [\phi\phi]_{0,J_1},[\phi\phi]_{0,J_2};
J_1-\dg n)}^{\mathrm{LS,\mrmII}}(u_{s1}^2u_{s2}u_{s3}u_{s4}^2u_{s5})^{h_\phi}
(u_{s5} (1-u_{s2}))^{\dg n}\, \mathcal{K}_{h_\phi+\dg n}(u_{s3}u_{s5} J_2^2). 
\nonumber 
\end{eqnarray}
The normalization $\nn_{( [\phi\phi]_{0,J_1},[\phi\phi]_{0,J_2};J_1-\dg n)}^{\CC,\mrmII} 
:= \nn_{(2h_\phi,2h_\phi,2h_\phi+J_1,2h_\phi+J_2;J_1-\dg n)}^{\CC,\mrmII}$ is given in 
eq.~\eqref{norm_block_case_II}.  Hence, there is nothing that prevents us from evaluating 
crossing symmetry in the regime $\LCL^{(4)}$,  without sending $u_{s2}$ to zero.  In this 
case, the crossing equation becomes
\begin{multline*}
C_{\phi\phi\phi}u_{s5}^{-2h_\phi}=
u_{s3}^{h_\phi}(1-z)^{h_\phi} \\
\times \sum_{\dg n=0}^\infty z^{\dg n} \int \frac{\dd^2\la}{16\sqrt{\la_1\la_2}} 
\nn_{( [\phi\phi]_{0,J_1},[\phi\phi]_{0,J_2};J_1-\dg n)}^{{\CC,\mrmII}}
P_{[\phi\phi]_{0,J_1}[\phi\phi]_{0,J_2}}^{(J_1-\dg n)} u_{s5}^{h_\phi+\dg n}
\mathcal{K}_{h_\phi+\dg n}(\la_1 u_{s5}+\la_2u_{s3}u_{s5}). 
\end{multline*}
 Homogeneity of the equation in $\ep_{15},\ep_{34}$ imposes $\mathcal{N}P \propto 
 \la_1^{2h_\phi+\dg n-1/2}\la_2^{h_\phi-1/2}$, so without loss of generality we can write
\begin{equation}
\frac{\nn_{( [\phi\phi]_{0,J_1},[\phi\phi]_{0,J_2};J_1-\dg n)}^{{\CC,\mrmII}}
P_{[\phi\phi]_{0,J_1}[\phi\phi]_{0,J_2}}^{(J_1-\dg n)} }{16\sqrt{\la_1\la_2}}=
C_{\phi\phi\phi} \frac{2 \la_1^{2h_\phi+\dg n-1} \la_2^{h_\phi-1}}{\Gamma(2h_\phi+\dg n)
\Gamma(h_\phi)\Gamma(2h_\phi)} a_{\dg n},
\end{equation}
for some series of coefficients $a_{\dg n}$.  Plugging this back into the crossing equation 
and using the double integral formula of appendix~\ref{app:double_integral} yields a sum 
rule for the $a$-coefficients:
\begin{equation}
1=  (1-z)^{h_\phi} \sum_{\dg n=0}^\infty  a_{\dg n} z^{\dg n}.
\end{equation}
The sum rule is easily solved by imposing $a_{\dg n}= (h_\phi)_{\dg n}/\dg n!$, such that the full solution is
\begin{equation}
\frac{\nn_{(2h_\phi,2h_\phi,2h_\phi+J_1,2h_\phi+J_2;J_1-\dg n)}^{\CC,\mrmII}
P_{[\phi\phi]_{0,J_1}[\phi\phi]_{0,J_2}}^{(J_1-\dg n)}}{16 J_1J_2} \sim 
C_{\phi\phi\phi} \frac{2J_1^{4h_\phi+2\dg n-2}J_2^{2h_\phi-2}}{\Gamma(h_\phi)
\Gamma(2h_\phi)^2} \frac{(h_\phi)_{\dg n}}{\dg n! (2h_\phi)_{\dg n}}.
\label{caseII_1phiexch}
\end{equation}
 The quantity $P$ on the left-hand side is 
given by the product 
\begin{equation} \label{eq:PtoCCC} 
P_{[\phi\phi]_{0,J_1}[\phi\phi]_{0,J_2}}^{(J_1-\dg n)} 
= C_{\phi\phi[\phi\phi]_{0,J_1}} 
C^{(J_1-\dg n)}_{[\phi\phi]_{0,J_2}\phi[\phi\phi]_{0,J_1}} 
C_{\phi\phi[\phi\phi]_{0,J_2}}\ . 
\end{equation}
Since the operator product coefficients $C_{\phi\phi[\phi\phi]_{0,J}}$ 
for large $J$ are known (see eq.~\eqref{eq:CphiphidtphiJ}), our new formula \eqref{caseII_1phiexch} 
allows us to compute the coefficients in the center of the expression
\eqref{eq:PtoCCC} for large $J_a$ but any finite integer value of the label
$\dg n$. We will discuss this formula further in the section~\ref{ssec:applications} and 
test it through a non-trivial example. But before that, we will first go 
through the other direct channel exchanges.

\paragraph{$[\phi\vert\mathbf{1}]$-exchange in the direct channel.} As we saw previously,  
the identity exchange in (34) forces the exchange of $\phi$ in the (12) OPE of the crossed 
channel.  As a result, the crossed channel block decomposition is a specific example of case 
II scaling where $\la_1=0+\orm(1)$ is fixed.  The corresponding solution to the crossing 
equation was already computed in the second paragraph of section \ref{ssec:5ptBS_CCsingle}.
Hence, the $[\phi\vert\mathbf{1}]$ does not provide any new insight into OPE coefficients. 

\subsubsection{No internal identity exchange in the direct channel}
Even in the relaxed lightcone limit $LCL_{\vec{\ep}}^{(4)}$,  the exchange of $\oo_\star$ in 
the (15) OPE of the direct channel will yield differing asymptotics in the $\ep_{12}\rightarrow 0$ 
limit depending on whether $h_\star<h_\phi$, $h_\star=h_\phi$ or $h_\star >h_\phi$.  As we determined 
in the previous section, these $X_{12}$ asymptotics affect the twist of operators in the (12) OPE of 
the crossed channel: $[\phi\oo_\star]$ is exchanged when $h_\star \leq h_\phi$,  $[\phi\phi]$ is 
exchanged when $h_\star \geq h_\phi$, and an anomalous dimension appears when $h_\star =h_\phi$.   
Each contribution will have a non-trivial dependence on the finite cross-ratio $u_{s2}$ corresponding 
to the lightcone blocks for $\oo_\star$ exchange --- the latter were  computed in full detail in 
section \ref{ssec:5pt_blocks_dc}.

\paragraph{$[\oo_\star\vert\phi]$-exchange in the direct channel.} If there is $\phi$ exchange in 
the (34) OPE of the direct channel, then for any $\oo_\star$ exchange in the (12) OPE, the left hand side of 
the crossing equation will take the form
\begin{equation}\label{cross15-2ext-2}
\DC = \dots+C_{\phi\phi\phi} 
\left(\frac{u_{s1} u_{s4}}{u_{s5}} \right)^{2 h_\phi}u_{s5}^{h_\star} 
\frac{C_{\phi\phi\oo_\star}^2}{\mathrm{B}_{\bar h_\star}} \log u_{s1} u_{s2}^{h_\phi} \,
\hypg{h_\star}{2h_\phi-h_\star}{2h_\star}(z) +...
\end{equation}
If we define $\gamma_0:=2 \mathrm{B}_{\bar h_\star}^{-1}C_{\phi\phi\oo_\star}^2 
\Gamma(2h_\phi-h_\star)^2 \Gamma(2h_\phi)^{-2}$, then it is easy to verify that the direct channel 
is reproduced by the perturbation $\frac{1}{2}\gamma_0\frac{\ds}{\ds h_1} (P\psi) = 
\gamma_0\log u_{s1}(P\psi)+\dots $ of the direct channel sum computed in eq.~\eqref{caseII_1phiexch} 
for $[\mathbf{1}\vert\phi]$ exchange, i.e.
\begin{equation}
\mathrm{DC}=\dots+\int \frac{\dd J_1\dd J_2}{4} \left(\log u_{s1}
\frac{\gamma_0}{2J_1^{2h_\star}}\right) P_{[\phi\phi]_{0,J_1}[\phi\phi]_{0,J_2}}^{(J-\dg n)} 
\psi_{(2h_\phi,2h_\phi+J_a;J_1-\dg n)}^{\mathrm{CC}}+\dots 
\label{caseII_logterm}
\end{equation}
Hence, the direct channel contribution from $[\oo_\star\vert\phi]$-exchange is nicely reproduced 
by crossed channel terms, but it does not provide any new information on OPE coefficients. 

\paragraph{$[\mathcal{O}_\star\vert\mathcal{O}_\star]$-exchange in the direct channel,
$h_\star< h_\phi$.} We can now look at the direct channel contribution given by
\begin{equation}
\mathrm{DC}=\dots + P_{\oo_\star\oo_\star}^{(0)} 
u_{s1}^{h_\star + h_\phi}   u^{h_\star - h_\phi}_{s3}
u_{s4}^{2h_\phi} u_{s5}^{h_\star - 2 h_\phi} (1-z)^{h_\star}z^{J_\star} 
\hypg{\bar h_\star}{\bar h_\star+ h_\star-h_\phi}{2\bar h_\star}(z)
 +\dots
\end{equation}
For this contribution, the crossing equation reduces to
\begin{align*}
&(u_3u_5)^{h_\star-2h_\phi}(1-z)^{h_\star-h_\phi} z^{J_\star} 
\hypg{\bar h_\star}{\bar h_\star+ h_\star-h_\phi}{2\bar h_\star}(z)=\\[2mm] 
&\sum_{\dg n=0}^\infty  \int \frac{\dd \la_1\dd \la_2}{16\sqrt{\la_1\la_2}} 
\nn_{( [\phi\oo_\star]_{0,J_1},[\phi\phi]_{0,J_2};J_1-\dg n)}^{{\CC,\mrmII}}
P_{[\phi\oo_\star]_{0,J_1}[\phi\phi]_{0,J_2}}^{(J_1-\dg n)}\, z^{\dg n}u_{5}^{h_\phi+\dg n} 
\mathcal{K}_{h_\phi+\dg n}(\la_1 u_5+\la_2u_3u_5),
\end{align*}
where the normalization $\nn_{( [\phi\oo_\star]_{0,J_1},[\phi\phi]_{0,J_2};J_1-\dg n)}^{{\CC,\mrmII}} 
:= \nn_{(h_\phi+h_\star,2h_\phi,h_\phi+h_\star+J_1,2h_\phi+J_2;J_1-\dg n)}^{{\CC,\mrmII}}$ is again to 
be found in eq.~\eqref{norm_block_case_II}.  Homogeneity of the crossing equation in 
$\ep_{15},\ep_{34}$ fixes the dependence of the OPE coefficients in $\la_1,\la_2$ up to a series 
in $\dg n$, which we express as
\begin{equation}
\frac{\nn_{( [\phi\oo_\star]_{0,J_1},[\phi\phi]_{0,J_2};J_1-\dg n)}^{{\CC,\mrmII}}
P_{[\phi\oo_\star]_{0,J_1}[\phi\phi]_{0,J_2}}^{(J_1-\dg n)}}{16 \sqrt{\la_1\la_2}} =
P_{\oo_\star\oo_\star}^{(0)}  \frac{2 \la_1^{h_\phi+\dg n-1}\la_2^{2h_\phi-h_\star-1}}
{\Gamma(h_\phi+\dg n)\Gamma(2h_\phi-h_\star)^2} b^{-}_{\dg n}.
\label{caseII_h*<hphipre}
\end{equation}
Plugging this form back into the crossing equation yields the sum rule
\begin{equation}
(1-z)^{h_\star}z^{J_\star} \hypg{\bar h_\star}{\bar h_\star+ h_\star-h_\phi}{2\bar h_\star}(z) 
=\sum_{\dg n=0}^\infty b^{-}_{\dg n} z^{\dg n}.
\end{equation}
The solution to this sum rule is immediate: the coefficients $b^-_{\dg n}$ on the right are 
coefficients of the function on the left in a power series expansion around $z=0$. By combining 
the previous two equations we obtain the following expression for the OPE coefficients 
\begin{eqnarray} \label{caseII_h*<hphi}
& & \hspace*{-8mm} 
\frac{\nn_{(h_\phi+h_\star,2h_\phi,h_\phi+h_\star+J_1,2h_\phi+J_2;J_1-\dg n)}^{{\CC,\mrmII}}
P_{[\phi\oo_\star]_{0,J_1}[\phi\phi]_{0,J_2}}^{(J_1-\dg n)}}{16 \sqrt{\la_1\la_2}} = 
  \\[2mm] 
& & \quad \quad = P_{\oo_\star\oo_\star}^{(0)}  \frac{2 \la_1^{h_\phi+\dg n-1}\la_2^{2h_\phi-h_\star-1}}
{\Gamma(h_\phi+\dg n)\Gamma(2h_\phi-h_\star)^2 \dg n!} 
\frac{\dd^{\dg n}}{\dd z^{\dg n}}  (1-z)^{h_\star}z^{J_\star} 
\hypg{\bar h_\star}{\bar h_\star+ h_\star-h_\phi}{2\bar h_\star}(z)\Bigl \vert_{z=0}.
\nonumber
\end{eqnarray}
This is a genuinely new result on OPE coefficients involving the double twist families 
$[\phi\oo_\star]$ and $[\phi \phi]$. We shall discuss its applications in the next subsection. 
But before we do so, let us analyze the final case in which $[\oo_\star\vert \oo_\star]$ is 
exchanged in the direct channel. 

\paragraph{$[\mathcal{O}_\star\vert\mathcal{O}_\star]$-exchange in the direct channel,
$h_\star>h_\phi$.} When 
$h_\star > h_\phi$, the relevant behavior of direct channel blocks is 
\begin{equation}
\psi_{\oo_\star\oo_\star;n}^{(15),(34)}(u_{si}) \stackrel{\LCL^{(4)}}{\sim}
(u_{s1}u_{s2})^{h_\phi}(u_{s3}u_{s5})^{h_\star}
g_{\oo_\star\oo_\star;n}^{\mathrm{fin}}(z), \quad z:=1-u_{s2}.
\end{equation}
An integral representation for the function $g^{\mathrm{fin}}$ was given in equation 
\eqref{eq:intgfin}. It is also possible to work out the following power series expansion, 
see eq.~\eqref{eq:sumgfin} in section~\ref{ssec:5pt_blocks_dc}. For $\oo_1 = \oo_\star = 
\oo_2$, it reads 
\begin{align}
g_{\oo_\star\oo_\star;n}^{\mathrm{fin}}(z) =& 
\frac{\Gamma(2\bar h_\star)\Gamma(h_{\star\phi}+n)}
{\Gamma(\bar h_\star)\Gamma(\bar h_\star+h_{\star\phi}+n)} \sum_{k=0}^\infty 
\frac{(2\bar{h}_\star - \bar{h}_\phi)_k}{k!} \frac{(\bar h_\star)_k}{(2\bar h_\star)_k} 
\frac{(h_{\star\phi}+n)_k}
{(h_{\star\phi}+n+\bar h_\star)_k} \label{eq:gfinpseries} \\[2mm] 
& \hypg{\bar h_\star-2h_{\phi}-n}{\bar h_\star}{2\bar h_\star+k}(z) \, 
\hypg{n-\bar h_\star - h_\star}{\bar h_\star}
{\bar h_\star+h_{\star\phi}+n+k}(1-z). \nonumber 
\end{align}
If we plug our Ansatz \eqref{ope_coeffs_case_II} for the operator product 
coefficients into the crossing equation, we obtain the following constraint:
\begin{equation}
\sum_{n=0}^{J_\star} P_{\oo_\star\oo_\star}^{(n)} 
g^{\mathrm{fin}}_{\oo_\star\oo_\star;n}(z)
=  \sum_{\dg n=0}^\infty b_{\dg n } z^{\dg n}.
\label{cse_finite_ts}
\end{equation}
This constraint now replaces the $z$-independent sum rule \eqref{sum_rule} found in the 
previous subsection, the latter of which is recovered for $z= 1-u_{s2} = 1$. Since the 
new equation \eqref{cse_finite_ts} contains a power series in $z$ on 
both sides, it suffices to completely determine the coefficients 
$b_{\delta n}$.  In conclusion, the operator product coefficients for 
two double-twist exchanges at subleading order are given by
\begin{eqnarray}\label{ope_coeffs_discrete_ts}
\frac{\nn_{(2h_\phi,2h_\phi,2h_\phi+J_1,2h_\phi+J_2;J_1-\dg n)}^{\CC,\mrmII}}
{16\sqrt{\la_1\la_2}}  P_{[\phi\phi]_{0,J_1}[\phi\phi]_{0,J_2}}^{(J_1-\dg n)} 
& \sim & \\[2mm] 
& &  \hspace*{-3cm} \sim \frac{2 \la_1^{h_\phi+\dg n-1}\la_2^{2h_\phi-h_\star-1}}
{\Gamma(h_\phi+\dg n) \Gamma(2h_\phi-h_\star)^2} 
\sum_{m=0}^{J_\star} P_{\oo_\star\oo_\star}^{(m)}
 \frac{\dd^{\dg n}g^{\mathrm{fin}}_{\oo_\star\oo_\star;m}}
 {\dg n!\,\dd z^{\dg n}} \Bigl\vert_{z=0}. 
\nonumber 
\end{eqnarray}
It is not difficult to evaluate arbitrary derivatives of the function 
$g^\textrm{fin}(z)$ at $z=0$ using the series expansion formula 
\eqref{eq:gfinpseries}. Recall that OPE coefficients on the right hand 
side are given by $P^{(m)}_{\oo_\star\oo_\star} = C^2_{\phi\phi\oo_\star} 
C^{(m)}_{\oo_\star \phi \oo_\star}$ with a label $m$ that enumerates 
non-trivial tensor structures in case $\oo_\star$ carries spin $J_\star
\neq 0$. Assuming this data is known, one can use our formula to compute 
the quantity $P$ on the left-hand side of the equation via 
the expression \eqref{norm_block_case_II} for the normalization of 
case~II lightcone blocks.

The last formula can be made even more explicit if the intermediate 
field $\oo_\star$ is scalar. In this case, the finite sum over $m$ consists 
of a single term with $m=0$ and the function $g^{\mathrm{fin}}(z)$ reduces 
to a Gauss hypergeometric series, 
\begin{equation}
 g_{\oo_\star\oo_\star;0}^{\mathrm{fin}}(z) = 
 \frac{\mathrm{B}_{h_\star,h_\phi} 
\mathrm{B}_{h_\star,h_\star-h_\phi}}{\mathrm{B}_{h_\star}^2}\hypg{h_\phi}
{h_\phi}{h_\star+h_\phi}(z).
\end{equation}
From this expression, it is straightforward to evaluate the derivatives 
of $g^\textrm{fin}$ that appear in eq.~\eqref{ope_coeffs_discrete_ts}. The 
net effect is encoded in the replacement  
\begin{equation}
 \sum_{m=0}^{J_\star} P_{\oo_\star\oo_\star}^{(m)}
 \frac{\dd^{\dg n}g^{\mathrm{fin}}_{\oo_\star\oo_\star;m}}
 {\dg n!\,\dd z^{\dg n}} \Bigl\vert_{z=0} 
 \ \mapsto \
P_{\oo_\star\oo_\star}^{(0)}\frac{\mathrm{B}_{h_\star,h_\phi} 
\mathrm{B}_{h_\star,h_\star-h_\phi}}{\mathrm{B}_{h_\star}^2} 
\frac{(h_\phi)_{\dg n}^2}{\dg n!\, (h_\star+h_\phi)_{\dg n}}. 
\label{ope_coeffs_discrete_ts_scalarO*}
\end{equation}
With this substitution rule, our new result \eqref{ope_coeffs_discrete_ts} on 
the contributions to double twist OPE coefficients that stem from $[\oo_\star
\oo_\star]$-exchange in the direct channel applies to the case of scalar 
$\oo_\star$. 

\paragraph{Summary.} We have now bootstrapped OPE coefficients of large-spin 
double-twist operators in the crossed channel as a function of direct-channel leading-twist 
exchanges $\oo_\star$ and their CFT data $P_{\oo_\star\oo_\star}^{(n)}$. We were able to 
obtain asymptotics of double-twist OPE coefficients in two regimes: either $n\gg J_1,J_2\gg 1$ 
(I) or $J_2\gg J_1,n\gg 1$ with $J_1-n=\dg n=0,1,2,\dots,\infty$ (II). The results in regime (I), 
derived in section~\ref{ssec:5ptBS_CC}, only apply to direct-channel exchanges with $h_\star\leq 
h_\phi$. The OPE coefficients in this regime were computed in \cite{Antunes:2021kmm} using different 
methods, and we find exact agreement between their results and ours. On the other hand, the results 
in regime (II), derived in section~\ref{ssec:discrete_tensor_structures},  apply 
for any $h_\star< 2h_\phi$ exchanges in the direct channel. It is actually 
the first time that a discrete tensor structure dependence for the OPE coefficients $C_{[\phi\phi]\phi [\phi\phi]}^{(n)}$,  $C_{[\phi\oo_\star]\phi [\phi\phi]}^{(n)}$ has been resolved.  This progress came 
about because we were able to control the lightcone blocks on both sides of the crossing symmetry 
equation with only four, rather than the usual five lightcone limits taken. While in principle, the 
relevant information on lightcone blocks resides in the integral formula and hence in the usual 
lightcone OPE, the lightcone Casimir operators played a crucial role in extracting this data. 
 
\subsection{Applications to specific models}
\label{ssec:applications}

To explore the predictions that follow from formulas~\eqref{caseII_1phiexch}, \eqref{caseII_h*<hphi} and 
\eqref{ope_coeffs_discrete_ts}, we would like to discuss possible applications, at least 
briefly. In the first subsection, we shall go through each of the new formulas and list a 
few cases to which they apply, mostly within the context of models in $d=3$. We shall then 
employ one of the concrete realizations in the second subsection to check our result 
\eqref{caseII_1phiexch} against leading order perturbative computations in scalar 
$\phi^3$ theory. 

\subsubsection{Possible leading-twist exchanges} 

Our results on OPE coefficients were obtained by solving the crossing equation for specific leading-twist exchanges in the direct 
channel. Which of these leading-twist exchanges actually appear depends on the specific model 
under investigation and the choice of the field $\phi$. We will review this case-by-case:
\begin{itemize}
\item $[\mathbf{1}\vert \phi]$-exchange only exists if $C_{\phi\phi\phi}\neq 0$ and entails 
$P_{[\phi\phi][\phi\phi]}\propto C_{\phi\phi\phi}$ in the crossed channel, see eq.\ 
\eqref{caseII_1phiexch}. Of course, in all models with $\mathbb{Z}_2$ symmetry, the 
OPE coefficients $C_{\phi\phi\phi}$ vanish if $\phi$ is $\mathbb{Z}_2$-charged. Hence, 
there are two cases where this contribution to the OPE coefficients does not vanish. On the 
one hand, we can consider $\mathbb{Z}_2$-charged operators in models that break the $\mathbb{Z}_2$ symmetry, like $\phi^3$ theory. Alternatively, our formula \eqref{caseII_1phiexch} applies to any 
$\Zs_2$-singlet operator, even in models with $\mathbb{Z}_2$ symmetry, such as the field $\phi
\equiv \ep$ in the $d=3$ Ising model or $\phi\equiv\varphi^2$ at the Wilson-Fisher 
fixed-point. The first realization within scalar $\phi^3$ theory provides a nice opportunity to check formula \eqref{caseII_1phiexch} against an independent perturbative calculation of the OPE coefficients at tree level. The latter are derived in 
appendix~\ref{app:phi_cubed} and they are compared with our bootstrap result in 
section~\ref{ssec:check_phi3} below. 
 
\item $[\oo_\star \vert \oo_\star]$-exchange with $0<h_\star<h_\phi$ entails 
$P_{[\phi\oo_\star][\phi\phi]} \propto P_{\oo_\star\oo_\star}^{(0)}$ in the crossed 
channel, see eq. \eqref{caseII_h*<hphi}. These exchanges can of course only appear 
when $\phi$ is \emph{not} the lowest dimension scalar of the theory. A typical example 
would be $(\phi,\oo_\star)=(\ep,T_{\mu\nu})$ in the Ising model, or more generally 
$\oo_\star=T_{\mu\nu}$ whenever $\Dg_\phi>d-2$. Since these cases involve external composite 
fields, performing explicit perturbative calculations of our formula \eqref{caseII_h*<hphi} 
would be significantly more difficult than for the case we mentioned at the end of the 
previous paragraph. This makes eq.~\eqref{caseII_h*<hphi} an interesting prediction 
of the multipoint bootstrap. 

\item $[\oo_\star \vert \oo_\star]$-exchange with $h_\star>h_\phi$ entails 
$P_{[\phi\phi][\phi\phi]}\propto P_{\oo_\star\oo_\star}^{(n)}$, as detailed in our 
formula \eqref{ope_coeffs_discrete_ts}. When $[\mathbf{1}\vert\phi]$-exchange 
contributes to the direct channel --- see our discussion in the first item --- the contributions 
from $[\oo_\star\vert \oo_\star]$-exchange correspond to subleading terms in the large spin expansion 
of the double-twist OPE coefficients, such as $\oo_\star=T_{\mu\nu}$ in $\phi^3$ theory or 
$(\phi,\oo_\star) = (\ep,\ep')$ in the Ising model. On the other hand, one can also find 
models in which the exchange of $[\oo_\star\vert\oo_\star]$ with  $h_\star>h_\phi$ is leading, 
like the Gross-Neveu-Yukawa models with $\phi\equiv \sigma$ the pseudoscalar and $\oo_\star 
\equiv T_{\mu\nu}$. In all of these cases, $\oo_\star$ must be the unique field with twist $h_\star$ in the spectrum. Such a twist gap is characteristic of strongly interacting CFTs, making formula~\eqref{ope_coeffs_discrete_ts} a non-trivial prediction of the five-point lightcone bootstrap that would be difficult to reproduce with perturbative methods. 
\end{itemize}

Before we conclude this short list of applications, we would like to briefly comment on 
an aspect that specifically applies to five-point functions in $d=3$-dimensional models. It is well known (see e.g.~\cite[Sec.~4.2.3]{Costa:2011mg}) that 
three-point tensor structures of two STTs and one scalar can only be parity-odd in $d=3$, whereas the results we have 
stated above only apply to parity-even tensor structures. Fortunately, it is not difficult to derive analogous expressions for parity-odd tensor structures. As demonstrated in 
appendix~\ref{app:parity-odd_blocks},  the parity-odd five-point blocks in either channel 
are obtained by a simple shift of parameters $(h_\phi,n)\rightarrow (h_\phi+1/2,n+1)$
\footnote{Here,  $h_\phi$ is to be understood as the dependence of the blocks on the 
conformal dimension of the external field when the quantum numbers $h_a,\bar h_a,n$ are 
kept arbitrary.}.  We can therefore perform the same analysis and bootstrap parity-odd 
OPE coefficients of double-twist operators.  There are only two qualitative differences: 
first,  odd parity excludes scalar exchange in either channel. Second, the direct channel 
asymptotics now depend on the sign of $h_\star-h_\phi+1/2$, such that $[\phi\oo_\star]$ 
only appears if $\frac{d-2}{2}\leq h_\star<h_\phi-1/2$,  which in turn requires 
$\Dg_\phi>d-1$. In the case where formula \eqref{ope_coeffs_discrete_ts} applies, i.e. $h_\star>h_\phi-1/2$, the 
shift of parameters leads to the following explicit result: 
 \begin{align}
& \hspace*{-15mm}\frac{\nn_{([\si\si]_{0,J_1}[\si\si]_{0,J_2};J_1-\dg n-1/2)}^{{\CC,\mrmII}}}
{16 J_1 J_2}  P_{[\si\si]_{0,J_1}[\si\si]_{0,J_2}}^{(J_1-\dg n,\mathrm{odd})} 
\sim \nonumber \\[2mm] 
&\frac{2 J_1^{2(h_\si+\dg n-3/2)}J_2^{2(2h_\phi-h_\star-1)}}
{\Gamma(h_\phi+\dg n-1/2) \Gamma(2h_\phi-h_\star)^2} 
\sum_{m=0}^{J_\star-1} P_{\oo_\star\oo_\star}^{(m,\mathrm{odd})}
 \frac{\dd^{\dg n}g^{\mathrm{fin},h_\phi\rightarrow h_\phi+1/2}_{\oo_\star\oo_\star;m+1}}
 {\dg n!\,\dd z^{\dg n}} \Bigl\vert_{z=0}.
\label{ope_coeffs_gny}
\end{align}
For pseudoscalar five-point functions, the comments on applications that we listed in the final item of our 
list will therefore apply after replacing the parity-even result~\eqref{ope_coeffs_discrete_ts} with the parity-odd result~\eqref{ope_coeffs_gny} above.

\subsubsection{Explicit checks in \texorpdfstring{$\phi^3$}{phi-cubed} theory at tree level} 
\label{ssec:check_phi3}
In this subsection, we perform explicit checks for the predicted leading large spin asymptotics of double-twist OPE coefficients that reproduce the identity exchange in the (15) OPE of the direct channel. This corresponds to eq.~\eqref{NPJphiJ} for case~I scaling and eq.~\eqref{caseII_1phiexch} for case~II scaling. We check that these expressions can be obtained from the exact OPE coefficients of $\phi^3$ theory at tree-level. At this first order in perturbation theory, the five-point correlator is disconnected and takes the form
\begin{equation}
\langle \phi(X_1)\dots\phi(X_5)\rangle = \langle \phi(X_1)\phi(X_5)\rangle\langle \phi(X_2)\phi(X_3)\phi(X_4)\rangle+\mathrm{perms}+\orm(g^2). 
\label{phi3_5ptc_treelevel}
\end{equation}
This applies to either the $\epsilon$-expansion around the six-dimensional conformal field theory, or the holographic description of a perturbative $\phi^3$ theory in AdS$_d$.  In appendix~\ref{app:phi_cubed},  we  compute the double-twist OPE coefficients in the OPE decomposition of this correlator and obtain 
\begin{equation}
P_{[\phi\phi]_{0,J_1}[\phi\phi]_{0,J_2}}^{(n)} = C_{\phi\phi\phi} \frac{1}{n!} \prod_{i=1}^2P_{[\phi\phi]_{0,J_i}}\frac{J_i!}{(J_i-n)!}  (\Dg_\phi/2)_{J_i-n} \, \mathcal{F}_{(\Dg_\phi;J_1,J_2)}^{(n)} +\orm(g^2),
\label{PJ1J2n_phi3_tree}
\end{equation}
where $P_{[\phi\phi]_{0,J}}=C^2_{\phi\phi[\phi\phi]_{0,J}}$ are the OPE coefficients of the GFF four-point function, given. in e.g. \cite[eq.~(11)]{Fitzpatrick:2012yx}, and 
\begin{align}
\mathcal{F}_{(\Dg_\phi;J_1,J_2)}^{(n)} &=\sum_{j=0}^n \binom{n}{j} \frac{(\Dg_\phi/2)_{n-j}}{(\Dg_\phi)_j(\Dg_\phi)_{J_1-j}(\Dg_\phi)_{J_2-j}} \\
&= \frac{(\Dg_\phi/2)_n}{(\Dg_\phi)_{J_1}(\Dg_\phi)_{J_2}}\, \tensor[_3]{F}{_2} \begin{bmatrix}
-n,  & 1-\Dg_\phi-J_1, & 1-\Dg_\phi-J_2 \\
1-\Dg_\phi/2-n, & \Dg_\phi &
\end{bmatrix} (1). 
\label{FJ1J2n_phi3_tree}
\end{align}

\paragraph{Comparison with case~I asymptotics. } In the case~I regime $(J_1^2,J_2^2,n)=\orm(\ep_{15}^{-1}\ep_{23}^{-1},\ep_{15}^{-1}\ep_{34}^{-1},\ep_{15}^{-1})$ with tensor structures larger than spin, we cannot directly compare the asymptotics of eq.~\eqref{PJphiJ} with the tree-level OPE coefficients in eq.~\eqref{PJ1J2n_phi3_tree} --- the derivation in the latter case requires $[\phi\phi]_{0,J}$ to be local operators with integer spin, in which case $n\leq J_i$. Nevertheless, we can meaningfully compare the two results by inverting the order of limits from $(\ep_{15},\ep_{34},\ep_{23})$ to $(\ep_{34},\ep_{23},\ep_{15})$ in eq.~\eqref{PJ1J2n_phi3_tree} to put the spins and tensor structures back into a physical regime. Indeed, given the explicit expression for the tree-level correlator in eq.~\eqref{phi3_5ptc_treelevel}, it is easy to compare the expansion in the two different orders of limits.  While the permutation makes the identity exchange in (15) subleading (compared to the identity in (34)), it will instead appear at next-to-leading order in the crossing equation of the tree-level correlator. In the limit $(J_1^2,J_2^2,n)=\orm(\ep_{23}^{-1},\ep_{34}^{-1},1)$ we find
\begin{equation}
\mathcal{F}_{(\Dg_\phi;J_1,J_2)}^{(n)} =\frac{1}{(\Dg_\phi)_{J_1}(\Dg_\phi)_{J_2}} \frac{(J_1J_2)^n}{n! (\Dg_\phi)_n}\left[1+\orm(J_1^{n-1}J_2^n,J_1^nJ_2^{n-1})\right].
\end{equation}
This result coincides with the large $J_1,J_2$ asymptotics predicted in \cite{Antunes:2021kmm}.  As a result,  isolating this term and taking the limit $J_1,J_2\gg n \gg 1$ in the ratio of $\Gamma$ functions indeed reproduces the lightcone bootstrap prediction of eq.~\eqref{PJphiJ} in case~I asymptotics. 
\paragraph{Retrieving case~II asymptotics.} In the case~II regime $(J_1,J_2,n)=\orm(\ep_{15}^{-1},\ep_{15}^{-1}\ep_{34}^{-1},\ep_{15}^{-1})$ with $J_1-n=\dg n=0,1,\dots,\infty$,  the spins and tensor structure label take physical values that allow for a direct check of eq.~\eqref{caseII_1phiexch} against eq.~\eqref{PJ1J2n_phi3_tree}.  In this regime, we find
\begin{equation}
\mathcal{F}^{(J_1-\dg n)}_{(\Dg_\phi;J_1,J_2)} = \frac{ 1+ \orm((J_2-J_1)^{-1})}{(\Dg_\phi)_{\dg n}(\Dg_\phi)_{J_2-J_1+\dg n}(\Dg_\phi)_{J_1-\dg n}}.
\end{equation}
After this approximation, we can again take the large $J_1,J_2$ limit in eq.~\eqref{PJ1J2n_phi3_tree} and retrieve the solution to the crossing equation in eq.~\eqref{caseII_1phiexch}.

\section{Outlook: Six-point comb channel lightcone bootstrap}
\label{sect:six_pt_outlook}

The goal of this final section is to provide a detailed  outlook on the second part 
of this work, in which we will apply the new methodology we have developed here to 
study triple-twist operators. While \cite{Antunes:2021kmm} explored crossing symmetry 
between two OPE channels with snowflake topology, our upcoming paper addresses the 
duality between a direct snowflake channel and a crossed channel of comb topology, 
see fig.~\ref{fig:6pt_cse_outlook}. The treatment of the crossed comb channel is made 
possible by the new technology we have -- to a large extent -- developed above. 
And it is the comb channel that will give us access to triple-twist data.

\subsection{Crossing a snowflake into a comb}
\begin{figure}[ht]
\centering
\includegraphics[scale=0.1]{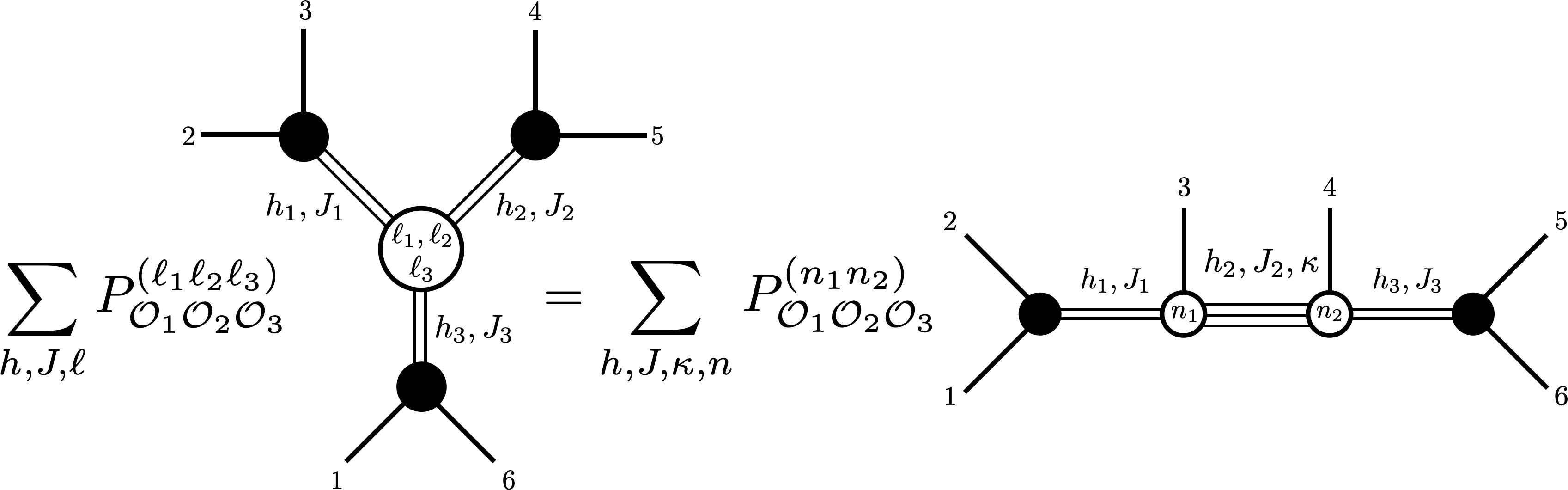}
\caption{Graphical representation of a planar six-point crossing symmetry equation from the snowflake channel $(61)(23)(45)$ to the comb $(12)3(4(56))$. Internal legs are labeled by half-twists $h$ and spin labels $J,\kappa$ (the lengths of the two rows of a mixed-symmetry tensor Young tableau), while tensor structure labels are $\ell_1,\ell_2,\ell_3$ at the central spinning vertex of the snowflake and $n_1,n_2$ at the two innermost vertices of the comb.}\label{fig:6pt_cse_outlook}
\end{figure}

In our forthcoming work~\cite{In_preparation}, we will further extend the methods and results of this paper to six points. Indeed, the setup of fig.~\ref{fig:5pt_cse_general} naturally generalizes to a six-point setup with the comb-channel expansion $(12)3(4(56))$ as the crossed channel, and a direct channel containing the $(16)$ OPE. In particular, the snowflake channel $(16)(23)(45)$ leads to the planar crossing equation depicted in fig.~\ref{fig:6pt_cse_outlook}. In the snowflake channel, all intermediate operators 
are STT with twist denoted by $h_a, a=1,2,3$ and spin $J_a, a = 1,2,3$. The tensor 
structures at the central vertex are parameterized by three integers $\ell_1,\ell_2,\ell_3$.
In the comb channel, the nine degrees of freedom in the sum over conformal blocks are divided among twists $h_1,h_2,h_3$, spins $J_1,J_2,\kappa_2,J_3$, and tensor structures $n_1,n_2$, where the new symbol $\kappa_2$ represents the length of the second row of the Young tableau associated with the mixed-symmetry tensor exchanged at the middle leg. Note that in $d=3$, where the scalar six-point function has one degree of freedom less, there is an extra relation between the tensor structure labels $\ell_1,\ell_2,\ell_3$ in the snowflake channel, while $\kappa_2=0$ in the comb channel. As in our treatment of five-point crossing, we are using the same symbols for the twists and STT spins of the intermediate fields in the direct and crossed channel. We trust that it 
will be clear from the context which set we are referring to. 

The sequence of lightcone limits in which we propose to analyze the 
crossing symmetry relations shown in fig.~\ref{fig:6pt_cse_outlook} 
is given by 
\begin{equation}
\LCL_{\vec{\ep}}:X_{16} \ll X_{23}, X_{45}  \ll X_{12},X_{56} \ll 
X_{46} \ll X_{34}  \ll  1.
\label{lcl_6pt}
\end{equation}
The first three limits we perform clearly expose leading-twist contributions 
in the direct snowflake channel. The remaining three limits favor leading-twist 
terms in the crossed channel. Note in particular that the limit $X_{45},
X_{56}\ll X_{46}\rightarrow 0$ projects to leading twist in the $(4(56))$ 
OPE of the comb channel. Through the final limit in which $X_{34}$ is sent 
to zero, we reach the kinematics of a null polygon in which $X_{i(i+1)} \ll 
1$ for all $i=1, \dots, 6$, with the additional lightcone constraint 
$X_{46} = 0$ imposed before we complete the null polygon. 

A convenient choice of cross-ratios that generalizes the snowflake cross-ratios
$u_{s1},\dots, u_{s5}$ we used in the context of five-point functions and that 
additionally preserves the $\Zs_2$ reflection symmetry of the  comb channel is 
given by
\begin{equation}
	\begin{gathered}\dutchcal{u}_1=\frac{u_1}{v_2}=\frac{X_{12}X_{35}}{X_{13}X_{25}}\,, \ \dutchcal{u}_2=\frac{u_2}{v_1v_3}=\frac{X_{13}X_{46}}{X_{14}X_{36}}\,, \ \dutchcal{u}_3=\frac{u_3}{v_2}=\frac{X_{24}X_{56}}{X_{25}X_{46}}\,,\\[2mm]
	v_1=\frac{X_{14}X_{23}}{X_{13}X_{24}}\,, \ v_2=\frac{X_{25}X_{34}}{X_{24}X_{35}}\,, \ v_3=\frac{X_{36}X_{45}}{X_{35}X_{46}}\,, \\[2mm]
	\mathcal{U}_1=\frac{U^{(5)}_1}{v_1v_2}=\frac{X_{15}X_{24}}{X_{14}X_{25}}\,, \ \mathcal{U}_2=\frac{U^{(5)}_2}{v_2v_3}=\frac{X_{26}X_{35}}{X_{25}X_{36}}\,, \  \mathcal{U}^6=\frac{U^{(6)}}{v_1v_2v_3}=\frac{X_{16}X_{24}X_{35}}{X_{14}X_{25}X_{36}}\,.
\end{gathered}
\label{eq:curlycrossratios}
\end{equation}
In terms of these cross-ratios, the lightcone regime $\LCL_{\vec{\ep}}$ we specified 
in eq.~\eqref{lcl_6pt} corresponds to  
\begin{equation}
\LCL_{\vec{\ep}}:\mathcal{U}^6 \ll v_1,v_3 \ll \dutchcal{u}_1, \dutchcal{u}_3 \ll \dutchcal{u}_2 \ll v_2 \ll 1.
\label{lcl_6pt_cr}
\end{equation}
Note that two cross-ratios $\mathcal{U}_1,\mathcal{U}_2$ do not contain any of the 
$X_{ij}$ that are sent to zero and hence they are left unconstrained in our limit. There is one less independent cross ratio in $d=3$ kinematics, which is in encoded in the relation $\mathcal{U}_1+\mathcal{U}_2=1$ at leading order in the lightcone limit \eqref{lcl_6pt_cr}. Finally, to fix our conventions for snowflake and comb channel blocks, we write down the corresponding conformal block decompositions below:
\begin{align*}
\langle \phi(X_1)\dots \phi(X_6)\rangle &= \left(X_{16}X_{23}X_{45}\right)^{-\Dg_\phi} \sum_{\oo_1,\oo_2,\oo_3,\ell_1,\ell_1,\ell_3} P_{\oo_1\oo_2\oo_3}^{(\ell_1\ell_2\ell_3)} \,\psi_{\oo_1\oo_2\oo_3;\ell_1\ell_2\ell_3}^{(16)(23)(45)}(\dutchcal{u}_a,v_a,\mathcal{U}_s,\mathcal{U}^6) \\
&=\left(\frac{X_{12}X_{34}X_{56}}{\sqrt{\dutchcal{u}_2 v_1 v_3}}\right)^{-\Dg_\phi} \sum_{\oo_1,\oo_2,\oo_3,n_1,n_2} P_{\oo_1\oo_2\oo_3}^{(n_1n_2)}\, \psi_{\oo_1\oo_2\oo_3;n_1n_2}^{(12)3(4(56))}(\dutchcal{u}_a,v_a,\mathcal{U}_s,\mathcal{U}^6).
\end{align*}
\medskip 

In order to analyze the crossing symmetry constraints we start with an explicit 
expansion of the direct channel. In the lightcone regime $\LCL_{\vec{\ep}}^{(6)}$
where six of the seven limits in eq.~\eqref{lcl_6pt} are performed, i.e. the 
cross-ratio $v_2$ is left unconstrained, the direct channel expansion reads 
\begin{align} \label{eq:6pt_dc_exp}
\sum_{\oo,\ell} P^{(\ell_1\ell_2\ell_3)}_{\oo_1\oo_2\oo_3} 
\psi^\DC_{\oo;\ell} & \sim  \Big[1+C_{\phi\phi \oo_\star}^2\, (\dutchcal{u}_2 v_1 v_3)^{h_\star} \hypg{\bar h_\star}{\bar h_\star}{2 \bar h_\star}(1-v_2) \\
&\qquad \,\, +C_{\phi\phi \oo_\star}^2\, (\frac{v_1\mathcal{U}^6}{\mathcal{U}_2})^{h_\star} \hypg{\bar h_\star}{\bar h_\star}{2 \bar h_\star}(1-\dutchcal{u}_1/\mathcal{U}_2) \nonumber \\
&\qquad \,\,+C_{\phi\phi \oo_\star}^2\, (\frac{v_3\mathcal{U}^6}{\mathcal{U}_1})^{h_\star} \hypg{\bar h_\star}{\bar h_\star}{2 \bar h_\star}(1-\dutchcal{u}_3/\mathcal{U}_1) \nonumber \\[2mm]
& \hspace*{-2cm} +\left(\dutchcal{u}_2v_1v_3\,\mathcal{U}^6\right)^{h_\star}\sum_{\ell} P_{\oo_\star \oo_\star \oo_\star}^{(\ell_1\ell_2\ell_3)} \left( g_{\oo_\star \oo_\star \oo_\star;\ell}\left(0,v_2\mathcal{U}_1,0,\dutchcal{u}_2,\mathcal{U}_1,\mathcal{U}_2\right)+\orm(X_{12,56})\right) \nonumber \\
&\hspace*{7.55cm}+\orm(X_{16}^{h>h_\star})\Big]. \nonumber
\end{align}
In writing the direct channel expansion we have dropped the prefactor $(X_{16}X_{23}X_{45})^{-\Dg_\phi}$ 
so that we obtain functions of the cross ratios only. On the fourth line,  we used 
the same conventions as in  \cite{Antunes:2021kmm} for six-point OPE coefficients and lightcone blocks, except 
that we replaced their label $k$ by $\oo$. The snowflake lightcone blocks $g_{\oo_1\oo_2\oo_3;\ell_1\ell_2\ell_3}(u_1,u_3,u_5,U_1,U_2,U_3)$ are given by the 
integral formula \cite[eq.~(22)]{Antunes:2021kmm}. The latter is a direct 
consequence of the lightcone OPE. Therefore, the direct channel expansions are 
under good control. 

In \cite{Antunes:2021kmm}, the leading terms of the direct channel expansion were 
reproduced from a crossed channel with snowflake topology. Here we address the same 
problem, but for a crossed channel of comb topology, i.e. the crossing equation we 
analyze takes the form 
 \medskip
\begin{equation} \label{eq:6pt_cse_snowflake}
\Big( \frac{\dutchcal{u}_1\sqrt{\dutchcal{u_2}}\dutchcal{u}_3v_2}{\mathcal{U}^6\sqrt{v_1v_3}} \Big)^{\Delta_\phi}\sum_{\oo_a;\ell_a} P^{(\ell_1,\ell_2,\ell_3)}_{\oo_1\oo_2,\oo_3} 
\psi^\DC_{\oo_a,\ell_a} = 
\sum_{\oo_a;n_\rho} P_{\oo_1\oo_2\oo_3}^{(n_1,n_2)} \, \
\psi^\CC_{\oo_a,n_\rho}
(\dutchcal{u}_i,v_i,\mathcal{U}_1,\mathcal{U}_2,\mathcal{U}^6). 
\end{equation} 
The prefactor on the left-hand side is $\Omega_{\DC} \Omega^{-1}_\CC$. The twists 
$(h_a)_{a=1}^3$ of the three intermediate operators that are exchanged at leading 
order in the crossed channel may be read off from the exponents of the cross-ratios
$(\dutchcal{u}_a)_{a=1}^3$ on the left-hand side of the crossing equation 
\eqref{eq:6pt_cse_snowflake}. The scaling of the spin labels $(J_1,J_2,J_3,\kappa)$ 
in the crossed channel, on the other hand, can be determined from the degree of 
singularity of the three second-order comb-channel Casimir operators 
$\mathcal{D}_{12}^2,\mathcal{D}_{456}^2,\mathcal{D}_{56}^2$ along with the 
fourth-order operator $\mathcal{D}_{456}^4$ for the middle leg. \medskip

The leading contribution in the direct channel expansion~\eqref{eq:6pt_dc_exp}, corresponding to an exchange of three identities, does not probe any new data in the 
crossed channel because it is reproduced by a sum over contributions of $(\oo_1;\oo_2;\oo_3)=([\phi\phi]_{0,J_1};\phi;[\phi\phi]_{0,J_3})$ at large $J_1,J_3$.  In particular, the exchange of $\phi$ at the middle leg follows from the exact diagonalization of $\mathcal{D}_{456}^2$ by the direct channel contribution with eigenvalue $\frac{1}{2}\Dg_\phi(\Dg_\phi-d)$.  \smallskip 

All of the remaining contributions of the direct channel expansion \eqref{eq:6pt_dc_exp}
are reproduced by exchanges of three double-twist operators $[\phi\phi]_{0,J_1}$, 
$[\phi\oo_\star]_{0,J_2,\kappa}$, $[\phi\phi]_{0,J_3}$ in the crossed channel.  
Therein,  the leading logarithmic singularity coming from the ${}_2 F_1(1-v_2)$ 
is reproduced by what we call ``case~I'' scaling, namely   
\begin{equation}
\mathrm{LS}_{\mathrm{I}}: \quad J_1^2=\orm(\ep_{16}^{-1}\ep_{23}^{-1}), \quad J_2^2=\orm(\ep_{16}^{-1}\ep_{34}^{-1}), \quad J_3^2=\orm(\ep_{16}^{-1}\ep_{45}^{-1}).
\label{eq:6pt_LS_I}
\end{equation} 
The remaining contributions are what we call ``case~II",  in that $J_2$ scales 
only with $X_{16}$:
\begin{equation}
\mathrm{LS}_{\mathrm{II}}: \quad J_1^2=\orm(\ep_{16}^{-1}\ep_{23}^{-1}), \quad
J_2^2=\orm(\ep_{16}^{-1}), \quad J_3^2=\orm(\ep_{16}^{-1}\ep_{45}^{-1}).
\label{eq:6pt_LS_II}
\end{equation}
Following our analysis of case~II scaling in the five-point crossing equation,  we expect
it will be necessary to solve the crossing equation at all orders in $X_{34}$ to completely
fix the double-twist data that appears in the crossed channel. 
\medskip

\paragraph{Explicit solution in case~I.}  After describing the general setup, 
we now want to anticipate a few of the new results. In particular, we state the lightcone
limit of the comb channel blocks in the regime $\LCL_{\vec{\ep}}$, see eq.~\eqref{lcl_6pt},
with scaling $\mathrm{LS}_{\mrmI}$ as specified in eq.~\eqref{eq:6pt_LS_I}, and in $d=3$, where the two unconstrained cross-ratios satisfy the extra relation $\mathcal{U}_1+\mathcal{U}_2=1$. These lightcone blocks take the form 
\begin{align}
\frac{\psi_{\oo;n}(\dutchcal{u}_a,v_a,\mathcal{U}_1,\mathcal{U}^6)}
{\prod_{a=1}^3 \dutchcal{u}_a^{h_a} (v_1v_2^2v_3)^{h_\phi}} 
\underset{\mathrm{LS}_{\mathrm{I}}}{\overset{\LCL_{\vec{\ep}}}{\sim}} \, 
\frac{\mathcal{N}^{\mathrm{I},3d}_{\oo;n} \,  \delta(n_2\mathcal{U}_1-n_1 (1-\mathcal{U}_1))}{\exp((n_1+n_2)\mathcal{U}^6 + \frac{\la_1v_1}{n_1} + \frac{\la_2v_2}{n_1+n_2}+\frac{\la_3v_3}{n_3})} ,
\label{eq:6pt_caseI_dgeq3}
\end{align}
as we shall prove in 
\cite{In_preparation}. Here, the scaling $n_1,n_2 = \orm(\ep_{16}^{-1})$ of the tensor structure labels is 
further corroborated by the asymptotics of the vertex operators, and the normalization 
is given by
\begin{equation}
\mathcal{N}^{\mathrm{I},3d}_{\oo;n} = \frac{1}{\sqrt{\pi}} 4^{\bar h_1+\bar h_2+\bar h_3-1/2}  \frac{\left(\frac{n_1}{e}\right)^{n_1}\left(\frac{n_2}{e}\right)^{n_2} (n_1n_2)^{h_\star} }{J_1^{n_1+h_\star}J_2^{n_1+n_2+2h_\phi-1/2}J_3^{n_2+h_\star}} (n_1+n_2)^{2h_\phi-1}.
\end{equation}
Plugging these blocks back into the crossing equation, we can reproduce the $\log v_2^{-1}$ 
term in the first line of the direct channel expansion eq.~\eqref{eq:6pt_dc_exp} if the OPE 
coefficients are given by 
\begin{equation}
P_{[\phi\phi]_{0,J_1}[\phi\oo_\star]_{0,J_2}[\phi\phi]_{0,J_3}}^{(n_1n_2)} \sim \frac{C_{\phi\phi\oo_\star}^2}{\mathrm{B}_{\bar h_\star}} \frac{4^{\frac{5}{2}-\sum_{a=1}^3 \bar h_a}}{\sqrt{\pi}}\frac{J_1^{2\Dg_\phi-h_\star-n_1-1}J_2^{\Dg_\phi-n_1-n_2-\frac{1}{2}}J_3^{2\Dg_\phi-h_\star-n_2-1}}{\Gamma(\Dg_\phi)\Gamma(\Dg_\phi-h_\star)^2 \, e^{n_1+n_2} n_1^{n_1+\Dg_\phi} n_2^{n_2+\Dg_\phi}}.
\end{equation}
Setting $J_3=J_1$, $n_1=n_2=n$ and dividing both sides by $C_{\phi\phi[\phi\phi]_{0,J_1}}^2$ in eq.~\eqref{eq:CphiphidtphiJ}, we obtain a new result for OPE coefficients of two double-twist 
operators,
\begin{equation}
C_{[\phi\phi]_{0,J_1} \phi [\phi\oo_\star]_{0,J_2}}^{(n)} \sim \frac{C_{\phi\phi\oo_\star}^2}{\mathrm{B}_{\bar h_\star}} \frac{4^{1-\bar h_1-\bar h_2}}{\sqrt{\pi}} \frac{\Gamma(\Dg_\phi)}{\Gamma(\Dg_\phi-h_\star)^2} \frac{J_1^{2\Dg_\phi-2h_\star-2n-\frac{1}{2}} J_2^{\Dg_\phi-2n-\frac{1}{2}}}{e^{2n} n^{2n+\Dg_\phi}}.
\end{equation}
The result holds at leading order for large spins $J_a$ and large tensor structure 
label $n$, but it does not assume any additional relations between these large 
quantum numbers. 
\medskip 

\paragraph{Higher order terms and case~II.}  On the first line of the direct channel expansion in
eq.~\eqref{eq:6pt_dc_exp}, the remaining non-divergent terms in the expansion of the four-point lightcone 
block ${}_2F_1(1-v_2)$ around $v_2=0$, should be reproduced in the crossed channel by a large spin 
sum over double-twist OPE coefficients with the case~II scaling defined by
eq.~\eqref{eq:6pt_LS_II}. Expanding both sides of the crossing symmetry equation 
\eqref{eq:6pt_cse_snowflake} in the same basis of functions of the cross-ratio 
$v_2$, we expect to solve the crossed channel OPE coefficients for the direct 
channel data similarly to how we solved for the OPE coefficients 
$P_{[\phi\phi][\phi\phi]}^{(J-\dg n)}$ with discrete dependence on the tensor structure $\dg n$ in
section~\ref{sec:5pt_bootstrap}. Next, we expect the four-point contributions from the second and third line to be reproduced by the OPE coefficients and anomalous dimensions of the double-twist operators $[\phi\phi]_{n_{1,3},J_{1,3}}$ at the left and right legs of the comb channel. Moving on to the fourth line of the direct 
channel expansion~\eqref{eq:6pt_dc_exp}, we will analyze the asymptotics 
of the lightcone block $g_{\oo_\star \oo_\star \oo_\star }(0,v_2 \mathcal{U}_1^5,0
,\dutchcal{u_2},\mathcal{U}_1^5,\mathcal{U}_1^5)$ in the limit where the cross-ratio
$\dutchcal{u_2} \ll 1$.  In comparison to the setup of
\cite{Bercini:2020msp,Bercini:2021jti,Antunes:2021kmm}, this amounts to taking one 
of the three origin limits ($U_1\rightarrow 0$ in their notation) before taking the 
last null polygon limit ($u_3\rightarrow 0$ in their notation).  Following a 
differential-operator-based analysis of these asymptotics in the spirit of our analysis
above, we will show that $g_{\oo_\star \oo_\star \oo_\star }=\orm(\log \dutchcal{u_2})$ 
at leading order in this limit.  This leading contribution on the second line can 
only be reproduced from a $\orm(J_2^{-2h_\star})$ correction to the anomalous 
dimension of the double-twist operators $[\phi\oo_\star]_{0,J_2,\kappa}$ in 
the crossed channel.  
 
\subsection{Triple-twist data and non-planar crossing equation}
\begin{figure}[t]
\centering
\includegraphics[scale=0.11]{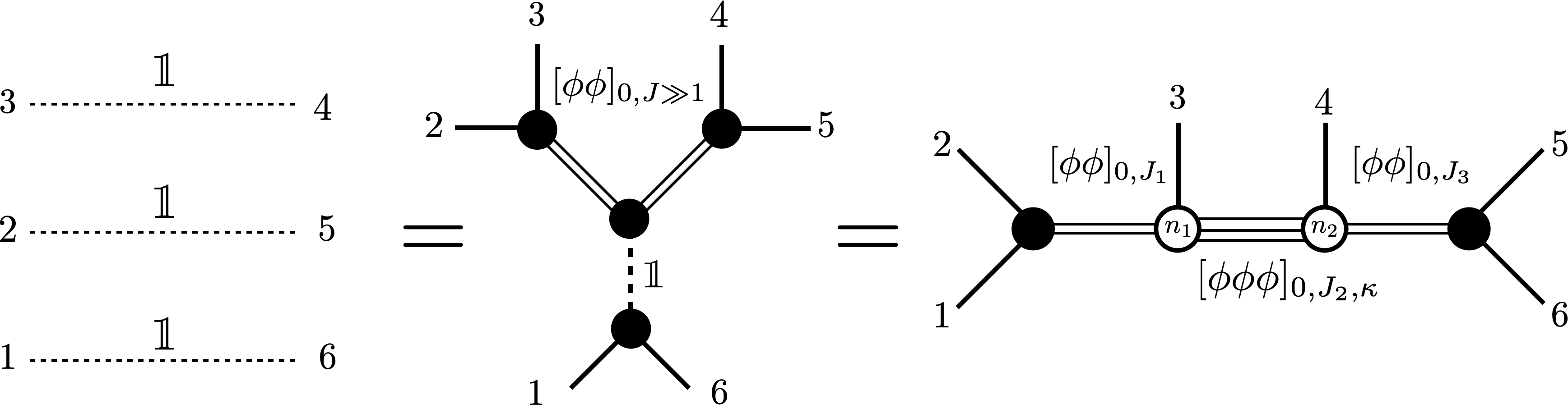}
\caption{Leading contribution to the six-point crossing equation with triple-twist exchange in the crossed channel. }\label{fig:triple_twist}
\end{figure}
In the crossing equation of fig.~\ref{fig:6pt_cse_outlook}, triple-twist exchanges appear in the crossed channel only if double-twist exchanges appear in the direct channel.  At the same time,  for identity exchange in the (16) OPE, a four-point crossing between (23)(45) and (25)(34) relates double-twist exchange in the (16)(23)(45) channel to three identity exchanges in the (16)(25)(34) channel, see fig.~\ref{fig:triple_twist}.  This observation motivates us to study a novel crossing equation where (16)(25)(34) is the direct channel. In this case, the relevant sequence of lightcone limits is given by
\begin{equation}
\LCL_{\vec{\ep}^{\,\prime}}: X_{16}\ll X_{34} \ll X_{12},X_{56} \ll X_{23},X_{45}, X_{46} \ll 1.
\label{lcl_6pt_nonplanar}
\end{equation}
If we only expand the direct channel explicitly in the first two limits $X_{16} \ll X_{34}\ll 1$, then the non-planar crossing equation takes the form
\begin{eqnarray}
\left( \frac{\dutchcal{u}_1\dutchcal{u_2}^{\frac{3}{2}}\dutchcal{u}_3 \sqrt{v_1v_3}}{\mathcal{U}^6} \right)^{\Delta_\phi} & & \left(1+v_2^{h_\star} \hypg{\bar h_\star}{\bar h_\star}{2\bar h_\star}(1-v_1 v_3 \dutchcal{u}_2)+\orm(X_{16}^{h>h_\star}) \right) \label{eq:6pt_cse_nonplanar}\\[2mm]
& & \hspace*{5cm} = \sum_{\Xi} P_{\Xi} 
\psi^{\CC}_{\Xi}(\dutchcal{u}_a,v_a,\mathcal{U}_s,\mathcal{U}^6),
\nonumber
\end{eqnarray}
where the crossed-channel sum is specified by
\begin{equation}
\Xi = \left([\phi\phi]_{0,J_1} [\phi\phi\phi]_{0,J_2,\kappa}[\phi\phi]_{0,J_3};n_1n_2 \right).
\end{equation}
Acting with the comb channel Casimirs at leading order in the lightcone limit \eqref{lcl_6pt_nonplanar},  we then find that the crossed channel sum is dominated by the large spin regime
\begin{equation}
\mathrm{LS}_{\mathrm{II}'} :J_1^2=\orm(\ep_{16}^{-1}), \quad J_2^2=\orm(\ep_{16}^{-1}\ep_{34}^{-1}), \quad J_3^2=\orm(\ep_{16}^{-1}).
\label{eq:6pt_LS_II_nonplanar}
\end{equation}
This ``case~II$'$" scaling regime is the closest analog to the five-point case~II blocks computed in section~\ref{sect:five_pt_crossed_blocks}.  From this point of view, one may expect that a solution to the crossing equation at all orders in $X_{23},X_{45}$ would further constrain the triple-twist OPE coefficients in the crossed channel, as we saw in section~\ref{sec:5pt_bootstrap} for double-twist OPE coefficients $P_{[\phi\phi][\phi\phi]}^{(J-\dg n)}$.  However, the direct channel contribution on the left-hand side of eq.~\eqref{eq:6pt_cse_nonplanar} is exactly the same for $X_{23},X_{45} \ll 1$ or $X_{23},X_{45}=\orm(1)$, which suggests that we may not obtain any new information by relaxing the latter limits.  We will address this question in our upcoming work~\cite{In_preparation}. \medskip

The next to leading $\orm(X_{34}^{h_\star})$ contribution to the non-planar crossing equation \eqref{eq:6pt_cse_nonplanar} in the direct channel is
\begin{equation}
v_2^{h_\star} \hypg{\bar h_\star}{\bar h_\star}{2\bar h_\star}(1-v_1v_3\dutchcal{u}_2) = v_2^{h_\star} \left(\frac{\log \dutchcal{u}_2^{-1}}{\mathrm{B}_{\bar h_\star}} + \orm( \dutchcal{u}_2^0)\right).
\end{equation}
In the crossed channel, we expect this logarithm to come from a correction to the anomalous dimension of $[\phi\phi\phi]_{0,J_2,\kappa}$.  
\subsection{Concluding remarks}
This work can be seen as a proof of concept for the multipoint lightcone bootstrap.  In particular, the notions of Casimir and vertex singularity of the direct channel provide an efficient criterion for the presence of universal data in the crossed channel.  The variation in the degree of Casimir/vertex singularity, which first appears non-trivially at $N>4$ points, translates into different scaling regimes in the spin and tensor structures of the multipoint data.  While subleading scaling limits raise the difficulty in computing the blocks (c.f.  case~I and case~II blocks in this paper),  the payoff is a higher resolution in the conformal field theory data. 

As we briefly discussed in this section,  the triple-twist data of the six-point comb channel appears only in subleading regimes,  where the spins and tensor structures do not all scale independently.  Solving in the full lightcone limit,  we expect the crossing equation to fix only certain averages of this data over a subset of spins and tensor structures.  It would be interesting to better understand how far these averages can be resolved by relaxing certain lightcone limits and, if not, to find a natural physical interpretation. 

We believe that our integrability-based approach to 
lightcone blocks can be generalized to a wide class of OPE channels and lightcone 
limits.  In this approach, the blocks in the relevant regime are computed by 
interpolating the OPE limit and the lightcone limit with the differential 
equations of the integrable system.  From this point of view, the computation is 
analogous to an integrable scattering problem for a basis of many-body quantum mechanical wave functions.  While its technical implementation may seem daunting, we are motivated by the simplicity of the differential equations in lightcone limits.  It would be interesting to investigate how the standard integrability machinery may adapt to and simplify in such limits,  particularly in determining the spectra of the vertex differential operators and relations between different bases of tensor structures.

\bigskip 

\noindent 
\textbf{Acknowledgements:} We are grateful to Ant\'onio Antunes, Carlos Bercini, Ilija Buri\'c, 
Miguel Costa, Aleix Gimenez-Grau, Vasco Gonçalves, Sebastian Harris, Murat Kolo$\breve{\mathrm{g}}$lu, Petr Kravchuk, Sylvain Lacroix, Pedro Liendo, Andreas Stergiou, Pedro Vieira, and Jo\~ao Vilas Boas
for useful discussions. This project received funding from the German Research 
Foundation DFG under Germany’s Excellence Strategy -- EXC 2121 Quantum Universe -- 
390833306 and from the European Union’s Horizon 2020 research and innovation 
programme under the MSC grant agreement No.764850 “SAGEX”. J.A.M. is funded by the Royal Society under grant URF$\backslash$R1$\backslash$211417.

\appendix
\section{Five point lightcone blocks}
\label{app:5pt_lightcone_blocks}
\subsection{Integral representation from lightcone OPE}
\label{app:5pt_integral_formula}
The lightcone OPE for two scalars can be written as
\begin{equation}
\phi_1(X_1)\phi_2(X_2) \stackrel{X_{12}\rightarrow 0}{\sim}\sum_{\oo_3}  \frac{X_{12}^{h_3-h_1-h_2}C_{\phi_1\phi_2\oo_3}}{\Rs^\times\mathrm{B}_{\bar h_3+h_{12},\bar h_3+h_{21}}} \int_{\Rs_+^2} \frac{\dd s_1}{s_1}s_1^{\bar h_3+h_{21}} \frac{\dd s_2}{s_2} s_2^{\bar h_3+h_{12}} \oo_3(X, Z),
\label{eq:lOPE}
\end{equation}
with $X:= s_1 X_1+s_2 X_2$, $X\wedge Z := X_1 \wedge X_2$.  We write the Euler Beta function as $\mathrm{B}_{a,b} :=\Gamma(a+b)^{-1}\Gamma(a)\Gamma(b)$ and use the shorthand notation ``$\Rs^\times := \int_0^\infty r^{-1} \dd r $" to denote the volume of the dilation group, which acts on the integration variables as $(s_1,s_2)\rightarrow (r s_1,r s_2)$.  Since the integrand is invariant under any such dilation,  we may factorize $\Rs^\times$ via a change of variables or by using the Faddeev-Popov method.  In particular, the more familiar formula of Ferrara et al.  is obtained by the change of variables $(s_1,s_2)=(r t,r(1-t))$, for $(r,t)\in \Rs_{>0} \times [0,1]$, 
\begin{equation*}
\phi_1(X_1)\phi_2(X_2)  \stackrel{X_{12}\rightarrow 0}{\sim} \sum_{\oo_3}  \frac{X_{12}^{h_3-h_1-h_2}C_{\phi_1\phi_2\oo_3}}{\mathrm{B}_{\bar h_3+h_{12},\bar h_3+h_{21}}} \!\int_{0}^1\!\! \frac{\dd t}{(t(1-t))^{1-\bar h_3}} (1/t-1)^{h_{12}} \oo_3(X_2-t(X_2-X_1), X_2-X_1).
\end{equation*}
Applying this OPE twice in a five-point function of identical scalars yields an integral formula for lightcone blocks,
\begin{align}
\psi_{\oo_1\oo_2;n}^{(12),(45)}(u_{si}) \stackrel{u_{s1,4}\rightarrow 0}{\sim} &X_{12}^{h_1} X_{45}^{h_2} (X_{15}X_{24}-X_{14}X_{25})^n  \int_{\Rs_+^4} \frac{\dd^4 \log s\,(s_1s_2)^{\bar h_1}(s_4s_5)^{\bar h_2}}{(\Rs^\times)^2\,\mathrm{B}_{\bar h_1}\mathrm{B}_{\bar h_2}} \label{five_point_LC_block}\\
&X_{a3}^{h_2-\bar h_1-h_\phi+n}X_{b3}^{h_1-\bar h_2-h_\phi+n} X_{ab}^{-\bar h_{12;\phi}} J_{a,3b}^{J_1-n}J_{b,3a}^{J_2-n},
\label{int_ts_5pt}
\end{align}
where $X_a = s_1X_1+s_2X_2$, $X_b = s_4X_4+s_5X_5$ and $X_a \wedge Z_a = X_1 \wedge X_2$, $X_b \wedge Z_b = X_5\wedge X_4$, such that e.g. $X_{a3}=s_1 X_{13}+s_2X_{23}$. To efficiently reduce this formula to cross-ratio space, we fix the gauge $X_i=X_i^{\star}$, with
\begin{equation}
X_2^{\star} = (1,0,\vct{0}), \quad X_3^{\star} = (0,1,\vct{0}), \quad X_{4} = (1,1,\vct{n}),
\end{equation}
in lightcone coordinates $\dd X^2 = -\dd X^+\dd X^-  + \dd \mathbf{X}^2$ and for some unit vector $\vct{n} \in S^{d-1}$, such that
\begin{align}
& X_{15}^{\star}=u_{s2}u_{s3} u_{s5}, \quad X_{14}^{\star}=u_{s2}, \quad X_{13}^{\star}=1, \quad X_{12}^{\star}=u_{s1}u_{s3} \label{1i5p} \\
& X_{25}^{\star}=u_{s3}, \quad X_{24}^{\star}=1, \quad X_{23}^{\star}=1, \\
& X_{35}^{\star}=1, \quad X_{34}^{\star}=1, \\
& X_{45}^{\star}=u_{s2}u_{s4}\label{4i5p}.
\end{align}
We then evaluate the integrand at $X_{ij}^\star =X_{i}^\star \cdot X_{j}^\star$ to obtain a function of the cross-ratios.
\subsubsection{Equivalence with solution to Casimir equations} In section~\ref{ssec:5pt_blocks_dc}, we derived an Euler integral representation of five point blocks directly from the second-order Casimir equations of blocks with OPE limit boundary conditions. This representation can also be retrieved directly from the above lightcone integral after expressing the $X_{ij}^\star (u_{si})$ in terms of OPE cross-ratios $\bar z_a,z_a,w$,  where
\begin{equation}
(u_{si})_{i=1}^5 = \left(\frac{z_1}{1-z_2} \bar z_1,1-z_1,1-z_2,\frac{z_2}{1-z_1}\bar z_2,1-\frac{(1-w) z_1z_2}{(1-z_1)(1-z_2)}\right)+\orm(\bar z_a).
\end{equation}
The inverse map is 
\begin{equation}
(\bar z_1,\bar z_2,z_1,z_2,w)= \left(\frac{u_{s1}u_{s3}}{(1-u_{s2})},\frac{u_{s2}u_{s4}}{(1-u_{s3})}, 1-u_{s2},1-u_{s3},1-\frac{u_{s2}u_{s3}(1-u_{s5})}{(1-u_{s2})(1-u_{s3})}\right)+\orm(u_{s1,4}).
\end{equation}
We then obtain
\begin{align}
&X_{a3}^\star = s_1+s_2, \quad X_{3b}^\star = s_4+s_5, \\
&X_{ab}^\star=(s_1+s_2)(s_4+s_5)(1-z_1 S_1-z_2 S_5+w z_1 z_2 S_1 S_5), \\
&J_{a,b3}^\star = z_1 (s_4+s_5)(1-w z_2 S_5), \quad J_{b,a3} ^\star = z_2(s_1+s_2)(1-w z_1 S_1),
\end{align}
where $S_1 := s_1/(s_1+s_2)$ and $S_5:=s_5/(s_4+s_5)$. If we plug this formula into eq.\ \eqref{int_ts_5pt} and change variables to $(s_1,s_2;s_4,s_5) = (r_1 t_1,r_1(1-t_1);r_2(1-t_2),r_2 t_2)$, $r_1,r_2 \in \Rs_{>0}$, $t_1,t_2 \in [0,1]$, then we obtain
\begin{equation}
\psi_{\oo_1\oo_2;n}(u_{si}(\bar z_a,z_a,w)) \stackrel{\bar z_a\rightarrow 0}{\sim} \prod_{a=1}^2 \bar z_a^{h_a} z_a^{\bar h_a} (1-w)^n \tilde{F}_{(h_a,\bar h_a;n)}(z_1,z_2,w),
\end{equation}
where $\tilde{F}$ corresponds precisely to the integral formula \eqref{int_5pt_from_cas} derived from the Casimir equations.
\subsubsection{Euler transformation and alternative representation}
When $h_a>h_\phi$,  the Euler integral formula \eqref{int_5pt_from_cas} does not converge in the limit $X_{(a+1)(a+2)}\rightarrow 0$, i.e.  when $u_{s(a+1)}\rightarrow 0$ at $u_{s(b+1)},u_{s5}$ fixed.  In this case,  it will be convenient to analyze the integral in the cross-ratios $(u_a,v_a,x)$, where
\begin{equation}
(u_{si})_{i=1}^5 = \left(\frac{u_1}{v_2},v_1,v_2,\frac{u_2}{v_1},1-x\right).
\end{equation}
The inverse map is
\begin{equation}
(u_1,u_2,v_1,v_2,x) = (u_{s1}u_{s3},u_{s2}u_{s4},u_{s2},u_{s3},1-u_{s5}).
\end{equation}
After the  change of variables $\tilde{s}_1 = s_1 v_1$, $\tilde{s}_5 = s_5 v_2$,  we may rewrite the tensor structures in the integrand as
\begin{align*}
&X_{ab}^\star = (\tilde{s}_1+s_2)(\tilde{s}_5+s_4)\left(1- x \tilde{S}_1 \tilde{S}_5\right), \\
&J_{a,3b}^\star= (1-v_1)(\tilde{s}_5+s_4) \left(1+ \frac{v_1 x}{1-v_1} \tilde{S}_5\right), \quad J_{b,a3}^\star= (1-v_2)(\tilde{s}_1+s_2) \left(1+ \frac{v_2x }{1-v_2} \tilde{S}_1\right),
\end{align*}
where $\tilde{S}_1=\tilde{s}_1/(\tilde{s}_1+s_2)$ and $\tilde{S}_5=\tilde{s}_5/(s_4+\tilde{s}_5)$. Plugging this into eq.\ \eqref{int_ts_5pt} then yields the result
\begin{equation}
\psi_{\oo_1\oo_2;n}(u_{si}(u_a,v_a,x)) \stackrel{u_a\rightarrow 0}{\sim} \prod_{a=1}^2 u_a^{h_a} (1-v_a)^{J_a-n} v_a^{-h_{b\phi}} x^n F_{\oo_1\oo_2;n}(v_a,x),
\end{equation}
where $F_{\oo_1\oo_2;n}$ is the integral defined in eq.\ \eqref{eq:Fint}.  The two integral representations are related by a generalized Euler transformation,
\begin{equation}
\tilde{F}_{\oo_1\oo_2;n}(1-v_1,1-v_2,w(v_1,v_2,x))= \prod_{a\neq b} v_a^{-(h_{b\phi}+n)} F_{\oo_1\oo_2;n}(v_1,v_2,x). 
\end{equation}
where the change of variables $w\leftrightarrow x$ is given by
\begin{align}
x(z_a,w)=\frac{z_1z_2(1-w)}{(1-z_1)(1-z_2)}, \quad w(v_a,x) =1-\frac{v_1 v_2x}{(1-v_1)(1-v_2)}.
\end{align}
For $a\neq b$ and $h_{b\phi}+n\neq 0$, this relation ensures that $F$ converges at $v_a=0$ when $f$ diverges and vice versa. 

\subsection{Expansion of lightcone blocks around decoupling limit}
The goal of this section is to expand blocks as a power series in $x$ around the decoupling limit $x=0$.  With applications to crossed channel blocks in mind, we will assume $h_1,h_2>h_\phi$ such that $F_{\oo_1\oo_2;n}(v_a,x)$ converges at $v_a=0$.  Setting $\tilde{s}_{1,5}\equiv s_{1,5}$ for simplicity, such that $X_a= \frac{s_1}{v_2} X_1+s_2X_2$ and $X_b=s_4X_4+\frac{s_5}{v_2}X_5$, we can express the integral as
\begin{align}
F_{\oo_1\oo_2;n}(v_a,x)= &\int_{\Rs_+^2} \frac{\dd^2 \log (s_1,s_2)\,(S_1S_2)^{\bar h_1}}{\Rs^\times\mathrm{B}_{\bar h_1}(1-(1-v_1)S_1)^{\bar h_1-h_{2\phi}-n}}\int_{\Rs_+^2}  \frac{\dd^2 \log (s_4,s_5)\,(S_4S_5)^{\bar h_2}}{\Rs^\times\mathrm{B}_{\bar h_2}(1-(1-v_2)S_4)^{\bar h_2-h_{1\phi}-n}}  \nonumber \\
&\left(1+\frac{v_1 x}{1-v_1}S_5 \right)^{J_1-n} \left(1+\frac{v_2 x}{1-v_2} S_1 \right)^{J_2-n} (1-x S_1S_5)^{-\bar h_{12;\phi}}.
\label{integral_TS}
\end{align}
To obtain a power series in $x$, we first expand the integrand and then express each coefficient as a differential operator acting on $F_{\oo_1\oo_2;n}(v_a,0)$. 
\subsubsection{Factorization in the decoupling limit}
Evaluating eq.\ \eqref{integral_TS} at $x=0$, we find a factorization into a product of two integrals,
\begin{equation}
F_{\oo_1\oo_2;n}(v_1,v_2,0) = \int_{\Rs_+^2} \frac{\dd^2 \log (s_1,s_2)\,(S_1S_2)^{\bar h_1}}{\Rs^\times\mathrm{B}_{\bar h_1}(1-(1-v_1)S_1)^{\bar h_1-h_{2\phi}-n}}\int_{\Rs_+^2}  \frac{\dd^2 \log (s_4,s_5)\,(S_4S_5)^{\bar h_2}}{\Rs^\times\mathrm{B}_{\bar h_2}(1-(1-v_2)S_4)^{\bar h_2-h_{1\phi}-n}}.
\end{equation}
Each integral is equal to one of the Gauss hypergeometric functions in eq.\ \eqref{dec_f_a}.  More generally,  the three-parameter Gauss hypergeometric function admits an integral representation
\begin{align}
&\hypg{a}{b}{c}(1-v) = \int_{\Rs_+ ^2} \frac{\dd s_1}{s_1}\frac{\dd s_2}{s_2}\frac{1}{\Rs^\times}\,\,\hygen{f}{a}{b}{c}(1-v;s_1,s_2), \\
&\hygen{f}{a}{b}{c}(1-v;s_1,s_2)=  \frac{s_1^{c-b} s_2^b}{\mathrm{B}_{b,c-b}}\frac{(s_1+s_2)^{a-c}}{(s_1+s_2v)^a},
\end{align}
which reduces to the Euler Beta integral representation after the change of variables $(s_1,s_2)=(r(1-t), rt)$,  $(r,t) \in \Rs_{>0} \times [0,1]$.  In this homogeneous integral representation, the Euler transformation of the Gauss hypergeometric follows from
\begin{equation}
\hygen{f}{a}{b}{c}(1-v;s_1 v,s_2)=  v^{c-a-b} \hygen{f}{c-a}{c-b}{c}(1-v;s_2,s_1).
\end{equation}

\subsubsection{Expansion around the decoupling limit}
We can now rewrite the integral in eq.~\eqref{integral_TS} as 
\begin{align}
F_{\oo_1\oo_2;n}(v_a,x) =\!\!\int_{\Rs_+^4}\!\frac{\dd^4\log s}{(\Rs^\times)^2}\, & \hygen{f}{\bar h_1-h_{2\phi}-n}{\bar h_1}{2\bar h_1}(1-v;s_1,s_2)\, \hygen{f}{\bar h_2-h_{1\phi}-n}{\bar h_2}{2\bar h_2}(1-v;s_5,s_4) \nonumber \\
& \left(1+\frac{v_1 x}{1-v_1}S_5 \right)^{J_1-n} \left(1+\frac{v_2 x}{1-v_2} S_1 \right)^{J_2-n} (1-x S_1S_5)^{-\bar h_{12;\phi}}
\end{align}
Since $0 \leq S_1,S_5 \leq 1$ for $s_i\in \Rs_+$ we can expand each factor into a binomial series when $0\leq x \leq 1$ and $0 \leq (1-v_a)^{-1}v_a x \leq 1$. In this case, we observe that all higher-order corrections to the integrand in $x$ are proportional to
\begin{align*}
x^{k+m_1+m_2} S_1^{k+m_2} \hygen{f}{\bar h_1-h_{2\phi}-n}{\bar h_1}{2\bar h_1}(1-v_1;s_1,s_2)S_5^{k+m_2} \hygen{f}{\bar h_2-h_{1\phi}-n}{\bar h_2}{2\bar h_2}(1-v_2;s_5,s_4),
\end{align*}
for some triplets of positive integers $k,m_1,m_2$. At the same time,  these integrand shifts $S_i^\nu$ are equivalent to
\begin{equation}
S_1^{\nu}\,\hygen{f}{a}{b}{c}(1-v;s_1,s_2) = \frac{(c-b)_\nu}{(c)_\nu} \hygen{f}{a}{b}{c+k}(1-v;s_1,s_2)
\end{equation}
The right-hand side lifts to the integral itself,  and to express it in a simple way we define an operator $\mathcal{S}_1^\nu$ that realizes this integrand transformation,
\begin{equation}
\mathcal{S}_1^\nu\cdot  \hypg{a}{b}{c}(1-v) := \frac{(c-b)_\nu}{(c)_\nu} \hypg{a}{b}{c+\nu}(1-v) .
\label{def_Sk}
\end{equation}
The operator $\mathcal{S}_1^\nu$ obviously depends on the parameters $a,b,c$ and the variable $v$, but we will not need to make this dependence explicit in future uses.  Using the explicit form of the integrand $\tensor[_2]{f}{_1}(z;s_1,s_2)$, it is not difficult to write this transformation as a differential operator,
\begin{align*}
\frac{(c-b)_\nu}{(c)_\nu} \hypg{a}{b}{c+\nu}(1-v) &=v^{c+\nu-a-b} \frac{(c-b)_\nu}{(c)_\nu} \hypg{c-a+\nu}{c-b+\nu}{c+\nu}(1-v) \\
&= v^{c+\nu-a-b} \frac{(-\ds_v)^\nu}{(c-a)_\nu} \hypg{c-a}{c-b}{c}(1-v)  \\
&= v^{c-a-b}\frac{v^\nu(-\ds_v)^\nu}{(c-a)_\nu} v^{a+b-c} \hypg{a}{b}{c}(1-v).
\end{align*}
Since the differential operator acting on the $\tensor[_2]{F}{_1}$ is homogeneous of degree zero in $v$, it can be written as a function of $\vt_v:= v\ds_v$. By inspecting its action on the eigenbasis of $\vt_v$, we find
\begin{equation*}
v^\nu (-\ds_v)^\nu \cdot v^n = (-n)(-n+1)\dots(-n+\nu)v^n \Rightarrow v^\nu (-\ds_v)^\nu= (-\vt_v)_\nu.
\end{equation*}
Altogether, we obtain two useful representations of $\mathcal{S}_1^\nu$, defined in eq.\ \eqref{def_Sk},  as a differential operator.
\begin{equation}
\mathrm{Acting} \,\, \mathrm{on} \,\,\,  \hypg{a}{b}{c}(1-v), \quad \mathcal{S}_1^\nu =v^{c-a-b}\frac{v^\nu(-\ds_v)^\nu}{(c-a)_\nu} v^{a+b-c} = \frac{(c-a-b-\vt_v)_\nu}{(c-a)_\nu}. \label{Sk_diff}
\end{equation}
Here, for the all orders expansion of lightcone blocks in $x$, we will use two such shift operators,
\begin{equation}
\mathcal{S}_a^\nu\cdot  \hypg{\bar h_a-h_{b\phi}-n}{\bar h_a}{2\bar h_a}(1-v_a):=\frac{(\bar h_a)_\nu}{(2\bar h_a)_\nu} \hypg{\bar h_a-h_{b\phi}-n}{\bar h_a}{2\bar h_a+\nu}(1-v_a), \quad a\neq b=1,2.
\label{Ska}
\end{equation}
The resulting formulas for $\mathcal{S}_1^\nu,\mathcal{S}_2^\nu$ as differential operators can then be read from eq.\ \eqref{Sk_diff} by inserting the corresponding values of $(a,b,c)$ in the $\tensor[_2]{F}{_1}$'s. We will also adopt the notation of eq.\ \eqref{dec_f_a} for the product of hypergeometrics at $x=0$, that is
\begin{equation}
f_a^{\mathrm{dec}}(v_a) := \hypg{\bar h_a-h_{b\phi}-n}{\bar h_a}{2\bar h_a}(1-v_a), \quad a\neq b=1,2.
\label{dec_f_a_app}
\end{equation}
Finally, with all of the necessary conventions listed in eq.~\eqref{Sk_diff}---\eqref{dec_f_a_app},  the all-orders expansion of five-point lightcone blocks around the decoupling limit $x=0$ can be expressed as
\begin{equation}
F_{\oo_1\oo_2;n}(v_a,x)=\left(1-x\, \mathcal{S}_1\mathcal{S}_2\right)^{-\bar h_{12;\phi}} \left(1+x \frac{v_1}{z_1}\mathcal{S}_2\right)^{J_1-n}\left(1+x\,\mathcal{S}_1 \frac{v_2}{z_2}\right)^{J_2-n} f_1^{\mathrm{dec}}(v_1) f_2^{\mathrm{dec}}(v_2).
\label{block_decsum_exact}
\end{equation}
In this formula, one should understand the three factors on the right-hand side as a binomial power series of the form $(1-x \la)^{-\Dg}= \sum_{k=0}^\infty \frac{x^k}{k!} (\Dg)_k \la^k$.  This culminates in a triple-sum power series expansion in $x$,
\begin{equation}
F_{\oo_1\oo_2;n}(v_a,x)=\sum_{k=0}^\infty \frac{(\bar h_{12;\phi})_k}{k!} \prod_{b\neq a=1} ^2\sum_{m_a=0}^{J_a-n} \binom{J_a-n}{m_a} \left(\frac{v_a}{z_a}\right)^{m_a} \mathcal{S}_a^{k+m_b} f_a^{\mathrm{dec}}(v_a)  \,  x^{k+m_1+m_2}.
 \label{rhs_block_decsum_exact}
\end{equation}
\subsubsection{Dual expansion}
While less useful for our purposes, note that an analogous expansion around the decoupling limit $\mathcal{X}:=1-w=0$ can be defined from the dual integral formula $\tilde{F}_{\oo_1\oo_2;n}(z_a,w)$ of eq.~\eqref{eq:Fint}, which can be rewritten as
\begin{align}
\tilde{F}_{\oo_1\oo_2;n}(z_a,1-\mathcal{X}) = &\int_{\Rs_+^4} \frac{\dd^4\log s}{(\Rs^\times)^2}\,  \hygen{f}{\bar h_1+h_{2\phi}+n}{\bar h_1}{2\bar h_1}(z_1;s_2,s_1)\, \hygen{f}{\bar h_2+h_{1\phi}+n}{\bar h_2}{2\bar h_2}(z_2;s_4,s_5) \nonumber \\
& \left(1+\frac{\mathcal{X} z_1 S_5}{1-z_1 S_5}\right)^{J_1-n}\left(1+\frac{\mathcal{X}z_2 S_1}{1-z_2 S_1}\right)^{J_2-n} \left(1-\frac{\mathcal{X} z_1z_2 S_1S_5}{(1-z_1 S_5)(1-z_2 S_1)}\right)^{-\bar h_{12;\phi}}.
\end{align}
If we now define
\begin{equation}
\tilde{f}_a^{\mathrm{dec}}(z_a) := (1-z_a)^{-(h_{b\phi}+n)} f_a^{\mathrm{dec}}(1-z_a) = \hypg{\bar h_a+h_{b\phi}+n}{\bar h_a}{2\bar h_a}(z_a),
\end{equation}
then all higher $\mathcal{X}$ corrections can be expressed in terms of the differential operators
\begin{equation}
\tilde{\mathcal{S}}_a^\nu \cdot \tilde{f}_a^{\mathrm{dec}}(z_a) := \frac{z_a^\nu\ds_{z_a}^\nu   \tilde{f}_a^{\mathrm{dec}}(z_a)}{(\bar h_a+h_{b\phi}+n)_\nu}, \quad a\neq b=1,2.
\label{tildeSk}
\end{equation}
In this case, we may express the expansion of $\tilde{F}$ around the decoupling limit as
\begin{equation}
\tilde{F}_{\oo_1\oo_2;n}(z_a,1-\mathcal{X})=\left(1-\mathcal{X} \tilde{\mathcal{S}}_1\tilde{\mathcal{S}}_2\right)^{-\bar h_{12;\phi}} \left(1+\mathcal{X} \tilde{\mathcal{S}}_2\right)^{J_1-n}\left(1+\mathcal{X}\tilde{\mathcal{S}}_1 \right)^{J_2-n} \tilde{f}_1^{\mathrm{dec}}(z_1) \tilde{f}_2^{\mathrm{dec}}(z_2).
\label{dec_exp_Fdual}
\end{equation}

\subsection{Application to direct channel blocks}
\label{app:direct_channel_5pt}
In our analysis of the direct channel, we will be interested in lightcone limits of the form
\begin{equation}
(z_a,w)=(1,1)+\orm(\ep_a) \iff (v_a,x) = (\orm(\ep_a),\orm(1)), \quad \ep_a\rightarrow 0.
\end{equation}
The asymptotics of blocks in this limit depend on the sign of $h_{b\phi}+n$: for $h_{b\phi}+n\geq 0$, these are easiest to derive from the expansion of $F_{\oo_1\oo_2;n}(v_a,x)$ in eq.~\eqref{rhs_block_decsum_exact}, while for $h_{b\phi}+n \leq 0$, these are easiest to derive from the expansion of $\tilde{F}_{\oo_1\oo_2;n}(z_a,w)$ in eq.~\eqref{dec_exp_Fdual}. 

\paragraph{Asymptotics of blocks with $h_{b\phi}+n< 0$.} Consider first the lightcone limit $z_a=1+\orm(\ep_a)$, $\mathcal{X}=1-w=\orm(\ep_a)$,  in the case where $h_{b\phi}+n<0$.  In this case,  $\tilde{f}_a^{\mathrm{dec}}(z_a)$ converges at $z_a=1$ and
\begin{equation}
\mathcal{X}^m \tilde{\mathcal{S}}_a^m \cdot  \tilde{f}_a^{\mathrm{dec}}(z_a) = \orm\left(\epsilon_a^{\mathrm{min}(\abs{h_{b\phi}+n},m)}\right), \quad \mathcal{X}^m \tilde{f}_a^{\mathrm{dec}}(z_a) = \orm\left(\epsilon_a^{m}\right), \quad \forall m \in \Zs_{>0}.
\end{equation}
As a result, the whole function $\tilde{F}$ converges in this double scaling limit, which coincides with the evaluation of $\tilde{F}$ at $z_a=1,w=1$. Taking $a=2$ without loss of generality,  we may thus write
\begin{equation}
\lim_{\ep_2\rightarrow 0} \tilde{F}_{\oo_1\oo_2;n}(z_1,1+\orm(\ep_2), 1+\orm(\ep_2)) = \tilde{f}_1^{\mathrm{dec}}(z_1).
\end{equation}
\paragraph{Asymptotics of blocks with $h_{b\phi}+n< 0$.} In this case, we know that $\tilde{f}_a^{\mathrm{dec}}(1-\ep_a)=f_a^{\mathrm{dec}}(\ep_a) = \mathrm{B}_{\bar h_2}^{-1}\log\ep_a^{-1}+\orm(\ep_a^0)$ admits a logarithmic divergence.  At the same time,  it is easy to check from the differential operator expression for $\tilde{\mathcal{S}}_a^\nu$ in eq.~\eqref{tildeSk} (respectively $\mathcal{S}_a^\nu$ in eq.~\eqref{Sk_diff}) that all higher powers in $1-w=\mathcal{X}$ (respectively $x$) in the expansion of $\tilde{F}$ (respectively $F$) around the decoupling limit are subleading, i.e. for any $m \in \Zs_{>0}$
\begin{align*}
\mathcal{X}^m \tilde{\mathcal{S}}_a^m \tilde{f}_a^{\mathrm{dec}} = \orm(\ep_a^0)&,  \quad \mathcal{X}^m \tilde{f}_a^{\mathrm{dec}} = \orm(\ep_a^m \log \ep_a), \\[2mm] 
\quad x^m \mathcal{S}_a^m f_a^{\mathrm{dec}} = \orm(\ep_a^0)&,  \quad \left(\frac{x v_a}{1-v_a} \right)^m \tilde{f}_a^{\mathrm{dec}} = \orm(\ep_a^m \log \ep_a).
\end{align*}
It follows that the leading asymptotics in the double scaling limit coincides with the leading asymptotics when $z_a\rightarrow 1$ in the decoupling limit $\mathcal{X}=0$. In concrete terms,  for $z_1=1-v_1$,
\begin{align}
\lim_{v_2\rightarrow 0} \frac{\tilde{F}_{\oo_1\oo_2;n}(z_1,1-v_2, 1+\orm(v_2))}{\log v_2^{-1}}  =\tilde{f}_1^{\mathrm{dec}}(z_1) \lim_{v_2\rightarrow 0} \frac{\tilde{f}_2^{\mathrm{dec}}(1-v_2)}{\log v_2^{-1}}  = \frac{\tilde{f}_1^{\mathrm{dec}}(z_1)}{\mathrm{B}_{\bar h_2}} \nonumber \\
=\lim_{v_2\rightarrow 0} \frac{F_{\oo_1\oo_2;n}(v_1,v_2, x)}{\log v_2^{-1}}  =f_1^{\mathrm{dec}}(v_1) \lim_{v_2\rightarrow 0} \frac{\tilde{f}_2^{\mathrm{dec}}(v_2)}{\log v_2^{-1}}  = \frac{f_1^{\mathrm{dec}}(v_1)}{\mathrm{B}_{\bar h_2}}.\nonumber 
\end{align}
\paragraph{Asymptotics of blocks with $h_{b\phi}+n>0$.} In this case, not only does $f_a^{\mathrm{dec}}(v_a)$ converge at $v_a=0$,  but 
\begin{equation}
\mathcal{S}_a^\nu\cdot f_a^{\mathrm{dec}}(v_a) = \frac{\mathrm{B}_{\bar h_a,h_{b\phi}+n}}{\mathrm{B}_{\bar h_a}} \frac{(2\bar h_a)_\nu (h_{b\phi}+n)_\nu}{(\bar h_a+h_{b\phi})_\nu (\bar h_a)_\nu}+\orm(v_a).
\end{equation}
We can thus write
\begin{align}
\lim_{v_a\rightarrow 0} F_{\oo_1\oo_2;n}(v_1,v_2,x) =&  \label{eq:Fint_one_v_to_0}  \\
 &\hspace*{-3cm}= \frac{\mathrm{B}_{\bar h_a,h_{b\phi}+n}}{\mathrm{B}_{\bar h_a}}\sum_{k=0}^\infty \frac{(\bar h_{12;\phi})_k}{k!} \sum_{m_b=0}^{J_b-n} \binom{J_b-n}{m_b} \left(\frac{v_b}{z_b}\right)^{m_a} \frac{(2\bar h_a)_{k+m_b} (h_{b\phi}+n)_{k+m_b}}{(\bar h_a+h_{b\phi})_{k+m_b} (\bar h_a)_{k+m_b}} \mathcal{S}_b^k f_b^{\mathrm{dec}}(v_b) x^{k+m_b}. \nonumber
\end{align}

\subsection{Application to crossed channel blocks}
\label{app:cross_channel}
In regimes relevant to the crossed channel OPE decomposition, the quantum numbers $\bar h_1,\bar h_2,n$ diverge by a specified $\ep$-scaling with the limits $v_{1,2}= \orm(\ep_{23,34})$ and $1-x=\orm(\ep_{15})$.  The spins $\bar h_1,\bar h_2$ are large numbers in all of these cases, and the tensor structure $n$ may also scale with $\ep_{15}$.  Using the differential operator representation \eqref{Sk_diff} of the $\mathcal{S}$-operators and their powers, we find
\begin{equation}
\mathcal{S}_a^k(\vt) = \bar h_a^{-k} (h_{b\phi}+n-\vt)_k \left(1+\orm(\bar h_a^{-1})\right).
\end{equation}
Consequently,  the two last factors in eq.\ \eqref{block_decsum_exact} lead to corrections that scale non-trivially with $J,v$,
\begin{align}
\left(1+x \frac{v_a}{z_a}\mathcal{S}_b\right)^{J_a-n} = 1+\orm \left(\frac{(J_a-n)v_a}{\bar h_b} \right), \quad a\neq b=1,2.
\end{align}
These corrections will be subleading in any regime where
\begin{equation}
\bar h_2^{-2}, v_2 = \orm(\ep_{34}), \quad \bar h_1 \ll \bar h_2.
\end{equation}
This approximation applies consistently to all large spin limits considered in the crossed channel $(12)3(45)$ that are relevant to the bootstrap analysis of section~\ref{sec:5pt_bootstrap}.

\paragraph{External scalar exchange in the crossed channel.}
Consider the special case
\begin{equation}
\oo_1 = \phi, \quad n=0, \quad \oo_2=[\phi\phi]_{0,J_2}, \quad J_2^2=\orm(\ep_{15}^{-1}\ep_{34}^{-1}).
\end{equation}
In this case, blocks satisfy the same Casimir equations \eqref{D12m10_eq},\eqref{D45m1m10_eq} as case~II blocks, but with $\la_1u_{s5}\rightarrow 0$\footnote{This implies that the blocks are in the kernel of \emph{both} $\mathcal{D}_{12}^{(-1,0,-1)}$ and $\mathcal{D}_{12}^{(-1,0,0)}$.} and $\dg n=0$.  We can thus write their asymptotics as
\begin{equation}
F_{\phi[\phi\phi]_{0,J_2};0}(u_{s2},u_{s3},1-u_{s5}) \sim\nn_{\phi [\phi\phi]_{0,J_2}}^\phi u_{s5}^{h_\phi}  \mathcal{K}_{h_\phi}(J_2^2 u_{s3} u_{s5}).
\label{block_34_id}
\end{equation}
In this case, however, the normalization $\nn^\phi$ cannot be computed in the same way as case~II blocks because we now have $h_1=h_\phi-n$ instead of $h_1>h_\phi-n$, meaning that the limit of the block at $u_{s3} J_2^2 \ll u_{s5}^{-1}$ has different asymptotics than $(v,x)\rightarrow (0,1)$ at finite $J_2$.  More specifically, the hypergeometrics in the decoupling limit simplify to
\begin{equation}
f_1^{\mathrm{dec}}(v_1) = 1, \quad f_2^{\mathrm{dec}}(v_2) = \hypg{\bar h_2}{\bar h_2}{2\bar h_2}(1-v_2).
\end{equation}
We can recognize on the right-hand side a four-point lightcone block, with logarithmic divergence as $v_2\rightarrow 0$.  Instead,  we will determine the normalization $\nn^\phi$ by direct computation, taking the lightcone blocks at the $(0,v_2,x)$ node\footnote{Since $f_1^{\mathrm{dec}}(v_1)=1$, the function $F$ is independent of $v_1$ in this case.} along the two green arrows in fig.~\ref{fig:lim_diag_5pt_gen} to the $(0,0,1)$ node where eq.\ \eqref{block_34_id} applies.  In this specific case, the expansion around the decoupling limit takes the form
\begin{equation}
F_{\phi[\phi\phi]_{0,J_2};0}(v_a,x) = \mathrm{B}_{\bar h_2}^{-1} \sum_{k=0}^\infty \frac{x^k}{k!} \frac{(h_\phi)_k}{(2h_\phi)_k} \frac{\Gamma(\bar h_2+k)^2}{\Gamma(2\bar h_2+k)} \,\hypg{\bar h_2}{\bar h_2}{2\bar h_2+k}(1-v_2).
\end{equation}
We begin with the first limit
\begin{equation}
\mathrm{LS}_{34}: \quad J_2^{-2},v_2 = \orm(\ep_{34}), \quad \ep_{34}\rightarrow 0.
\end{equation}
In this regime, the Gauss hypergeometric functions reduce to modified Bessel functions following the identity
\begin{equation}
\lim_{\bar h\rightarrow \infty} \frac{\Gamma(\bar h+k)^2}{\Gamma(2\bar h+k)} \, \, \hypg{\bar h}{\bar h}{2\bar h+k}(1-\bar h^{-2} y) =2 \mathcal{K}_{-k}(y).
\end{equation}
We thus obtain
\begin{equation}
F_{\phi[\phi\phi]_{0,J_2};0}(v_a,x)  \stackrel{\mathrm{LS}_{34}}{\sim} 2 \mathrm{B}_{\bar h_2}^{-1} \sum_{k=0}^\infty \frac{x^k}{k!} \frac{(h_\phi)_k}{(2h_\phi)_k} \mathcal{K}_{-k}(v_2 J_2^2).
\label{phi_exch_k}
\end{equation}
At this stage, we refrained from applying the Stirling formula to the Beta function $\mathrm{B}_{\bar h} = \Gamma(2\bar h)^{-1}\Gamma(\bar h)^2$ in order to keep the above expression more compact.  Now,  to reach the limiting form of eq.~\eqref{block_34_id},  we take the second large spin limit to the $(0,0,1)$ node,
\begin{equation}
\mathrm{LS}_{34,15}: \quad v_2J_2^2, 1-x= \orm(\ep_{15}^{-1}), \quad \ep_{15}\rightarrow 0.
\end{equation}
In this limit,  the sum \eqref{phi_exch_k} is dominated by the regime $k=\orm(\ep_{15}^{-1})$.  Approximating the sum over $k=\orm(\ep_{15}^{-1})$ by an integral and using the large $k$ formulas
\begin{align}
(\Dg)_k &=  \sqrt{\frac{\pi}{k}} \left(\frac{k}{e}\right)^{k+\Dg} \Gamma(\Dg)^{-1}\left(1+\orm(k^{-1})\right),  \quad x^k = e^{-k(1-x)} \left(1+\orm(k^{-1}) \right),\\
\mathcal{K}_k(k x) &= \frac{1}{2} \sqrt{\frac{\pi}{k}} \left(\frac{k}{e}\right)^{k} e^{-x} \left(1+\orm(k^{-\frac{1}{2}})\right),
\end{align}
we find
\begin{equation}
F_{\phi[\phi\phi]_{0,J_2};0}^{(12),(45)}  \stackrel{\mathrm{LS}_{34,51}}{\sim} \frac{2 \Gamma(2h_\phi)}{\mathrm{B}_{\bar h_2} \Gamma(h_\phi)}\, \frac{1}{2}\int_{\orm(1)}^{\infty} \frac{\dd k}{k^{1+h_\phi}} e^{-k(1-x)-\frac{v_2 J_2^2}{k}}.
\end{equation}
After a change of variables $t=k(1-x)$, we can identify the integral representation \eqref{bessel_clifford_integral} of the Bessel-Clifford function, with the ratio of Gamma functions on the left providing the desired normalization of eq.~\eqref{block_34_id}:
\begin{equation}
\nn_{\phi [\phi\phi]_{0,J_2}}^\phi = 2 \frac{\Gamma(2h_\phi)}{\Gamma(h_\phi)} \lim_{\bar h_2\rightarrow \infty} \Gamma(2\bar h_2) \Gamma(\bar h_2)^{-2}= \frac{ \Gamma(2h_\phi)}{\Gamma(h_\phi)} 4^{\bar h_2} \sqrt{\frac{\bar h_2}{\pi}}.
\end{equation}
\section{Double integral of a Bessel Function}
\label{app:double_integral}
\subsection{Modified Bessel-Clifford function}
\label{app:bessel_clifford}
For any $\ag \in \Rs$, the modified Bessel-Clifford function $x\mapsto \mathcal{K}_{\ag}(x)$ is defined as the solution to the differential equation
\begin{equation}
\ds_x(x\ds_x+\ag)\mathcal{K}_{\ag}(x) = \mathcal{K}_{\ag}(x),
\end{equation}
with $x\rightarrow 0$ asymptotics depending on the sign of $\ag$,
\begin{equation}
\mathcal{K}_\ag(x) \stackrel{x\rightarrow 0}{\sim} \begin{cases} &\frac{1}{2}\Gamma(-\ag), \quad \ag <0, \\
&-\frac{1}{2} \log x, \quad \ag=0 \\
& \frac{1}{2} \Gamma(\ag) x^{-\ag}, \quad \ag >0.  \end{cases}
\end{equation}
It is related to the modified Bessel function $K_\ag(y)$ by the relation
\begin{equation}
 \mathcal{K}_\ag(x) =x^{-\ag/2} K_\ag(2\sqrt{x}) \iff K_{\ag}(y) = \left(\frac{y}{2}\right)^\ag \mathcal{K}_\ag\left(\frac{y^2}{4} \right).
\end{equation}
This solution admits an integral representation,
\begin{equation}
\mathcal{K}_\ag(x) = \frac{1}{2}\int_0^\infty \frac{\dd t}{t^{1+\ag}} e^{-\left(t+t^{-1}x\right)}.
\label{bessel_clifford_integral}
\end{equation}
Applying the change of variables $t'=t^{-1} x$ in this formula then implies the Euler transformation
\begin{equation}
\mathcal{K}_\ag(x) = x^{-\ag} \mathcal{K}_{-\ag}(x). 
\end{equation}
This Bessel-Clifford function controls the asymptotics of a class of Gauss hypergeometric functions by virtue of the identity
\begin{equation}
\lim_{\bar h\rightarrow \infty} \frac{\Gamma(\bar h+c-a)\Gamma(\bar h+c-b)}{2\, \Gamma(2\bar h+c)}\, \hypg{\bar h+a}{\bar h+b}{2\bar h+c}\left(1-\bar h^{-2}x\right) = \mathcal{K}_{a+b-c}(x).
\end{equation}
\subsection{The double integral}
In this appendix, we will prove the identity~\eqref{eq:double_integral} by direct computation of a generalization of the integral on its left-hand side, namely
\begin{equation}
I(\ag,\bg) := \int_{\Rs_+^2} \frac{\dd x \dd y}{x y} x^\ag y^\bg \mathcal{K}_{\gamma}(x+y).
\end{equation}
Since $x,y\geq 0$, it is natural to intuit the argument of the Bessel-Clifford function $\mathcal{K}_\gamma$ as the square radius of a circle in the plane. To substantiate this intuition, we make the change of variables
\begin{equation}
x = r^2 \cos^2 \theta, \quad y = r^2 \sin^2 \theta, \quad (r,\theta) \in \Rs_+ \times [0,\pi/2).
\end{equation}
The variables $(r,\theta)$ can be understood as polar coordinates on the plane, and the domain of $\theta$ ensures a bijection with the upper right quadrant $(\sqrt{x}, \sqrt{y}) \in\Rs_+^2$. The measure transforms as
\begin{align*}
\frac{\dd x \dd y}{x y} = 4 \frac{\dd (\sqrt{x}) \dd (\sqrt{y})}{\sqrt{xy}} = 4 \frac{r \dd r \dd \theta}{r^2 \cos\theta\sin\theta} = 2 \frac{\dd (r^2)}{r^2} \frac{\dd \theta}{\cos\theta\sin\theta}.
\end{align*}
The double integral thereby factorizes in polar coordinates and takes the form
\begin{equation}
I(\ag,\bg)= 2 \int_0^\infty \frac{\dd (r^2)}{r^2} (r^2)^{\ag+\bg} \mathcal{K}_{\gamma}(r^2) \int_0^{\pi/2} \dd \theta\, \cos^{2\ag-1}\theta \sin^{2\bg-1}\theta.
\end{equation}
Using $\mathcal{K}_{\gamma}(r^2) = r^{-\gamma} K_\gamma(2 r)$, we retrieve two known integrals of special functions for $z=2r$ and $\theta$, namely
\begin{align*}
2\int_0^\infty \frac{\dd z}{z} \left(\frac{z}{2}\right)^{2\ag+2\bg-\gamma}K_{\gamma}(z) = \frac{1}{2}\Gamma(\ag+\bg)\Gamma(\ag+\bg-\gamma), \quad \mathrm{Re}(2\ag+2\bg-\gamma)> \abs{\mathrm{Re}(\gamma)} \\\int_0^{\pi/2} \dd \theta\, \cos^{2\ag-1}\theta \sin^{2\bg-1}\theta = \frac{1}{2} \frac{\Gamma(\ag)\Gamma(\bg)}{\Gamma(\ag+\bg)}, \quad \mathrm{Re}(\ag),\mathrm{Re}(\bg)>0.
\end{align*}
Putting everything together, this gives us
\begin{equation}
    \int_{\Rs_+^2} \frac{\dd x \dd y}{x y} x^\ag y^\bg \mathcal{K}_{\gamma}(x+y)=\frac12 \Gamma(\alpha)\Gamma(\beta)\Gamma(\alpha+\beta-\gamma).
\end{equation}
When applied to the bootstrap analysis around eq.~\eqref{eq:double_integral}, we have $\ag=\gamma=h_\phi+\dg n$ and $\bg=2h_\phi-h_\star$, while in the case~II bootstrap analysis of section~\ref{sssec:singleid_CaseII} we have $\alpha=2h_\phi+\dg n$, $\beta=h_\phi$, and $\gamma=h_\phi+\delta n$, with parameters always satisfying $0<h_\phi< h_\star< 2 h_\phi$ and $\dg n>0$. We conclude that in both cases the integral converges, and our ans\"atze for the OPE coefficients in eqs.\ \eqref{ope_coeffs_case_II} and~\eqref{caseII_1phiexch} are consistent.

\section{Comments on tensor structure larger than spin}
\label{app:tensor_structure_larger}
In our construction of crossed channel five-point lightcone blocks we re-scaled the spins $J_1$ 
and $J_2$ of the exchanged fields as well as the tensor structure $n$. For case~II blocks the 
scaling behavior was consistent with the usual condition $n \leq \min(J_1,J_2)$. But in 
the study of case~I blocks, $n$ was sent to $\infty$ faster than $J_1,J_2$. Here we want to 
explain why this can make sense when the spins become continuous.  Let us consider the STT-STT-scalar 
three-point vertex operator we introduced in \cite{Buric:2021ttm}. Using the same notations as 
in the main text, this operator is given by
\begin{equation}
    \frac{1}{2}\textit{str}\left[\mathcal{T}_1^3 \mathcal{T}_3\right] \,,
\end{equation}
where we are considering legs 1 and 2 to be associated with STTs with spin $J_1$, $J_2$, and 
leg 3 to be associated with a scalar field. We know that the eigenfunctions of this vertex operator 
are a basis of the space of tensor structures that can also be spanned by simple powers of the 
cross-ratio $\mathcal{X}$ defined in eq.~\eqref{3pt_stt_scalar_stt}. This means that we can recast the action of the vertex operator in 
terms of the basis $\mathcal{X}^n$, which acquires the following form:
\begin{equation}
    \mathcal{V}\mathcal{X}^n= c_{-1} \mathcal{X}^{n-1}+c_0 \mathcal{X}^{n}+ c_1 
    \mathcal{X}^{n+1} + c_2 \mathcal{X}^{n+2}\,.
\end{equation}
In particular, we have
\begin{gather}
    c_{-1}=- n \left(n+\frac{d-4}{2}\right) \left(2 n+\Delta _1-\Delta _2-\Delta _3-J_1-J_2\right) 
    \left(2 n-\Delta _1+\Delta _2-\Delta _3-J_1-J_2\right),\\[2mm]
    \begin{split}c_1=4 \left(n-J_1\right) \left(n-J_2\right) 
    \Bigl[\frac{d}{2}\left(\Delta _1+\Delta _2+\Delta _3-n+1-d\right)+n(2 J_1+2 J_2 -3 n+\Delta _3-2)\\[2mm]
    -\Delta _1 \Delta _2-\Delta _3+J_1-J_1 J_2+J_2-1\Bigr]\,,
    \end{split}\\[2mm]
    c_{2}= 4 \left(n-J_1\right) \left(n-J_1+1\right) \left(n-J_2\right) \left(n-J_2+1\right).
\end{gather}
From here it is easy to understand the usual dimension of the space of tensor structures for integer 
spins. In fact, the space of powers $\mathcal{X}^n$ with $0\le n \le \min(J_1,J_2)$ is closed under 
the action of the vertex operator: $c_{-1}$ stops lowering the order of $\mathcal{X}^n$ at $n=0$, 
while $c_2$ and $c_1$ will stop raising powers respectively at $n=\min(J_1,J_2)-1$ and 
$n=\min(J_1,J_2)$.

When taking large spin limits, however, we are often replacing the summation over spins in the OPE 
with integrals. In other words, a continuum limit is taken for the spins, which implies on the other 
hand that the space of tensor structures ceases to be finite. In fact, if we assume the tensor 
structure $\mathcal{X}^0$ to be part of the space of allowed tensor structures, repeated action of 
$\mathcal{V}$ on it will always produce higher powers of $\mathcal{X}$, as the zeros of the power 
raising coefficients, $n=J_i$, $n=J_i-1$, will never be reached for non-integer spins. Conversely, if assume the tensor structure $\mathcal{X}^{\min(J_1,J_2)}$ to be allowed in our space, the vertex operator action shows us that the $n$ spectrum is not bounded from below, allowing for the label $n$ to scale faster (in absolute value) than the spins $J_i$.
With this argument, we can conclude that as long as the continuum limit for the spins $J_i$ can be taken, 
the tensor structure $n$ is allowed to scale faster than the spin $J_i$.

\section{First-order computation of OPE coefficients in \texorpdfstring{$\phi^3$}{phi-cubed} theory}
\label{app:phi_cubed}
In this appendix, we will determine the OPE coefficients of $\phi^3$ theory at first order in perturbation theory in the cubic coupling $g$, i.e.  at tree level. This can be thought of as being either the $d=6-\epsilon$ expansion or the holographic description of a perturbative $\phi^3$ theory in AdS$_d$. To obtain these OPE coefficients, we perform a direct computation of the following three-point function:
\begin{equation}
\langle [\phi\phi]_{0,J_1}(P_1,Z_1)\phi(X_3)[\phi\phi]_{0,J_2}(P_2,Z_2)\rangle =  \Omega_{J_1J_2} \sum_{n=0}^{\mathrm{min}(J_1,J_2)} C^{(n)}_{[\phi\phi]_{0,J_1}\phi[\phi\phi]_{0,J_2}} \xx^n,
\label{eq:appD_three_pt_target}
\end{equation}
where we define
\begin{equation}
\Omega_{J_1J_2}:= X_{13}^{-h_\phi}X_{23}^{-h_\phi}X_{12}^{-3 h_\phi} \frac{J_{1,23}^{J_1}J_{2,13}^{J_2}}{(X_{13}X_{23}X_{12})^{J_1+J_2}}, \quad \xx =\frac{H_{12} X_{13}X_{23}}{J_{1,23}J_{2,13}}.
\end{equation}
Since the double-twist operators $[\phi\phi]_{0,J}$ are not renormalized at leading order, we can make use of explicit formulas relating them to bilinear forms of the field $\phi$. As a result, we can relate the above three-point function to the tree-level five-point function of the field $\phi$, which takes the form
\begin{equation}
\langle \phi(X_1)\dots\phi(X_5)\rangle = C_{\phi\phi\phi} \, X_{15}^{-2h_\phi} \left(X_{23}X_{34}X_{24}\right)^{-h_\phi}+\mathrm{perms}+\orm(g^2),
\label{5pt_phi3_tree}
\end{equation}
where $C_{\phi\phi\phi}=\orm(g)$. Using an efficient expression for the bilinear forms derived in section~\ref{app:gen_funct_2twist}, we compute the three-point function and the resulting OPE coefficients of eq.~\eqref{PJ1J2n_phi3_tree} in section \ref{app:derivation_opecoeffs}.
\subsection{Parameterization and normalization of GFF double-twist operators}
\label{app:gen_funct_2twist}
The extensive literature on perturbative conformal field theories contains  many different (but physically equivalent) expressions for double-twist operators $[\phi\phi]_{n,J}$ in generalized free field theory. Our approach, which applies to leading-twist operators $[\phi\phi]_{0,J}$, is similar to previous works of Derkachov and Manashov --- see e.g. \cite{Derkachov:1995zr} for the original paper or \cite{Derkachov:2010zza} for a more recent review. 

\paragraph{Generating function of double-twist operators.} Double-twist operators and their descendants can always be expressed as linear combinations of products of derivatives of $\phi$. In the leading-twist sector $\Dg-J=2\Dg_\phi$,  these linear combinations can be expressed as
\begin{equation}
(Z\ds_P)^M[\phi\phi]_{0,J}(P,Z) = \psi_{J,M}(\ds_{\ag_1},\ds_{\ag_2})\phi(P+\ag_1 Z)\phi(P+\ag_2 Z) \vert_{\ag_i=0},
\label{phiphi_to_psi}
\end{equation}
where $\psi_{J,M}(\ag_1,\ag_2)$ is a homogeneous polynomial of degree $J+M$.  The right hand side of this formula is to be understood as the result of a limiting procedure.  Indeed,  the operator product $\phi(X_1)\phi(X_2)$ is divergent when $X_{12}\rightarrow 0$ due to the identity contribution in the GFF OPE:
\begin{equation*}
\phi(X_1)\phi(X_2)=X_{12}^{-\Dg_\phi}\, \mathbf{1} + \orm(X_{12}^0).
\end{equation*}
Instead, we define the operator product  in eq.~\eqref{phiphi_to_psi} as the lightcone limit of the OPE with the identity contribution subtracted.  More specifically, if $X_1,X_2$ are generic and $P_1,P_2$ are two lightlike separated points, then we define the GFF operator product at lightlike separation by
\begin{equation}
\phi(P_1)\phi(P_2):= \lim_{X_1,X_2\rightarrow 0} \left(\phi(X_1+P_1)\phi(X_2+P_2)-X_{12}^{-\Dg_\phi}\, \mathbf{1}\right).
\label{ope_gff_lightlike}
\end{equation}
In this case, the derivative $\ds_{\ag_i}$ acts as the operator $Z\ds_P$ on the $i$'th insertion of the field $\phi$, which increases both the spin and the conformal dimension by one.  Similarly, the action of $(Z\ds_P)^M$ on the primary $[\phi\phi]_{0,J}$ does not change the twist, so the corresponding descendants remain at leading twist. 

Given the definition \eqref{ope_gff_lightlike}, the product in eq.\ \eqref{phiphi_to_psi} is finite and can therefore be expanded as a power series around $\ag_1,\ag_2=0$.  We can organize this power series into leading-twist primaries and descendants:
\begin{equation}
\phi(P+\ag_1 Z)\phi(P+\ag_2 Z) = \sum_{J,M=0}^\infty \hat{\psi}_{J,M}(\ag_1,\ag_2) (Z\ds_P)^M[\phi\phi]_{0,J}(P,Z),
\label{phiphi_to_psihat}
\end{equation}
where $\hat{\psi}_{J,M}(\ag_1,\ag_2)$ is a homogeneous polynomial of degree $J+M$.  The above expansion agrees with eq.~\eqref{phiphi_to_psi} if
\begin{equation}
(\psi_{J,M},\hat{\psi}_{J',M'}):= \psi_{J,M}(\ds_{\ag_1},\ds_{\ag_2})\hat{\psi}_{J',M'}(\ag_1,\ag_2) \vert_{\ag_i=0}= \dg_{JJ'}\dg_{MM'},
\label{pairing_psi_psihat}
\end{equation}
which defines a duality relation between the two spaces of polynomials.  The polynomials $\psi,\hat{\psi}$ are further constrained by the action of the conformal subalgebra that preserves leading-twist fields. The latter is isomorphic to $\mathfrak{sl}(2)$,  and its representation on the space of fields spanned by $\{(Z\ds_P)^M [\phi\phi]_{0,J}\}_{M=0}^\infty$ is given by the generators $Z\ds_P, Z\ds_Z-P\ds_P,P\ds_Z$.  By virtue of the relation \eqref{phiphi_to_psihat}, \eqref{pairing_psi_psihat}, the spaces spanned by $\{\psi_{J,M}\}_M$ and $\{\hat{\psi}_{J,M}\}_M$ also transform in a representation of $\mathfrak{sl}(2)$.  These representations are all isomorphic to lowest or highest weight representations, and the explicit action of the raising/lowering operators on the polynomials $\psi,\hat{\psi}$ is
\begin{align*}
& \sum_{i=1}^2 (\ag_i^2\ds_{\ag_i} +2 \ag_i \Dg_\phi)\psi_{J,0}(\ag_1,\ag_2) =0, \quad \psi_{J,M+1}(\ag_1,\ag_2) = (\ag_1+\ag_2) \psi_{J,M}(\ag_1,\ag_2), \\
& \sum_{i=1}^2 \ds_{\ag_i} \hat{\psi}_{J,0}(\ag_1,\ag_2)=0,\quad \hat{\psi}_{J,M+1} (\ag_1,\ag_2) = \sum_{i=1}^2 (\ag_i^2\ds_{\ag_i} +2 \ag_i \Dg_\phi)  \hat{\psi}_{J,M}(\ag_1,\ag_2).
\end{align*}
The same relations can be found in \cite[Sec.~5]{Derkachov:2010zza}. The second-order differential equation satisfied by $\psi_{J,0}$ has a unique polynomial solution, which specifies $[\phi\phi]_{0,J}$ up to a multiplicative constant (by plugging the solution into eq.~\eqref{phiphi_to_psi}).  However, the corresponding expression is tedious to manipulate when computing correlation functions of $[\phi\phi]_{0,J}$.  Instead, we will exploit the simple functional form of $\hat{\psi}_{J,0}$:
\begin{equation}
\hat{\psi}_{J,0}(\la \ag_1,\la\ag_2)=\la^J\hat{\psi}_{J,0}(\ag_1,\ag_2), \quad ( \ds_{\ag_1}+\ds_{\ag_2}) \hat{\psi}_{J,0}(\ag_1,\ag_2)=0  \Longrightarrow \hat{\psi}_{J,0}(\ag_1,\ag_2) = C_J(\ag_1-\ag_2)^J. 
\end{equation}
Next,  we use an integral transform to relate $\psi$ to $\hat{\psi}$:
\begin{equation}
\psi_{J,0}(\ds_{\ag_1},\ds_{\ag_2}) = \prod_{i=1}^2 \int_C \frac{\dd t_i}{2\pi \mathrm{i}}  \frac{\Gamma(\Dg_\phi)}{t_i^{\Dg_\phi}} e^{t_i} \hat{\psi}_{J,0}\left(t_1^{-1}\ds_{\ag_1},t_2^{-1}\ds_{\ag_2}\right).
\label{psihat_to_psi}
\end{equation}
Here,  $C$ is a Hankel contour going from $t_i=-\infty-\mathrm{i}\ep$ to $t_i=-\infty+\mathrm{i}\ep$ after winding once around $t_i=0$.  The Hankel contour is commonly used to study the Gamma function away from $\mathrm{Re}(z)>0$. In fact,  for any non-integer $z$ (see \cite[eq.~(5.9.2)]{dlmf}):
\begin{equation}
\frac{1}{\Gamma(z)} = \int_C \frac{\dd t}{2\pi\mathrm{i}}  \frac{e^t}{t^z}. 
\end{equation}
Note that the inverse transform can be found in \cite[eq.~(2.14)]{Derkachov:1995zr}:
\begin{equation}
\hat{\psi}_{J,0}(\ag_1,\ag_2)= \prod_{i=1}^2 \int_0^\infty \frac{\dd t_i}{t_i} \frac{t_i^{\Dg_\phi}}{\Gamma(\Dg_\phi)} e^{-t_i} \psi_{J,0}(t_1 \ag_1,t_2\ag_2). 
\end{equation}

Assuming $\Dg_\phi$ non-integer, we can expand $\hat{\psi}_{J,0}$ in eq.~\eqref{psihat_to_psi} into monomials to obtain
\begin{equation}
\psi_{J,0}(\ds_{\ag_1},\ds_{\ag_2}) = C_J \sum_{k=0}^J \binom{J}{k} (-1)^k \frac{\ds_{\ag_1}^k}{(\Dg_\phi)_k} \frac{\ds_{\ag_2}^{J-k}}{(\Dg_\phi)_{J-k}}. 
\end{equation}
This combination of derivatives indeed reproduces a well-known form of double-twist operators that is commonly used in the literature.  In this appendix, with applications to three-point functions of double-twist operators in mind, we instead introduce a generating function
\begin{equation}
\sum_{J=0}^\infty \frac{s^J}{J!} \frac{\psi_{J,0}(\ds_{\ag_1},\ds_{\ag_2})}{C_J} = \prod_{i=1}^2 \int_C \frac{\dd t}{2\pi \mathrm{i}}  \frac{\Gamma(\Dg_\phi)}{t_i^{\Dg_\phi}} e^{t_i} e^{s \left(\frac{\ds_{\ag_1}}{t_1}-\frac{\ds_{\ag_2}}{t_2}\right)}.
\end{equation}
To equate the left and right hand sides, we of course needed to permute the sum over spins with the integration over $t_1,t_2$ ––– this applies as long as the sum over spins converges.  Plugging the above generating function of $\psi_{J,0}$'s back into eq.\ \eqref{phiphi_to_psi}, we obtain a generating function of $[\phi\phi]_{0,J}$'s:
\begin{equation}
\sum_{J=0}^\infty \frac{s^J}{J!} C_J^{-1} [\phi\phi]_{0,J}(P,Z) =    \frac{\Gamma(\Dg_\phi)^2}{(2\pi \mathrm{i})^2} \int_{C\times C}  \dd t_1\dd t_2\, e^{t_1+t_2} \phi(t_1P+sZ)\phi(t_2P-sZ).
\label{generating_function_phphiJ}
\end{equation}
In this expression, the invariance of the primaries under gauge transformations $Z\rightarrow Z+\beta P$ on the left hand side translates to the invariance of the integral under shifts $(t_1,t_2)\rightarrow (t_1+s\beta,t_2-s\beta)$ on the right hand side --- this is the main benefit of introducing the generating function. 
\paragraph{Fixing the normalization.} To derive normalized three-point functions of $[\phi\phi]_{0,J}$ from the generating function in eq.~\eqref{generating_function_phphiJ}, we first need to fix the undetermined constant $C_J$.  We can derive this constant from the normalized three-point function
\begin{equation}
\langle [\phi\phi]_{0,J}(P_1,Z_1) \phi(X_3)\phi(X_4)\rangle = C_{\phi\phi[\phi\phi]_{0,J}}(X_{13}X_{14})^{-\Dg_\phi} \left(\frac{J_{1,34}}{X_{13}X_{14}}\right)^J,
\label{3pt_standard}
\end{equation}
where
\begin{equation}
C^2_{\phi\phi[\phi\phi]_{0,J}} = \frac{(1+(-)^J)}{J!} \frac{(\Dg_\phi)_J^2}{(2\Dg_\phi+J-1)_J}=\frac{(1+(-)^J)}{J!} \frac{\Gamma(\Dg_\phi+J)^2}{\Gamma(\Dg_\phi)^2} \frac{\Gamma(2\Dg_\phi-1+J)}{\Gamma(2\Dg_\phi-1+2J)}.
\label{4pt_gff_opec}
\end{equation}
Inserting the generating function,  we obtain
\begin{multline*}
\sum_{J=0}^\infty \frac{s^J}{J! C_J} \langle [\phi\phi]_{0,J}(P_1,Z_1) \phi(X_3)\phi(X_4)\rangle  = \\
\frac{\Gamma(\Dg_\phi)^2}{(2\pi \mathrm{i})^2} \int_C  \dd t_1\dd t_2 e^{t_1+t_2}\langle  \phi(t_1P_1+sZ_1)\phi(t_2P_1-sZ_1) \phi(X_3)\phi(X_4)\rangle.
\end{multline*}
Now, given the definition of the GFF operator product in eq.~\eqref{ope_gff_lightlike}, the four-point function on the right hand side will only involve two Wick contractions, namely
\begin{align*}
 \langle \phi(t_1P_1+sZ_1)\phi(t_2P_1-sZ_1) \phi(X_3)\phi(X_4)\rangle = \sum_{3\leq i\neq j\leq 4} \left(t_1 X_{1i}+s Z_1\cdot X_i\right)^{-\Dg_\phi}\left(t_2 X_{1j}-s Z_1\cdot X_j\right)^{-\Dg_\phi}.
\end{align*}
After the gauge transformation/change of variables
\begin{align*}
(t_1',t_2') = \left(t_1+ s \frac{Z_1\cdot X_3}{X_{13}},t_2- s \frac{Z_1\cdot X_3}{X_{13}}\right),
\end{align*}
we obtain 
\begin{multline*}
 \langle \phi(t_1P_1+sZ_1)\phi(t_2P_1-sZ_1) \phi(X_3)\phi(X_4)\rangle =\\
 (X_{13}X_{14})^{-\Dg_\phi}\left[t_1'^{-\Dg_\phi}\left(t_2'
+s \frac{J_{1,34}}{X_{13}X_{14}}\right)^{-\Dg_\phi}+t_2'^{-\Dg_\phi}\left(t_1'
-s \frac{J_{1,34}}{X_{13}X_{14}}\right)^{-\Dg_\phi}  \right].
\end{multline*}
The integrand can be expanded into a power series in $s$,  and after integrating over $t_1',t_2'$ we obtain
\begin{equation}
\langle [\phi\phi]_{0,J}(P_1,Z_1) \phi(X_3)\phi(X_4)\rangle = C_J (1+(-1)^J)(X_{13}X_{14})^{-\Dg_\phi} \left(\frac{J_{1,34}}{X_{13}X_{14}}\right)^J.
\end{equation}
To retrieve the normalization of eq.~\eqref{3pt_standard}, we must set $C_J=\frac{1}{2}C_{\phi\phi[\phi\phi]_{0,J}}$.

\subsection{Application to OPE coefficients of two double-twist operators}
\label{app:derivation_opecoeffs}
The computation of $\langle [\phi\phi]_{0,J_1}(P_1,Z_1)\phi(X_3)[\phi\phi]_{0,J_2}(P_2,Z_2)\rangle$ can be done similarly to the computation of $\langle [\phi\phi]_{0,J_1}(P_1,Z_1)\phi(X_3)\phi(X_4)\rangle$ in section \ref{app:gen_funct_2twist}. First of all, we have
\begin{align*}
&\sum_{J_1,J_2=0}^\infty \frac{4 s_1^{J_1}s_2^{J_2}}{J_1!J_2!} \frac{\langle [\phi\phi]_{0,J_1}(P_1,Z_1)\phi(X_3)[\phi\phi]_{0,J_2}(P_2,Z_2)\rangle}{C_{\phi\phi[\phi\phi]_{0,J_1}}C_{\phi\phi[\phi\phi]_{0,J_2}}} = \\
&\prod_{i=1,2,4,5} \frac{\Gamma(\Dg_\phi)}{2\pi\mathrm{i}} \int_C \dd t_i \, e^{t_i} \langle \phi(t_1P_1+s_1Z_1) \phi(t_2P_1-s_1Z_1)\phi(X_3) \phi(t_4P_2-s_2Z_2)\phi(t_5P_2+s_2Z_2)\rangle.
\end{align*}
We can now insert the tree-level form of the five-point function in eq.~\eqref{5pt_phi3_tree}, with all Wick contractions of the first two and last two fields subtracted following eq.~\eqref{ope_gff_lightlike}. The correlator simplifies after the gauge transformation/change of variables
\begin{equation*}
(t_1',t_2',t_4',t_5') = \left(t_1+s_1\frac{Z_1\cdot P_1}{X_{12}}, t_2-s_1\frac{Z_1\cdot P_1}{X_{12}}, t_4-s_2\frac{Z_2\cdot P_1}{X_{12}}, t_5+s_2\frac{Z_2\cdot P_1}{X_{12}} \right),
\end{equation*}
in terms of which the five-point correlator takes the form
\begin{align*}
\langle &\phi(t_1P_1+s_1Z_1) \phi(t_2P_1-s_1Z_1)\phi(X_3) \phi(t_4P_2-s_2Z_2)\phi(t_5P_2+s_2Z_2)\rangle =C_{\phi\phi\phi}(X_{12}^3 X_{13}X_{23})^{-h_\phi} \\
& \left(t_1't_5' \!+\! s_1s_2 \frac{H_{12}}{X_{12}^2}\right)^{-2h_\phi}\! \left(t_2't_4' \!+\! s_1s_2 \frac{H_{12}}{X_{12}^2}\right)^{-h_\phi}\!\left(t_2' \!+ \!s_1\frac{J_{1,23}}{X_{12}X_{13}}\right)^{-\frac{1}{2}\Dg_\phi}\!\left(t_4' \!+\! s_2\frac{J_{2,13}}{X_{12}X_{23}}\right)^{-h_\phi}\! + \mathrm{perms},
\end{align*}
where $h_\phi:=\Dg_\phi/2$ and
\begin{equation}
\mathrm{perms}= [t_1'\leftrightarrow t_2', s_1\rightarrow -s_1]+[t_4'\leftrightarrow t_5', s_2\rightarrow -s_2]+[t_1'\leftrightarrow t_2', s_1\rightarrow -s_1,t_4'\leftrightarrow t_5', s_2\rightarrow -s_2]. 
\end{equation}
Since the integral is symmetric under the above permutations of $t_i'$,  the four terms differ only by the signs of $s_1,s_2$.  Therefore,  in a power series expansion in $s_1,s_2$,  the three permutations will be related to the first one by multiplicative factors of $(-1)^{J_1}$, $(-1)^{J_2}$ and $(-1)^{J_1+J_2}$ respectively, such that the total three-point function is the integral of the first term multiplied by $(1+(-1)^{J_1})(1+(-1)^{J_2})$.  

Note that $H_{12}$ always appears with a factor of $s_1s_2$ in these expressions --- this ensures that we are within the physical range $n\leq \min(J_1,J_2)$ when expanding a three-point function in the $n$-basis.  The OPE coefficients $C^{(n)}_{[\phi\phi]_{0,J_1}\phi[\phi\phi]_{0,J_2}}$ in this basis can therefore be obtained from the coefficient of $s_1^{J_1-n}s_2^{J_2-n} (s_1s_2H_{12})^n$ in a series expansion around $s_1,s_2,H_{12}=0$.  After dividing out by the kinematical prefactor, we can write this coefficient for even $J_1,J_2$ as
\begin{equation*}
\frac{\ds_{s_1}^{J_1-n}\ds_{s_2}^{J_2-n}\ds_{H_{12}}^{n} \langle \phi\phi\phi\phi\phi\rangle\vert_{s_1,s_2,H_{12}=0}}{4 n! \Om_{J_1J_2} (\xx/H_{12})^n} = \prod_{i=1}^2 (h_\phi)_{J_i-n} \sum_{j=0}^n \frac{(2h_\phi)_j (h_\phi)_{n-j}}{j!(n-j)!} \frac{(t_1't_5')^{-2h_\phi-j}}{t_2'^{2h_\phi+J_1-j}t_4'^{2h_\phi+J_2-j}}.
\end{equation*}
Each of the $j=0,\dots,n$ summands is proportional to a product of powers $t_i'^{-(2h_\phi+M)}$, where $M$ is a non-negative integer, each of which integrates to $1/(2h_\phi)_M$.  We therefore obtain
\begin{equation}
\prod_{i=1}^2 \frac{(J_i-n)!}{J_i!} \frac{C^{(n)}_{[\phi\phi]_{0,J_1}\phi[\phi\phi]_{0,J_2}}}{C_{\phi\phi[\phi\phi]_{0,J_1}}C_{\phi\phi[\phi\phi]_{0,J_2}}} =  \prod_{i=1}^2 (h_\phi)_{J_i-n} \sum_{j=0}^n \frac{(h_\phi)_{n-j}}{j!(n-j)!}\frac{1}{(2h_\phi)_j (2h_\phi)_{J_1-j}(2h_\phi)_{J_2-j}}.
\label{eq.}
\end{equation}
Multiplying the left hand side by $P_{[\phi\phi]_{0,J_1}} P_{[\phi\phi]_{0,J_2}}$ reproduces the product of OPE coefficients entering the five-point function, and we retrieve the formula in eq.~\eqref{PJ1J2n_phi3_tree}.

\section{Lightcone blocks with parity-odd tensor structures}
\label{app:parity-odd_blocks}
In section~\ref{sect:lightcone_blocks_five_pt}, our derivation of the lightcone blocks for the direct and crossed channels has been operated under the implicit assumption that the tensor structures at the central vertex of the OPE channel are even under parity transformations. While this is certainly true for CFTs in $d\ge 4$, in the special case of three-dimensional CFTs one can also have tensor structures that are odd under parity transformations. As reviewed in~\cite[Section~3.2]{Buric:2021ttm}, these tensor structures arise because of the possibility of constructing three-point invariants using the totally-antisymmetric symbol $\varepsilon$. For a three-point function $\expval{\mathcal{O}_1(X_1,Z_1)\mathcal{O}_2(X_2,Z_2)\phi(X_3)}$ expressed in embedding space coordinates and polarizations, one can see that
\begin{equation}
    \left(\varepsilon_{ABCDE}X_1^A X_2^B X_3^C Z_1^D Z_2^E\right)^2= 2\, \mathcal{X} (1-\mathcal{X})\frac{J_{1,23}^2 J_{2,31}^2}{X_{12}X_{23}X_{13}}
\end{equation}
which implies that the presence of the parity-odd tensor structures $\varepsilon_{ABCDE}X_1^A X_2^B X_3^C Z_1^D Z_2^E$ manifests in cross-ratio space as the presence of the factor $\sqrt{\mathcal{X}(1-\mathcal{X})}$. 

With this in mind, we can aim to understand what changes does the presence of these tensor structures imply for the expressions of five-point lightcone blocks we derived in section~\ref{sect:lightcone_blocks_five_pt}.
In the OPE limit, we have that $\sqrt{\mathcal{X}(1-\mathcal{X})}=\sqrt{w(1-w)}$, so the presence of parity-odd can be imposed as a correction to the OPE-limit asymptotics of the five-point blocks. For the direct channel, this tells us that the expression of blocks for odd tensor structures will change from eq.~\eqref{eq:DCLCL2} to
\begin{equation} \label{eq:DCLCL2-odd}
\psi^{\mathrm{DC-odd}}_{(h_a,\bar h_a;n)}(u_{si}(\bar z_a,z_a,w)) \stackrel{\textit{LCL}^{(2)}}{\sim} 
\prod_{a=1}^2 \bar z_a^{h_a}z_a^{\bar h_a} (1-w)^n \sqrt{w(1-w)}
\tilde{F}_{(h_a,\bar h_a;n)}^{\mathrm{(odd)}}(z_1,z_2,w)\,,
\end{equation} 
where the yet to be specified function $\tilde{F}_{(h_a,\bar h_a;n)}^{\mathrm{(odd)}}$ must asymptote to a constant in the OPE limit.

To compute the form of $\tilde{F}_{(h_a,\bar h_a;n)}^{\mathrm{(odd)}}$, we can use the Casimir differential equations to express this object in terms of the parity-even solution $\tilde{F}$. In fact, denoting the differential operators~\eqref{eq:Da_for_Ftilde} that act on $\tilde{F}$ as $\mathcal{D}_a(h_\phi;h_a,\bar h_a;n)$, one can directly check that the operators acting on $\tilde{F}^{\mathrm{(odd)}}$ correspond to
\begin{equation}
    \mathcal{D}_a^{\mathrm{odd}}=\mathcal{D}_a\!\left({h_\phi+\frac12;h_a,\bar{h}_a;n+1}\right) 
\end{equation}
which implies that their eigenfunctions must be equal once one performs the same shift in parameters. The same result applies to the crossed channel once a prefactor analogous to that in eq.~\eqref{eq:DCLCL2-odd} is extracted, so this constitutes for both cases a simple recipe that we can use to avoid computing from scratch the conformal blocks with parity-odd tensor structures.

For the direct channel, following this recipe means we just need to correct eq.~\eqref{eq:5ptLCLdc} to
\begin{equation}
\psi^{\mathrm{DC-odd}\, (0)}_{(h_a,\bar h_a;n)}(u_{si}) \
\stackrel{{\LCL}_{\vec{\ep}}}{\sim} \nn^{\DC\, (0)}_{\left({h_\phi+\frac12;h_a,\bar{h}_a;n+1}\right)}\,
(u_{s1} u_{s2})^{h_\phi} u_{s5}^{h_1} u_{s3}^{h_2}\,
f_{n+h_{2\phi}+\frac12}(u_{s2})f_{n+h_{1\phi}+\frac12}(u_{s1}).
\label{eq:5ptLCLdc-odd}
\end{equation}

For the crossed channel, we can once more the distinguish the solutions for two types of scaling of eigenvalues: the case~I blocks with scaling~\eqref{case_I_regime} are such that the correction to eq.~\eqref{sol_5pt_I}
\begin{equation}
\psi^{\mathrm{CC-odd},\mathrm{I} }_{(h_a,\bar h_a;n)}(u_{si})
\stackrel{\LCL_{\vec{\epsilon}}}{\sim}
\nn^{\CC,\mathrm{I}}_{\left({h_\phi+\frac12;h_a,\bar{h}_a;n+1}\right)}
 u_{s1}^{h_1} u_{s4}^{h_2} (u_{s2}u_{s3})^{h_\phi}
 e^{-(n+1) u_{s5}-\frac{\bar h_1^2 u_{s2}+ \bar h_2^2 u_{s3}}{n+1}}
\label{sol_5pt_I-odd}
\end{equation}
is negligible due to $n+1\sim n+\dutchcal{O}(\epsilon_{15}^0)$, while the case~II solution~\eqref{blocks_case_II} becomes in the presence of odd tensor structures
\begin{multline}
\hspace*{-3mm}\psi^{\mathrm{CC-odd},\mathrm{II}}_{(h_a,\bar h_a;J_1-\dg n)}(u_{si}) \stackrel{{\LCL}_{\vec{\epsilon}}^{(4)}}
{\sim} \nn^{\CC, \mathrm{II}}_{\left({h_\phi+\frac12;h_a,\bar{h}_a;n+1}\right)} u_{s1}^{h_1} u_{s4}^{h_2}
(u_{s2}u_{s3})^{h_\phi}\\
(1-u_{s2})^{\dg n-\frac12}u_{s5}^{h_\phi + \dg n-\frac12}
\mathcal{K}_{h_\phi+\dg n-\frac12} \left(\bar h^2_1 u_{s5} +
\bar h_2^2 u_{s3} u_{s5}\right).
\label{blocks_case_II-odd}
\end{multline}


\end{document}